\documentclass[12pt,twoside]{report}

\usepackage{natbib}
\usepackage{suthesis-2e}
\usepackage{graphicx}
\usepackage{amsmath}
\usepackage{amssymb}
\usepackage{comment}
\usepackage{pifont}
\usepackage{mathabx}
\usepackage{url}

\usepackage[hmargin=1.5in,vmargin=1in]{geometry}

\def\g{\gamma}
\def\G{\Gamma}

\newcommand{\be}{\begin{equation}}
\newcommand{\ee}{\end{equation}}
\newcommand{\ba}{\begin{eqnarray}}
\newcommand{\ea}{\end{eqnarray}}
\newcommand{\ep}{E_p}
\newcommand{\eiso}{{\cal E}_{\rm iso}}
\newcommand{\egamma}{{\cal E}_\gamma}
\newcommand{\tjet}{\theta_{\rm jet}}
\newcommand\ie{{\it i.e.\,}}
\newcommand\eg{{\it e.g.\,}}
\newcommand{\epo}{E_p^{\rm obs}}

\newcommand{\gray}{$\gamma$-ray}
\newcommand{\Tz}[1]{{$T_0+#1$\,s}}

\DeclareGraphicsExtensions{.jpg,.mps,.pdf,.png,.gif}

\newenvironment{packed_enum}{
\begin{itemize}
  \setlength{\itemsep}{0.5mm}
  \setlength{\parskip}{0pt}
  \setlength{\parsep}{0pt}
}{\end{itemize}}

\begin{document}

	\title{Gamma-Ray Burst Observations at High-Energy with the $Fermi$ Large Area Telescope}
	\author{Aur\'elien Philippe Bouvier}
	\date{\today}
	\dept{Physics}
	\principaladviser{Tsuneyoshi Kamae}
	\firstreader{Hiroyasu Tajima}
	\secondreader{Stefan Funk}
	\thirdreader{Vah\'e Petrosian}
 
	\beforepreface  	
\prefacesection{Preface}

\indent 
The {\it Fermi Gamma-ray Space Telescope (Fermi)} mission was launched by NASA from Cape Canaveral on June $11^{th}$, 2008. The Large Area Telescope (LAT), the primary instrument onboard, is an imaging, wide field-of-view (FoV), high-energy $\g$-ray telescope, covering the energy range from $\sim 20$ MeV to more than 300 GeV. With an expected five to ten years time of operation, the mission aims at acquiring a deeper understanding of astrophysical objects producing high-energy radiation such as supermassive black holes in Active Galactic Nuclei (AGNs), supernovae remnants, Gamma-Ray Bursts (GRBs), Pulsars, microquasars and novae (recently discovered as \gray{} emitter \cite{Abdo_novae}). On this list, Gamma-Ray Bursts stand out as the most violent, the most energetic and the most distant objects ever detected by humans.

This thesis focuses on the observations of Gamma-Ray Bursts during the first year and a half of operation (up to January 2010) of the $Fermi$ LAT instrument which opened up a new era in the study of the high-energy $\g$-ray sky\footnote{However, blazars (which are a class of bright AGNs with their jet pointed toward us) detected by the LAT will also be used in combination with GRBs in the analysis presented in chapter \ref{EBL}.}. Through the detection of high-energy radiations from GRBs, one can learn about the nature of these objects but also use them as a tool to constrain the amount of radiation present in the Universe as well as the fundamental nature of light propagation.

The first two chapters introduce the reader to the state of GRB observations and their theoretical interpretations before the launch of $Fermi$.
Follows in chapter \ref{chap:introEBL} an introduction to the use of $\g$-ray sources to constrain the Infrared-Optical-Ultraviolet Extragalactic Background Light (EBL). Chapter \ref{Fermi} provides a detailed description of the $Fermi$ LAT and GBM instruments that are relevant for GRB observations.
Chapter \ref{event_optimization} summarizes a work performed in the course of my thesis to improve the LAT performance specifically for the observation of transient objects through the definition and optimization of new LAT event selections.
And chapter \ref{detLATGRB} describes the standard procedure that is used for the detection and analysis of GRBs with the LAT instrument.

A comprehensive and detailed characterization of the high-energy prompt emission of LAT GRBs during the first year and a half of operation will be given in chapter \ref{HEprompt}.
We provide in the chapter a comparison of the LAT GRB detection rate with predictions based on pre-$Fermi$ observations of the sub-MeV emission by the BATSE detectors onboard the {\it Compton Gamma-Ray Observatory (CGRO)}. This will allow us to derive few general conclusions on the behavior and energetics of the GRB population in the high-energy regime.
In its first year and a half, $Fermi$ LAT detected 14 GRBs among which 4 are extremely bright in the LAT energy range which allowed us to probe the spectral behavior of these sources with unprecedented precision. A number of interesting features in the high-energy prompt emission have been revealed through their analyses. These include delay of the high-energy photons with respect to the sub-MeV emission, an additional spectral component dominating over the so-called 'Band' function in the $Fermi$ energy range and finally a curvature in the spectrum above $\sim 1$ GeV observed in one of the bright LAT GRBs. 
We provide a brief discussion of the possible theoretical interpretations for these features which are still a source of much debate in the GRB community. Constraints on the bulk Lorentz factor of the GRB jet are also discussed at the end of this chapter.

The following two chapters focus on the use of GRBs with very high-energy emission (typically $\gtrsim 10$ GeV) as a tool to constrain interesting physics. The first constraint has to do with the nature of light, and specifically the speed at which it travels in vacuum which is considered as a fundamental constant in special relativity. However, some models of quantum gravity predict that Lorentz Invariance (namely the independence of the speed of light with energy) would break down for high energy photons due to the intrinsic nature of the quantum 'foam' in which light is traveling in these models.
Thanks to the very sharp onset of GRB emissions, their extreme cosmological distances and the detection of very high-energy events by the LAT, it is possible to set significant limits on the linear dependence of the speed of light with energy and thus constrain these quantum gravity models. Details of this analysis are presented in chapter \ref{lorentz}.

In chapter \ref{EBL} we constrain the content of the Optical-Ultraviolet radiation in the Universe - the so-called Extragalactic Background Light (EBL) - which is difficult to observe directly due to the presence of foreground zodiacal and galactic light.
Through the LAT observations of GRBs as well as blazars, it is possible to probe for the effect of the optical-Ultraviolet EBL on the \gray{} spectrum of bright extragalactic sources. Indeed, high-energy \gray{s} are known to interact with the EBL through the pair creation process ($\g \g \rightarrow e^+ e^-$) when they carry sufficient energy (typically $\gtrsim 10$ GeV). We found no such EBL-related signature in the LAT data and exclude some EBL models which predict an opaque Universe inconsistent with the $Fermi$ LAT data. In particular, the constraints we find significantly reject the Stecker's 'baseline' and 'fast evolution' models \citep{Stecker06}. Our investigations on this topic are described in full details in chapter \ref{EBL}.

Finally, Chapter \ref{Conclusion} will conclude and provide prospects for the future of high energy observations of Gamma-Ray Bursts with $Fermi$ as well as with next generations instruments and how it might help bringing us closer to a unified and complete picture of these GRB phenomenon.

	\prefacesection{Acknowledgements}

I view my time spent as a graduate student in California as one of the happiest of my life. Many reasons for this spring to my mind but my work as a PhD student at KIPAC - of which this manuscript is the product - is certainly high on the list. Although the path to a PhD can be made of times where you wonder why you didn't choose surf instructor as a profession, it is also made of most pleasurable rewards: understanding a little bit better every day how the universe works, getting a kick at pinning down a difficult (or not so difficult) problem, always learning new skills, discussing interesting ideas with friends and colleagues, seeing oneself slowly becoming a scientist with its wonderful approach to problem solving, or simply providing you with some sense of accomplishment in your life. I here would like to express my most sincere feelings toward those of you who played a significant role during my time here at Stanford.

\vspace{0.3cm}

The scientific method has for a long time been a foggy concept in my mind. Now that my PhD research taught me its inner-working, I realize that this is probably the most valuable lesson I learned during those years. If only for that, I would certainly do it all over again. Without a doubt, my mentor Pr. Hiro Tajima was at the front-line of this educational process. By patiently guiding me through meaningful investigations and with his acute sense of priorities in conducting scientific research, he has been and remains a great inspiration to me. His trust in my scientific work as well as his strong support within the collaboration gave me confidence in my abilities as a scientist. And as I was acquiring more experience, he always gave me the freedom to venture into particular topics I was interested in. He also shared with me his enthusiasm and knowledge in the world of instrumentation which has become a great interest of mine that I will pursue during my postdoc. For all of that Hiro, thank you very much!

\vspace{0.3cm}

My research would simply not be without an amazing group of scientists that decided to make the $Fermi$ telescope a reality and a success. The GRB group in particular was a very fertile environment for me to grow in and I thank all of you I had the chance to collaborate with: Fred, Nicola, Jim, Valerie, Michael, Dan, Yoni and a special thought to my graduate and postdoc colleagues and friends: Sylain, Vlasios, Elena, Vero, Michael, Alexander, Johan, Masa, Francesco. I'd like to thank Fred and Nicola in particular for all the care and time they invested to help me find a postdoc position, I'm really touched. A big thank you also to the 'EBL team' which I truly enjoyed working with: Anita, Andrew, Luis, Silvia and Soeb.

\vspace{0.3cm}

The KIPAC institute is an extremely rich environment to conduct research in astrophysics and I feel lucky and proud having been part of it for a little while. Many people greatly contributed to my maturation as a scientist in this place. My advisor Pr. Tune Kamae provided me with very valuable scientific and academic advice as well as the freedom necessary for me to find an area of research I was most interested in. Pr. Vah\'e Petrosian introduced me to the topic of Gamma-Ray Bursts which would become my dream and nightmare for the years to come, and has always been very helpful in sharing his immense knowledge in this field. For the support and scientific help they provided me with, I'd like to thank Jim, Seth and Stefan. Finally, a big thumb up to the whole graduate and postdoc gang which brings so much life to this place: Herman, Keith, Josh, Warrit, Martin, Rudy, Alex, Rolf, Marco, Peter, Masaaki, Dan, Justin, Shizuka, Simona, Aurora, Norbert, Taka, Luigi, Yvonne, Ping, Bijan. You are what I will miss the most in this place. A special thought to David Paneque who besides being a top-notch scientist I had the privilege to work with, is most and for-all a friend who introduced me to the best drug ever: adrenaline rush!

\vspace{0.3cm}

Finally, I would certainly not have been able to carry out all these years of research without the love and friendship of my closest ones. My mum and dad, who gave me a beautiful childhood and view of the world, and my siblings Mat, Vic and Lea: I miss not having you closer to me. Alice who brightens my life every single day! Thomas for being my best friend throughout those years. And all of you I shared so many good times with in the bay area: David M., Alex, Pierre S., Charles H. D., Negin, Dan B., Harper \& co., Pierre G., Magali, David P., Shizuka, Sarah, Dima, Mr. George, Anna, Kevin, Fabio, Julien, Maki, Fab, Cristo... and many of you that stayed close despite time and distance. Thanks for walking part of the journey with me!

\vspace{0.3cm}

\begin{center}
{\it Don't walk in front of me, I may not follow. 

Don't walk behind me, I may not lead. 

Walk beside me and be my friend. 

- Albert Camus -}
\end{center}

\vspace{10cm}

\begin{flushright}
{\it To the spirit of Rimini...}
\end{flushright}

\newpage
\afterpreface

%	\tableofcontents
 
%	\include{Chapters/test}
	\chapter{\label{chap:GRB}Gamma-Ray Bursts}

This chapter aims at introducing the reader to the field of Gamma-Ray Bursts (GRBs) with a brief introduction to these objects in section \ref{intro_grb} to set the stage. Section \ref{sec:GRBs_History} provides a historical overview of GRB observations in the pre-$Fermi$ era. And section \ref{sec:GRB_Hosts} summarizes the theoretical interpretation that were based on these observations.

\section{Introduction}
\label{intro_grb}

Gamma-Ray Bursts are the brightest and the most distant explosions ever observed in our universe. 
They emit brief and intense electromagnetic radiation in the kilo-electron volt (keV) to Mega-electron volt (MeV) energy range at a rate of about one per day for the typical GRB instrument sensitivities. 
GRBs are now known to be distributed uniformly on the celestial sphere to cosmological distances,
they outshine all other sources in the $\gamma$-ray sky during their short period of emission. The energy released by a single GRB as keV/MeV $\gamma$-rays is comparable to the total energy released by a supernovae explosion over many months. 

Early observations have detected what is now referred to as the prompt emission which is brief (milliseconds to minutes), highly variable
(in time scales of ms to tens of seconds), non-thermal, and observed
mostly in the keV/MeV energy range. The observed prompt emission in the $\g$-ray energy band is believed to be produced by electrons accelerated during collisionless shocks inside highly collimated relativistic jets \cite{Narayan:92}. For the majority of GRBs, the prompt emission is followed by
a smoothly decaying and long-lasting ``afterglow'', observed in
longer wavelengths (from X-rays to optical), and believed to be produced by the deceleration
of a relativistic jet in the surrounding interstellar or circumburst
medium. GRBs of long durations (typically $\gtrsim\,2s$) are believed to be produced by the collapse of the cores of massive spinning stars while the progenitor of shorter duration GRBs (typically $\lesssim\,2s$) is less certain, although mergers of compact binaries (neutron star-neutron star or neutron star-black hole) seem to be a likely candidate.

GRBs are extremely bright objects and are expected to occur and be detectable out to redshifts greater than $z\simeq 10$ and possibly up to $z\simeq 15-20$, while Active Galactic Nuclei (AGNs) occur only out to redshift $z\sim7$ \citep{Lamb:01}. GRBs are therefore unique tools for probing
the very distant (or very young) universe as early in time as the
epoch of reionization. In particular, GRBs have a great potential to become a new probe into the evolution history of the universe (star formation history, 
metallicity at different redshifts), its large-scale structure, and
the properties of the earliest generations of stars.

The afterglow of GRBs passes through multiple regions filled with
gas or radiation fields around their progenitor before it reaches the earth: the circumstellar
medium (CSM) around the GRB progenitor, the star-forming region surrounding
the GRB (HII Region, $H_{2}$ Cloud), the ambient interstellar medium (ISM)
of the host galaxy, the baryonic halo of the galaxy (Halo gas), 
and the intergalactic medium (IGM). 
Spectroscopic studies on the absorption features imprinted on the
GRB afterglow at each of these regions provide unique information
regarding the regions' chemical and nuclear composition and density. 
GRBs are also believed to be intense sources of neutrinos, cosmic
rays (up to ultra-high energies $10^{20}\,$ eV), and gravitational waves. Observations
of these emissions can potentially answer questions in astrophysics, 
particle physics, and general relativity. 

The diverse and intriguing properties of GRBs make them the focus
of intense scientific research and debate, 
and the observational target of multiple instruments. 
Despite the fact that we have known about GRBs for almost forty years, they continue to spark scientific interest and their origin and mechanism remain to be known.
Table \ref{tab:GRB Experiments} gives a non-exhaustive list of past and current instruments which measure the keV to TeV gamma-ray emission of GRBs.
Some future instruments that promise new insights into GRBs are: SVOM (Space-based multi-band astronomical Variable Objects Monitor), a fast repointing $\gamma$-ray, X-ray and optical instrument with a ground segment allowing optical observation of the prompt emission and redshift measurement (launch scheduled for the end of 2014), HAWC (High Altitude Water Cherenkov) the successor of MILAGRO currently in construction, or CTA (Cherenkov Telescope Array) the next generation air Cherenkov telescopes (target first light before the end of this decade).
Further experiments hope to detect neutrino emission from those objects (ANTARES, IceCube...) and gravitational-wave emission (VIRGO, LIGO, LISA...) which would be a by-product of compact binaries mergers.

\begin{table}[htbp]
   \centering
   \begin{tabular}{|p{2.5cm}|p{5.8cm}|p{2.2cm}|p{1.7cm}|} % Column formatting, @{} suppresses leading/trailing space
   \hline
      	Instrument    & Observing energy range & Period of operation & GRBs detected  \\
      	\hline
      	Vela 3-6 	& 200 keV - 1 MeV &1967 - 1979 & $\sim 100$  \\
   	\hline
      	CGRO	& 50-300 keV (BATSE) 	& 1991- 2000 & 2704  	 \\
      			& 20 MeV - 30 GeV (EGRET) 	&   &  5  	 \\
   	\hline
      	BeppoSAX 	& 2 - 600 keV & 1996 - 2002 & 944 \\
   	\hline
      	Milagrito-Milagro 	& 100 GeV - 100 TeV & 1997 - now & 1? \\
   	\hline
      	HETE II 	& 2 - 400 keV & 2000 - now & 80	 \\
   	\hline
      	Integral	& 15 keV - 10 MeV & 2002 - now & $\sim 600$  \\
   	\hline
      	Swift 	& 15 - 150 keV & 2004 - now & $\sim 600$  	 \\
   	\hline
	MAGIC 	& 50 GeV - 30 TeV &  2004 - now & 0	 \\
   	\hline
      	AGILE 	& 10 keV - 700 keV (SA+MCAL) & 2007 - now & $\sim 130$	 \\
      			& 30 MeV - 30 GeV (GRID) & 	 & 3 \\
   	\hline
      	$Fermi$ 	& 10 keV - 40 MeV (GBM) & 2008 - now & $\sim 400$	 \\
      			& 20 MeV - 300 GeV (LAT) &	 & $\sim 14$	 \\
	\hline	\end{tabular}
   \caption{List of past and operating instruments that have claimed GRB detection along with the approximate numbers of detected GRBs detected as of January 2010.}
   \label{tab:GRB Experiments}
\end{table}

The remainder of this chapter will give an overview of our current knowledge regarding GRBs.
A historical overview of GRB observations will be given in section
\ref{sec:GRBs_History}, and the currently accepted
model for the progenitor and mechanism of GRBs will be presented in
section \ref{sec:GRB_Hosts}.

\section{\textmd{\label{sec:GRBs_History}}History of GRB Observations}

\subsection{The first years (1967-1991)}

\begin{figure}[ht]
\begin{centering}
\includegraphics[width=0.75\columnwidth]{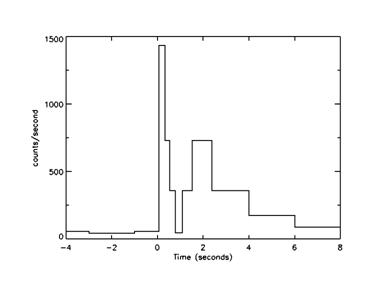}
\par\end{centering}
\caption{\label{firstGRB} Lightcurve of the first GRB detected by the military satellites Vela 3 and 4. The discovery of the detection of brief and sudden $\gamma$-ray flashes not originating from Earth was made public only 7 years later \cite{Klebesadel:73}.
Credit: R. Klebesadel, I. Strong \& R. Olson}
\end{figure}

GRBs were serendipitously discovered in 1967 by the U.S. Vela satellites, 
operated by Los Alamos National Laboratory, 
whose purpose was to monitor from space violations of the nuclear-test ban
treaty by looking at brief X-ray and $\gamma$-ray flashes, signatures of nuclear explosions. The Vela satellites carried omnidirectional $\gamma$-ray detectors which detected bursts of $\gamma$-rays soon after they were launched. These bursts were eventually identified as coming from space. 
The detection of the first GRBs was immediately classified and not made public until seven years later \citep{Klebesadel:73}, when
sixteen GRB detections were reported in the $0.2-1.5$ MeV energy range. GRB 670702 is known as the first GRB detected by humans (Fig. \ref{firstGRB}).
After the Vela satellites, a number of instruments were built to detect
GRBs. However, the number of bursts detected was small and the angular resolution was very poor making understanding their origin extremely difficult. 

\subsection{BATSE (1991-2000)}

A breakthrough in the field of GRB research happened with the numerous
GRB detections from the Burst And Transient Source Explorer (BATSE), which flew, with other instruments, on board the Compton Gamma-Ray Observatory
 (CGRO). BATSE, which operated from 1991 to 2000, detected $\g$-rays in the $15\,\mathrm{keV}-2\,$ MeV energy
range, had a wide field of view ($4\pi\,\mathrm{str}$ minus 30\% because of
Earth obscuration), and a moderate angular resolution ($\sim4^{o}$). 
It detected 2704 GRBs \citep{Paciesas:99}, significantly
more than the total number of GRBs in the pre-existing catalog
(a few hundred). In combination with the Energetic Gamma-Ray Experiment
Telescope (EGRET), a gamma-ray detector also aboard the CGRO, GRB observations
were extended in the $15\,keV-30\,$ GeV energy range.

Before BATSE, the distance scale of GRBs was unknown. The
scientific community was divided among theories predicting
distance scales ranging from our own solar system to the edges of the known
universe. Even though GRBs were observed uniformly on the sky, it was believed that
the reason for this uniform-in-space distribution instead of the pancake shape
of our galaxy was that the pre-BATSE instruments were not sensitive enough
to see the galactic structure of the Milky Way.

\begin{figure}[ht]
\begin{centering}
\includegraphics[width=0.93\columnwidth]{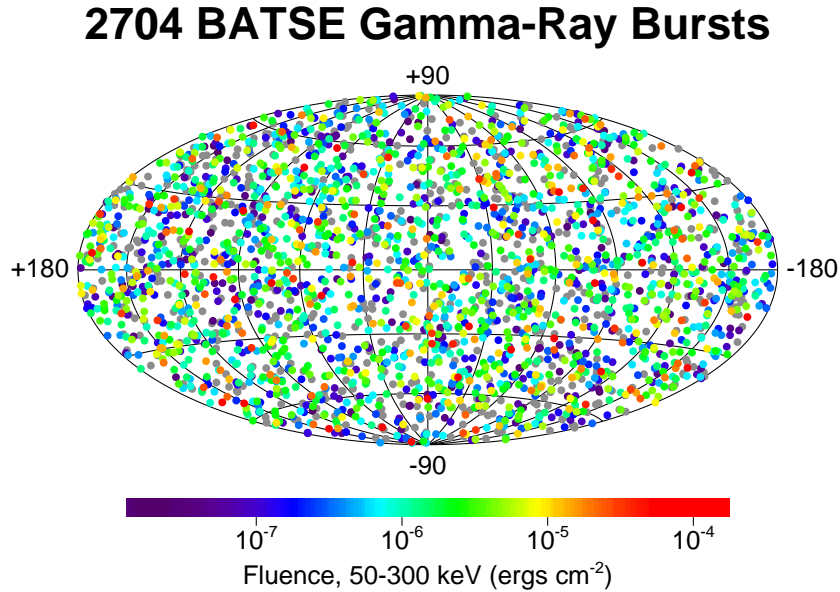}
\par\end{centering}
\caption{\label{fig:GRB_Locations}Locations of all 2704 GRBs detected
by BATSE in galactic coordinates. The isotropic distribution
of GRB localizations suggested that GRBs are most likely extragalactic sources. Source: \cite{URL_BATSE_Locations}}
\end{figure}

BATSE's improved spatial sensitivity proved that GRBs are isotropically distributed on the celestial sphere
(Fig. \ref{fig:GRB_Locations}) ruling out possible correlations with the local distribution of stellar or gaseous mass (our galaxy, the LMC, M31, globular clusters, the Virgo cluster, etc...) which is not isotropic.
This piece of evidence strongly suggested a cosmological origin for GRBs although an extended halo around our galaxy could still generate a uniform distribution similar to the one observed.

\begin{figure}[ht]
\begin{centering}
\includegraphics[width=1\columnwidth]{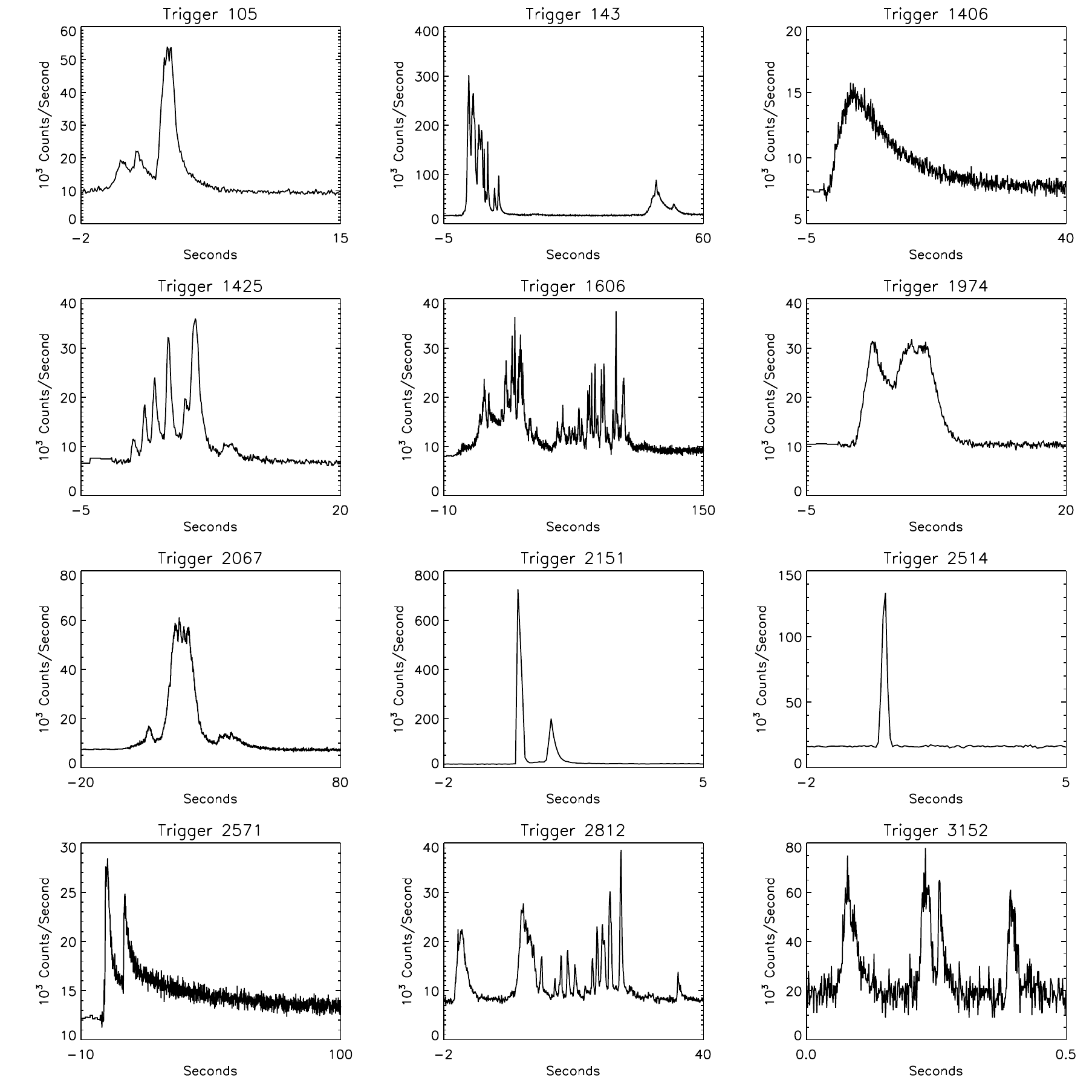}
\par\end{centering}
\caption{\label{fig:GRB_TimeCurves}Diverse light curves of GRBs detected
by BATSE. Source: J. T. Bonnell (NASA/GSFC)}
\end{figure}

Key GRB properties have emerged from the extensive dataset of BATSE-detected GRBs. 
The light curves of GRBs showed great diversity,
from smooth, fast rise, and quasi-exponential decays, 
to curves with many peaks and with a high variability, ranging
from timescales of milliseconds to minutes (Fig. \ref{fig:GRB_TimeCurves}). 

\begin{figure}[ht]
\begin{centering}
\includegraphics[width=0.7\columnwidth]{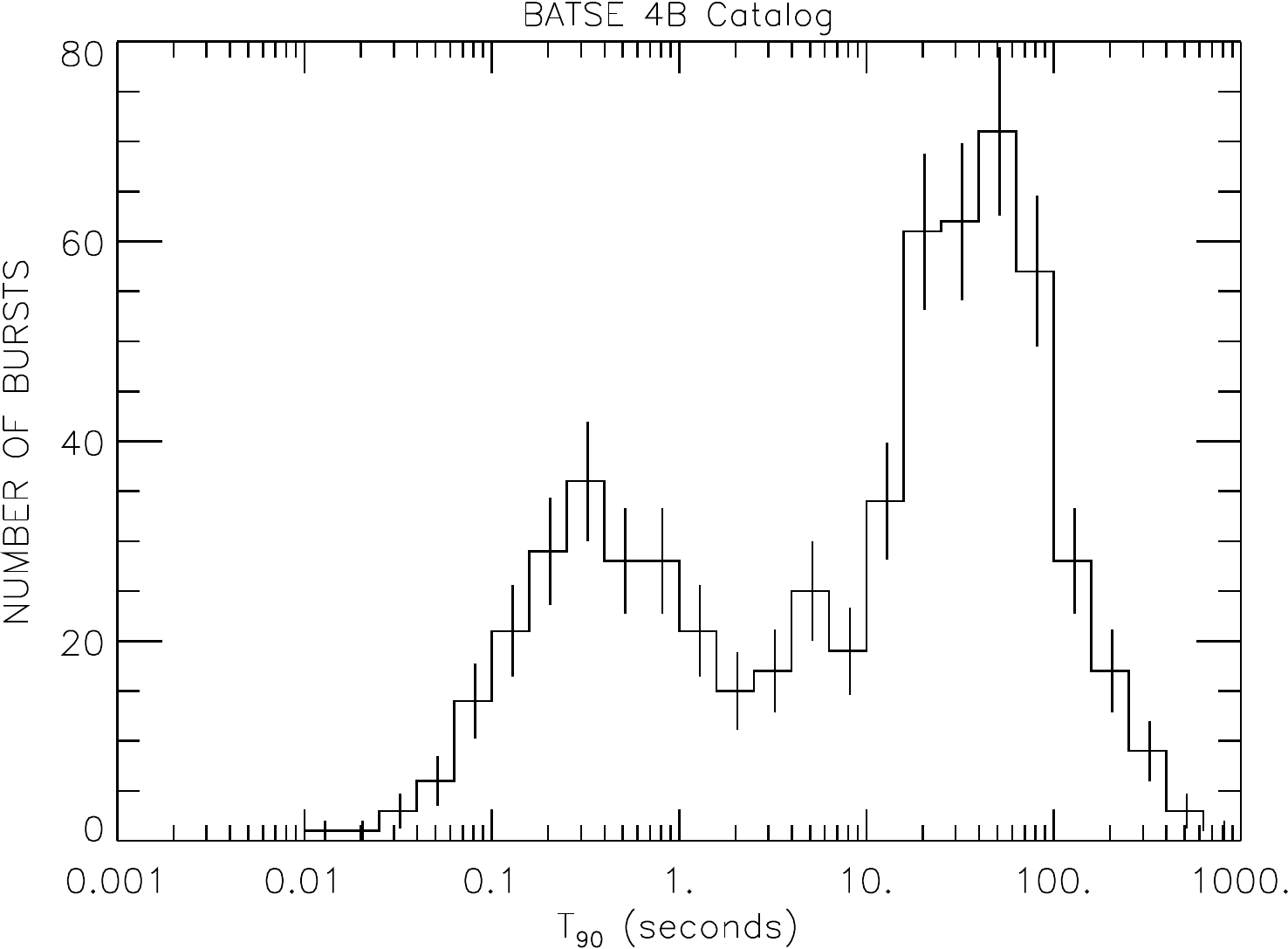}
\par\end{centering}
\caption{\label{fig:GRB_T90}Duration ($T_{90}$) distribution of 
the GRBs detected by BATSE. The bimodal distribution is indicative of two populations of bursts separated around $T_{90}\simeq2\,s$. Source: \cite{URL_BATSE_Durations}}
\end{figure}

\begin{figure}[ht]
\begin{centering}
\includegraphics[width=0.7\columnwidth]{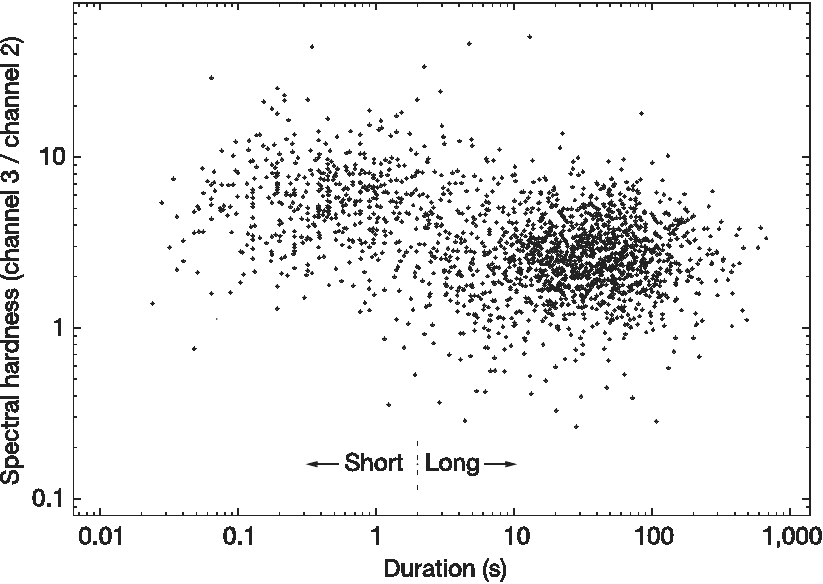}
\par\end{centering}
\caption{\label{fig:GRB_HardnessRatio}Possible correlation between the hardness ratio and the duration $(T_{90})$
of BATSE GRBs. Short bursts ($T_{90}\lesssim2\,s$) have higher hardness 
ratios than long bursts ($T_{90}\gtrsim2\,s$), suggesting that short and long bursts constitute two separate populations, 
probably originating from different progenitors. Source : \cite{Hjorth:05} }
\end{figure}

The duration of GRBs is usually described by the $T_{90}$ parameter, 
which is equal to the time over which the burst emits from 5\% to
95\% of its measured counts. The $T_{90}$ distribution of GRBs
(Fig. \ref{fig:GRB_T90}) spans a long range of durations and seems to show a bimodal shape. This led to the phenomenological division of GRBs into two categories:
``short bursts'' having $T_{90}\lesssim2\,s$, and ``long bursts''
having $T_{90}\gtrsim2\,s$ with short bursts constituting $\sim30\%$
of the BATSE sample and quite different spectral properties between these two categories. 
BATSE measured the fluence of a burst in different channels, each
one corresponding to a different energy range. The ``Hardness Ratio,''
defined as the ratio of the fluence in channel 3 ($100-300\,$ keV)
over the fluence in channel 2 ($50-100\,$ keV), was a measure of the spectral
hardness of a burst. Short bursts were found to have on average larger hardness 
ratios than long bursts, as shown on Figure \ref{fig:GRB_HardnessRatio}. 
The existence of two populations of bursts was suggestive of the existence of two
kinds of progenitors and inner engines during the BATSE era.

\begin{figure}[ht]
\begin{centering}
\includegraphics[width=0.8\columnwidth]{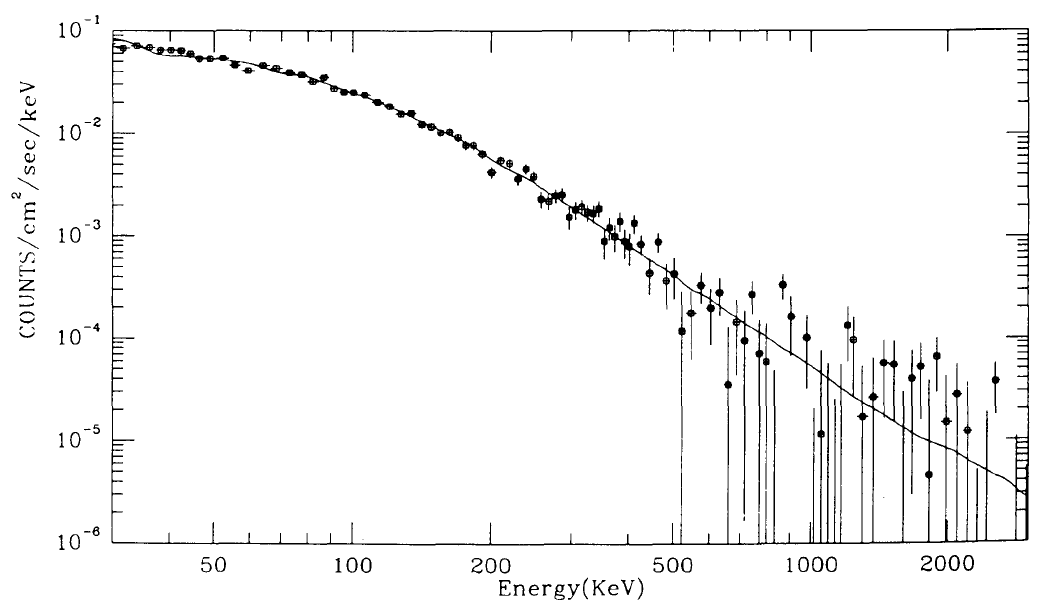}
\par\end{centering}
\caption{\label{fig:GRB_ASpectrum} Band function fit to GRB911127 with the following fit parameters $\alpha=-0.967\pm0.022$ and $\beta=-2.427\pm0.07$. The Band function was found to be a very good phenomenological description of the prompt GRB spectrum in the sub-MeV regime. Source: \cite{Band:93} }
\end{figure}

The measured $\gamma$-ray spectra are non-thermal, and evolve with time from hard to soft. 
The time-averaged spectrum is found to follow an \textit{ad hoc} function, 
called the ``Band function'' $S(E)$ \cite{Band:93}, 
given by:

\begin{equation}
S(E)=A\times\begin{cases}
\left(\frac{E}{100keV}\right)^{\alpha}e^{-\frac{E}{E_{0}}} & E\le(\alpha-\beta)E_{0}\\
\left(\frac{(\alpha-\beta)E_{0}}{100keV}\right)^{\alpha-\beta}\left(\frac{E}{100keV}\right)^{\beta}e^{\beta-\alpha} & E\ge(\alpha-\beta)E_{0},\end{cases}\label{eq:GRB_BandFunction}
\end{equation}

where $\alpha$ and $\beta$ are the low- and high-energy spectral indices,
respectively. The
parameters were estimated by measurements of bright BATSE bursts to
be on average $\alpha\simeq-1$, $\beta\simeq-2. 25$, and $E_{0}\simeq250$ keV \cite{Preece:00}. The spectral energy distribution $\nu\,F_{\nu}$
peaks at the peak energy $E_{p}\equiv(2-\alpha)E_{0}$. 
Figure \ref{fig:GRB_ASpectrum} shows an example
of a spectral fit using the Band function. 
It is generally believed that this prompt emission spectrum originates from from synchrotron emission produced by electrons accelerated to a power-law distribution of energy within internal shocks. However this explanation faces some difficulties that still remain to be solved. An example of such difficulty is the fact that synchrotron emission cannot produce a spectral index harder than $-\frac{2}{3}$ (refered to as the {\it line of death}) but BATSE observed a substantial fraction of its bursts with $\alpha> -\frac{2}{3}$ which is a puzzle that still remain to be solved (instrumental or physical explanation have been proposed without reaching any consensus).

\subsection{BeppoSAX \& HETE-2 (1996-2007)}

\begin{figure}[ht]
 \begin{centering}
 \includegraphics[width=0.8\columnwidth]{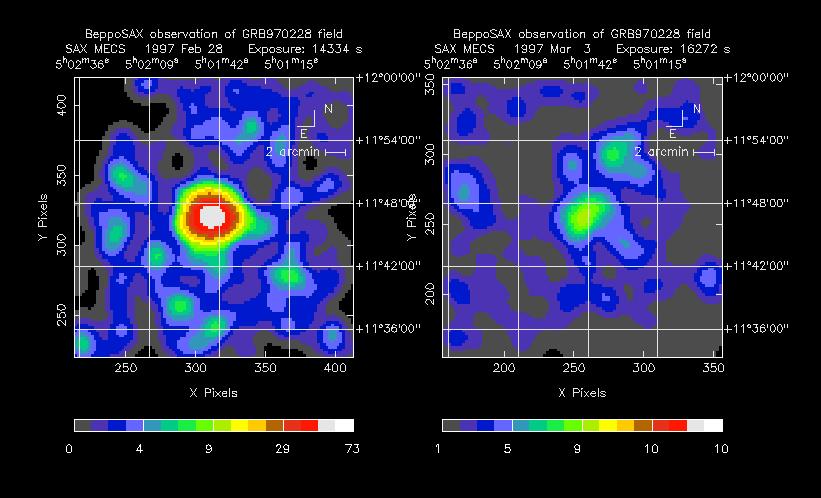}
 \par\end{centering}
 
 \caption{\label{grb970228} Localization of GRB 970228 observed by BeppoSAX. 
The left image is the X-ray emission in the $\g$-ray instrument error box taken $\sim 8$ hours after trigger while the right image is taken $\sim  3$ days after trigger.
A fainting source was clearly detected at the GRB location which is the first detection of an X-ray afterglow. Through the subsequent measurement of the host galaxy redshift, this was the first conclusive evidence of the extragalactic nature of long GRBs. }
 \end{figure}

The angular resolution of gamma-ray instruments such as BATSE was still too coarse for optical or X-ray telescopes (which have a small field-of-view necessitating few arc-minute localizations) to search for a burst counterpart which was theoretically predicted at the time \cite{Rees:92,Meszaros:97}.
This in particular prevented the identification of the progenitors and the environments of GRBs, as well as the measurement of their distances for which optical spectroscopy is needed. As a result, a verification of the
cosmological origins of GRBs was still lacking. A breakthrough happened
in early 1997, with the Dutch/Italian satellite BeppoSAX
(1996-2002) which was equipped with a coded aperture hard X-ray camera and could localize GRBs with a precision of $<10$ arcmin, and a wide field X-ray telescope aboard the satellite to follow-up by repointing of the satellite (within $\sim 3-4$ hours). This allowed the detection of the first fading X-ray emission from a long burst: GRB 970228 (Fig. \ref{grb970228}). After processing data for a few
hours, GRBs localization are accurate enough for ground-based follow-up observations
at optical, radio, and other wavelengths. These observations
initially identified a fading optical counterpart, and, after
the burst had faded, long-duration deep imaging identified a very
distant ($z=0.498$) host galaxy at the location of the burst. This observation
was the first conclusive piece of evidence that long GRBs are cosmological
sources. BeppoSAX also paved the way for identifications of more host galaxies
and for redshift determinations through spectroscopy of the GRB host galaxy, 
and settled the distance-scale argument for long GRBs \cite{Metzger:97}.

The distance measurement of GRBs gave access to intrinsic properties of their sources for the first time. In particular the immense energy release in the form of $\gamma$-rays. 
The typical $\gamma$-ray fluences measured by BATSE are of the order of $10^{-5}\,\mathrm{erg}/cm^{2}$ which lead to an isotropically-emitted energy of $E_{iso}\simeq10^{53}\,\mathrm{erg}$. The total $\g$-ray fluence can be significantly smaller if the emission is beamed as seems to be the case (see section \ref{sub:fireball_model} for further details).

After BeppoSAX, the High-Energy Transient Explorer (HETE-2) (2000-2007)
\cite{URL_HETE2} performed more high-quality afterglow observations and helped identify
a new class of sources called `X-Ray Flashes', similar to
softer GRBs identified earlier by BeppoSAX. HETE-2 also made the
first observations of long GRBs spatially associated with Type Ic supernovae (see
subsection \ref{sub:Progenitors}). 

By 2005, afterglows had been detected from about fifty long GRBs, but there were
no such detections for short GRBs. The afterglows of short
GRBs were hard to detect because the detectors had to achieve precise
localization using a smaller number of photons.
The afterglow from a short GRB was detected first by the Swift satellite (described next) which has a higher sensitivity and fast slewing (re-pointing) capabilities.

\subsection{Swift (2004 - present)}

In the instruments described in the previous section, there was an $\sim8$ hour or longer delay
between the initial burst detection and the start of the optical follow-up
observations.
A new satellite, called Swift \cite{URL_SWIFT}, that can observe the afterglow of GRBs rapidly after its detection, was launched in 2004. Swift has a $\gamma$-ray detector combined
with a wide field X-ray and an optical/ultraviolet telescope, 
and is capable of repointing automatically in seconds. It can localize afterglows with arcsec
accuracy a minute or so after the burst and facilitates follow-up observations in gamma-ray, X-ray, and optical wavelengths. 

\begin{figure}[ht]
\begin{centering}
\includegraphics[width=0.8\columnwidth]{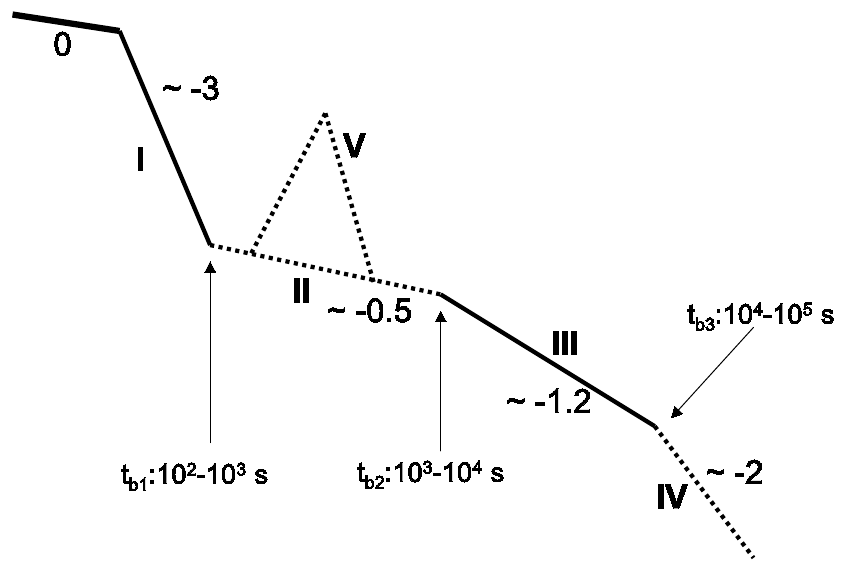}
\par\end{centering}
\caption{\label{fig:GRB_Afterglow_structure}Sketch of an afterglow light curve
based on Swift's observations. Phase ``0,'' 
corresponds to the end of the prompt emission. Four power-law light-curve
segments together with a flaring component are identified in the afterglow
phase. The components marked with solid lines are the features common to most long GRBs, 
while the ones marked with dashed lines are observed in only a fraction
of GRBs. The typical spectral indices of the power-law decay
are shown for each segment. The break between regions III and IV occurs
simultaneously for all observed frequencies (achromatic break) and is related
to the geometry of the GRB relativistic jets. Source:
\cite{Zhang:06}}
\end{figure}

Swift's capabilities have enabled us to study the transition between the
energetic and chaotic prompt emission, and the smoothly decaying
afterglow in lower energy bands. Figure \ref{fig:GRB_Afterglow_structure} gives a synthetic sketch of typical afterglow transition behavior observed by Swift \cite{Zhang:06}: a fast decay (phase I) followed by a plateau (phase II) which in a few thousands seconds reaches the typical afterglow decay with index $\sim -1.2$ already observed pre-Swift (phase III) and finally a possible spectral break (transition from phase III to phase IV) which provides support to
the collimated-emission model of GRBs and allowed to significantly
constraint the energetics of GRBs (see subsection \ref{sub:fireball_model}). 
Furthermore, it provided, for the first time, observations of the
afterglows of short bursts, which led to redshift
measurements and verified the cosmological origin of short GRBs.

Swift's observations of short GRBs showed that, unlike long GRBs, 
they usually originate from regions with a low star-formation rate. 
This suggested that short GRBs are related to old stellar populations, possibly
from mergers of compact-object binaries (i.e., neutron star-neutron
star or neutron star-black hole). 
Furthermore, even though supernova features such as red bumps and late-time rebrightening were detected in the afterglows of most long GRBs close enough to allow such a detection (few close-by  long GRBs do not show such feature and are still a puzzle to the community: GRB 980425 \cite{Kulkarni:98} and GRB 030329 \cite{Hjorth:03}), there was no
evidence of such features in the afterglows of short GRBs. These observations strengthened
the case for long and short GRBs having different kinds of progenitors: massive stars for long GRBs versus compact-object binaries, the leading candidate for short GRBs. An important piece of evidence strengthening the merger scenario was the discovery of significant spatial offsets of short GRB afterglow with respect to their supposed host. This can be easily interpreted as the escape of a binary system from the host galaxy via an initial kick applied after one (or both) of the companions underwent supernova explosion and the time needed for the two objects to loose enough gravitational energy before a final catastrophic merger. This time offset is long enough (on order of few tens of Megayears) for the system to significantly escape its host galaxy.

The GRB afterglows, as observed by Swift, decayed on a power law and
progressively softened from X-rays, to optical, to radio. As of 2010, Swift detected more than 500 bursts in $\gamma$-rays, with almost all
of them having an X-ray afterglow. 

\begin{figure}[ht]
\begin{centering}
\includegraphics[width=0.8\columnwidth]{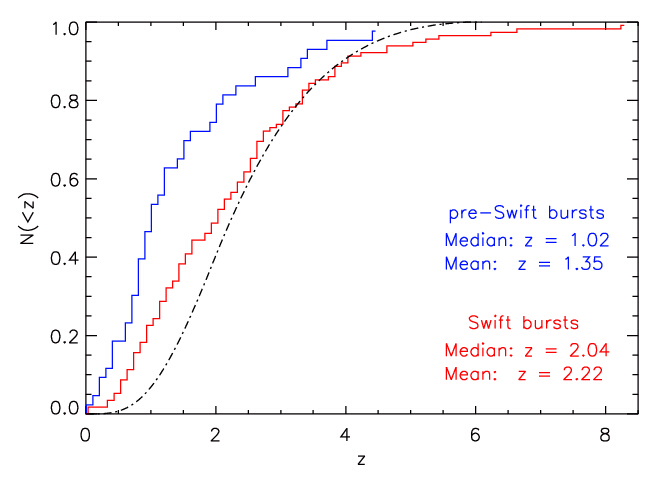}
\par\end{centering}
\caption{\label{fig:GRB_Redshifts}Redshift distributions of GRBs detected
by Swift (up to April 2010) and of GRBs detected by pre-Swift satellites. The average
redshift of the Swift sample (red, 115 bursts with mean $\sim 2.2$) is higher than redshift of the previous observations (blue, 44 bursts with mean $\sim 1.35$) because of the greater sensitivity of Swift to distant GRBs. Dot-dashed curve is the expectation of the redshift distribution of GRBs predicted by a simple model (see \cite{Jakobsson:06} for details). Source: \cite{z_dist_swift}}
\end{figure}

\begin{figure}[ht]
\begin{centering}
\includegraphics[width=0.8\columnwidth]{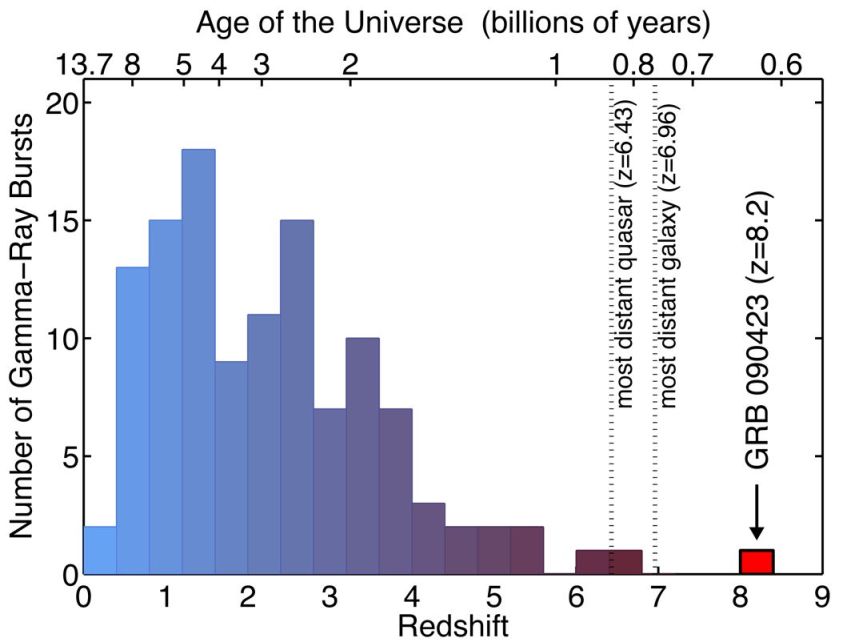}
\par\end{centering}
\caption{\label{fig:highest_redshift} Redshif distributions of GRBs detected up to the $23^{rd}$ of April 2009, when the highest redshift GRBs was discovered at $z = 8.2$. Credit: Edo Berger (Harvard/CfA)}
\end{figure}

Swift is sensitive to a lower energy range ($15-150\,$ keV) and to
bursts of longer durations than other detectors. 
Therefore, it is more sensitive to GRBs of higher redshifts, since the signal from such GRBs is
more redshifted and time dilated. Due to its increased sensitivity to distant GRBs, 
Swift observed GRB 090423, the most distant GRB ever observed. 
GRB 090423 had a redshift of $z=8.2$, and when it exploded the 
age of the universe was less than 700 million years old (figure \ref{fig:highest_redshift}).
The redshift distribution of Swift GRBs and pre-Swift GRBs is shown in Fig. \ref{fig:GRB_Redshifts}. 
As can be seen, Swift GRBs are on average more distant than pre-Swift GRBs. 

%\FloatBarrier
\section{\label{sec:GRB_Hosts}The GRB Model}

Despite more than 40 years of observation, the basic processes underlying the GRB phenomena are still poorly understood. In this section I will describe the theoretical model that has been derived to interpret GRB observations. As a warning, let me mention that this model is neither complete nor self-consistent and I will do my best to stress areas that are source of controversy. For an extensive review on the physics of
GRBs, the progenitors and hosts of GRBs, see \cite{Piran:04, Meszaros:06}. 

\subsection{\label{sub:Progenitors}Progenitors of GRBs}

\subsubsection{Long GRBs}

The light curves of the prompt emission show a variability of 
milliseconds to many minutes. The shortest variability ($\Delta t \sim 1$ ms) imposes a strong upper limit via causality argument on the size of the inner engine: $R \lesssim c \Delta t / (1+z) \sim 10^7$ cm $\sim 100$ km. An upper limit on the mass of the object can also be obtained by requiring that the size of the engine is at least the Schwarzschild radius $R_{Sh} = 2 GM / c^2$: $R>R_{Sh} \Rightarrow M < c^3 \Delta t / (2 (1+z) G) \sim 10^{35} g \sim 100 M_{\Sun}$. Therefore, the object(s) responsible for GRB events has to be a stellar compact object with stellar black holes and neutron stars being natural candidates.
The fact that the burst duration is usually longer than the variability suggests
a prolonged and intermittent inner-engine activity in two or three
different simultaneous time scales. In particular, an explosive model
that releases the energy at once is disfavored as it would require interactions with a clumpy environment which is quite inefficient at emitting $\g$-rays and would pose a serious challenge to the overall energetics. 

We also note that: $L_{Edd} \sim 1.5 \times 10^{38} \frac{M}{M_{\Sun}} \lesssim 1.5 \times 10^{40}$ ergs.s$^{-1}$ where $L_{Edd}$ is the Eddigton luminosity above which gravitation is dominated by radiation pressure which pushes matter outward. However the isotropically-radiated luminosity is many orders of magnitude higher than this value ($L \gtrsim 10^{50}$ erg.s$^{-1}$) which means that GRBs must be an explosive phenomena.
All the above suggest that the inner engine
of GRBs consists of a massive stellar object, most likely a newborn black
hole, with a massive ($mass\gtrsim0.1\,M_{\odot})$ disk accreting onto
it. The accretion explains the prolonged activity and the different
time scales, and the black hole satisfies the size and energy requirements. 

There are multiple observational pieces of evidence that suggest that
not all GRBs are the same, and that there are different kinds of
progenitors and inner engines. Specifically, the duration and the hardness ratio-duration
distributions (Figs. \ref{fig:GRB_T90} and \ref{fig:GRB_HardnessRatio}) 
show that there are two kinds of bursts: short-hard
bursts and long-soft bursts. 

Long bursts are observed in spiral galaxies displaying more star forming activity than an average spiral galaxy and when high resolution is granted, they are usually located in regions where high mass stars are formed and is where massive stars, with their short lifetime, die.  Additionally, supernova-emission spectra were detected superimposed on the afterglows of a few close-by long GRBs (\cite{Soderberg:06,Kulkarni:98,Hjorth:03}). 
In 1998, the optical telescope ROTSE discovered a transient emission 
coincident in space and time with BeppoSAX/BATSE long GRB 980425 \cite{Akerlof:99}. 
The location, spectrum and light curve of the optical transient led
to its identification as a very luminous Type Ic supernova%
\footnote{A Type Ic supernova has no hydrogen in its spectrum and lacks strong
lines of He I and Si II. } (SN 1998bw) \cite{Galama:98, Kulkarni:98}. This
detection was a first of its kind, and suggested that long GRBs are
related to supernovae, and therefore to the deaths of massive stars. 
Because GRB 980425 was very subenergetic comparing to other GRBs (isotropic
energy emitted was $\sim8\times10^{47}\,\mathrm{erg}$ instead of the usual
$10^{51}-10^{54}\,\mathrm{erg}$), the supernova-GRB connection was initially called into question. 
However, a few years later, a similar event happened. Emission from
a supernova (SN2003dh \cite{Hjorth:03}) was detected on the
afterglow of long GRB 030329. This time, the associated GRB
had a normal isotropic energy. In addition to those events, there have also been red
emission `bumps' superimposed on the afterglows of GRBs, with
color, timing, and brightness consistent with the emission of a Type
Ic supernova similar to SN 1998bw (see \cite{Woosley:06}
and references therein). 
All the above suggest that a good fraction of long GRBs are probably related to the death of massive stars. 
The massive star involved is most likely
a Wolf-Rayet star\footnote{Massive stars (Mass$>20\,M_{\odot}$) that rapidly lose their outer envelope by means of a very strong stellar wind (before supernova explosion).},
given that absorption features in the afterglow of long
GRBs \cite{Moeller:02} were explained by the presence
of the fast-moving wind of such a star.
For some unknown reason, the collapse of
the core of the Wolf-Rayet star creates a GRB\footnote{It is believed that most, if not all, long
GRBs are accompanied with a Type Ic supernova} instead
of just a type Ic supernova. 
The specific conditions that lead to the creation of a GRB is one of the open questions
of the field. Observational and theoretical evidence imply that high
rotational speeds, high progenitor masses, and regions of low
metallicity \cite{Fynbo:03, MacFayden:99} favor
the creation of GRBs. 
Some of the differences between short and long GRBs come from the fact that the engine of long GRBs operates at the center of a collapsing star, 
therefore it is covered by the mantle of the star, while the engine
of short GRBs is more or less exposed. 

\subsubsection{Short GRBs}

Short duration bursts are also observed in star forming galaxies but with lower star formation activity and lower metallicity than in the case of long GRB hosts. When a good localization is available, short bursts are primarily observed in regions with low or no star formation, therefore they are likely to be related to old stellar populations. Also for some of them, the afterglow position is observed with a larger offset with respect to the supposed host galaxy.
This suggests that these bursts could be the result of mergers
of compact binaries, such as neutron star-neutron star or neutron
star-black hole which live long and escape far from the star forming region where they were born. The binary loses rotational energy through the
emission of gravitational radiation and eventually merges, forming
a black hole and an accretion disk surrounding it which is a similar system as obtained from the collapse of the core of a massive stars. A similar emission model can therefore apply to short and long burst after this stage.
Another piece of evidence indicating that short burst are not associated with the death of massive stars is the absence of an associated supernovae in any deep long-duration observations of the
optical afterglows of short bursts where such an afterglow should have been detected \cite{Bloom:06, Hjorth:05, Fox:05}.

According to the generally accepted model of the progenitor and the emission mechanism
of GRBs, GRBs start with a cataclysmic event, such as the merger of two
compact objects or the collapse of the core of a rotating massive
star, followed by the creation of a rapidly spinning black hole and an accreting
envelope around it. This model, called the ``Collapsar model'',
was initially proposed to explain long GRBs \cite{MacFayden:99}. However, it
was realized that the mergers of compact-object binaries that create short GRBs
also produce a black hole-accretion disk system similar to the one in the collapsar model.

The collapse of the accreting material that is near the plane perpendicular to the rotation axis of the envelope is somewhat
inhibited by the strong centrifugal forces. Most of the accretion
happens through two funnels that form on the rotation axis of the black
hole (on the axis of rotation). The sources for the deposited energy are
the gravitational and rotational energy of the accreting envelope
and the spinning black hole which are converted to large amounts of kinetic energy ($\sim10^{50}\,\mathrm{erg/s}$) near the polar region possibly through neutrino-driven
winds \cite{Narayan:92}, or magneto-hydrodynamic processes
\cite{Blandford:77}, magnetic instabilities in the
disk \cite{Blandford:82}. The relative
contribution of each source (envelope or black hole) is unknown and depends on which energy-transfer
mechanism is more efficient. It is more likely that the largest fraction
is supplied by the gravitational energy of the envelope. 

Outward radiation and matter pressure gradually build up at the poles; however, they are initially
smaller than the pressure from the in-falling material. A point
is reached, at which the matter density over the poles and the accretion
rate are reduced to a level where they cannot counter-balance
the outward pressure. At that point an explosion occurs. The subsequent stages of the phenomena comprise the magneto-hydrodynamic development of the ejected plasma, its interaction with the external medium as well as the dominant radiation mechanisms involved throughout this evoution. A model that has been fairly successful to describe pre-$Fermi$ data is the so-called `fireball' model which we will describe after some general description of the ultra-relativistic plasma outflow.

While the general picture described by the collapsar model
is accepted by the scientific community, there is little consensus
regarding some of its details. The inner engine of GRBs has not been observed until now, so we can make only indirect inferences about its nature.
As a result, there is still uncertainty regarding many aspects of
the model, such as how exactly the jets are formed; which mechanism
transfers energy from the inner engine to the jets; the baryonic load
of the jets; the jets' bulk Lorentz factor; which physical processes
are involved in the internal shocks; what specific circumstances
lead to the creation of a GRB instead of just a supernova, etc...

In the following, I will give a brief review of some of the observed
GRB properties along with their physical explanation when such an explanation exists. I will also mention, where applicable, how these properties are related to
the collapsar model. 

\subsection{\label{sub:fireball_model}The GRB plasma outflow}
 
 After the initial explosion of the progenitor, a hot baryon-loaded $e^{-}, e^{+}, \gamma$ plasma (also called the `fireball') pushes outwards through the layers of the stellar envelope. 
Matter and pressure gradients and magnetic fields collimate the
outflow, until it finally manages to erupt from the surface of the
object and break free in the form of two opposite narrow jets of half-opening
angle of order of a few degrees.

 It was rapidly realized in the BATSE era that the plasma responsible for the $\gamma$-ray emission of GRB must be ultra-relativistic. The next section describes this argument in more detail.
 
 \subsubsection{Relativistic expansion}

The GRB fireball has a high radiation density, so photon pairs of center
of mass energy $\geq2\,m_{e}\,c^{2}$ should readily create
$e^{-}e^{+}$ pairs, instead of escaping from the fireball. A calculation using
typical values yields an optical depth $\tau_{\gamma\gamma}\sim10^{15}$
\cite{Piran:97}. In such a case, the emitted spectrum should
be thermal and should not contain an MeV or higher-energy component. 
This is the source of the so-called `Compactness problem',
since the observed spectrum is a power law and extends up to energies of at least tens
of GeVs, with no indication of a cutoff for long GRBs and up to tens of MeV for short GRBs. 

This paradox can be solved if the radiating material is moving with relativistic
velocities towards us. In such
a case, the observed GeV/MeV photons actually have lower energies
in the fireball frame of reference. Therefore, the
optical depth of the fireball for the observed photons is actually
lower, since there is now a smaller number of photon pairs with a
center of mass energy over the pair creation threshold ($2m_{e}c^{2}$). If we assume that the photon energies inside the fireball are distributed
on a power law $I_{o}E^{-\alpha}$, then this effect will decrease the
opacity by factor $\Gamma^{-2\alpha}$, where $\Gamma$ is the bulk Lorentz factor
of the fireball.
Furthermore, because of relativistic contraction, the apparent linear dimension along the line of sight of the source
moving towards us will be smaller by a factor of $\Gamma^{2}$ than
its real size.
The combined effect is
that the optical depth is actually lower by a factor of $\Gamma^{-2\alpha-2}$ \cite{Lithwick:01}
than what it would be for a non-relativistic jet, thus solving the paradox. 
Based on the above considerations and the amount of detected MeV/GeV
radiation from GRBs, lower limits on the bulk Lorentz factor of $\Gamma\gtrsim15$
were placed for short GRBs \cite{Nakar:07} and
$\Gamma\gtrsim100$ for long GRBs \cite{Lithwick:01}. 

Another piece of evidence supporting the case for relativistic motion
of the ejecta comes from the fact that estimates of the size of the
afterglow two weeks after the burst, independently provided by
radio scintillation \cite{Goodman:97} and lower-frequency self
absorption \cite{Katz:97}, lead to relativistic velocities for the afterglow expansion of the burst.

As a consequence of the ultra-relativistic velocities inferred for the jet, the baryon loading of the fireball plasma must be small ($M_{b}c^{2}\ll E_0$, where $M_{B}$ is the total mass of the baryons, and $E_0$ is the
total energy of the fireball).

\subsubsection{Energetics and collimated emission}

The afterglow light curves of long GRBs exhibit achromatic spectral breaks (Fig. \ref{fig:GRB_Afterglow_structure}) that
can be explained by assuming that the geometry of the ejecta
is conical (on two opposite jets) instead of spherical. Because the fireball is moving with relativistic
velocities, its emission is beamed. Consider an observer that is inside
the projection of the emission cone of the fireball. When the bulk
Lorentz factor of the fireball is very high, the relativistically-beamed
radiation will be emitted in a very narrow cone. As a result, the
observer will not be able to see the emission from all parts of the fireball and radiation from the conical boundary of the fireball will not be visible
by the observer. As the GRB progresses, the surface area of the fireball cone
expands (as $\propto t^{2}$), and the emitted radiation density drops
inversely proportional to the area, causing a gradual decrease in the observed brightness
of the burst. Because of the expansion, the bulk Lorentz
factor is reduced, and the relativistic beaming becomes wider. As a
result, a larger fraction of the surface of the fireball will come
into the field of view of the observer (middle picture), which compensates the
decline of the observed GRB brightness (now $\propto t^{-1. 2}$
instead of $\propto t^{-2}$). Eventually, all of the surface
of the burst becomes visible to the observer.
The decay rate of the burst's brightness now depends only on the expansion
of the fireball's surface and becomes proportional to $t^{-2}$. This
transition, appearing as an achromatic break in the afterglow light curve, 
has been observed on many GRBs. For the GRB afterglows with no observed jet
breaks, it is usually assumed that the breaks happened at a tim after the observation has stopped (corresponding to jets with wide angles).
However for a few burst with extremely long follow up observations, the non-detection of any jet break remains an unsolved puzzle.

\begin{figure}[ht]
\begin{centering}
\includegraphics[width=0.8\columnwidth]{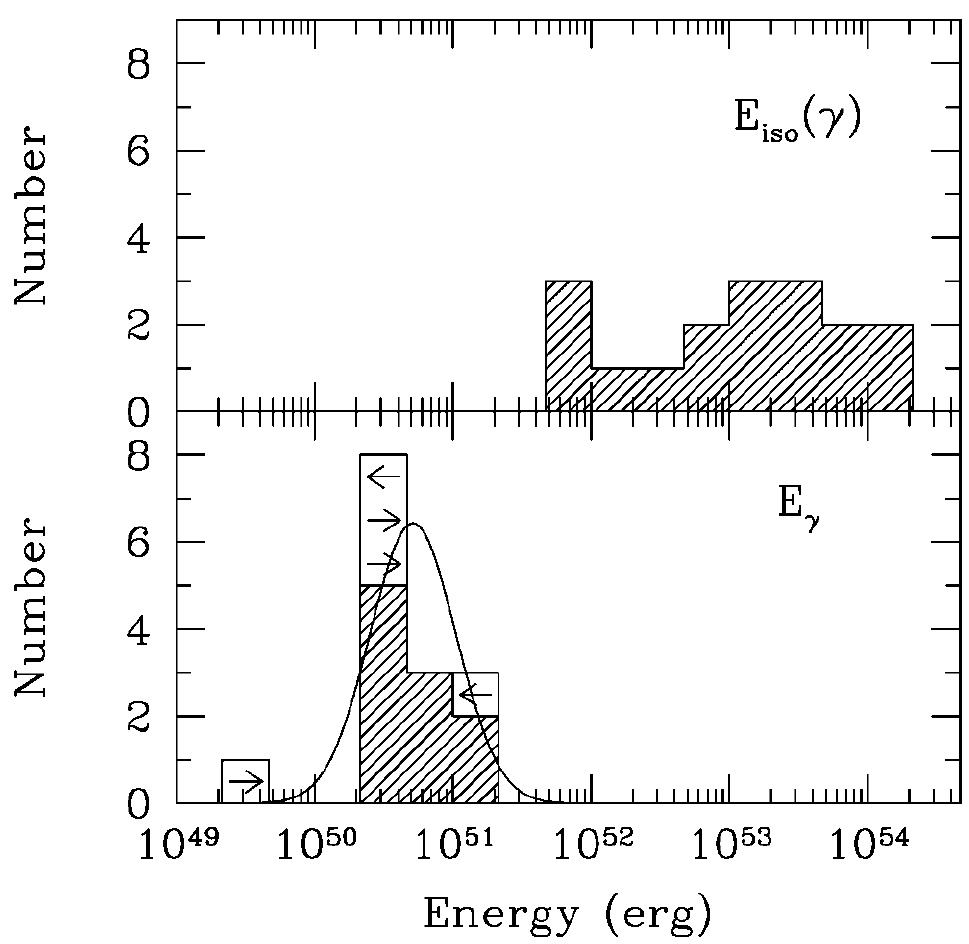}
\par\end{centering}
\caption{\label{fig:GRB_CorrectedEnergy}
\textit{Top}: distribution of the isotropic-equivalent energy output in $\g$-rays for a sample of GRBs with known redshifts. \textit{Bottom:} distribution of the $\g$-ray energy output corrected for the collimation of the jet (collimation angle has been deduced of late time achromatic break). Arrows indicate lower or upper limits to the collimation-corrected energy (when a jet break has not been clearly identified). Source: \cite{Frail:01}}
\end{figure}

The typical GRB gamma-ray fluence on the Earth is of the order of $10^{-5}\,\mathrm{erg/cm}^{2}$. 
If we assume an isotropic emission from GRBs at cosmological distance of $z=2$, this fluence would result to an isotropically-emitted energy
of $E_{iso}\simeq10^{53}\,\mathrm{erg}$. Such an energy emission is considerably
higher than the electromagnetic energy release from a typical supernova ($10^{51}\,\mathrm{erg}$
in few months or $10^{49}\,\mathrm{erg}$ in hundreds of seconds), and a better explanation is needed. However, if the emission is confined in a narrow cone,
the energy requirement will be reduced. 
If the emission actually happened in a solid angle $\Delta\Omega$, 
then the true amount of emitted energy (corrected for beaming) is 
\begin{eqnarray*}
E_{\gamma} & = & 2\,E_{iso}\,\Delta\Omega/4\pi
  =  2\,E_{iso}\,\frac{1-cos(\theta_{jet})}{2}
  \simeq  E_{iso}\,\frac{\theta_{jet}^{2}}{2}
\end{eqnarray*}
where $\theta_{jet}$ is the half opening angle of the emission cone. 
Frail \textit{et al.} \cite{Frail:01} estimated $\theta_{jet}$
for a sample of GRBs, based on the occurrence time of the achromatic break in
their afterglow curves. Based on $\theta_{jet}$, they calculated the true amount of emitted
energy from the isotropic-equivalent amount. Their result (Fig. \ref{fig:GRB_CorrectedEnergy})
showed that even though the isotropic-equivalent emitted energy spans a wide energy range
$(4\times10^{52}-2\times10^{54}\,\mathrm{erg})$, the true amount of emitted energy spans a considerably
narrower energy range centered at $\sim3\times10^{50}\,\mathrm{erg}$. This shows that the energy emission
of GRBs is comparable to that of supernovae.
The fact that the emission is confined within a cone also increases
the implied rate of GRBs by the same factor ($\simeq\theta_{jet}^{2}$), since only GRBs with their
emission cones pointing to the Earth are detected. 

The analysis by Frail et al. \cite{Frail:01}, that the energy release by GRBs might be narrowly concentrated, led people to think that GRBs might have a standard energy reservoir and thus be used as standard candles to constraint cosmological evolution. However, jet breaks are not observed in many GRBs (due to lack of optical afterglow and considerable structure in the X-ray afterglow or simply absence of any sign of jet break up to the latest times of observation. This has put the whole idea of jet breaks and calculating $E_{\gamma}$ into question. The discovery of correlations between a distance independent parameter (or one which scales simply with redshift) and a distance dependent parameters have brought hope in the field although these correlations are subject to wide statistical scatter and more importantly strong bias which are not easy to get rid of. Examples of such correlations are the lag-luminosity (lag refers to the time offset between two different energy band) or variability luminosity relations \citep{Norris:00,Norris:02} and the correlation between $E_{peak}$ and $E_{iso}$ \cite{Amati:02} or $E_{gamma}$ \cite{Ghirlanda:04}.
During my Ph.D, I worked on a paper adressing some of the contentious issues in this field as well as presenting a statistical technique properly taking simple instrumental bias into account which allows us to derive GRB luminosity function and density rate from a sample of bursts with measured redshift \cite[see Appendix \ref{GRB_cosmology}]{Petrosian:09}.

\subsubsection{The `fireball' model}

\begin{figure}[ht]
\begin{centering}
\includegraphics[width=0.95\columnwidth]{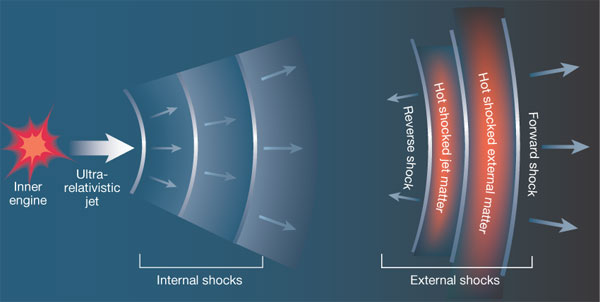}
\par\end{centering}
\caption{\label{fig:GRB_Model} Sketch showing the different steps involved in the `fireball' model with internal shocks producing the $\gamma$-ray prompt emission and external shock with the interstellar medium or the star wind responsible for the afterglow phase observed in radio, optical, X-ray, $\gamma$-ray. Source: \cite{Piran:03}}
\end{figure}

We now describe the `fireball' model which is illustrated in Figure \ref{fig:GRB_Model}.
As discussed above, the ultimate source of GRB is associated with a catastrophic energy release in stellar mass objects. Most of this energy goes into neutrinos and gravitational waves while a significantly smaller fraction (0.1\%-1\%) goes into a high temperature fireball ($k T \gtrsim$ MeV) consisting of $e^{\pm}$, $\gamma$-rays and baryons.
Neither neutrino nor gravitational waves have been detected from a GRB, only the electromagnetic output of the fireball, which despite being a small fraction of the total energy release, still constitute a formidable energy output much more intense than any other explosive event in the universe besides the Big Bang.

In order for the jet to reach ultra-relativistic velocities, a small baryon loading is required to keep the kinetic energy of the jet at a reasonable level.
For very low baryon loading, the radiation may decouple from the fireball electrons before reaching the photospheric radius. This would result in most of the radiation energy not being converted to kinetic energy prior to radiation decoupling and therefore most of the fireball energy would escape in the form of thermal radiation. As a consequence, the final bulk Lorentz factor must be within the range $10^2 \lesssim \Gamma \lesssim 10^3$ in order to allow for the production of the observed non-thermal $\gamma$-ray spectrum \citep{Waxman:03}.

In the first stages following the ejection of the jet (preburst),
the density of the jet is very high, and most of the radiation produced in it is readily absorbed and only a thermal component is emitted at the surface of the hot plasma. As it expands, the optical depth is reduced, and non-thermal radiation can escape from it. The radius over which such non-thermal radiation can escape is usually referred to as the photospheric radius.
Collision of the relativistic plasma with the circumstellar medium \citep{Rees:92,Meszaros:97} and internal collisions within the ejecta itself \citep{Narayan:92, Meszaros:94, Paczynski:94}  were proposed as possible dissipation processes.

Most GRBs show variability on time scales much shorter ($10^2$ up to $10^5$ times shorter) than the total GRB duration. Such variability is hard to explain in models where the energy dissipation is due to external shocks \citep{Sari:97}. Therefore, it is believed that internal collisions are responsible for the prompt emission of $\gamma$-rays.
The internal
shocks happen inside the jet and between shells of material moving
at different velocities. Such shells can be created if the 
energy-deposition mechanism is intermittent at the source for example. During these shocks, 
the jet's electrons are accelerated to ultra-relativistic 
velocities (probably via first-order Fermi mechanism \citep{Fermi:49}) and emit synchrotron radiation. 
Each peak of the prompt light curve is considered to be created by such an internal shock. 
The ejected shells of different Lorentz factor $\Delta \Gamma \sim \Gamma$ are initially separated by $c \Delta T$ and they catch up with each other at an internal shock (or dissipation) radius $r_{dis} \sim c \Delta T \Gamma^2 \sim 3 \times 10^{14} \Delta T_{1ms} \Gamma_2^2$ cm.

%# mention photospheric model (Ryde et al.) right here

As the fireball expands, it drives a relativistic shock into the surrounding environment filled by interstellar medium, star wind, etc.... This interaction smoothly and slowly decelerates the jet. At a typical deceleration radius $r_{dec} \sim (3 E_0 / 4 \pi n_{ext} m_p c^2 \Gamma^2)^{1/3} \sim 5 \times 10^{16} E_{53}^{1/3} n_0^{-1/3} \Gamma_{2.5}^{-2/3}$ cm, the initial bulk Lorentz factor has decreased to approximately half its original value ($n_{ext}$ is the matter density of the external medium).
 Similarly to internal shocks, relativistic particles emit non-thermal radiation
observed as an afterglow. And the afterglow first peaks in gamma rays and gradually softens to longer wavelengths, down to radio as the jet is 
attenuated by the circumburst medium. 
Synchrotron emission by relativistic electrons has been quite successful to explain the temporal and spectral properties of GRB afterglows \cite{Granot:02}.

\subsubsection{$\g$-ray emission mechanisms}

According to the fireball model, the observed radiation is produced by energetic particles produced
in internal or external shock fronts. 
In these shocks, energy is transferred to the electrons in the jet through a diffusive
shock acceleration mechanism refered as the `first order Fermi mechanism' \cite{Fermi:49} in which 
magnetic field irregularities in the shocks keep scattering the particles back and forth multiple times. In ultra-relativistic shocks, electrons gain an 
amount of energy of the order of $\Gamma^{2}_{sh}$ during the first crossing, where $\Gamma_{sh}$ is the Lorentz factor
of the shock front measured in the rest frame of the jet while
subsequent crossings are less efficient, with gains of the order of unity \cite{Achterberg:01}. 
During these shocks, the electrons are accelerated to ultra-relativistic 
velocities ($\gamma_{e}$ up to $\sim1000$) and emit synchrotron radiation. 

The shocks may also accelerate protons. However, the power of the synchrotron emission from protons
is approximately $(m_e/m_p)^2 \sim 3 \times 10^{-7}$) of the power from the electrons for the same Lorentz factor. Additionally, the baryon loading\footnote{the baryon loading is simply the amount of baryons in the jet.} of the jets is believed to be low making a hadronic component in the keV-MeV range unlikely. Therefore the detected radiation is believed to be produced by electrons.

\begin{figure}[ht!]
\begin{centering}
\includegraphics[width=0.8\columnwidth]{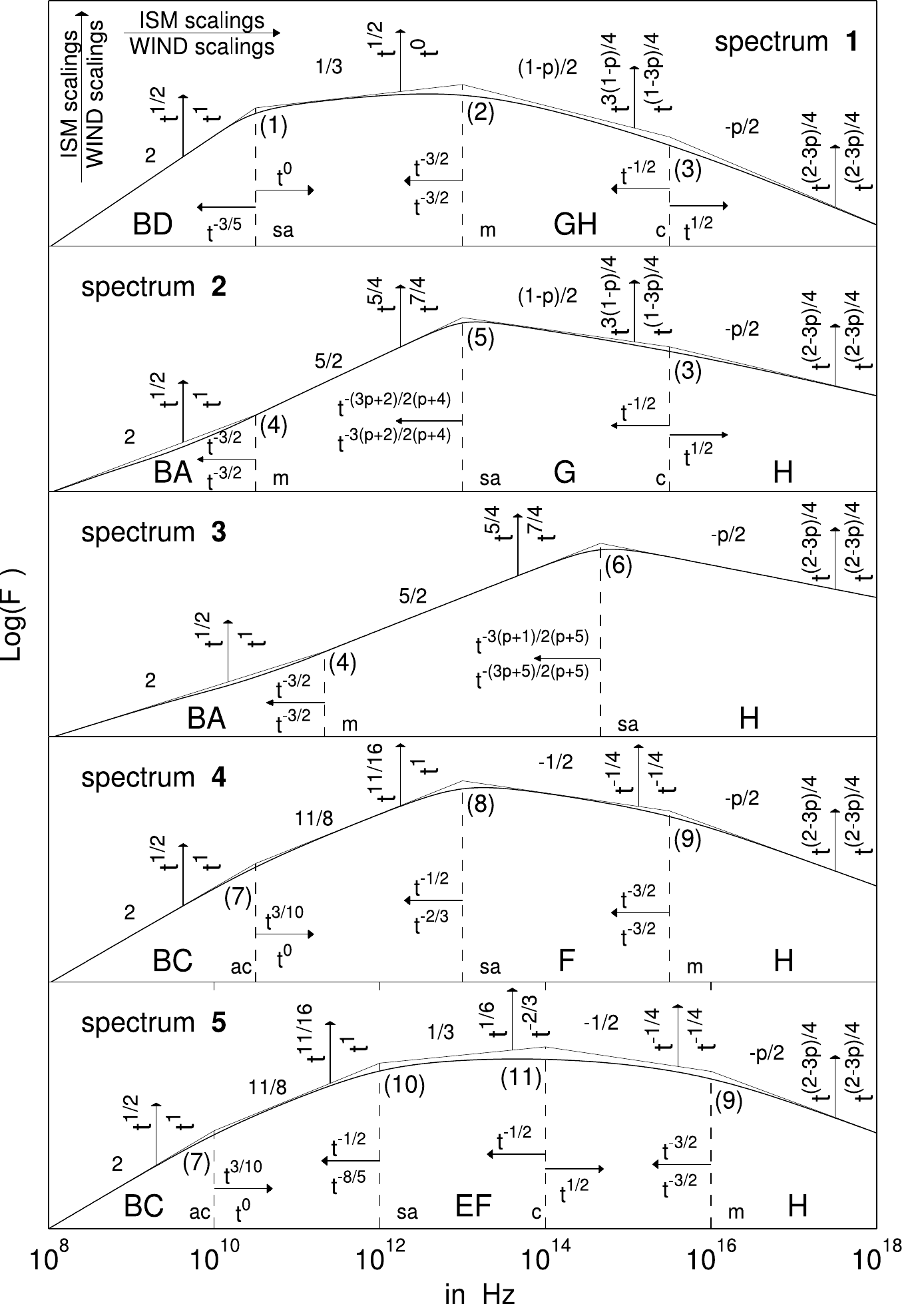}
\par\end{centering}
\caption{\label{fig:GRB_SyncSpectra}Different possible broadband synchrotron spectra from a relativistic blast wave that accelerates electrons to a power-law distribution of energies ($N_{\gamma}\propto\gamma^{-p}$, with $\gamma$ the Lorentz factor of the electrons).
The break frequencies are shown by vertical dashed lines for different sets of physical conditions corresponding to different orderings of the break frequencies: the minimal synchrotron frequency of the least energetic electron $\nu_{m}$, the self-absorption frequency $\nu_{sa}$, and the typical synchrotron frequency of an electron whose cooling time equals the dynamical time of the system $\nu_{c}$. The similarity of these spectra to the spectrum of the prompt emission from GRBs implies that the latter is most likely synchrotron radiation. 
Source: \cite{Granot:02}. }
\end{figure}

While the predictions of the synchrotron
model (Fig. \ref{fig:GRB_SyncSpectra}) are in reasonable agreement with afterglow observations \cite{Wijers:99, Panaitescu:02}, 
there are some inconsistencies between its predictions and the observed spectral slopes during the prompt emission \cite{Preece:02}. 
Alternative models for the emission in internal shocks include synchrotron
self-Compton \cite{Waxman:97, Ghisellini:99} and
inverse Compton scattering of external light \cite{Shaviv:95} similarly to the emission mechanism for blazar jets.

In the next chapter, we will come back to these emission mechanisms in conjunction of the high-energy emission observed by the $Fermi$ Large Area Telescope.

	\chapter{\label{chap:VHE}High-energy $\g$-ray Emission from GRBs}

\section{Introduction}

Different energy ranges of the electromagnetic spectrum provide unique information regarding the environment and mechanism of GRBs. Afterglow emission has been observed from radio to $\gamma$-rays while the prompt emission was mostly observed in the $\g$-ray sub-MeV regime with very scarce observation in the $E>20$ MeV band during the pre-$Fermi$ era.
GRBs are a candidate source of Ultra-High Energy Cosmic Rays (UHECRs). Acceleration of elementary particles to such high energies, inevitably, will produce extremely energetic $\g$-rays up to TeV energies and beyond.
This chapter will present the physical processes involved in the generation and absorption
of High-Energy\footnote{I will hereby use the word High-Energy as meaning emission above 100 MeV} (HE) photons in GRBs, and will provide insight into the emission region and mechanisms.

Section \ref{HE_Observations} will present 
the status of HE observations of prompt and afterglow emission from GRBs in the pre-$Fermi$ era. Then, section \ref{sub:HE_Emission}
will give an overview of the processes that can generate HE photons in GRBs. Lastly,
section \ref{sub:HE_Absorption} will describe the processes that can absorb part of that radiation 
at the site of the burst. 

\section{\label{HE_Observations}Pre-$Fermi$ observations of HE Emission from GRBs}

Before the advent of the $Fermi$ mission, constraints on the origin of the high-energy emission from GRBs were quite limited due to both the small number of bursts with firm
high-energy detection and even in the case of detection, the small number of $\g$-rays that were observed.  High-energy emission from GRBs was first
observed by the Energetic Gamma-Ray Experiment Telescope (EGRET,
covering the energy range from $30\;$MeV to $30\;$GeV) onboard
CGRO. Emission above $100\;$MeV was detected in five cases by the EGRET spark chamber: 
GRBs~910503, 910601, 930131, 940217 and 940301 \citep{Dingus:95}. 
One of these sources, GRB~930131, had high-energy emission that was consistent with an
extrapolation from its spectrum obtained with BATSE between 25~keV --
4~MeV~\citep{Sommer:94}.  

\begin{figure}[bt]
\includegraphics[width=1\columnwidth]{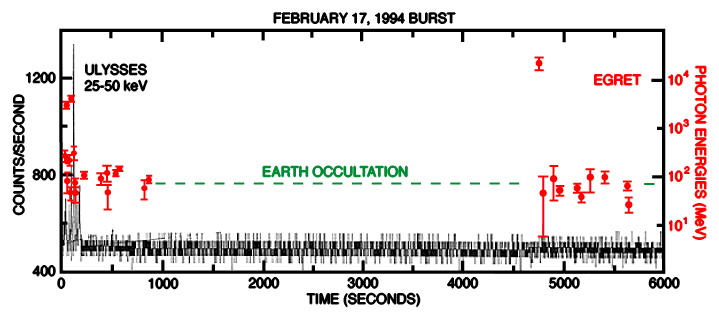}
\caption{\label{fig:GRB_Egret18Gev} Ulysses and EGRET observations of GRB 970217.
Notice in particular the $18\,GeV$ photon detected $\sim90\,min$ by EGRET after the GRB onset. Source: \cite{Hurley:94}}
\end{figure}

\begin{figure}[bt]
\begin{centering}
\includegraphics[width=0.8\columnwidth]{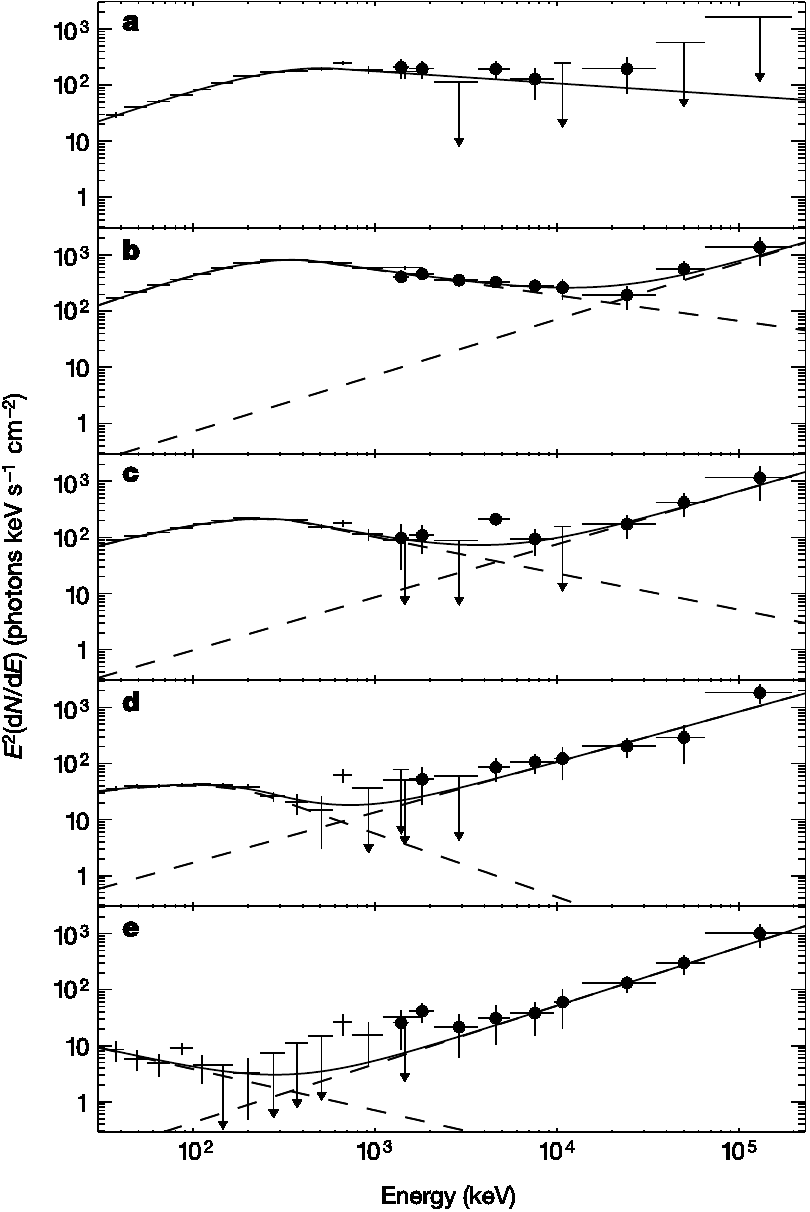}
\par\end{centering}
\caption{\label{fig:GRB_Magdas}Combined spectrum of GRB 941017 measured
by BATSE and EGRET at different times relative to the BATSE trigger,
a ($-18,14\,s$), b ($14,47\,s$), c ($47,80\,s$), d ($80,113\,s$), and e ($113,211\,s$). 
An additional component is clearly observed at high energy and stays roughly constant with a hard index ($\sim 1$) while the sub-MeV emission is decreasing.
Source: \cite{Gonzalez:03}}
\end{figure}

Evidence for an additional
high-energy component up to $200\;$MeV with a different temporal
behavior to the low-energy component was discovered in GRB~941017
\citep[in EGRET's calorimeter TASC; ][]{Gonzalez:03}. The high-energy emission for this GRB lasted more
than 200 seconds with a single spectral component being ruled out.
The extra-component appeared $\sim10-20\,s$ after the main burst and had
a roughly constant flux, while the lower-energy component decayed
by three orders of magnitude (see figure \ref{fig:GRB_Magdas}). The higher-energy component also had
a hard and rising spectral slope (index $\sim1.0)$. Some time after the
main burst ($\sim150\,s)$, it contained more energy than the lower-energy
peak ($30\,keV-2\,MeV$).
And GRB~940217 showed high-energy emission which lasted up to $\sim$$90$
minutes after the BATSE GRB trigger, including an 18 GeV photon
at $\sim$$75$ minutes post-trigger
\cite{Hurley:94} (see figure \ref{fig:GRB_Egret18Gev}). 

More recently, the GRID instrument onboard 
Astro-rivelatore Gamma a Immagini LEggero (AGILE) detected 10
high-energy events with energies up to $300\;$MeV from GRB 080514B, in
coincidence with its lower energy emission, with a significance of
$3.0\;\sigma$ \citep{Giuliani:08}.
The Fermi observatory is expected to drastically improve GRB observations at high energy thanks to its main instrument, the Large Area Telescope which has unprecedented effective area, energy resolution, time resolution and field-of-view in the 20 MeV to 300 GeV energy range. A pre-launch analysis \cite{Band:09} estimated that the LAT would approximately detect 10 bursts per year with few of them having more than 100 photons detected allowing precise time-resolved spectroscopy of the prompt emission to be performed (see \ref{overviewLATGRBs} for a comparison with actual detection rate after launch).

There is no conclusive detection at Very High Energies\footnote{I will hereby use the word Very High-Energy as meaning emission above 100 GeV} (VHE) as of August 2010.
The most sensitive ground-based detectors Imaging Atmospheric Cherenkov Telescopes (IACTs)
have a narrow field of view ($\lesssim3^{o}$), a small duty cycles (less than $10\%$) and have slewing limitation to catch the prompt emission (MAGIC telescope which has a dedicated light structure for fast slewing takes on order of a hundred seconds to be on target). Given these constraints, a detection of prompt VHE emission with current IACTs is unlikely. 
Water Cherenkov Telescopes, on the other hand have wide
field of view ($\sim2\,sr$) and a high duty cycle ($>90\%)$ which allows observation of the GRB prompt emission although their sensitivity is much lower than IATCs. 
Milagrito, the prototype of Milagro, was used to search for $\g$-rays in coincidence with 56 BATSE GRBs that were in its field of view. It
detected an excess in coincidence in time and in the error box
of GRB 970417 with a post-trials probability of $1.5\times10^{-3}$ (or
$3\sigma$) \cite{Atkins:00}.
However, the statistical significance was not high enough for a definitive
detection to be claimed.

\section{\label{sub:HE_Emission}High-energy emission Processes}

Various radiation mechanisms are capable of producing high energy photons in GRBs.
Leptons and hadrons are certainly present within the ultra-relativistic plasma and may be responsible for the HE emission observed. Distinguishing between a leptonic and a hadronic origin of the HE emission is a key issue still awaiting to be resolved.
Because the exact particle composition and the plasma condition at the
acceleration site of a burst are not very well known, the expected emission due to the different
processes is only moderately constrained.
We will now review the leptonic (section \ref{subsec:LeptonicEmission}) and hadronic (section \ref{subsec:HadronicEmission}) processes capable of producing HE emission in GRBs.

\subsection{\label{subsec:LeptonicEmission}Leptonic emission}

It is believed that the GRB jet converts its macroscopic kinetic energy into non-thermal energy via internal and external shocks which accelerate electrons to ultra-relativistic energies. These accelerated electrons can then dissipate their energy via two
processes. The first is synchrotron radiation, believed to be responsible
for the observed $keV/MeV$ and lower-energy emission from GRBs. The
second process is inverse Compton scattering, in which the ultra-relativistic
electrons upscatter low-energy photons to higher energies. The energy
of a photon that underwent inverse Compton scattering is \cite{Fan:08}\begin{equation}
E_{ic}\simeq\frac{2\Gamma}{1+z}\frac{\gamma_{e}^{'2}E_{s\gamma}^{'}}{1+g},\label{eq:GRB_ICENergy}\end{equation}
where $\Gamma$ is the bulk Lorentz factor of the fireball; $z$
is the redshift of the burst; $\gamma_{e}^{'2}$ is the Lorentz factor
of the electron that caused the inverse Compton scattering; $E_{s\gamma}^{'}$
is the initial energy of the seed photon that underwent Inverse Compton
scattering, and $g\equiv\gamma_{e}^{'}E_{s\gamma}^{'}/m_{e}c^{2}$.
The primed parameters refer to the
fireball frame of reference, and the non-primed parameters refer to the observer frame
of reference. In the Thomson limit ($g\ll1)$, equation \ref{eq:GRB_ICENergy}
becomes \begin{equation}
E_{ic,Thomson}\simeq\Gamma\gamma_{e}^{'2}E_{s\gamma}^{'}.\label{eq:GRB_ICEnergy_Thomson}\end{equation}
As can be seen, the upscattered energetic photons will have on average an
energy $\gamma_{e}^{2}$ times higher than the target photons. If
the energy of the seed photons is high ($g\gg1$) (for example, if
they have already underwent one inverse Compton scattering), then
we are in the Klein-Nishina regime, and the cross section of inverse Compton scattering becomes much lower than the Thompson limit.

During internal shocks, the synchrotron photons generated by the electron
population can undergo inverse Compton scattering by that same electron
population. This process is called {}``Synchrotron Self-Compton''
(SSC), with {}``Self'' referring to the electrons that both produce
and upscatter the radiation. The typical Lorentz factor of the internal
shock electrons is $\gamma_{e}^{2}\sim10^{3}$ (in the fireball's
rest frame). Therefore, a typical synchrotron photon of $\sim300\,keV$
energy (from our reference frame) will be upscattered to an energy
$\sim10^{6}$ times higher, equal to a few hundred $GeV$. This process
is believed to produce a second $GeV/TeV$ peak at the GRB spectra, similar to
the one observed in blazar spectra. X-ray flare photons ($E\sim10\,keV)$ 
can also be upscattered to $GeV/TeV$ energies
whether X-ray emission is produced by the synchrotron radiation of late internal shocks \cite{Guetta:03}
or by shocks between slowly moving and fast moving matter ejected simultaneously
during the onset of the prompt emission (refreshed shocks) \cite{Galli:08}.

Synchrotron emission and inverse Compton scattering both contribute to cooling of the electrons. The cooling time through synchrotron emission is $t_{syn}=6\pi\,m_{e}\,c/(\sigma_{T}\,B^{2}\,\gamma_{e})$,
where $\sigma_{\tau}$ is the Thomson cross section, and $B$ is the
magnetic field. The cooling time through inverse Compton scattering
can be written as $t_{IC}=t_{syn}/Y$, where $Y$ is {} 'Compton
Y parameter', given by \cite{Sari:96}
\begin{equation}
\label{VHEGRB_ComptonY}
Y=\begin{cases}
\frac{\epsilon_{e}}{\epsilon_{B}} & ,\,\,\frac{\epsilon_{e}}{\epsilon_{B}}\ll1\\
\sqrt{\frac{\epsilon_{e}}{\epsilon_{B}}} & ,\,\,\frac{\epsilon_{e}}{\epsilon_{B}}\gg1,\end{cases}
\end{equation}
where $\epsilon_{e}$ and $\epsilon_{B}$ are the fractions of the
shocked material's energy carried by electrons and the magnetic field,
respectively. Depending on the relative magnitudes of $\epsilon_{e}$
and $\epsilon_{B}$, cooling either through synchrotron emission or through
inverse Compton scattering dominates. Cooling through inverse Compton
scattering is only important for $\epsilon_{e}>\epsilon_{B}$. 

\begin{figure}[bt]
\begin{centering}
\includegraphics[width=0.8\columnwidth]{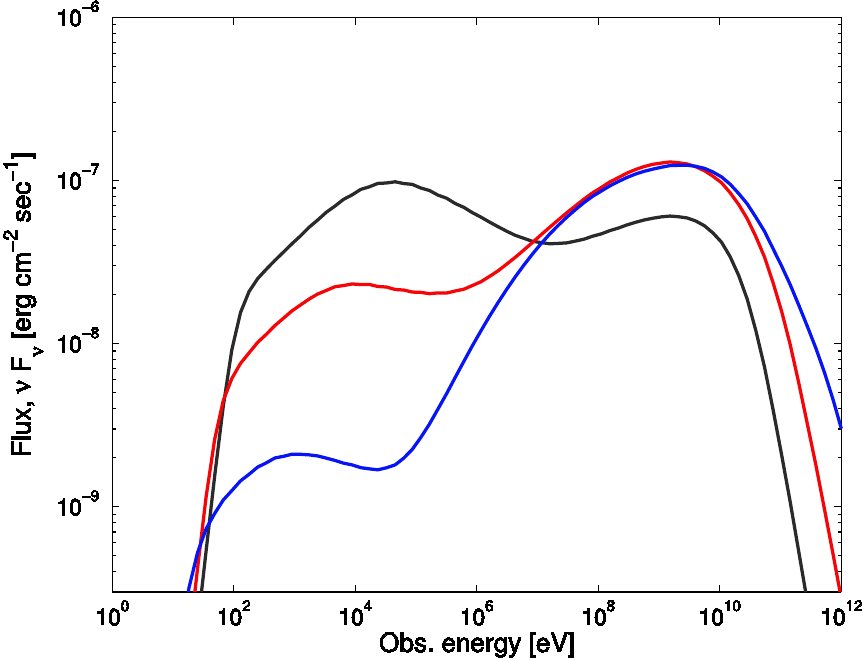}
\par\end{centering}

\caption{\label{fig:VHE_DiffEB}Predictions for synchrotron and synchrotron self-compton radiation produced at internal
shocks for different fractions of the jet's energy carried by the
magnetic field $\epsilon_{B}$: black $\epsilon_{B}=0.33$, red $\epsilon_{B}=10^{-2}$,
blue $\epsilon_{B}=10^{-4}$. Here $\Gamma=600$, $\epsilon_{e}=0.316$,
$\Delta{}t=10^{-3}$, $l'=0.8$, and the GRB has a redshift $z=1$.
The calculation is for a low-opacity fireball. The higher $\epsilon_{B}$ is, the smaller
the fraction of the electrons' energy dissipated through inverse Compton
scattering. The fraction of energy emitted at high energy (as opposed to the well-known sub-MeV behavior) is therefore heavily dependent on the amount of energy deposited into the electrons ($\epsilon_e$) compare to the amount of energy in the magnetic field ($\epsilon_{B}$) if the scenario of leptonic emission. Source: \cite{Pe'er:04}}
\end{figure}

Pe'er and Waxman \cite{Pe'er:04} calculated the leptonic
emission from internal shocks inside the GRB fireball. Their time-dependent
numerical calculations included all the relevant physical processes:
synchrotron emission, synchrotron self-compton, inverse Compton scattering, $e^{-}e^{+}$ pair production and annihilation,
and the evolution of high-energy electromagnetic cascades. 
Figure \ref{fig:VHE_DiffEB} shows the effect
of the ratio $\epsilon_{e}/\epsilon_{B}$ on the resulting spectral
energy distribution. The first peak in the figure comes from synchrotron
emission, and the second higher-energy peak comes from inverse Compton
scattering (SSC). As can be seen, the higher $\epsilon_{B}$ is, the
larger the amount of energy dissipated by synchrotron emission. 
Note that we here consider a fireball relatively transparent to HE energy photon. The effect of opacity of the emitting region will be discussed in section \ref{sub:HE_Absorption}

Inverse compton scattering operates whenever a population of accelerated electrons encounters a photon radiation field. Models have therefore also be put forward where the nature of these two population vary. Ultra-relativistic electrons can be produced through internal or external shocks. 
For Inverse Compton scattering, the radiation field, may come from the photospheric emission the jet emits at an early stage. 

I now moves to hadronic processes that could produce high energy GRB emission.

\subsection{\label{subsec:HadronicEmission}Hadronic emission}

The protons of the GRB fireball are most likely accelerated to relativistic
energies (up to $10^{20}\,eV$?) and emit synchrotron radiation at high energies \cite{Vietri:97, Waxman:95}.
However, synchrotron emission from protons is weak, smaller by a factor
$(m_{e}/m_{p})^{2} \sim 3 \times 10^6$ than the synchrotron emission by electrons. 

In GRB internal and external shocks, various hadronic processes can create
neutral pions that later decay to higher energy gamma rays. These
processes are: \begin{eqnarray*}
p+\gamma\rightarrow\Delta\rightarrow\pi^{o}+p,\\
p+X\rightarrow p+X+\pi^{0}\\
\end{eqnarray*}

where X can be a neutron or a proton. The produced pions will be moving relativistically along the rest
of the fireball towards the observer. For that reason,
the energy of the pions and their decay photons will be 
relativistically boosted to higher energies by a factor of $\Gamma$, as observed by our
reference frame. 

Charged pions also produce higher energy photons via synchrotron emission
of their decay electrons \cite{Gupta:07,Fan:08}.
These pions are produced by processes such as \begin{eqnarray*}
p+X\rightarrow X+n+\pi^{+/-},\\
p+\gamma\rightarrow\Delta\rightarrow\pi^{+}+n,
\end{eqnarray*}
and decay through\begin{eqnarray*}
\pi^{+} & \rightarrow & \mu^{+}+\nu_{\mu}\rightarrow e^{+}+\nu_{e}+\bar{\nu}_{\mu}+\nu_{\mu}\\
\pi^{-} & \rightarrow & \mu^{-}+\bar{\nu}_{\mu}\rightarrow e^{-}+\bar{\nu}_{e}+\nu_{\mu}+\bar{\nu}_{\mu}.\end{eqnarray*}
The energetic decay electrons and positrons emit synchrotron
radiation at high energies \cite{Fan:08}. However, the energy
radiated by this process is expected to be smaller by orders of magnitude
than the synchrotron component by primary electrons, unless they are not efficiently accelerated or have lost their energy by synchrotron or inverse Compton cooling \cite{Gupta:07,Fan:08}.

\section{\label{sub:HE_Absorption}Internal-Absorption Processes}

High-energy $\g$-rays living within a high density fireball are expected to be attenuated through various processes that contribute to the opacity of the GRB fireball, such as Compton scattering and pair creation.
The dominant process is pair creation after the scattering of the
high energy ($E>>1\,MeV)$ photons with lower-energy photons of the fireball
($\gamma\gamma\rightarrow e^{-}e^{+}$). In that process, a photon
with energy $\epsilon_{\gamma}$ can annihilate with another photon
of energy $\gtrsim(m_{e}c^{2})^{2}/2\epsilon_{\gamma}$, where $m_{e}$
is the electron mass, creating an electron-positron pair. It is likely
that the opacity at the GRB site varies from burst to burst, depending
on the local conditions, making it difficult to predict.

\begin{figure}[bt]
\begin{centering}
\includegraphics[width=0.8\columnwidth]{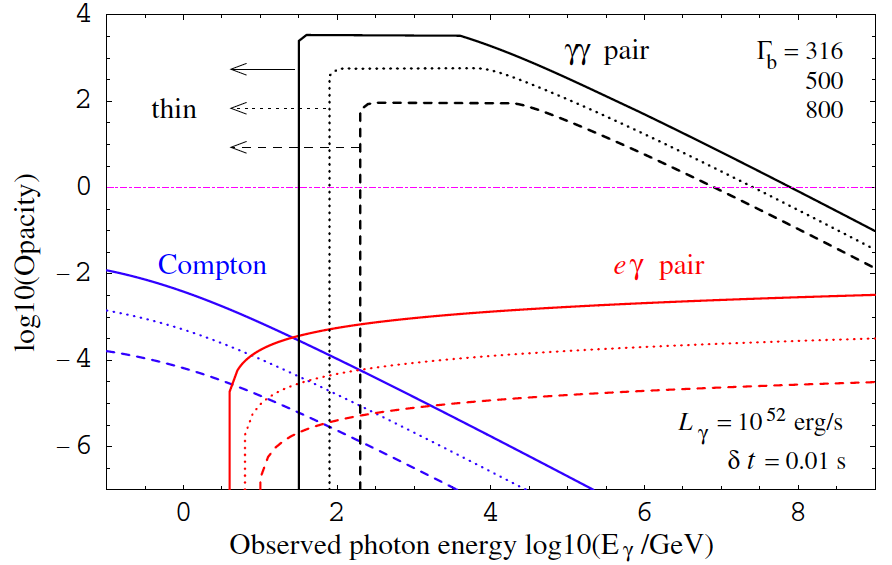}
\par\end{centering}
\caption{\label{Optical_depth} Opacities of high energy photons in the GRB internal shocks for various processes with the dominant $\g-\g$ pair production process which make the fireball optically thick for specific bulk Lorentz factor and energy regime. Photons below the $e^{\pm}$ pair production threshold, in the comoving frame, escape freely both below 10-100 GeV and above 10 PeV. When other parameters are fixed, the opacity decreases with increasing $\Gamma$ ($= 316, 500, 800$ for solid, dotted and dashed curves respectively). The GRB fireball becomes completely optically thin to photons of all energy for a very high bulk Lorentz factor ($\Gamma >> 1000$). \citep{Razzaque:07}}
\end{figure}

The opacity of the emitting region depends on the radiation density.
A high radiation density provides an abundance of lower-energy target
photons with which the higher-energy photons can be absorbed. The radiation
density is proportional to the luminosity of the emitting region,
and inversely proportional to its size. An estimate of the size of
the emitting region is obtained through the variability time scale
of the prompt emission light curve. Each spike in the prompt light
curve is indicative of the thickness of one internal shock. Therefore, there is a relationship between the spatial ($\Delta{}R$) and temporal ($\Delta{}t$) width. If the fireball is
moving towards us with a bulk Lorentz factor $\Gamma$, then $\Delta{}R=\Gamma\,c\,\Delta{}t$,
with $\Delta{}R$ measured in the burst frame. Using $\Delta{}R$ and
the photon luminosity produced from the shock $L$, the ``comoving
compactness'' parameter $l'$ can be calculated as $l'=\Delta{}R\, n_{\gamma}^{'}\,\sigma_{T}$,
where $n_{\gamma}^{'}=\epsilon_{e}L/(4\pi\,m_{e}\,c^{3}\,\Gamma^{2}\,r_{i}^{2})$
is the comoving number density of photons with an energy $E_{ph}$ that
exceeds the electron's rest mass $E_{ph}\ge{}m_{e}\,c^{2}$; $\epsilon_{e}$
is the fraction of the post-shock thermal energy carried by the electrons;
$r_{i}\simeq2\,\Gamma^{2}\,c\,\Delta{}T$ is the radial distance of the shock
from the center of the system; and $\sigma_{T}$ is the Thomson cross
section \cite{Pe'er:04}. 
The compactness parameter gives
a measure of the opacity of the prompt emission region with high-compactness conditions which will result in a suppressed HE emission.  Figure. \ref{Optical_depth} \citep{Razzaque:07} represents the $\g-\g$ and $e-\g$ pair production and Compton scattering opacity as a function of energy and bulk Lorentz factor $\Gamma$.

A direct measure of the bulk Lorentz factor could therefore be achieved through a clear identification of a cut-off in the spectrum associated with this internal absorption process. But a spectral cut-off has not been identified in the $pre-Fermi$ era. Even if such cut-off had been observed, it remains a real challenge to determine its origin. Internal absorption, roll-over in the emitting particle energy distribution or external absorption such as interaction with the extragalactic background light are all viable candidates to explain such feature.
However the observation of high energy $\g$-rays in the GRB emission may be used to set a lower limit on the bulk Lorentz factor given a measure (or upper limit) of the time variability of the burst and and estimation of the spectral behavior during the prompt emission.
Such technique was applied by \citep{Lithwick:01} on a few BATSE bursts although assumptions on the redshift and/or the highest photon from the source had to be made which were not justified.
In section \ref{bulk_lorentz_factor}, we will describe the reliable constraints that the $Fermi$ data allows us to set on the bulk Lorentz factor of a few jet observed by the LAT instrument which had a follow-up redshift measurement.

	\chapter{\label{chap:introEBL} EBL absorption of the $\g$-ray emission from GRBs and AGNs}

Density and spectrum of the Extragalactic Background Light (EBL) in the early Universe carry important aspects of the evolution of matter: star formation history, absorption and re-emission of light by dust, chemical evolution, galaxy formation... Unfortunately its direct measurement is hindered by the much stronger foreground zodiacal and Galactic light.
Bright $\g$-ray emitters observed up to large cosmological distances can be used to derive indirect information on the integrated flux of Extragalctic Background Light between the source and the solar system. Gamma-Ray Bursts along with Active Galactic Nuclei (AGNs), blazars in particular, satisfy this condition as they are extremely luminous in the $\g$-ray band and have been detected up to very high redshifts (GRBs: $z \sim 8.2$; AGNs: $z \sim 5.4 $).
Indirect clue of what column density of infrared, optical and ultraviolet EBL can be obtained since very high-energy (typically above 10 GeV) $\g$-rays interact with these lower energy photons via pair production. As a consequence $\g$-ray emission is absorbed during its propagation throughout extragalactic space until it reaches the Earth. 

The purpose of this chapter is to present the different models of the
EBL, and the effects of EBL absorption on the high-energy emission from $\g$-ray emitters. 
Section \ref{sec:IR_EBL} provides a general description of the EBL. Section \ref{sub:IR_Models} gives an overview of the currently available models and how they compare with the observational constraints.
Finally, section \ref{sub:IRAbs_Effects} 
presents the effects of EBL absorption on the high-energy emission from GRBs and blazars. 

\section{\label{sec:IR_EBL}The IR-optical-UV EBL}

Energetic $\g$-ray photons are subject to absorption by
production of electron-positron ($e^-e^+$) pairs while interacting with
low energy cosmic background
photons~\citep{Nishikov:61,Gould:66,Fazio:70} if above the interaction
threshold: 

\begin{eqnarray}
\epsilon_{\rm thr}=(2 m_e c^2)^2/(2E(1-\mu))
\label{eqEBL}
\end{eqnarray}

where
$\epsilon$ and $E$ denote the energies of the background photon and
$\g$ ray, respectively, in the comoving frame of the interaction,
$m_ec^2$ is the rest mass electron energy, and $\mu = cos(\theta)$ where $\theta$ is
the interaction angle between the two photons.  Because of the sharply peaked cross section
close to the threshold, most interactions are centered around
$\epsilon^*\approx 0.8(E/{\rm TeV})^{-1}$eV for a smooth broadband
spectrum.  Thus, the extragalactic background light (EBL) at UV
through infrared wavelengths constitutes the main source of opacity for
$\g$-rays from extragalactic sources (Active Galactic Nuclei: AGN and GRBs).
The effect of absorption of HE $\g$-rays is then reflected in
an energy- and redshift dependent softening of the observed spectrum
from a distant $\g$-ray source.
%%(see, e.g., \citet{axions,Essey} for an alternative explanation).  
The observation, or absence, of such
spectral features at HEs, for a source at redshift $z$ can be used to
constrain the $\g\g\to e^+e^-$ pair production opacity,
$\tau_{\g\g}(E,z)$.

\begin{figure}[ht]
\begin{centering}
\includegraphics[width=0.6\columnwidth]{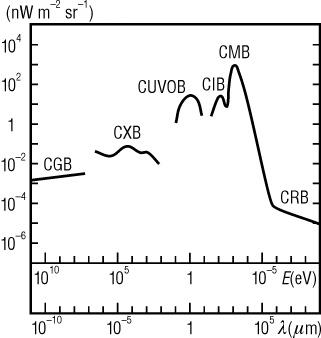}
\par\end{centering}

\caption{\label{fig:GRBSensi_SED_EBL} Schematic background radiation spectrum as a function
of wavelength. The optical-infrared-UV EBL consists of two kinds of light: redshifted
starlight  at near infrared-optical-UV (and optically bright AGN to a lesser extent) and starlight absorbed and remitted by dust in the infrared. The infrared to UV background light is the second most intense source of radiation background after the Cosmic Microwave Background (CMB). Source: Wikipedia}
\end{figure}

The EBL is the accumulated radiation from structure formation and its
cosmological evolution.  Figure \ref{fig:GRBSensi_SED_EBL} shows radiation background at different energy. As one can see, the infrared-optical-UV background light is the second most intense source of radiation background after the Cosmic Microwave Background (CMB).
Knowledge of its intensity variation with time
would probe models of galaxy and star formation.  The intensity of the EBL
from the near-IR to ultraviolet is thought to be dominated by direct
starlight emission out to large redshifts, and to a lesser extent by
optically bright AGN.  At longer wavelengths the infrared background
is produced by thermal radiation from dust which is heated by
starlight, and also emission from polycyclic aromatic
hydrocarbons~\citep[see e.g.][]{driver:08}.

\section{\label{sub:IR_Models}EBL Models}

Direct measurements of the EBL is difficult due to contamination by
foreground zodiacal and Galactic light~\citep[e.g.,][]{Hauser:01}, and
galaxy counts result in a lower limit since the number of unresolved
sources is unknown~\citep[e.g.,][]{Madau:00} and direct measurements can set loose upper limits. 
However, the density of the EBL at different redshifts cannot be
constrained by measuring the cumulative energy output only. For that
reason, there are multiple models that try to calculate the density
of the EBL at various redshifts and wavelengths. These models approach
the problem with different methods, degrees of complexity, observational
constraints and, inputs\footnote{The following information on the various models was mostly based on
the detailed overviews in \cite{Hauser:01,Reyes:07}.}: 

\begin{itemize}

\item The backward-evolution models \cite{Stecker:01,Stecker:06b} 
extrapolate present-day data or template spectra of local galaxies
to higher redshifts. They are simple and easily verifiable
since they predict quantities that can be compared with
observations. However, they do not include known processes occurring
in galaxies such as star formation, and re-emission of radiated
power by dust. 

\item The forward-evolution models predict the temporal evolution of galaxies
as a function of time, starting at the onset of star formation. 
In general, they have been proven successful
in fitting the spectra of individual galaxies, galaxy number counts
in specific bands, and the general characteristics of the EBL. However, they
do not include galaxy interactions and stochastic changes in
the star formation rate.

\item The semi-analytical models \cite{Primack:05} adopt the approach of the forward-evolution
models, but they also rely on simulations of structure formation.
This way, they can make predictions about the observable characteristics
of galaxies and the intensity and spectrum of the EBL. These models
take into account multiple physical processes, such as the cooling
of gas that falls in the halos, the star formation, and the feedback
mechanisms that modulate the star formation efficiency. In spite of
their general successes, there remain some discrepancies between their predictions
and observations. The origins of these discrepancies are difficult
to trace because of the inherent complexity of the models and the
multitude of parameters needed by them. 

\item The EBL provides an integrated over-time view of the energy release
by a wide variety of physical processes and systems that have populated
the universe. So, it is expected to be dependent mostly on the global
characteristics of cosmic history. Thus, chemical-evolution
models deal with the history of a few of the globally-averaged properties
of the universe instead of trying to model the complex processes that
determine galaxy formation, evolution, and emission. The main advantages
of these models are their global nature, their intrinsic simplicity,
and the fact that they do not require detailed knowledge of specific
processes involved in the evolution of galaxies. They provide a picture
of the evolution of the mean number density of stars, interstellar gas, metals,
and radiation averaged over the entire population of galaxies. They
have been successful in reproducing the generic spectral shape of
the EBL, but they fall short of some UV-optical and near-infrared measurements. 

\end{itemize}

In general, most models predict similar cosmic infrared background
spectra from $\sim5-1000\,\mu{}m$, mostly because they use similar cosmic
star formation histories. Backward evolution models assume a rising
SFR up to $z\sim1-1.5$ with a nearly constant rate at earlier times.
Forward-evolution and semi-analytical models try to reproduce the
same SFR in order to fit number counts or comoving spectral luminosity
densities at different redshifts. Larger differences are in the
predictions regarding the UV-optical spectral range of the EBL. Backward-evolution
models do not include the physical processes that link the cosmic
infrared background and the UV-optical part of the spectrum. Some
of them try to amend this by incorporating template spectra. Other
models, naturally arrive at a doubly-peaked EBL because they explicitly
include the absorption of starlight and the following re-emission
by dust. 

A number of EBL models
have been developed over the last two
decades~\citep[e.g.,][]{Salamon:98, Stecker:06, Kneiske:02, Kneiske:04,
Primack:05, Gilmore:09, Franceschini:08, Razzaque:09, Finke:09}.
However large scatter in available EBL data does not constrain these
models strongly.

Primack \textit{et al.} \cite{Primack:05}, use a semi-analytical
model, which in general predicts lower optical depths for nearby sources
$z\lesssim2$ than the other models. Stecker \textit{et al.} \cite{Stecker:01,Stecker:06b}
use a backward evolution model that has been frequently updated using
new data. Their model predicts a large UV photon density and consequently
a higher gamma-ray opacity at high redshifts. Kneiske \textit{et al.} \cite{Kneiske:04}
use a chemical-evolution model for the UV-optical part of the
EBL and, based on recent
deep galaxy surveys, a backwards-evolution for the infrared part.

\begin{figure}[ht]
\begin{centering}
\includegraphics[width=0.9\columnwidth]{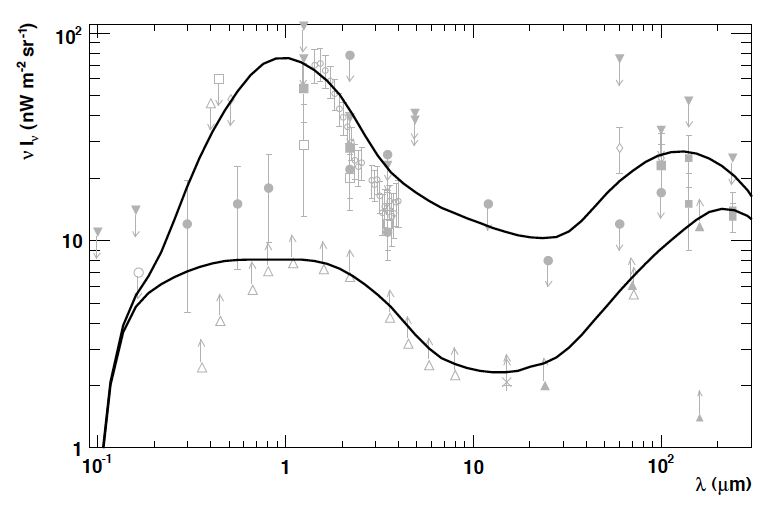}
\par\end{centering}
\caption{\label{fig:EBLmodels} EBL measurements and lower/upper limits (gray symbols) from various experiments. Solid curves represent minimum and maximum shape for the EBL spectrum which is reasonable given the observational constraints \citep{Mazin:07}. This illustrates the huge uncertainty in estimating the EBL content of the Universe. Source: \citep{Mazin:08}}
\end{figure}

Figure \ref{fig:EBLmodels} shows various observational measurements and lower/upper limits from various experiments. Minimum and maximum shape of the EBL spectrum that stays in reasonable agreement with the observational constraints are also plotted.
One can notice that the predictions differ significantly, sometimes by more than a factor five. Further observational constraints are therefore necessary in order to get a better handle on the total EBL content in the universe.

\section{\label{sub:IRAbs_Effects}Constraints on the EBL via $\g$-ray observations}

\begin{figure}[ht]
\begin{centering}
\includegraphics[width=0.75\columnwidth]{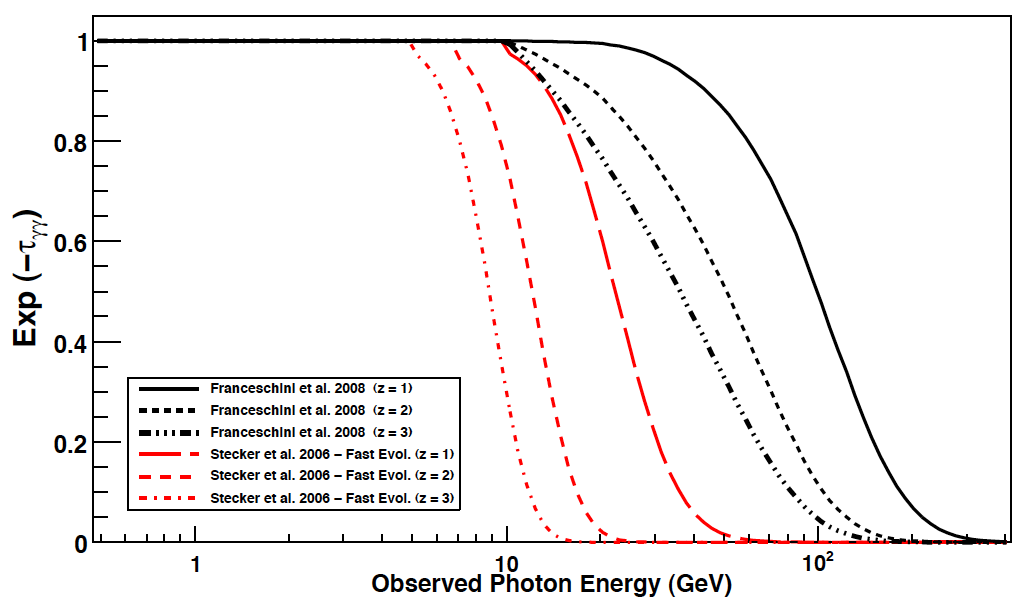}
\par\end{centering}
\caption{\label{fig:tau_vs_energy}Attenuation as a function of observed gamma-ray energy for the EBL models of 
\citep {Franceschini08} and \citep{Stecker06}. These models predict the minimum and maximum absorption 
of all models in the literature, and thus illustrate the range of optical depths predicted at very high energy.}
\end{figure}

Let us first describe the effects of the EBL absorption on the $\g$-ray spectra from 
distant sources.
Figure \ref{fig:tau_vs_energy} shows the attenuation factors ($e^{-\tau(E,z)})$
versus photon energy and GRB redshift for different EBL models.
As can be seen, Primack's 2004 model predicts less absorption,
while the Stecker models predict more. 
Also, the attenuation is positively correlated with the redshift
of the source and the photon energy. 

The primary extragalactic $\g$-ray point sources are considered to be
blazars, which are galaxies with relativistic jets directed along our
line of sight; and GRBs.
GRBs have not been used to contrain EBL absorption during
the pre-{\em Fermi} era mainly because of a lack of sensitivity to transient
objects above 10 GeV as well as the absence systematic redshift measurement during the EGRET era.
The sensitivity of EGRET decreased significantly above 10
GeV, and the field-of-view of TeV instruments is too small to
catch the prompt phase where most of the HE emission occurs. 

Because the e-folding cutoff energy, $E(\tau_{\g\g}=1)$, from $\g\g$ pair production in
$\g$-ray source spectra decreases with redshift, TeV $\g$-ray
telescopes are limited to probing EBL absorption at low redshift due
to their high detection thresholds (typically at 100~GeV).
Ground-based $\g$-ray telescopes have detected 35 extragalactic
sources to date\footnote{e.g., http://www.mpi-hd.mpg.de/hfm/HESS/pages/home/sources/, http://www.mppmu.mpg.de/~rwagner/sources/}, mostly of the high-synchrotron peaked (HSP) BL Lacertae objects
type.  The most distant sources seen from the ground with a confirmed
redshift are the flat spectrum radio quasar (FSRQ) 3C~279 at $z=0.536$ \citep{Albert:08}
and PKS~1510-089 at $z=0.36$ \citep{Wagner:10}.  Observations
of the closest sources at multi-TeV energies have been effective in
placing limits on the local EBL at mid-IR wavelengths, while spectra
of more distant sources generally do not extend above 1 TeV, and
therefore probe the optical and near-IR starlight peak of the
intervening EBL.

The starting point for constraining the EBL intensity from observation
of TeV $\g$-rays from distant blazars with atmospheric Cherenkov
telescopes~\citep[e.g.,][]{Stecker:93, Schroedter:05, Costamante:04,
Aharonian:06a, Mazin:07, Albert:08, Finke:09b} is the
assumption of a reasonable intrinsic blazar spectrum, which, in the case of a power law, $dN/dE \propto
E^{-\Gamma_{int}}$ for example, that is not harder than a pre-specified
minimum value, e.g., $\Gamma_{int}\geq\Gamma_{min}=0.67$ or 1.5. Upper limits on the
EBL intensity are obtained when the reconstructed intrinsic spectral
index from the observed spectrum, $\Gamma_{obs}$, presumably softened by EBL absorption of
VHE $\g$-rays, is required to not fall below $\Gamma_{int}$.  The
minimum value of $\G$ has been a matter of much debate, being reasoned
to be $\G_{int}=1.5$ by \cite{Aharonian:06a} from simple shock
acceleration theory and from the observed spectral energy distribution (SED) properties of blazars, while \cite{stecker:07} argued for harder values
under specific conditions based based on more detailed shock acceleration simulations.
It was suggested that a spectral index as hard as
$\Gamma_{int}=0.67$ was possible in a single-zone leptonic model if
the underlying electron spectrum responsible for inverse-Compton
emission had a sharp lower-energy cutoff \cite{Katarzynski:06}.  Finally we note that
Compton scattering of the cosmic microwave background radiation by
extended jets could lead to harder observed VHE $\g$-ray spectra \cite{Boettcher:08}, and that internal absorption could, in some cases, lead to harder spectra in the TeV range as well \cite{Aharonian:08}.

	\chapter{The $Fermi$ Gamma-ray Space Telescope}
\label{Fermi}

The $Fermi$ observatory continuously observes the entire celestial sphere in the gamma-ray band, with unprecedented sensitivity and improved angular resolution.
Covering the 20 MeV to
$\geq 300$ GeV energy range, its main instrument, the Large Area Telescope (LAT)\footnote{we refer the reader to \cite{FERMI} for a detailed description of the instrument.}, has a large detecting area, an
imaging capability over a large FOV, and the time resolution and low deadtime sufficient
to study transient phenomena. The LAT also provides active background discrimination
and rejection against the large fluxes of cosmic rays, and Earth albedo gamma rays. The Fermi Gamma-ray Burst Monitor
(GBM) onboard $Fermi$ provides spectral and temporal observations in the 10 keV to 40 MeV
energy band, detects and localizes
transient objects (GRBs as well as other transient emissions such as Soft-Gamma ray Repeaters, Terrestrial Gamma-ray Flashes...), and alerts the observatory control unit of a GRB trigger for a more refined search in the LAT instrument. $Fermi$ can autonomously repoint to its
observing plan to observe strong GRBs during and after the low-energy gamma-ray
emission, and provides rapid notification to the science community. The next sections
will describe the instruments in greater detail.

The primary data downlink and uplink path between the spacecraft and the ground is through the
Tracking and Data Relay Satellite System (TDRSS).\footnote{See
http://msl.jpl.nasa.gov/Programs/tdrss.html} Science and housekeeping data are
downlinked at Ku-band frequencies $\sim 10-11$ times per day at 40 Mbps during seven to
eight minute real time telemetry contacts. During these downlinks there is a direct 4 kbps
uplink rate available at S-band frequencies. 
The time between these downlinks, the transmission time through
TDRSS and the processing at the LAT Instrument Science and
Operations Center (LISOC) result in a latency of $\sim 6-12$ hours
between an observation and the availability of the
resulting LAT data for astrophysical analysis. 
Burst alerts and spacecraft alarms can be
downlinked in the S-band (variable data rates) at all times. 
When GBM or LAT software triggers, messages are sent to the ground through TDRSS
with a $\sim$15~s latency.
S-band uplink and downlink
(no science data) directly to ground stations is possible as a backup to TDRSS. Time and
spacecraft position are provided by an onboard GPS system.

\section{Mission timeline}

$Fermi$ was launched in June $11^{th}$, 2008, from Cape Canaveral by a Delta 2920H-10 (also
known as a Delta II 'Heavy') into an initial orbit of $\sim 565$ km altitude at a 25.6 degree
inclination with an eccentricity $<0.01$. The orbital period is 96.5 minutes, and has a
precession period of 53.4 days (so the RA and Dec of the orbit poles trace a 25.6 degree
circle on the sky every 53.4 days). The mission design lifetime is a minimum of 5 years,
with a goal of 10 years.

After launch the mission consists of three phases: a $\sim 2$ month on-orbit initial checkout
(Phase 0), a one year science verification period during which a full sky survey will be
performed (Phase 1), and then at least four years of operations determined by the
scientific goals and requirements of guest investigations (Phase 2). There is one cycle of
guest investigations during the verification and sky survey phase, and annual guest
investigation cycles during Phase 2. The GBM data were publicly released during
Phase 1 while the LAT data were released only in Phase 2, i.e., about 14 months after 
FermiÕs launch.

\section{Observing Modes}

The LAT and GBM have very wide FOVs, and the observatory is very flexible in the
direction in which it can point. An observational constraint is to avoid pointing at or near
the Earth to maximize the detection of astrophysical photons. The Earth's limb is indeed a very strong source of albedo gamma-rays which the LAT may  occasionally observe for instrument
calibration. Orientation requirements for the LAT's cooling radiators, the battery
radiators, and the observatory solar panels also impose engineering constraints,
particularly during slewing maneuvers. No science data is taken while the observatory is
transiting the South Atlantic Anomaly (SAA) since the instruments lower the voltage on
their photomultiplier tubes (PMTs). The SAA is a region over the South Atlantic with a
high density of charged particles that are trapped by the configuration of the Earth's
magnetic field. In a 25.6 degree inclination orbit and at Fermi's altitude, SAA outages
cost $\sim15\%$ of the LAT's and GBMÕs potential observing time.

The $Fermi$ spacecraft operates in a number of observing modes. Transitions between
modes may be commanded from the ground or by the spacecraft. Based on data from the GBM, 
an autonomous repoint request can be sent to the spacecraft in order to
change the observing mode to monitor the location of a GRB (or other short timescale
transient) in or near the LAT's FOV. 
This mode keeps the earth out of the FOV; the
default Earth Avoidance Angle (defined as the minimum angle between the LAT axis and
the Earth's limb) is 30 degrees. When the target is unocculted but within the Earth
Avoidance Angle of the Earth's limb, the spacecraft will keep the target in the LAT's
FOV while keeping the Earth out of the LAT's FOV. The observatory may observe a
secondary target when the Earth occults the primary target.
After a pre-determined time the spacecraft will
return to the scheduled mode. Currently the dwell time for such autonomous repoints is
five hours. The pointing accuracy is $\leq 2$ degrees ($1 \sigma$ goal of $\leq 0.5$ degrees), with a
pointing knowledge of $\leq 10$ arcsec (goal $\leq 5$ arcsec).

In survey mode, which predominate during most of the mission (e.g., $>80\%$
of the observing time), the LAT's pointing is relative to the zenith (the direction away
from the Earth), and therefore changes constantly relative to the sky. Uniformity of
exposure is achieved by "rocking" the pointing perpendicular to the orbital motion. The
default profile during the first year of the mission rocked the instrument axis 35 degrees 
north for one orbit, then 35 degrees south for one orbit, resulting in a two-orbit periodicity. 
In order to optimize the uniformity of sky coverage, the rocking profile was changed in September $2^{nd}$ 2009 to a 50 degree rocking angle. The maximum rocking angle (in case of sun maneuver for example) is set to 60 degrees. This observing mode provides uniform sky coverage when averaged over 2 orbits ($\sim 3$ hours) with the
equivalent of 30 minutes of on axis exposure for each part of the sky.

In the next 2 sections, we will describe the main characteristics of the LAT and GBM detectors which are relevant to GRB observations. The LAT is described in greater depth in \citep{FERMI} and the GBM in \citep{Meegan:09}.

\section{The Large Area Telescope (LAT)}
\label{sec:LAT_Description}

\subsection{Instrument capabilities}

The LAT's principal objective is high sensitivity gamma-ray observations of celestial
sources in the energy range from $\sim 20$ MeV to $\geq 300$ GeV. Table \ref{LAT capabilities} lists the main characteristics of the LAT instruments: wide FOV, large effective area, good energy
resolution and good angular resolution.

\begin{table}[htbp]
%   \centering
   \begin{tabular}{|p{3.5cm}|p{9.5cm}|} % Column formatting, @{} suppresses leading/trailing space
   \hline
      	Characteristics    & Capability  \\
      	\hline
      	Energy Range 		& $\sim 20$ MeV to $\geq 300$ GeV	 \\
   	\hline
	Energy Resolution 	& 10\% below 100 MeV, 5\% around 1 GeV and 20\% above 10 GeV \\
	\hline
   	Effective Area 		& $\geq 8,000$ cm$^{2}$ ($\geq 9,000$ cm$^{2}$) maximum effective area at normal incidence for the diffuse (transient) class. See figure \ref{irf diffuse} \\
      	\hline
	Single Photon Angular Resolution & $\leq 0.15^{\circ}$, on-axis, 68\% space angle containment radius for $E\geq10$ GeV;  $\leq3.5^{\circ}$, on-axis, 68\% space angle containment radius for E=100 MeV \\
      	\hline
	Field of View & $\sim 3$ str \\
      	\hline
	Time Accuracy 		& $\leq 10 \mu s$, relative to spacecraft time \\
      	\hline
	Dead Time 	& $\leq 100 \mu s$ per trigger \\
      \hline
      \end{tabular}
   \caption{LAT instrument characteristics}
   \label{LAT capabilities}
\end{table}

Key points of the LAT design for GRB observations are: 

\begin{itemize}
\item Wide Field-Of-View: $\sim 3$ str if the LAT FoV is defined as the direction where the effective area is at least one-third of the peak effective area.With this definition, $\sim 25 \%$ of the sky is observable at all time (with variation of the effective area over the FoV)
\item Large effective area: improved sensitivity to faint bursts and possibility for time resolved spectral analysis of the bright bursts
\item Good energy resolution: improved spectral analysis compared to EGRET
\item Point Spread Function: good GRB localization especially for hard bursts; we estimate that two or three photons above 1 GeV will localize a bursts to $\sim 5$~arcminutes
\item Short dead time: study of the sub-structure of the GRB pulses; typically of the order of milliseconds (compared to $\gtrsim 100$ ms dead time for EGRET) %\citet{Walker:00}, with a time resolution that has never before been accessible at GeV energies.
\end{itemize}

\subsection{Detection methodology}

\begin{figure}
\begin{center}
\includegraphics[ width=.6\linewidth, keepaspectratio]{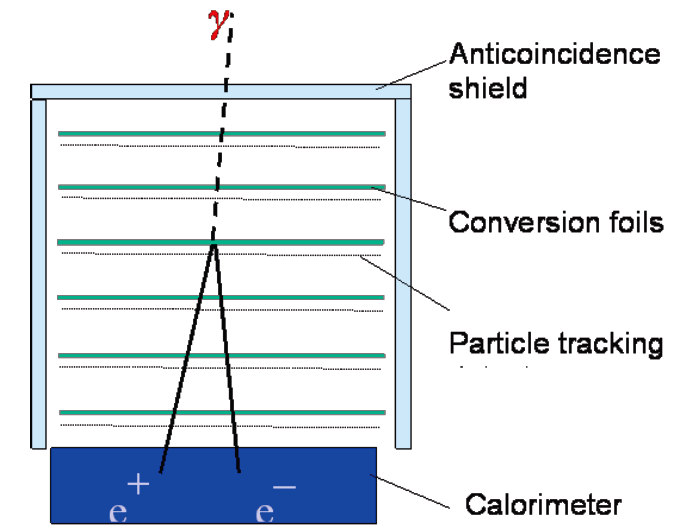}
\end{center}
\caption{Photon conversion in the LAT}
\label{photon conversion}
\end{figure}

The LAT is a 'pair conversion' $\g$-ray detector. Figure \ref{photon conversion} gives a schematic of the pair conversion process in the LAT instrument. Gamma rays penetrate into the detector and
interact with a high Z converter material, in this case tungsten, to produce an electron-positron
pair. Since the gamma-ray energy is much larger than the rest mass of the
electron and positron, both members of the pair continue predominantly in the direction
of the incident gamma ray. The passage of the electron and positron through the detector
is tracked by components that are sensitive to the passage of charged particles, in this
case of the LAT silicon strip detectors. At the bottom of the detector is a calorimeter array
made of CsI(Tl) logs where electromagnetic showers develop and that aims at recording the total energy deposited and therefore allowing a good estimation of the energy of the particle.

Charged particles incident on the LAT also interact in the LAT, resulting in multiple
charged particle tracks. To veto these charged particles, which are `background' for an
astrophysical telescope, the LAT is surrounded by an anticoincidence detector (ACD),
plastic scintillation tiles that scintillate when traversed by a charged particle. Very high energy
gamma rays ($E>> 1$ GeV) produces $\sim$ thousands of electrons, positrons and $\g$-rays some of which exit the LAT through the ACD (`backsplash'); the EGRET
detector on CGRO had a monolithic ACD that vetoed the instrument whenever such $e^-/e^+/\g$ hit the ACD.
The LAT ACD is segmented, and only when such a backsplash hits an ACD tile on the
path of the incoming $\g$-ray is the event vetoed. The segmentation also results in a more uniform anti-coincidence threshold over the ACD. This segmentation of the LAT ACD increases the LAT's
effective area for high-energy gamma rays dramatically relative to EGRET (whose sensitivity went down past a few GeV).

The output from the LAT consists of the pulse-height signals produced as charged
particles deposit energy in different parts of the tracker and calorimeter. By combining
the pulse heights with the x-y coordinates of each pair of silicon strip detectors, one can
reconstruct the particle trajectory and energy losses. The analysis both onboard and on
the ground reconstructs the tracks of the charged particles from these data, and then
characterizes the interaction that produced the charged particles; this analysis can
distinguish between events resulting from photons and background, determine the
incident direction and estimate the energy.

\subsection{Detector Structure}

\begin{figure}
\begin{center}
\includegraphics[ width=.6\linewidth, keepaspectratio]{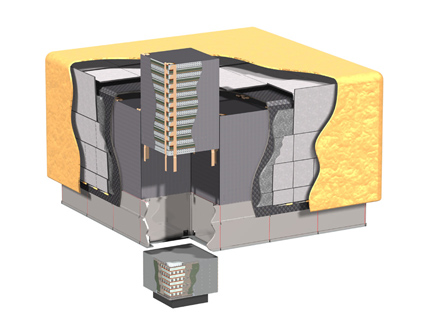}
\end{center}
\caption{View of the Fermi LAT instrument with the Anti-Coincidence Detector (yellow) "opened" and the tracker and calorimeter part of one tower separated for more clarity.}
\label{fermi inst}
\end{figure}

The LAT consists of an array of 16 tracker (TKR) modules, 16 calorimeter (CAL)
modules, and a segmented ACD. The TKR and CAL modules are mounted to the
instrument grid structure. Figure \ref{fermi inst} shows a schematic of the different part of the LAT instrument.

Each TKR module consists of 18 pairs of X and Y planes. Each XY pair has an array of
silicon-strip detectors (SSDs) for charged particle detection. The SSDs in each plane actually consist of two planes of silicon strips,
one running in the x and the other in the y direction, thereby localizing the passage of a
charged particle. The first 12 planes
have 0.035 radiation length thick tungsten plates in front of the SSD pairs, the next 4 planes
have 0.18 radiation length thick tungsten plate, and the last 2 planes, immediately in
front of the CAL, do not have tungsten plates. A radiation length is defined as the length
in a specific material in which an energetic electron loses $1-e^{-1}$ of its energy by
bremsstrahlung.

Gamma rays incident from within the LAT's FOV either convert
into an electron-positron pair in one of the TKR's tungsten plates or Compton scatter on an electron in the TKR. The initial directions of
the electron and positron are determined from their tracks recorded by the SSD planes
following the conversion point. Multiple scattering in the first few tungsten layers results in an
angular deflection that is a fundamental limit to the low energy angular resolution.
Although less frequent, bremsstrahlung by $e^+/e^-$ just converted in the TKR can offset the incident $\g$-ray direction.
Cosmic rays also interact within the TKR modules. Reconstruction of the interactions
from the tracks identify the type of particle as well as its energy and incident direction.

Each CAL module consists of 8 planes (4 X planes and 4 Y planes) of 12 CsI(Tl) crystals each. The crystals are read
out by two PIN diodes at each end. The CAL's segmentation and read-out provide precise
three-dimensional localization of the particle shower in the CAL. At normal incidence the
CAL's depth is 8.5 radiation lengths. The CAL is a total absorption calorimeter with excellent energy resolution.

The ACD is composed of plastic scintillator segmented into tiles, and supplemented with
fiber ribbons, and the plastic scintillators are read out by waveshifting fibers connected to one PMT at each end.

\subsection{Onboard and Ground Processing}
\label{processing}

The LAT's Data Acquisition System (DAQ) performs preliminary cuts on events within
the LAT, to reduce the rate of background events that is telemetered to the ground.
The DAQ processes the captured event data into a data stream with an average bit rate of
1.2 Mbps for the LAT. The DAQ also performs: command, control, and instrument
monitoring; housekeeping; and power switching. Onboard processing can be modified by
uploading new software, if necessary.

The astrophysical photons of primary interest are a tiny fraction of the particles that
penetrate into the LAT's TKR. The LAT performs on-board analysis cuts that reduces the $\sim 2.5$ kHz event rate that trigger the TKR to $\sim 400$ Hz event rate that is sent to the ground for further analysis; of these $\sim 400$ Hz only $\sim 2-5$ Hz are
astrophysical photons. The data for an event that passes the on-board analysis cuts is
stored in a packet with a time stamp and details of the signals from the various LAT
components. Because the number of signals for a given event varies, the data packets
have variable length. These data packets describing each event are the LAT's primary
data product. The LAT transfers these packets to the spacecraft's solid state recorder
(SSR) for subsequent transmission to the ground.

\begin{figure}
\begin{center} $
\begin{array}{cc}
\includegraphics[ width=.48\linewidth, keepaspectratio]{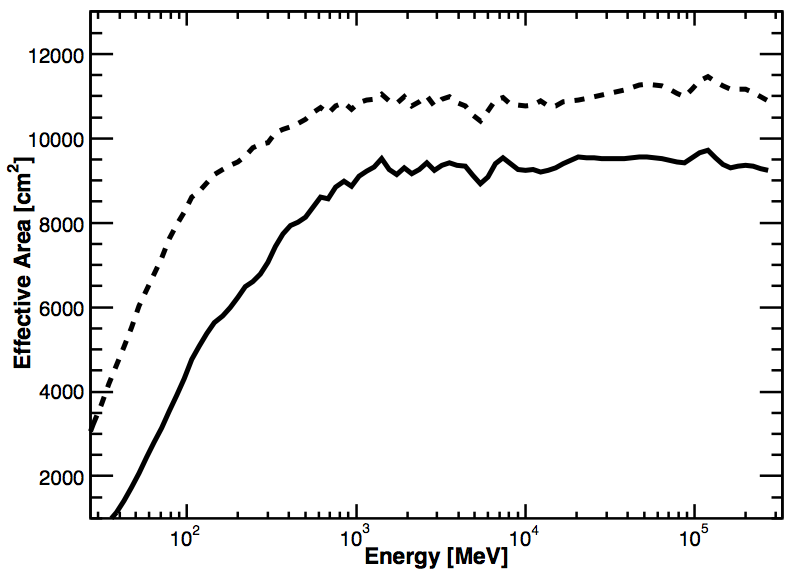} &
\includegraphics[ width=.48\linewidth, keepaspectratio]{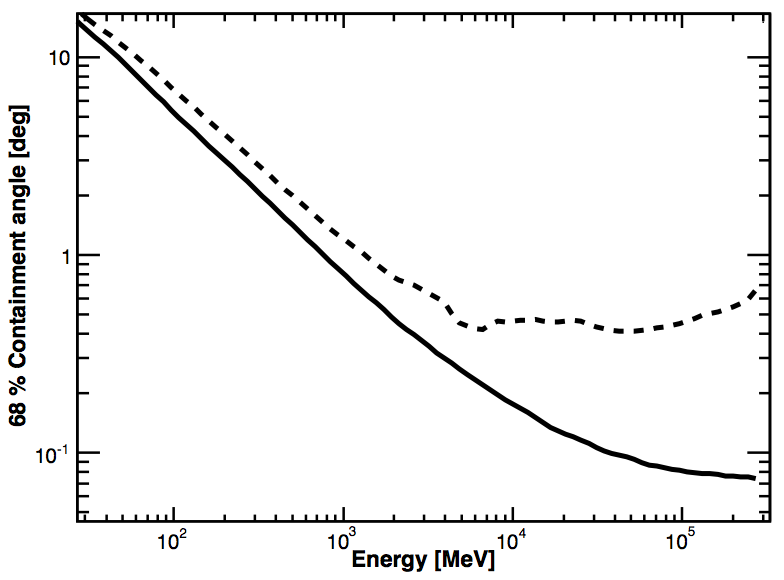}
\end{array} $
\end{center}
\caption{Comparison of the LAT onboard (dashed) and onground 'transient' class (solid) effective areas and point-spread-functions}
\label{onboardVSonground}
\end{figure}

The onboard flight software also performs event
reconstructions for a burst trigger.  Because of the
available computer resources, the onboard event selection
is not as discriminating as the on-ground event selection,
and therefore the onboard burst trigger is not as sensitive
because the astrophysical photons are diluted by a larger
background flux. Similarly, larger localization
uncertainties result from the larger onboard PSF.
Comparison of onboard and on-ground \emph{transient} effective area and PSF can be seen in figure \ref{onboardVSonground}

The data telemetered down from $Fermi$ enters the $Fermi$ ground system through the
Mission Operations Center (MOC) housed at GSFC. The MOC 'cleans up' the telemetry, monitors the spacecraft through the housekeeping portion of the telemetry. The resulting product, called Level 0 data, is transmitted to the other ground system elements. The MOC also sends commands to the
spacecraft, particularly a weekly load of observing and operational commands.

Each instrument team maintains an Instrument Operations Center (IOC). The LAT Instrument
Science and Operations Center (LISOC) is located at SLAC in Palo Alto, CA.
The IOC receive the 'cleaned up' telemetry, monitor their
detectors through the housekeeping portion of the telemetry, process the science data, and
transmit the resulting science data products to the FSSC. The LAT science data
processing is quite extensive, starting with event reconstruction from the 'hits' in different
parts of the LAT and ending with a characterization of these events. The data that leaves is called Level 1 data.

The data telemetered to the ground consists of the signals
from different parts of the LAT; from these signals the
ground software must `reconstruct' the events and filter
out events that are unlikely to be gamma-rays. Therefore,
the Instrument Response Functions (IRFs) depend not only on
the hardware but also on the reconstruction and event
selection software.  For the same set of reconstructed
events trade-offs in the event selection between retaining
gamma rays and rejecting background result in different
event classes.  The original three standard event
classes released to the scientific community are appropriate for
different scientific analyses (as their names suggest): the \emph{transient}, \emph{source} and
\emph{diffuse} event classes. Figure \ref{irf diffuse} shows the effective area and PSF of the \emph{diffuse} event class.
Less severe cuts increase the photon signal (and hence the
effective area) at the expense of an increase in the
non-photon background and a degradation of the PSF and the
energy resolution.
A significant part of my thesis was devoted to the optimization of event class for transient sources as will be described in details in chapter \ref{event_optimization}.

\begin{figure}
\begin{center}
\includegraphics[ width=1.\linewidth, keepaspectratio]{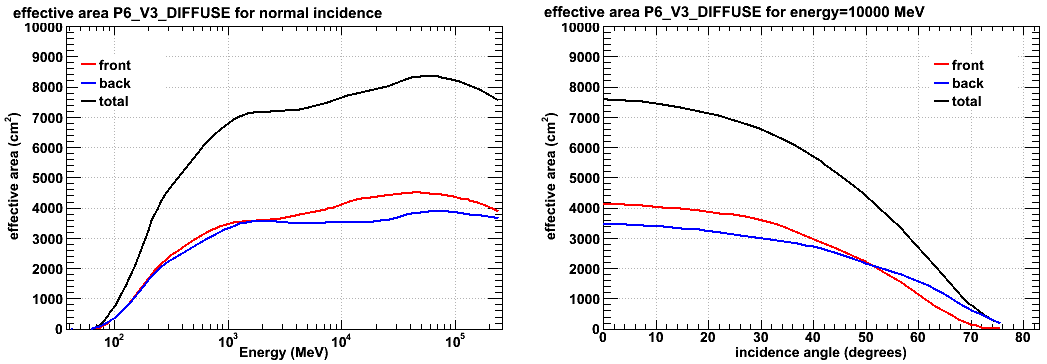}
\end{center}
\caption{Effective Area: the plot on the left is for normal incidence photons (defined here as $cos(\theta)>0.975$); the one on the right is for 10 GeV photons as a function of incidence angle. Blue and red curves represent the back and front effective area while black curve is the total instrument effective area (back+front).}
\label{irf diffuse}
\end{figure}

\begin{figure}
\begin{center}
\includegraphics[ width=.75\linewidth, keepaspectratio]{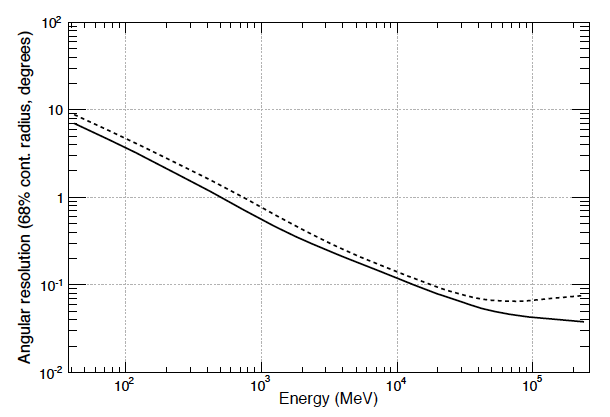}
\end{center}
\caption{Point-spread-function of the \emph{diffuse} class at incidence angle (solid) and $60^{\circ}$ off-axis (dashed).}
\label{irf diffuse}
\end{figure}

In addition to the search for GRB onboard the LAT and  manual follow-up analysis by duty scientists, there is also automated processing of the full science data. This processing performs an independent search for transient events in the LAT data, to greater sensitivity than is possible onboard, and also performs a counterpart search for all GRB detected within the LAT FoV.

\subsection{The LAT Collaboration}

The Fermi LAT collaboration includes scientists from Stanford University, including
SLAC/KIPAC (PI: Prof. P. Michelson); GSFC (Project Scientist: J. McEnery); University of
California at Santa Cruz; Naval Research Laboratory; University of Washington; Sonoma
State University; Ohio State University; University of Denver; Perdue University Ð
Calumet. in the United States; Stockholm University and Royal Institute
of Technology in Sweden; Commissariat a l'Energie Atomique, Departement
d'Astrophysique, Saclay; Institut National de Physique Nuclearie et de Physique des
Particules; and Centre dÕEtude Spatiale des Rayonnements in France; Instituto Nazionale
di Fisica Nucleare; Agenzia Spaziale Italiana; Instituto di Fisica Cosmica, CNR; and
Universita e Politecnico di Bari in Italy; Hiroshima University; Institute of Space and
Astronautical Science; and the Tokyo Institute of Technology, the University of Tokyo and Waseda University in Japan; and Institut de
Ciencies de lÕEspai in Sapin. In addition, 29 institutions world-wide host Affiliated
Scientists.

\section{$Fermi$ Gamma-ray Burst Monitor (GBM)}
\label{sec:GBM_Description}

The Fermi GBM\footnote{see \cite{Meegan:09} for a detailed description of the instrument.} provides simultaneous low-energy spectral and temporal measurements
for all GRBs within the LAT FOV. The combined GBM and LAT effective energy range
spans more than 7 energy decades from 10 keV to 300 GeV. The GBM extends the
energy coverage from below the typical GRB spectral break at $\sim100$ keV to above the
LAT's low-energy cutoff. However we will see that the LAT data at low-energy are currently difficult to use for spectral analysis making inter-calibration of the two instruments difficult at the moment. Furthermore, the GBM's sensitivity and FOV commensurate with the LAT's to ensure that many bursts will
have simultaneous low-energy and high-energy measurements with statistical
significance. The GBM also assists the LAT in detecting and localizing GRBs rapidly by
providing prompt notification to the ground of a burst trigger. Finally, the GBM provides
coarse GRB locations over a wide FOV that can be used to repoint the LAT on
particularly interesting bursts (both inside and outside the LAT FOV) for gamma-ray
afterglow observations. These locations are also sent to the ground with $\sim 15$ sec delay to notify external follow-up observers.

Using detection criteria similar to those of CGRO's Burst And Transient Source
Experiment (BATSE), the GBM burst detection rate is reaching $\sim 250$ bursts per year, well above the pre-launch. This improvement is mostly due to an increase in the number of time windows that are searched.... blablabla
In addition to measuring low-energy spectra below the LAT threshold, the GBM
significantly improves the constraints on high-energy spectral behavior compared to
those of the LAT alone. The combination of GBM and LAT data therefore provides a
powerful tool to study GRB spectra and their underlying physics.

\subsection{Detector Capabilities}

See table \ref{GBM capabilities}.

\begin{table}[htbp]
%   \centering
   \begin{tabular}{|p{8.5cm}|p{4.5cm}|} % Column formatting, @{} suppresses leading/trailing space
   \hline
      	Characteristics    & Capability  \\
      	\hline
	Low Energy Limit & $\sim 8$ keV \\
      	\hline
	High Energy Limit & $\sim 40$ MeV \\
      	\hline
	Energy Resolution (FWHM, 0.1-1 MeV)  & $\sim 12\%$ at 511 keV \\
      	\hline
	Field of View (co-aligned with LAT FOV) & $\sim 9.5$ sr \\
      	\hline
	Time Accuracy (relative to spacecraft time) & $\sim 2$ microseconds \\
      	\hline
	Average Dead Time & $\leq 2$ microsecond/count \\
      	\hline
	Burst Sensitivity (peak 50-300 keV flux for $5 \sigma$ detection in ph cm$^{-2}$.s$^{-1}$) & $\sim 0.6$ ph.cm.$^{-2}$.s$^{-1}$ \\
      	\hline
	Localization ($1\sigma$ systematic error radius) & $\leq15^{\circ}$ (onboard); $\leq5^{\circ}$ (on ground) \\
      	\hline
	Burst Alert Time Delay (time from burst trigger to spacecraft notification used to notify ground or LAT)  & $\leq 2$ s \\
      \hline
      \end{tabular}
   \caption{GBM instrument characteristics}
   \label{GBM capabilities}
\end{table}
	
\subsection{Hardware}

\begin{figure}
\begin{center}
\includegraphics[ width=.9\linewidth, keepaspectratio]{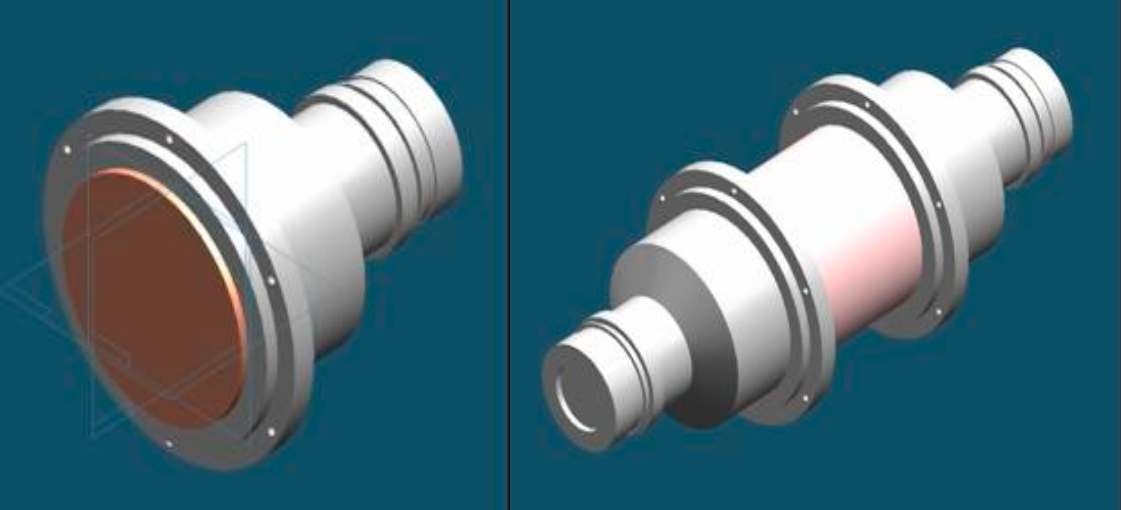}
\end{center}
\caption{a view of the NaI (left) and BGO (right) detectors}
\label{GBM detectors}
\end{figure}

\begin{figure}
\begin{center}
\includegraphics[ width=.6\linewidth, keepaspectratio]{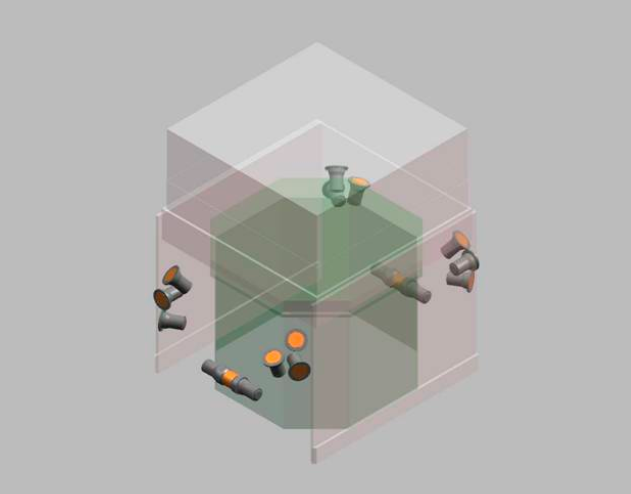}
\end{center}
\caption{GBM detectors placements around the satellite}
\label{GBM placement}
\end{figure}

To achieve the required GBM performance, the design and technology borrow heavily
from previous GRB instruments, particularly from BATSE. Like BATSE, the GBM uses
two types of cylindrical crystal scintillation detectors, whose light is read out by PMTs.
An array of 12 sodium iodide (NaI) detectors (0.5 in. thick, 5 in. diameter) covers the
lower end of the energy range up to $\sim1$ MeV. The GBM triggers on signals in the
NaI detectors. Each NaI detector consists of the crystal, an aluminum housing, a thin
beryllium entrance window on the front face, and a 5 in. diameter PMT assembly (including a
pre-amplifier) on the other. These detectors are distributed around the Fermi spacecraft
with different orientations to provide the required sensitivity, directionality and FOV. The cosine-like
angular variation of the effective area of the thin NaI detectors are used to localize burst sources by comparing
rates from detectors with different viewing angles. To cover higher energies, the GBM
also includes two 5 in. thick, 5 in. diameter bismuth germanate oxide (BGO) detectors. The
combination of the BGO detectors' high-density (7.1 g cm$^{-3}$) and large effective Z ($\sim 63$)
results in good stopping power overlapping with the low end of the LAT energy range at $\sim 20$ MeV.
Two BGO detectors are placed on opposite sides of the Fermi spacecraft to provide high energy
spectral capability over approximately the same FOV as the NaI detectors. For
redundancy and improve energy resolution, each BGO detector has two PMTs located at opposite ends of the crystal.

The signals from all 14 GBM detectors are collected by a central Data Processing Unit
(DPU). This unit digitizes and time-tags the detectors' pulse height signals, packages the
resulting data into several different types for transmission to the ground (via the Fermi
observatory), and performs various data processing tasks such as autonomous burst
triggering. In addition, the DPU is the sole means of controlling and monitoring the
instrument. For example, the DPU controls the PMTs' power supply to maintain their
gain.

\subsection{Data Types}

There are three basic types of GBM science data: (1) continuous data consisting of the
count rates from each detector in different energy and time integration bins;
(2) trigger data containing lists of individually time-tagged pulse heights from selected
detectors for periods before and after each on-board trigger; and (3) Alert Telemetry sent
down to the ground immediately after a burst containing computed data from a burst
trigger, such as intensity, location, and classification. The Burst Alert, the first packet of
the Alert Telemetry calculated by the on-board algorithm, arrives at the Gamma-ray burst Coordinates Network (GCN))\footnote{See http://gcn.gsfc.nasa.gov/} within
15 seconds of the burst detection. Alerts originating in the GBM are also sent to the LAT
to aid in LAT GRB detection and repointing decisions. The remaining data types are
transmitted via the scheduled Ku-band contacts. The GBM produces an average of 1.4
Gbits/day, with a minimum of 1.2 Gbits/day and a maximum allocated rate of 2.2
Gbits/day.

\subsection{The Fermi GBM Collaboration}

The Fermi GBM collaboration includes scientists from the Marshall Space Flight Center
(PI: Dr. C. A. Meegan), the Max Planck Institute for Extraterrestrial Physics (MPE; Co-
PI: Dr. J. Greiner), the University of Alabama in Huntsville, the Universities Space
Research Association (USRA), and Los Alamos National Laboratory. The Marshall,
University of Alabama, and USRA scientists are housed at the National Space Science
and Technology Center (NSSTC) in Huntsville.
       	\chapter{Optimization of event selection for GRB science}
\label{event_optimization}

The vast majority of instrument triggers and subsequently downlinked
data are background events caused by charged particles as well as
earth albedo \gray{s}.  The task of the hardware trigger is to
reduce data volume the online CPU can process while retaining most of the astronomical $\g$-ray events.  Subsequently the onboard filter eliminates background events further without
sacrificing celestial \gray{} events such that the resulting data can
be transmitted to the ground within the available bandwidth.   
These first two steps reduce the event rate from $2-4$ kHz to $\sim 400$ Hz.
The final task is for the analysis on the ground to distinguish between
background events and \gray{} events and minimize the impact of
backgrounds on \gray{} science.  The combination of these three elements (trigger, onboard filter and ground selection)
reduces the background by a factor of almost $10^6$ while preserving
efficiency for \gray{s} exceeding 75\%.  For reference, the average
cosmic \gray{} event rate in the LAT is $\sim$2 Hz.

As we will see in details below, the ground filtering algorithm defines several event classes optimized for different science analyses. This chapter describes the effort carried out to re-visit and further optimize the analysis of transient sources such as Gamma-Ray Bursts. Section \ref{intro_lat_sel} introduces the various LAT selection as defined before launch. Section \ref{gen_opt} provides a general approach to optimize event selection for specific science cases. Finally section \ref{opt_evt_class} summarizes our method and findings concerning the optimization of the LAT event classes for the analysis of transient sources.

\section{Introduction to LAT event selections}
\label{intro_lat_sel}

The on-ground algorithm, from the low-level information of each event, not only reconstructs the best direction and best energy for the event but also estimates the accuracy of the direction and energy reconstruction as well as the probability for the event to be a \gray{} (and not a cosmic-ray background).
These estimates are based on classification tree (CT) type analyses which generate parameters representative of the accuracy of the direction and energy measurement as well as the chance probability for the event to be a real $\g$-ray\footnote{Classification trees sort events through different branches where a selection process is taking place at each node. By training classification trees on Monte-Carlo data, one can assign probabilities for different branches to select $\g$-rays as opposed to charged particles and also determine the goodness of the direction and energy reconstruction on those events.}. These parameters can then be used as knobs by the various event selection to optimize the event sample for different science objectives (analyses of diffuse emission, point sources, transient objects...).

The background rejection is by far the most challenging of all the
reconstruction analysis tasks.   This is due to the large phase space
covered by the LAT and the very low $\g$-ray to cosmic-ray ratio in the
incoming data ($\sim$1:300 for down-linked data).  The first task is to
eliminate the vast majority of the charged particle flux that enters
within the FoV using the ACD in conjunction with the
found tracks.  
One cannot simply demand that there are no signal
from the ACD because high-energy \gray{s} can generate a considerable
amount of back splash
that can trigger several ACD
tiles\footnote{Back splash is caused when the shower of a high-energy event produces hard X-ray that scatter back into ACD tiles (usually the bottom tiles of the ACD).}.  Consequently only tiles which are not too distant from the reconstructed
event direction are considered to establish a veto by the presence of a signal in
excess of $\sim$$1/4$ that of a minimum ionization event.

Rejecting cosmic-ray backgrounds involve the detailed
topology of the events within the tracker and the overall match of the
shower profiles in 3D in both the tracker and the calorimeter.    The
tracker provides a clear picture of the early shower development.   For
example the identification of a 2-tracks vertex immediately reduces the
background contamination by about an order of magnitude.   However a
majority of events above 1 GeV do not contain such a recognizable
2-tracks vertex due to the small opening angle of the $e^+ e^-$ pair along the
incoming \gray{} direction.   The observation of a significant
number of extra hits in close proximity to the track(s) (i.e. consistent with electromagnetic shower development) indicates they
are electrons and hence from the conversion of a \gray{} while the
presence of unassociated hits or tracks are a strong indicator of
background.   These as well as other considerations are used for
training background rejection CTs.

The final discriminator of background is the identification of an
electromagnetic shower.   Considerations such as how well the shower axis reconstructed in the tracker points to the calorimeter centroid, how well the directional
information from the calorimeter matches that of the track found in
the tracker, as well as the width and longitudinal shower profile in
the various layers of the calorimeter, are important in discriminating
background.  Again the information from the reconstruction is used
to train CTs and the resulting probabilities are used to eliminate
backgrounds.

The broad range of LAT observations and analysis, from GRBs to
extended diffuse radiation, leads to different optimizations of the
event selections and different rates of residual backgrounds.  For
example, in analysis of a GRB, the extreme brightness of those objects in a relatively small region of the sky (they are point sources) and in a very short time window allow the background rejection
cuts to be relaxed relative to an analysis of a diffuse source
covering a large portion of the sky.   
And thus, the background rejection analysis has been constructed to allow analysis classes to be
optimized for specific science topics.

Basically three analysis classes have been defined based on Monte-Carlo, the
backgrounds in orbit, knowledge of the \gray{} sky,
and the performance of the LAT.  
Common to all
of these analysis classes is the rejection of the charged-particle
backgrounds entering within the field of view.  The classes are
differentiated by an increasingly tighter requirement that the
candidate photon events in both the tracker and the calorimeter behave
as expected for \gray{} induced electromagnetic showers.  The loosest
cuts apply to the so-called 'transient' class, for which the background rejection
was set to allow a background rate of $\lesssim$2 Hz, which would result in no more
than one background event every 5 sec inside a $10^\circ$ radius about a
source.  Next, the 'source' class was designed so that the residual background
contamination was similar to that expected from the extragalactic
\gray{} background flux over the entire field of view.  Finally, the
'diffuse' class has the best background rejection at the cost of a lower effective area and was designed such
that harsher cuts would not significantly improve the signal to noise.
The diffuse cut has also the best PSF which makes it optimal for studies of spatially extended structure (diffuse emission, supernovae remnants...).
Note that these three analysis
classes are hierarchical; that is all events in the 'diffuse' class are
contained in the 'source' class and all events in the 'source' class are
in the 'transient' class.

\begin{figure}
\begin{center}
\includegraphics[ width=.8\linewidth, keepaspectratio]{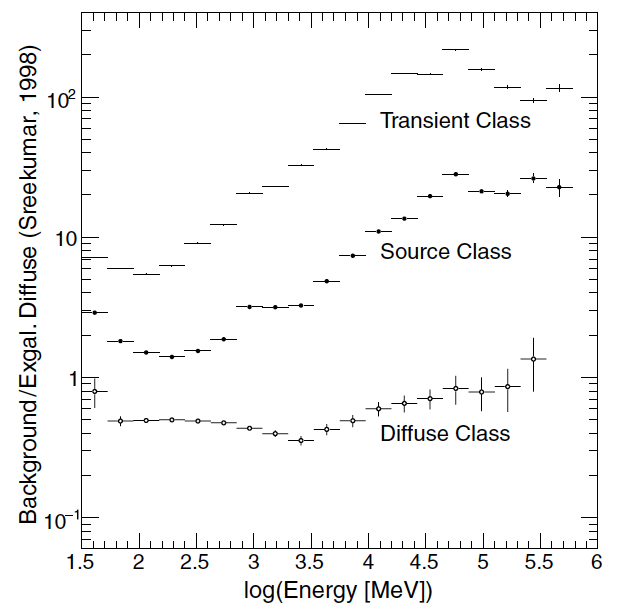}
\end{center}
\caption{Ratio of the cosmic-ray induced residual background to the extragalactic diffuse background 
inferred from EGRET observations \citep{Sreekumar:98} for each of the three 
analysis classes (transient, source, diffuse) in the Pass6 framework.  The integral EGRET diffuse flux is $1.45\times10^{-7}$ ph cm$^{-2}$ 
s$^{-1}$ sr$^{-1}$ above 100 MeV.}
\label{bkg_rate}
\end{figure}

The above optimization was first done pre-launch using a background model and Monte-Carlo and beam test data for the $\g$-ray instrument response function. Once in orbit, it was possible to further optimize the even selection based on the actual event rate detected by the LAT instrument.
Pre-launch studies as well as some first-year in-flight corrections led to the set of event selection and instrument response function which were released to the public under the name of Pass6\_V3 (on August 2009).
The residuals of cosmic-ray induced background events for the three Pass6\_V3 analysis classes are compared to the isotropic unresolved extragalactic $\g$-ray level in Figure \ref{bkg_rate}.  For the Diffuse class, the resulting rejection factor
is $\sim$1:$10^6$ at some energies (e.g., $\sim$10 GeV) while retaining $>$80\%
efficiency for retaining \gray{} events.

Part of my thesis work focused on the improvement of a LAT event selection for the detection and analysis of transient sources in the Pass6\_V3 framework (section \ref{opt_evt_class}).
A new set of event selection and corresponding instrument response function is under development within the LAT collaboration and will be released under the name of Pass7 in 2011. How my optimized event selection for transient sources fit within the Pass7 framework will also be described and compared to the characteristics in Pass6\_V3.

\section{Optimization of event classes for GRB science}
\label{gen_opt}

\subsection{Generalities}

This section is meant as a general introduction to the topic of event selection optimization that is useful to consider before going into the details of the analyses we performed. Ideally, one would like to retain all the $\g$-ray signal from the source of interest while rejecting the totality of the background events. This is of course not possible in practice and what one is reduced to do is maximize the $\g$-ray efficiency $Eff$, which is the ratio of the number of $\g$-rays retained after selection to the incoming $\g$-rays hitting the plane of your detector, while minimizing the backround rate $B$ in your detector. Both of these parameters, $Eff$ and $B$, are dependent on several variables (energy, spacecraft position, orientation, angle of incidence, conversion layer...) but for sake of simplicity we will for now assume that we are dealing with a simple counting $\g$-ray detector where $Eff$ and $B$ are unique numbers.

Let us define $S$ as the counting rate in the detector of the source alone (without any background) and $S_0$ the source rate actually incoming onto the detector plane so that $S = Eff \times S_0$.
The measurement of $S$ is normally carried out by counting the source plus background (at an average rate of $S+B$) for a time $T_{S+B}$ (on-source time interval) and then measuring background alone for a time $T_B$ (off-source time interval). The net rate due to the source alone is then:

$$ S = \frac{N_1}{T_{S+B}} - \frac{N_2}{T_B}$$

\noindent
where $N_1$ and $N_2$ are the total counts for each measure.
Since we are dealing with poissonian noise for our experiment, the uncertainty on S is then given by:

\begin{eqnarray*} 
\sigma_S &=& \sqrt{\left(\frac{\sigma_{N_1}}{T_{S+B}}\right)^2 + \left(\frac{\sigma_{N_2}}{T_{B}}\right)^2} \\
	&=& \sqrt{\frac{N_1}{T_{S+B}^2} + \frac{N_2}{T_{B}^2}} \\
	&=& \sqrt{\frac{S+B}{T_{S+B}} + \frac{B}{T_{B}}}
\end{eqnarray*}

In our case, the time for observing the transient $\g$-ray signal $T_{S+B}$ is determined by the nature of the transitory object and it is not in our capacity to change it. $T_B$ can also be considered as a given and much larger than $T_{S+B}$ since the procedure that we use to estimate background level makes use of the whole LAT data set accumulated over the mission lifetime (see Appendix \ref{bkg_est} for details). Furthermore, we note that only statistical uncertainties are included above when the backgournd estimation is actually dominated by systematics uncertainies which have non-trivial dependence with changing event selections. In the end, it is a reasonable assumption to assume that the fluctuation of the background during the on-source measurement will dominate over the statistical uncertainty of the background estimation ($\frac{B}{T_{S+B}} >> \frac{B}{T_B}$) as well as the systematic uncertainty of the background estimation which is estimated at 10-15\% (see Appendix \ref{bkg_est}) which remains true for on-source region with $\lesssim 40$ background events (always the case during GRB prompt emission for exemple). In the end:

$$ \sigma_S \backsimeq \sqrt{\frac{S+B}{T_{S+B}}}$$

We can now define two relevant Figure-Of-Merits (FOM) that we would like to maximize for the detection and spectral analysis\footnote{although we have put ourself in the simple case of a counting experiment, a detector measuring the energy of each event can easily be seen as multiple simple counting detectors operating in different energy bands which makes the above discussion relevant.} of $\g$-ray transient observations:

\begin{packed_enum}
\item {\bf Detection:} the significance of a signal can in first order be approximated by $S / \sigma_B$ where as we mentioned above, $\sigma_B$ is dominated by statistical uncertainty of the background emission during the on-source measurement.  A good FOM to maximize for such analysis would therefore be $S / \sqrt{B}$.
\item {\bf Spectral analysis:} the relative size of the spectral error bars with respect to the actual source flux measured which is proportianal to $\sigma_S / S \propto \sqrt{S+B} / S$. A good FOM to maximize for such analysis would therefore be $S / \sqrt{S+B}$.
\end{packed_enum}

\subsection{Method for optimizating event classes for GRB science}
\label{opt_evt_class}

Most of the events detected by the LAT instrument are background cosmic rays that we would like to separate from sources of $\gamma$-ray signal which we are interested in. Ideally, one would like to remove the totality of the background contamination while keeping the entire incoming gamma-ray signal. Unfortunately, this is not possible in practice and any real-world experiment suffers at some level from background contamination and loss of signal misinterpretated as background. Furthermore efficiency at detecting signal and background contamination usually work against each other. When reducing the former, one immediately increases the latter and vice versa. Therefore, a compromise between a high $\g$ efficiency and a low background level needs to be found. And this compromise of course strongly depends on the type of science that one is interested in.

The broad range of LAT observations and analyses, from GRBs
to extended diffuse radiation, lead to the definition of various
event selections with different rates of residual backgrounds (misclassified cosmic-rays) and $\g$-ray efficiency. During the pre-launch studies, three different event selections were defined \citep{FERMI}: the 'diffuse' class with the lowest background contamination and the best point-spread-function for the study of faint and diffuse sources ({\it i.e.} diffuse gamma-ray emission), the 'source' class with slightly better gamma efficiency and slightly increased background rate (optimized at the time for point source study although post-launch analysis found the 'diffuse' class to be more adapted for point source analysis in Pass6\_V3 IRFs), and the 'transient' class with the highest gamma efficiency at the cost of high background contamination (optimized at the time for transient events).

In the case of GRB observations of the prompt emission, the extreme brightness of those event on short time scale allow the background rejection cuts to be relaxed relative to an analysis of a diffuse source covering a large portion of the sky over longer periods of time. 
High energy extended emission from GRBs is another story as it is usually faint and therefore necessitates a purer event class such as the 'diffuse' class to detect it. The 'diffuse' class also has a much narrower PSF which makes it ideal for precise localization when plentiful of LAT 'diffuse' events were detected by the instrument in coincidence with a GRB trigger.

For object with a high $\g$-ray flux, a significant improvement in signal-to-noise ratio can be obtained by increasing the effective area while keeping the background rate at a reasonable level. The so-called 'transient' event class was developed for this specific purpose and is used for burst detection and localization when not enough events are detected in the diffuse class for an accurate localization determination. In the course of my thesis, I developed a new event selection to further improve the detection and spectral analysis of GRBs. Spectrum of astrophysical sources usually fall off rapidely (with a typical high energy index of $\sim -2.3$ for GRBs observed in the pre-$Fermi$ era). This means that most of the \gray{} photons are detected in the lower part of the LAT energy range. For this reason, the low energy effective area of the LAT is of crucial importance to improve the $\g$ efficiency for GRB detection. Since the LAT effective area falls off rapidly at low energy even for the 'transient' class, a small increase of the effective area in this region could have a substantial effect (see figure \ref{aeff_P6P7} for the Pass6\_V3 and Pass7 effective area of the 'transient' class).

We developed a more relaxed set of cuts than the 'transient' class in order to enhance the LAT effective area at the expense of background contamination. We estimated the improvements through a Figure-of-Merit based analysis which we will describe below. Our approach to develop new set of cuts was simple: we incorporated 'transient' class events by default and new events were added based on physically motivated cuts. Each LAT event is described by a set of merit variable which describe different properties of the interection in the detector for this particular event. Contrary to the development of the 'transient' class which used heavy classification tree training with simulated data, we simply used some physical intuition in order to define new set of cuts enhancing the $\g$ efficiency while keeping the background rate at a reasonable level. 

One cut selection we developped is called 'S3' and is defined with the following logic:

\vspace{0.5cm}
\noindent
S3 == [transient] $||$ ([Onboard filter cuts] \&\& AcdTileCount==0 \&\& VtxAngle$>0$ \nonumber \&\& CTBTkrCoreCalDoca$<150$)) \nonumber
\vspace{0.5cm}

with:

\begin{itemize}

\item  '[transient]': set of cuts applied to obtain the transient events. The '$||$' condition implies that the S3 selection is based on top of the transient selection (all transient events are also S3 events).

\item  '[Onboard filter cuts]': set of cuts applied onboard the spacecraft in order to reduce the event rate to a reasonable downlink rate for the Ku-Band antenna.

\item 'AcdTileCount': number of ACD tiles fired. When this parameter is set to zero, it means the ACD did not report any activity in coincidence with the event detection. Because the ACD efficiency is quite high (99.97\%), most of the charged particle having this parameter set to zero went through a 'crack' of the instrument or came from the back of the detector where the ACD shield is not present.

\item 'VtxAngle': angle between the two tracks converging at the vertex point. This variable has positive value when two tracks are reconstructed (otherwise a zero value) which is a signature of pair conversion (cosmic rays only produce one track).

\item 'CTBCoreCalDoca': distance between the projected vertex (or track if only one track) and the energy centroid in the calorimeter, evaluated at the z-plane of the centroid. This condition is requiring the vertex projection to be reasonably aligned with the energy deposition in the calorimeter. Note that events not leaving any energy in the calorimeter will not be included in the 'S3' selection.

\end{itemize}

Based on our discussion in section \ref{gen_opt}, we define the following Figure-of-Merit for spectral analysis:

\begin{itemize}

\item {\bf Spectral analysis:}

\begin{eqnarray}
FoM_{spectral} &=& \frac{Signal}{\sqrt{Signal+Background}} \nonumber \\
	&=& \frac{(S_{trans} \Delta T-B_{trans} \Delta T) \times \frac{Eff_{cut}}{Eff_{trans}}}{\sqrt{(S_{trans} \Delta T-B_{trans} \Delta T) \times \frac{Eff_{cut}}{Eff_{trans}} + B_{cut} \Delta T}} \nonumber \\
	&=& \sqrt{\Delta T} \times (S_{trans}-B_{trans}) \times \frac{ \frac{Eff_{cut}}{Eff_{trans}}}{\sqrt{(S_{trans}-B_{trans}) \times \frac{Eff_{cut}}{Eff_{trans}} + B_{cut}}} \nonumber
\end{eqnarray}

%\item {\bf Detection:} 
%
%\begin{eqnarray}
%FoM_{detection} &=& \frac{Signal}{\sqrt{Background}} \nonumber \\
%	&=& \frac{(S_{trans} \Delta T-B_{trans} \Delta T) \times \frac{Eff_{cut}}{Eff_{trans}}}{\sqrt{B_{cut} \Delta T}} \nonumber \\
%	&=& \sqrt{\Delta T} \times \frac{S_{trans}-B_{trans}}{Eff_{trans}} \times \frac{Eff_{cut}}{\sqrt{B_{cut}}} \nonumber
%\end{eqnarray}

\end{itemize}

$S_{trans}$, $B_{trans}$ and $Eff_{trans}$ are signal event rate, background rate and $\g$ efficiency for the transient selection respectively. $\Delta T$ is the approximate duration during which the burst emit.
What we are really interested in is the relative values of this FoM for different event selections. For that purpose, notice that $FoM_{cut1}/FoM_{cut0}$ is independent of the duration $\Delta T$ and only depends on the flux and high-energy index of the source. 
We will therefore look at $FoM_{spectral}$ as a function of $S_{trans}$, the transient event rate which is a direct observable, as well as the source high-energy index.

To compute this FoM, we need to compute two quantities for each event selection:

\begin{itemize}

\item The $\g$ efficiency ($Eff_{cut}$) which can be defined as the ratio of events passing the cut with the total number of events going through the geometrical area of the detector. However, for each cut we will only compute the ratio of the $\g$ efficiency with the $\g$ efficiency of the transient cut since this is all that matters to compare the FoMs of various cuts. This value will be computed from the Instrument Response Functions of each cut which is derived directly from Monte-Carlo simulations of $\g$-rays interacting with the LAT instrument. 

\item The background rate ($B_{cut}$) which will be derived directly from 500 orbits ($\sim 1$ month of operation) of real data which provides enough statistics for the purpose of this study. Note that our definition of background in this analysis includes all events that are not associated with the GRB and therefore includes residual cosmic-rays as well as $\g$-rays originating from the galactic and extragalactic background as well as from point sources.

\end{itemize}

These quantities will be computed as a function of energy as different energy threshold migh apply for the different analyses we are interested in.
In the following section, we provide the results of this Figure-of-Merit based analysis.

\section{Comparison of S3 with other classes}

This analysis was initially developped under the Pass6 environment, meaning that the new event class S3 was designed to significantly improve the analysis of transient objects with respect to the Pass6\_V3 `transient' class. However this new event selection will only be included in the next Pass7 environment which will be the next official public release and involves substantial re-definition of the `standard' event classes including the `transient' class. We will therefore also provide a comparison of the differences in S3 characteristics between the Pass6 and Pass7 environments.

\begin{figure}
\begin{center} $
\begin{array}{cc}
\includegraphics[ width=.5\linewidth, keepaspectratio]{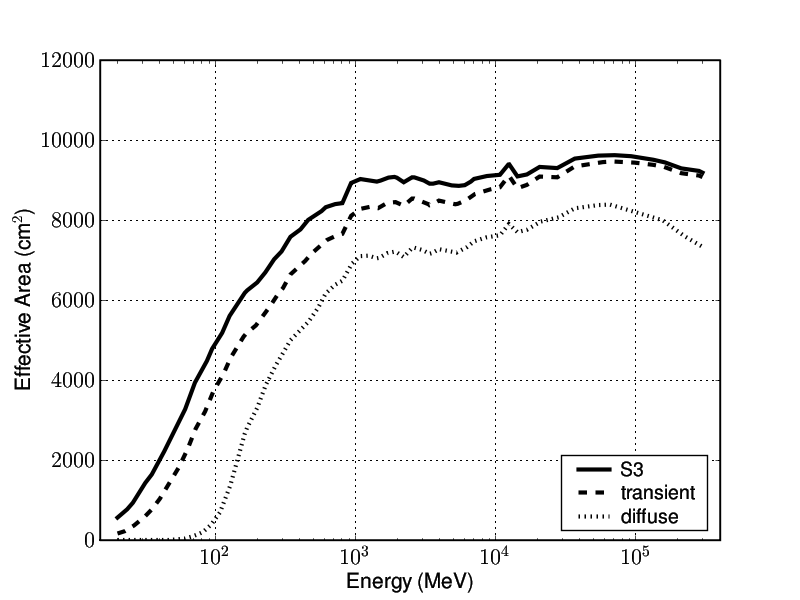}&
\includegraphics[ width=.5\linewidth, keepaspectratio]{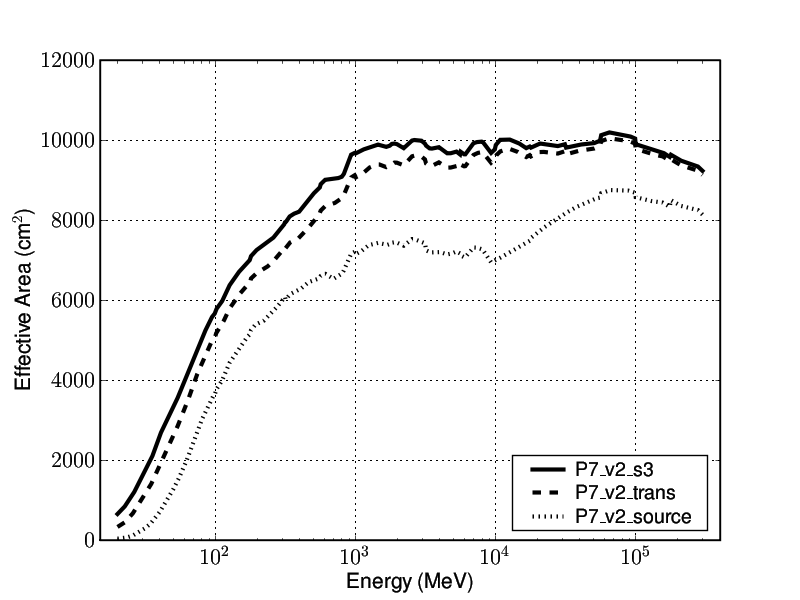}
\end{array} $
\end{center}
\caption{On-axis effective area of the Pass6 diffuse, transient and S3 class (left panel) and of the Pass7 source, transient and S3 class (right panel).}
\label{aeff_P6P7}
\end{figure}

The on-axis effective area of the Pass6 diffuse, transient and S3 cuts are shown in the left panel of figure \ref{aeff_P6P7}. The right panel shows the same event selections for Pass7 although the 'diffuse' class in Pass6 has by then been renamed as the 'source' class in Pass7 - the `source' class was initially intended for point source analysis while the `diffuse' class was more for diffuse emission analysis, but post-launch analyses with the Pass6\_V3 `source' class found the 'diffuse' class more suitable for point source analysis (the initial intent of the event class name will be restored with Pass7).

In Pass6\_V3 `S3' brings a significant improvement in effective area over the Pass6\_V3`transient' class.
Pure effective area increase is on the order of $\sim 25 \%$ at 100 MeV and $\sim 10\%$ at 1 GeV.
However this improvement in effective area is much reduced in Pass7 with $\sim 10 \%$ improvement at 100 MeV and $\sim 5\%$ at 1 GeV. This is due to the `transient' class that has been significantly relaxed in the Pass7 environment, therefore filling the gap that separated it from the S3 class in Pass6.

\begin{figure}
\begin{center}
\includegraphics[ width=.7\linewidth, keepaspectratio]{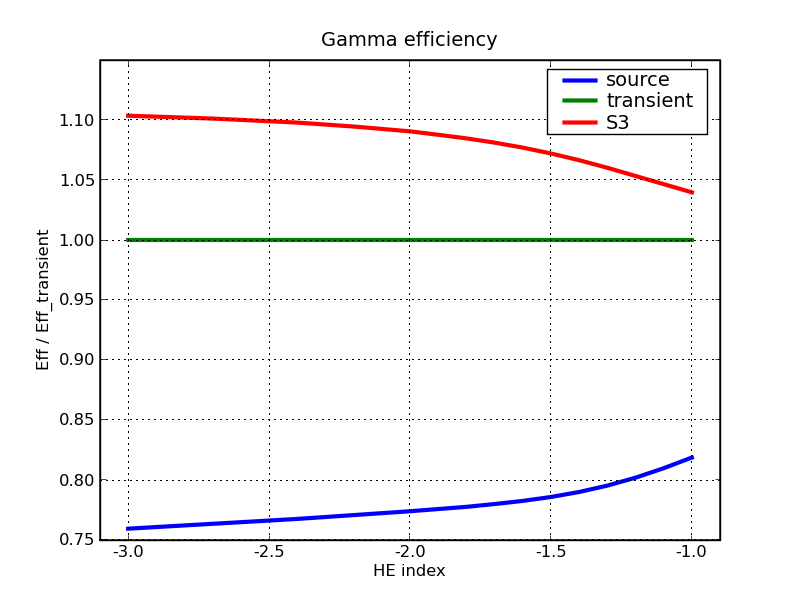}
\end{center}
\caption{Pass7 $\g$ efficiency above 100 MeV for different selections as a function of the high-energy index. Soft bursts benefit more from loose event selections at low energy while hard bursts benefit is mostly from the high energy effective area which shows smaller differences between the different classes thus the smaller difference in $\g$ efficiency.}
\label{gamma_eff}
\end{figure}

From the knowledge of the effective area, one can derive the $\g$ efficiency as a function of the high-energy index of the source in the LAT energy range. Figure \ref{gamma_eff} shows the Pass7 $\g$ efficiency above 100 MeV for the `source', `transient' and `S3' selection selections as a function of high-energy index.
As one could expect, soft bursts will benefit more from loose event selections at low energy while hard bursts benefit is mostly from the high energy effective area which shows smaller differences between the different classes thus the smaller difference in $\g$ efficiency.
%We note that the dependence of the $\g$ efficiency with  theta, the angle between the GRB and the LAT boresight, is small.

\begin{figure}
\begin{center} $
\begin{array}{cc}
\includegraphics[ width=.495\linewidth, keepaspectratio]{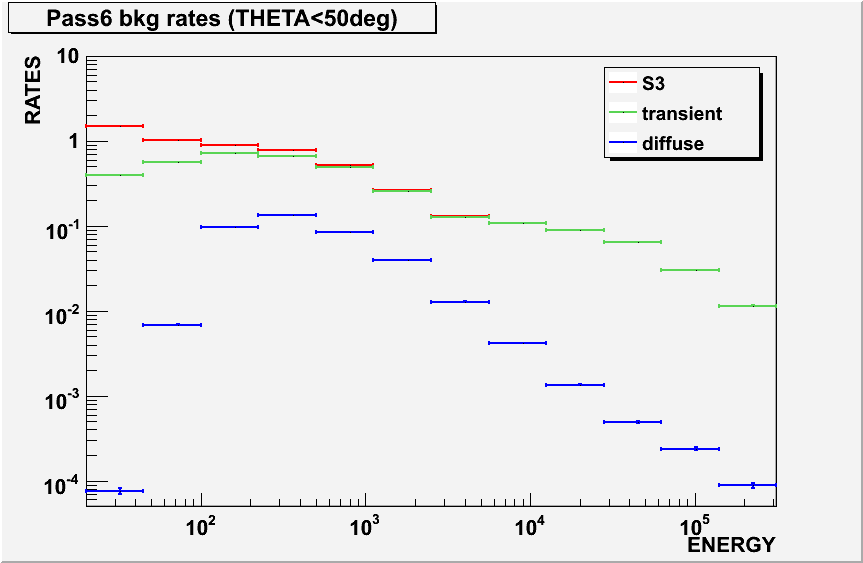}&
\includegraphics[ width=.475\linewidth, keepaspectratio]{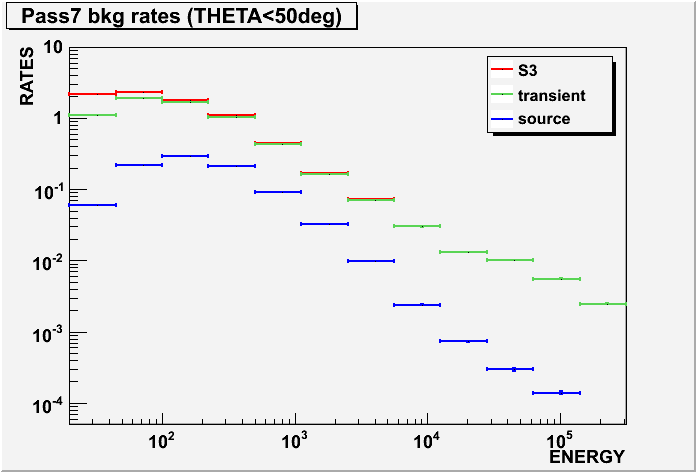}
\end{array} $
\end{center}
\caption{Background rates as a function of energy for Pass6 (left) and Pass7 (right). Note that despite the higher high energy effective area, the high energy background is much reduced in the transient/S3 class making Pass7 a lot cleaner in the high energy regime. Notice also that the transient and S3 event selections have very similar background rates above $\sim 100$ MeV, in particular for Pass7, so that the increase in effective area for the S3 class comes at a very little cost in terms of background contamination above 100 MeV}
\label{bkg_rates}
\end{figure}

The next thing we computed is background rates as a function of energy which is plotted in figure \ref{bkg_rates} for both Pass6 and Pass7. The increase of effective area in Pass7 has a visible cost in background rates up to a factor of two higher at energies less than 100 MeV. Note however that the high energy background is much reduced in the transient class although the effective area is higher making Pass7 a lot cleaner in the high energy regime.
One can also notice that the transient and S3 event selections have very similar background rates above $\sim 100$ MeV, in particular for Pass7, so that the increase in effective area for the S3 class comes at a very little cost in terms of background contamination above 100 MeV which makes this event class promising for spectral analysis using $> 100$ MeV events.

\begin{figure}
\begin{center} $
\begin{array}{cc}
\includegraphics[ width=.5\linewidth, keepaspectratio]{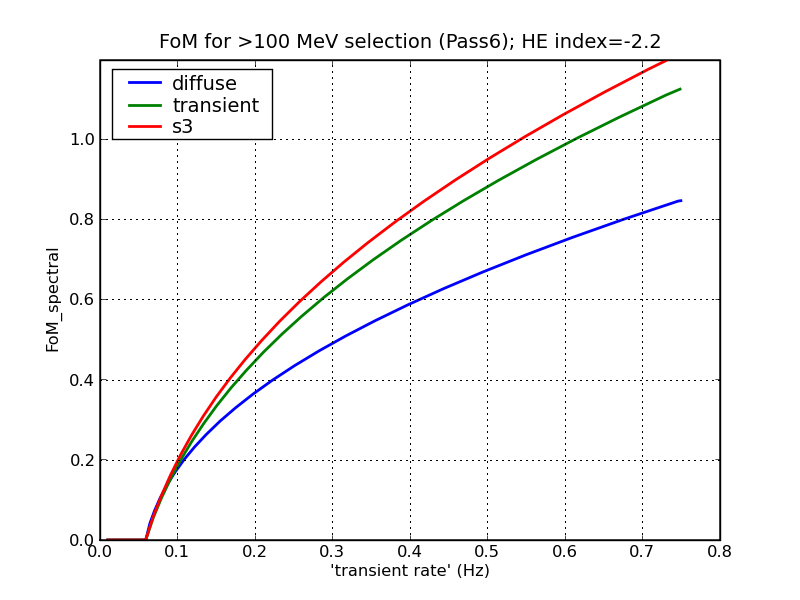}&
\includegraphics[ width=.5\linewidth, keepaspectratio]{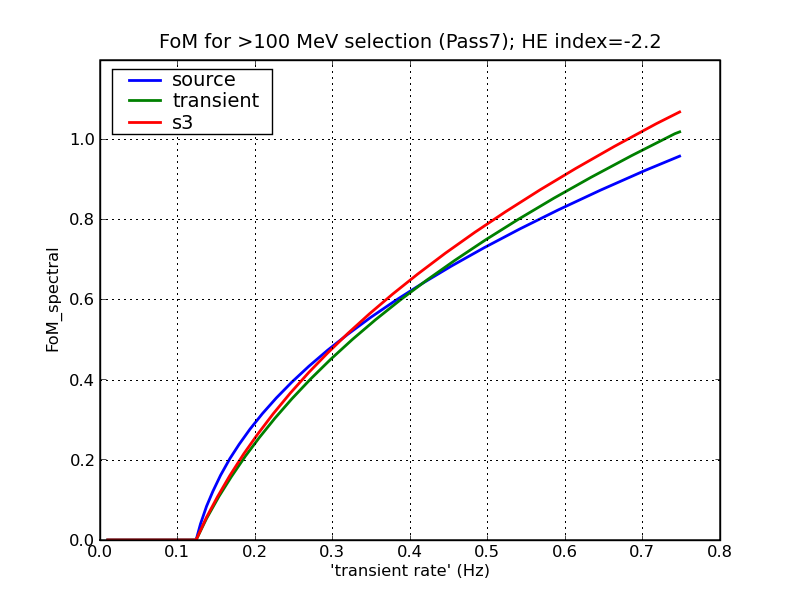}
\end{array} $
\end{center}
\caption{Relative values of the spectral Figure-of-Merit defined in section \ref{opt_evt_class} ({\it left}: Pass6; {\it right}: Pass7). A high energy index of -2.2 was assumed for this plot but the results remain similar for other values of high energy index. 
For Pass7, the source class is to be preferred for a source below a transient rate of $\sim 0.2$ Hz ($\pm 0.1$ when considering other high energy indices) while above this value the S3 class has the highest FoM for bright sources with an increase of $5-10 \%$ for Pass7 and $\sim 15-20\%$ for Pass6 (the transition is hard to see on the figure as it occurs at lower signal rate of $\sim 0.02$ Hz).
The S3 improvement was higher in the Pass6 framework compare to the transient class since this class has been much relaxed in the Pass7 framework. We note that the y-axis scale is arbitrary as only the relative value of $FoM_{spectral}$ matters to compare different cuts (the absolute value depends on the specific burst characteristics)..}
\label{FoM}
\end{figure}

Armed with this knowledge of the $\g$ efficiency and the background rates, it is now possible to compare the Figure-of-Merit for the different cuts. The relative values of the spectral FoM defined in section \ref{opt_evt_class} is shown in figure \ref{FoM} for an assumed high energy index of -2.2.
The cleanest event selections in the Pass6 (Pass6\_V3\_DIFFUSE) and Pass7 (P7\_v2\_source) environments are found to have the highest FoM for faint sources analyses as one could expect. Looser event classes are preferred for bright sources however, with the S3 selections bringing the most improvement in $FoM_{spectral}$. For Pass7, the transition regime between the purest class and S3 gives a background subtracted transient rate $>100$ MeV of $\sim 0.2$ Hz (note that this value is a direct observable a each source when the background contamination is well known). Above this value, the S3 class has the highest FoM for bright sources with an increase of $\sim 15-20\%$ for Pass6 (the transition is hard to see on the figure as it occurs at lower signal rate of $\sim 0.02$ Hz) and $5-10 \%$ for Pass7.
The results only changes slightly as a function of high energy index with the transition signal rate shifting from $\sim 0.1$ Hz to $\sim 0.3$ Hz as the index changes from -3.0 to -1.5. We also checked any dependence with the angle to the LAT boresight theta and that these results remain almost unaffected by a change in theta (which affects both the background rate and the $\g$ efficiency).
In the end, S3 is to be prefered over `transient' to perform spectral analysis above 100 MeV as long as the source flux is above the transition threshold which is the case for almost all LAT detected GRBs. We note however that the improvement brought by the S3 class over the `transient' class was reduced when moving to the Pass7 environment as the Pass7 `transient' class was much relaxed making the difference with S3 in terms of effective area and background rate small.
We also mention that a direct comparison of the Pass6 and Pass7 plot is not directly applicable for these plots as the transient background rate is different in each environment, requiring a shift along the x-axis to allow comparison.

As a final note, we mention that the Pass6 analysis led to the use of the 'S3' event class for the spectral analysis of GRB 080825C \cite{Abdo_080825C:09} as a clear improvement in FoM was found in the Pass6 environment. However, it was then decided to wait for the official public release of S3 in Pass7 (in 2011) in order to use it in further analysis.

%\begin{figure}
%\begin{center}
%\includegraphics[ width=.8\linewidth, keepaspectratio]{Plots/Event_optimization/FoM_detection_Pass7_s2s3_s3_trans_source_diff.png}
%\end{center}
%\caption{}
%\label{}
%\end{figure}

%\section{Optimization of the Region Of Interest}
%\label{ROI}
%
%Preparation of data, response and bkg data set based on this PSF optimization
%
%- optimization of ROI size
%- need for a proper spatial selection that takes into account energy dependent PSF
%
%Finally, the event
%selection makes use of the spatial information around the best LAT
%localization. The LAT point-spread-function (PSF) has a strong dependence with energy as well as with the conversion point in the tracker. LAT events are thus separated into FRONT (conversion in the upper part of the tracker) and BACK (conversion in the lower part of the tracker) events \cite{atw09} for which separate response functions are provided.
%The region of interest (ROI) 
%considered in this analysis is energy dependent and based on the $95\%$
%containment radius ($PSF95$) and the 95\% LAT error localization ($Err95$):  
%
%$$ ROI(E) = \sqrt{PSF95(E)^2 + Err95^2} $$
%
%To avoid large background contamination, a maximum size is set at 10 and 12 deg for FRONT and BACK events respectively.
%In the particular case of GRB 080825C: $ROI(E<200\mbox{ MeV})$ is set to this maximum size and $ROI(\sim 500 \mbox{ MeV}) = 2.9 \mbox{ and } 4.0$ degrees for FRONT and BACK events respectively.

      	\chapter{Detection procedure for a LAT GRB}
\label{detLATGRB}

\indent
The $Fermi$ satellite is monitoring a large fraction of the gamma-ray sky continuously\footnote{except during its passage in the South Atlantic Anomaly region which account for 17\% of the duty cycle.} making it an optimal instrument for observing transient sources which will by definition display significant emission only for short time periods, at unpredictable time and  region in the sky.
In this section, we will describe standard procedures developed by the LAT/GBM team in order to detect, localize and announce significant emission of a GRB in the LAT energy range.
Fermi LAT bursts are searched both on-board and on the ground through different methods that will be described below. For completeness, we will also give a brief description of the time-sequence for a GRB burst detection and announcement to the GCN community.

\section{GBM Burst Detection}\label{sec:GBM_Detection}

Onboard the $Fermi$ observatory the GBM
monitors count rates of the NaI detectors and look for a statistically
significant increase. The GBM triggers when two or more NaI detectors trigger.  The number of counts in an energy band $\Delta E$
over a time bin $\Delta t$ is compared to the expected number of background
counts in this $\Delta E$--$\Delta t$ bin; the background is
estimated from the rate in a 17 sec time window prior to the time bin being tested.  The GBM
trigger uses the twelve NaI detectors in the following energy band $\Delta E$=50--300~keV, and accumulation time from 16~ms to
4.096~s with various trigger threshold to avoid false triggers due to background fluctuations: 7.5 sigma for 16 and 32 ms, 5.0 sigma for 64 and 128 ms, 4.5 sigma for 256 ms to 4.096 s. 
Other energy bands where used on short time window in particular for Soft Gamma-ray Repeaters (20-50~keV) and Terrestrial Gamma-ray Flashes ($\geq 100$ keV and $\geq 300$ keV). Each time scale uses two separate search window offset by half the accumulation time in order to avoid splitting the burst counts in two separate search windows. This and the additional time scales used\footnote{BATSE trigger had one energy band---usually $\Delta E$=50--300~keV---and three time bins---$\Delta t=$0.064,
0.256, and 1.024~s.} led the GBM instrument to detect more bursts than initially predicted based on simple estimation based on BATSE detection rate (see \citep{Meegan:09} for more details).

When the GBM triggers it sends a series of burst alert
packets through the spacecraft and TDRSS to the Earth. Some
of these burst packets, including the burst location
calculated onboard, are also sent to the LAT unit to assist
in the LAT onboard GRB detection.  Burst localizations are
calculated by comparing the rates in the different
detectors since the effective area of each detectors is strongly dependent on the burst position in the FoV.
The continuous GBM data that are routinely telemetered to the ground
can also be searched for bursts that did not pass the GBM
onboard trigger criteria.  The GBM counting rate in
8 different energy channels with 0.256~s resolution and in 128~energy channels
with 4.096~s resolution.

Finally bright and hard GBM bursts (as well as onboad LAT detection) will result in automated repointing of the spacecraft in order to keep the GRB within the LAT FoV as practically feasible (the Earth albedo and the sun position imposes restrictions on the spacecraft pointing) for improved efficiency at detecting high-energy afterglow. The repointing is currently operational for the following 5 hours after the burst trigger time.

\section{Onboard GRB Detection in the LAT}\label{sec:LAT_Onboard}

The LAT flight software can detect bursts, localize them,
and report their positions to the ground through the burst
alert telemetry.  The rapid notification of ground-based
telescopes through GCN is intended for multi-wavelength
afterglow observations of high energy
GRBs.  The onboard burst trigger is described in greater details in \cite{Kuehn:07}.

The onboard processing for GRB detection in the LAT can be subdivided into three steps:  initial event
filtering; event track reconstruction; and finally burst
detection and localization. In the first step all
events---photons and charged particles---that trigger the
LAT hardware are filtered to remove events that are of no
further scientific interest (onboard gamma filter). The events that survive this
first filtering constitute the science data stream that is
downlinked to the ground for further processing. These
events are also fed into the second step of the onboard
burst processing pathway.

The second step is a crude reconstruction for all the events in the science data stream. The onboard reconstruction is less accurate, resulting in a larger PSF that obtained in the on-ground reconstruction and a poorer background rejection.

The rate of events that pass the onboard gamma filter is $\sim$400~Hz.  The event rate that events
are sent to the onboard burst trigger, which requires
3-dimensional tracks, is $\sim$120~Hz.  For comparison the on-ground
processing creates a transient event class with a rate of
$\sim$2~Hz (see section \ref{processing}).
This difference in background rate is the main difference between the on-board and on-ground sensitivity to GRB emission.

The third step in the burst processing is recognition as a burst,
which considers the events that have passed
of the first two steps, and thus have been assigned arrival times,
energies and positions in the sky. It will search for
temporal and spatial event clustering using an unbinned likelihood method.
The GRB alert trigger has two tiers.

The first Tier operates on a set of $N$ events (curent setting: $N=40-200$). All of these $N$ events are individually considered as the seed position for a potential GRB. In the current setting, all events within $\theta_0=17^{\circ}$ around the seed position are being considered as it is approximately the 68\%
containment radius of the onboard 3D tracks at low event energies.
A spatial and temporal probability are then calculated as follow:

\begin{equation}\label{eqn:spatialProb}
P_{S} =  \sum_{i=1}^{M}\, |\mathrm{log}_{10}(p_{s_i})| =
   \sum_{i=1}^{M} \Bigg{|} \mathrm{log}_{10}\bigg{(}\,\frac{1-\mathrm{cos}
   (\theta_{i})}{1-\mathrm{cos}(\theta_m)}\bigg{)}\Bigg{|}\; .
\end{equation}
\begin{equation}\label{eqn:temporalProb}
P_T = \sum_{i=1}^{M} | \mathrm{log}_{10}(p_{t_i})| = \sum_{i=1}^{M}
  \bigg{|} \mathrm{log}_{10}(1-e^{-r_t \Delta T_i})\, \bigg{|} \;.
\end{equation}

where $r_t$ is the rate at which events occur within the area of the cluster. 

And the trigger criterion is:
\begin{equation}
\xi  P_T  + P_S > \Theta
\label{eqn:combined_log_prob}
\end{equation}

where $\xi$ is an adjustable parameter that assigns
relative weights to the spatial and temporal clustering,
and $\Theta$ is the threshold (current setting: $\xi = 1$ and $\Theta = 150$).

The onboard burst localization algorithm uses a weighted
average of the positions of the cluster's events where the
weighting is the inverse of the angular distance of an
event from the burst position. Since the purpose of the
algorithm is to find the burst position, the averaging must
be iterated, with the weighting used in one step calculated
from the position from the previous step.  The initial
location is the unweighted average of the events positions.
The convergence criterion is a change of 1~arcmin between
iterations (with a maximum of 10 iterations).  The position
uncertainty depends on the number and energies of events,
but the goal is an uncertainty less that 1$^\circ$. Using
Monte Carlo simulations, this methodology was found to be
superior to others that were tried.

Tier 2 is very similar to tier 1 in principle although it will run with slightly different parameter:  tier 1 localization is the only seed position used, $N=500$, $\theta_0=10^{\circ}$, $\xi=1$, $\Theta=150$. If this value of $\Theta$ is exceeded, a tier 2 trigger results and the cluster events
are run through the localization algorithm. The resulting
trigger time, burst location and number of counts in four
energy bands are then sent to the ground through the burst
alert telemetry.

The parameters used by the onboard burst detection and
localization software are affected by the actual event
rates, and are bound to change as we learn more about onboard event rates. Currently the thresholds are set to keep the false trigger rate at an acceptable level. The falso positive level being set at $\lesssim 0.3$ GRB/year which would have (this configuration was only recently implemented) onboard triggered 5 of the 14 GRBs detected during the first year and a half of $Fermi$ operation.

\section{On-ground Automated Science processing}

A burst detection algorithm is applied on the ground
to all LAT counts after the events are reconstructed and
classified to detect bursts that were not detected by the
onboard algorithm. The
ground-based search is performed after each satellite
downlink and subsequent data processing. The ground-based blind search algorithm
is very similar to the onboard algorithm described in the
previous section, but will benefit from the full
ground-based event reconstruction and background rejection
techniques that are applied to produce the LAT counts used
for astrophysical analysis. Therefore for these data, the particle
background rates is lower than the onboard rates by at
least two orders-of-magnitude. Furthermore, the
reconstructed photon directions and energies is more
accurate than the onboard quantities.

The first stage of the
ground-based algorithm is applied to consecutive sets of 20
to 100 events. 
Based on reference distribution for the null-case derived from in-flight data, the burst detection threshold was set at 5$\sigma$.
If a candidate burst is found in the
ground-based analysis, counts from a time range bracketing
the trigger time undergoes further processing to determine
the significance of the burst.  If the burst is
sufficiently significant, it is localized and its spectrum
is analyzed.  These analyses use the unbinned maximum
likelihood method that is applied to LAT point sources.
The Automatic Science Processing software has blindly detected 6 GRBs up to January 2010: GRB 080916C, GRB 081024B, GRB 090217, GRB 090510, GRB 090902B and GRB 090926A.

\section{Human-in-the-loop analysis}

For every bursts detected by the GBM, the onboard or onground algorithm, or any other instruments (e.g. {\it Swift}), a Burst Advocate (BA) will look for GRB signal in the LAT. A LAT BA is on duty at all time in order to permorm a prompt analysis of the burst candidate and possibly send a circular to the Gamma-Ray bursts Coordinates Network (GCN). This procedure involving a human interaction is the most robust approach to convincingly announce the detection of a GRB by the LAT instrument because of the very unique observing conditions of each burst, such as if the burst is detected after a repointing of the instrument (which makes the background determination a challenge). We here describe the standart procedure for the Burst Advocate developped by the LAT team.

While the LAT BA is on shift, he/she is responsible to look for any GRB alert coming through a GCN notice or circular. If such alert is emitted, he/she will be in charge of analysing the LAT data and estimating the likelihood of a possible emission in the LAT energy range. Because the data processing has a 6 to 10 hours latency and it takes some time to perform the human analysis and make sure it is reliable, the detection of a LAT burst via a human-in-the-loop interaction is usually not announced less than $\sim 10$ hours after a burst trigger.

\begin{figure}
\begin{center}
\includegraphics[ width=.8\linewidth, keepaspectratio]{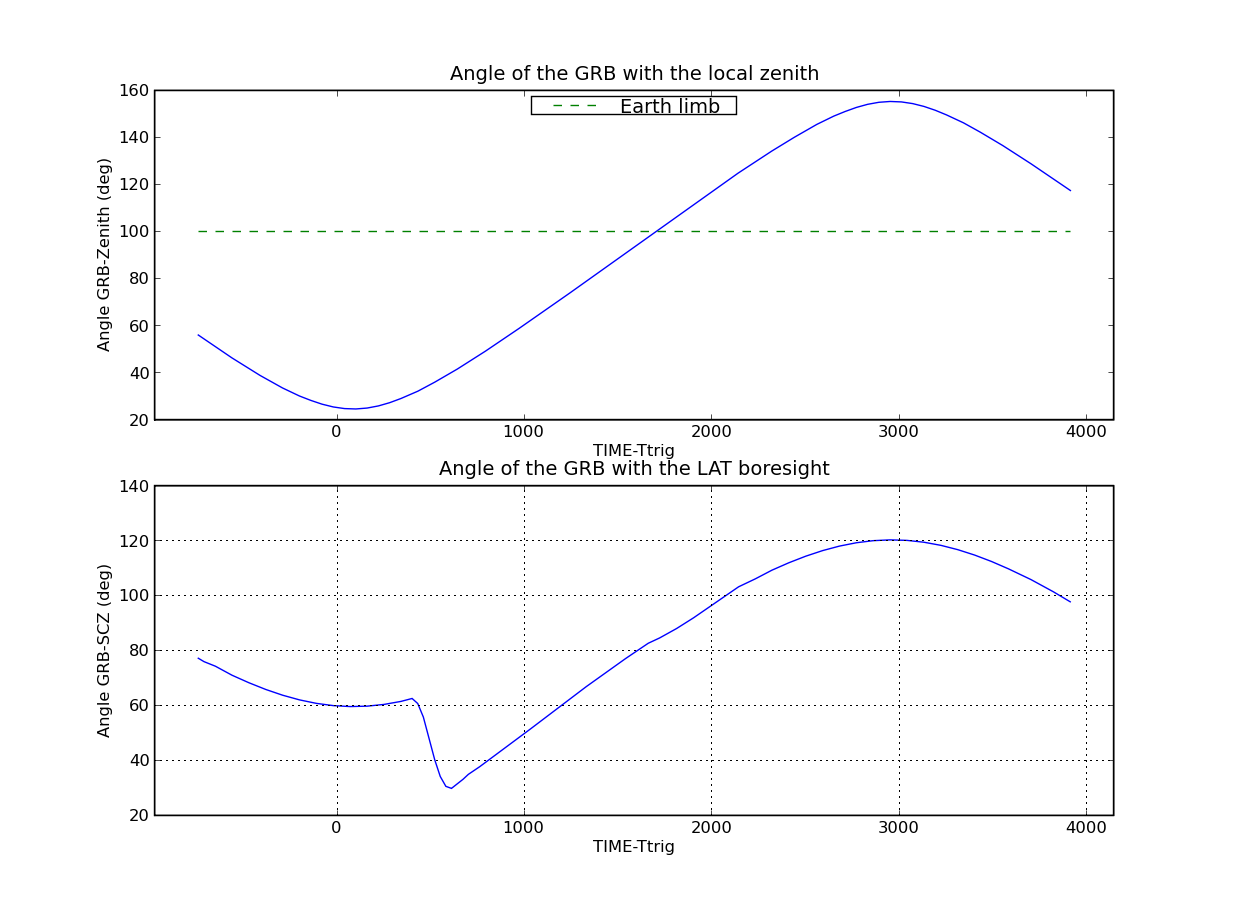}
\end{center}
\caption{Example of angles of the GBM best localization with the local earth zenith (top) and with the LAT boresight (bottom). These plots help determine when a burst location is above the earth horizon, not too contaminated by earth albedo, and within the LAT FoV after the burst trigger.}
\label{Angle plots}
\end{figure}

Since the GBM and the LAT FoV perfectly overlap by design, most of the burst alerts with a localization that falls within the LAT FoV are generated by the GBM instrument.
A first crucial information to derive for a particular GBM or {\it Swift} burst is the temporal variation of the LAT exposure at the burst location from the time it triggered. Figure \ref{Angle plots} is an example of how the angle of the GBM best localization with respect to the local earth zenith (top) and the LAT boresight (bottom) vary as a function of time.

At $\theta \sim 60^{\circ}$ the LAT effective area is still $\sim \frac{1}{3}$ of that for $\theta \sim 0^{\circ}$ while it drops to $\sim \frac{1}{10}$ of the incident angle value for $\theta \sim 70^{\circ}$ (see fig. \ref{irf diffuse}). Therefore bursts  with $\theta \leq 70^{\circ}$ have non-negligeable exposure during their prompt emission and are interesting candidates to look at. Bursts with high angle to the LAT boresight during the prompt emission have usually very low chance to display any emission in the LAT using a standart analysis. For example, the LAT team announced the detection of GRB 081215, a bursts with $\theta \sim 86^{\circ}$ which was detected using a non-standart even selection \cite{McEnery:08}.

\begin{figure}
\begin{center} $
\begin{array}{cc}
\includegraphics[ width=.48\linewidth, keepaspectratio]{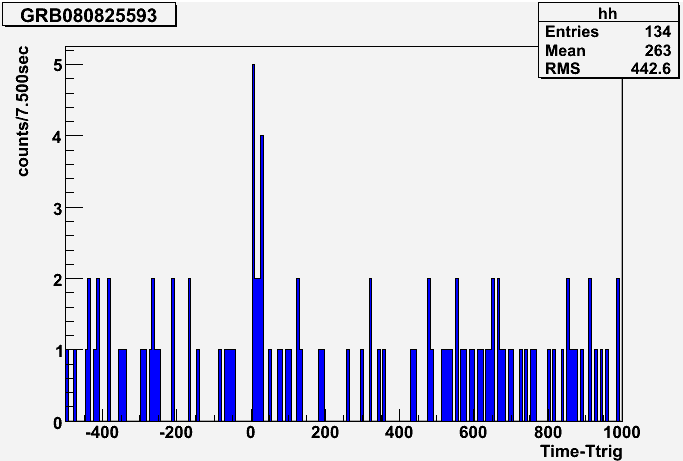} &
\includegraphics[ width=.48\linewidth, keepaspectratio]{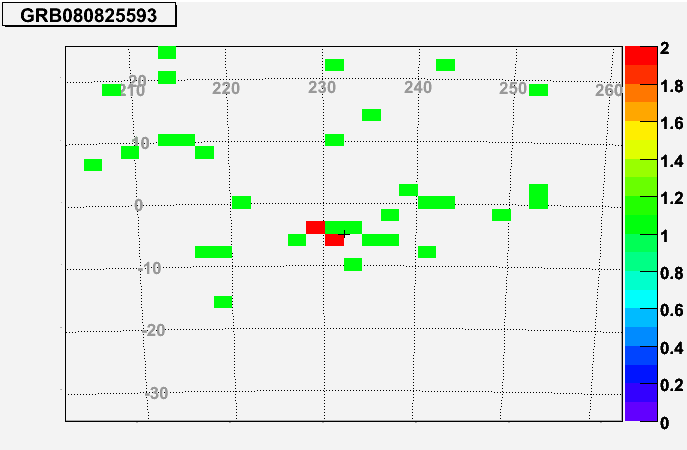}
\end{array} $
\end{center}
\caption{Quick look at GRB 080825C lightcurve (left panel) and count map (right panel) around the GBM localization (black cross) and trigger time. The presence of an increase count rate and a clustering of events around the GBM localization is visible by eye although a detailed analysis is necessary in order to claim for a significant detection of this burst by the LAT.}
\label{LC and cmap}
\end{figure}

Because the background rate is quite small for ROI less than $15^{\circ}$ (typically less than 0.1-0.5 Hz), it is possible for the BA to detect a possible significant LAT emission by simply performing a 'by-eye' analysis and looking at temporal and spatial distribution of events detected around the burst trigger and localization. Fig. \ref{LC and cmap} shows an example of such lightcurve and count map for GRB 080825C which was the first LAT GRB detected with a significance $>5 \sigma$. The increased count rate and a clustering of events is visible in the plots but an estimation of the exact significance of the detection requires very careful analysis that will be presented in the next section.

\section{Estimation of background and significance of detection}
\label{significance}

An accurate estimation of the background level at the time of the GRB trigger and around the location of the burst is crucial to estimate significance of the detection. 
One approach is the so-called 'off-source' technique which determines background from a surrounding spatial region or time window with respect to the region where the signal is detected (so-called 'on-source' region). However the rate of transient events is a strong function of the satellite position in the Earth coordinate (strong correlation with geomagnetic latitude), the angle of the GRB location to the LAT boresight, the position of the burst in the sky. 
A strong limitation for the 'off-source' technique is the faithfulness of the 'off-source' region to represent the background in the 'on-source' region. This is usually fairly easy for high counting rate instrument which have smoothly varying background over time and/or space.
However, the LAT background level changes rapidly and the event rate is very low which prevents us to estimate the background level during the 'on-source' interval reliably (see below for an exception with the case of GRB 080825C though).

To circumvent the difficulty in background estimation, we adopt another method which relies on past data at the orbit location to estimate the background level in the 'on-source' region. The basic idea is to bin the data into specific observing environment (geomagnetic latitude, angle to the LAT boresight, Earth zenith angle...) and accumulate enough statistics for any possible conditions a GRB might be observed in. We refer the interested reader to Appendix \ref{bkg_est} where the details of the method we use is provided.
Once the background spectrum is accurately predicted, one can compute the significance for any temporal increase or spatial clustering of events. In order to assess the significance of a GRB observation, 
we use four independent statistical methods which are described below and which we compare in the end.

The first method computes the probability of the null hypothesis being
true (the probability that the observed number of counts in the on-source
region is due to a background fluctuation) in a frequentist approach
where the background uncertainty is treated in a semi-Bayesian way
\cite{Conrad:03}. Given a certain estimated number of background counts B
during the on-source interval, we compute the probability of the
actual on-source measurement $N_{\rm on}$ being consistent with this
value. $P_{sup}(N_{\rm on},B)$ expresses the probability of the on-source
measurement being equal or larger than $N_{\rm on}$ when only statistical
fluctuations are considered:

\begin{equation}
P_{sup}(N_{\rm on},B) = \sum_{N=N_{\rm on}}^{\infty}  \frac{e^{-B} \times B^N}{N!}.
\end{equation}

Because of systematic uncertainties in the background estimation, each possible value for the number of background counts is
weighted by a Gaussian distribution with a mean of $B_{\rm est}$ and
a standard deviation of $0.15\times B_{\rm est}$ (since we estimate
our systematics to be about 15\%, see Appendix \ref{bkg_est}): ${\rm Gauss}(B)$. We then integrate
over all the possible values of B to compute a weighted probability:

\begin{equation}
P_{\rm null}=
\frac {\int_0^{\infty} dB \mbox{ } {\rm Gauss}(B) \times P_{\rm sup}(N_{\rm on},B)} 
{\int_0^{\infty} dB \mbox{ } {\rm Gauss}(B)}
\end{equation}

The second method is fully Bayesian. A GRB detection by the LAT is analyzed as an on-source/off-source
observation in the time domain. This method requires a reliable 'off-source' data. This was possible only for GRB 080825C since the inclination angle 
of the best source position with respect to the instrument boresight moved slowly during a $>500$ sec time interval before the trigger time.
The spacecraft motion was
favorable for GRB~080825C, allowing an unusually long off-source
interval to determine the background rate.  For other GRBs,
spacecraft motion causes the background in the LAT to vary more
quickly, limiting the time for which an off-source interval measures
the same background as that of the on-source interval, and therefore
limiting the applicability of this method.

The Bayesian method assumes that the counts during the background
interval are due to a Poissonian process with rate $b$. To evaluate the
probability that a source has been detected during the on-source
interval, the method compares two hypotheses for the Poisson rate
during the on-source interval: that observed counts are due to the
same background rate $b$, or that they are due to a background plus
source rate $b+s$.  This second hypothesis is insufficiently specified
to quantitatively solve the problem: one must specify some plausible
range for the source rate $s$ using prior data (i.e., not using the
LAT observations). We produce a `reasonable' estimate for the LAT source counts by extrapolating the time-integrated spectral 
fit of the GBM data to the LAT energy range.  The photon model from the GBM fit is
convolved with the LAT response function to predict the number of LAT
counts.  A real LAT observation could have a smaller value than this
estimate because of a statistical fluctuation, or because of a
spectral break between the GBM and LAT energy range. Alternatively, a real
LAT observation could actually exceed this estimate for the maximum
rate if there were an additional and distinct spectral component to the one found 
with the GBM data alone. Nevertheless, this is a
reasonable `prior' estimate for the maximum counts expected in the LAT.  Under the
assumptions described, both \cite{Loredo:92} and \cite{Gregory:05} give
analytic solutions for the probability $P$ that the source is
detected.

A third method computes the significance with a fully frequentist
method using the on-source/off-source approach as described by
\cite[equation (17) of][]{Li:83}. And the last method uses an unbinned likelihood analysis of the LAT data which takes into account the energy-dependent PSF in event-by-event basis. In the case of GRBs where high energy events (having the smallest point-spread-function) have been detected, this method is expected to provide the highest significance since it fully takes into account the spatial information for each event.

It should be noted that because such search for LAT excess is performed on all GRBs triggered by the GBM and other instruments (when the burst is in the LAT FoV), it is important to consider multi-trials in our analysis. For independent searches as is the case here, the post-trials probability threshold for obtaining a $5 \sigma$ result is $P_{post-trial} = 1- (1- P_{5 \sigma})^{1/N}$ where N is the number of trials and $P_{5 \sigma}$ the $5 \sigma$ probability threshold for a single search ($\sim 5.7 \times 10^{-7}$). We naturally applied this correction for all the LAT search we performed.

\section{LAT GCN circular}

Once the LAT detects a GRB and its localization deemed reliable, the BA with help from other members of the Fermi GRB group writes a GCN circular containing relevant information for follow-up observations. The circular usually contains the following information on the burst:

\begin{itemize}
\item Time window where a significant emission in the LAT is observed.
\item Significance of the detection
\item Best LAT localization with its error (crucial for afterglow follow-up)
\item Any additional information that might be relevant to prompt for follow-up observations (unusual high intensity, highest energies observed, high energy afterglow...)
\item Point of contact for the burst (usually the BA)
\end{itemize}

For the detailed procedure used for localization and high energy afterglow, we refer the interested reader to \cite{Abdo_080825C:09, Pelassa:10}.

It is the prime responsibility of a BA to deliver this LAT circular to the GRB community in the shortest time possible. However, mostly due to latency of satellite downlink , data processing and the necessary time to perform a decent LAT analysis, the emission of such circular is usually delayed by at least 10 hours after the initial burst alert. Here is a brief outline of the sequence of events leading (potentially) to the announcement of a LAT GRB detection through the GCN network\footnote{timing sequence only intends to show the average trend but might actually not be representative for certain specific bursts.}:

\begin{itemize}
\item {\bf $\mathbf{T_0}$:} GBM (or {\it Swift}) GCN alert (with possible Automatic Repoint Request of the spacecraft)
\begin{itemize}
\item {\bf $\mathbf{T_0 + 1-3} \mbox{hours}$:} GBM human-in-the-loop localization provided to LAT BA (few degrees accuracy typically)
\end{itemize}
\item {\bf $\mathbf{T_0 + 6-10} \mbox{hours}$:} LAT data available
\begin{itemize}
\item {\bf $\mathbf{\sim 30-60} \mbox{minutes later}$:} Automatic Science Processing + human-in-the-loop LAT localization (accuracy depends on burst brightness and hardness: arcmin to degree)
\item {\bf $\mathbf{\sim 1-2} \mbox{hours later}$:} {\it Swift} Target-of-Opportunity (ToO) request (if accuracy $\lesssim 10$ arcmin)
\end{itemize}
\item {\bf $\mathbf{T_0 + 10-15} \mbox{hours}$:} First LAT GCN circular
\item {\bf $\mathbf{T_0 + 24-48} \mbox{hours}$:} Second LAT GCN circular
\end{itemize}

Finally, we mention that we also published LAT GCN circulars for bright hard GBM bursts which triggered an autonomous repointing of the spacecraft and which were not detected by the LAT instrument (e.g. GRB 091010, GRB 091024). For those we provided upper limits in various time intervals.

       	\chapter{High energy prompt emission of GRBs detected by the LAT}
\label{HEprompt}

The $Fermi$ LAT unique sensitivity and low deadtime is opening a whole new era to explore the high energy emission of Gamma-Ray Bursts. In this section, we will focus more particularly on the high energy prompt emission which in the pre-$Fermi$ era is fairly unknown territory on the observational side with only few low statistics detection by the EGRET and AGILE instruments.
For a couple of bright bursts, more than hundred photons have been detected by the LAT which allows very precise investigation of the temporal and spectral characteristics of the prompt emission of these bursts over a wide energy range, typically from tens of keVs up to tens of GeVs. 
The LAT also detects with somewhat lower significance bursts which were until known unaccessible to any high energy instrument. A rate of $\sim 10$ GRBs/year detected by the LAT is therefore steadily increasing the population of GRBs detected at high energy allowing us to spot repeating features in various bursts which is key to eventually derive a common theoretical framework for the interpretation of those objects.
We provide in this section a detailed description of the key features of the prompt emission of LAT detected GRBs up to January 2010. Section \ref{overviewLATGRBs} gives an overview of general characteristics of LAT detected GRBs as well as a discussion of LAT detection rates. Section \ref{features} describes one of the main features observed in LAT GRBs: a delay of the high-energy emission with respect to the sub-MeV emission. Section \ref{spec} provides a detailed description of the time-resolved spectroscopy performed on the prompt emission of LAT detected bursts that have sufficient statistics for such analysis. Finally, section \ref{discussion} briefly discusses the possible theoretical interpretation of the various spectro-temporal features observed on LAT GRBs (sub-section \ref{interpretation}) as well as the type of constraints $Fermi$ data allow us to put on the bulk Lorentz factor (sub-section \ref{bulk_lorentz_factor}).

\section{Overview of LAT detected GRBs}
\label{overviewLATGRBs}

After one year and a half of operation (up to January $22^{nd}$ 2010), GBM detected 375 bursts (detection rate of $\sim 0.7 $ GRB/day) among which 193 where in the LAT field-of-view - which is arbitrarily defined as $\Theta<75^{\circ}$ where $\Theta$ is the angle between the GRB best localization and the LAT boresight. In total, 14 of those GRBs were detected by the LAT instrument with a significance of more than $5 \sigma$ (detection rate of$\sim 0.75$ GRB/month). All those LAT detection were reported to the scientific community via one or more GCN circular.
Table \ref{LATGRBtable} provides general information on these LAT detected GRBs.

\begin{table}[htbp]
%   \centering
%   \begin{tabular}{|p{3cm}|p{2cm}|p{2cm}|p{2cm}|p{2cm}|} % Column formatting, @{} suppresses leading/trailing space
   \begin{tabular}{|p{1.5cm}|p{1.2cm}|p{1cm}|p{1.3cm}|p{1.2cm}|p{1.0cm}|p{1.cm}|p{0.8cm}|p{0.9cm}|} % Column formatting, @{} suppresses leading/trailing space
      \hline
      	GRB    & Sig. & $\Theta$  & $\Delta T$ (sec) [type] & Events $>100$ MeV & Events $>1$ GeV & $E_{max}$ (GeV)& z& GCN circulars \\
      	\hline
      	080825C 		& $\gtrsim 6\sigma$ & $60^{\circ}$ & $\sim35$ [long] & $\sim 10$ & 0 & $\sim 0.6$ &  - & \cite{Bouvier:08}\\
      	\hline
	080916C 		& $\gtrsim 35\sigma$ & $49^{\circ}$ & $\sim80$ [long] & 145 & 13 & $\sim 14$ & 4.3 & \cite{Tajima:08} \\
      	\hline
	081024B 		& $\gtrsim 10\sigma$ & $19^{\circ}$ & $\sim1$ [short] & $\sim 10$ & 2 & $\sim 3$ & - & \cite{Omodei:08}\\
      	\hline
	081215A 		& $\gtrsim 8\sigma^{\dagger}$ & $86^{\circ}$ & ** [long] & - & - & - & - & \cite{McEnery:08} \\
      	\hline
	090217 		& $\gtrsim 8\sigma$ & $34^{\circ}$ & $\sim30$ [long] & $\sim 10$ & 1 &	 $\sim 1$ & - & \cite{Ohno:09}\\
      	\hline
	090323 		& $\gtrsim 5\sigma^{\dagger\dagger}$ & $57^{\circ}$ * & ** [long] & $\sim 20$ & $\gtrsim 1$	& $\sim 7$ & 3.6 & \cite{Ohno2:09}\\
      	\hline
	090328 		& $\gtrsim 8\sigma^{\dagger\dagger}$ & $65^{\circ}$ * & ** [long] & $\sim 20$ & $\gtrsim 1$ & $\sim 24$ & 0.7 &  \cite{McEnery2:09} \cite{Cutini:09}\\
      	\hline
	090510 		&  $\gtrsim 45\sigma$ & $14^{\circ}$ * & $\sim 2$ [short] & $\sim 150$ & $\sim 20$ & $\sim 31$ & 0.9 & \cite{Ohno3:09} \cite{Omodei2:09}\\
      	\hline
	090626 		&  $\gtrsim 8\sigma$ & $18^{\circ}$ * & ** [long] & $\sim 20$ & 1 & $\sim 2$ & - & \cite{Piron:09} \\
      	\hline
	090902B 		&  $\gtrsim 40\sigma$ & $51^{\circ}$ * & $\sim20$ [long] & $\sim 200$  & $\sim 30$ & $\sim 33$ & 1.8 & \cite{dePalma:09} \cite{dePalma2:09}\\
      	\hline
	090926A 		&  $\gtrsim 40\sigma$ & $48^{\circ}$ * & $\sim20$ [long] & $\sim 150$ & $\sim 50$ & $\sim 20$ &  2.1 & \cite{Uehara:09} \cite{Bissaldi:09}\\
      	\hline
	091003A 		&  $\gtrsim 10\sigma$ & $12^{\circ}$ * & $\sim21$ [long] & $\sim 20$ & 2 & $\sim 3$ & 0.9 & \cite{McEnery3:09}\\
      	\hline
	091031A 		&  $\gtrsim 6\sigma$& $22^{\circ}$ & $\sim35$ [long] & $\sim 20$ & 0 & $\sim 0.7$ & - & \cite{dePalma:09}\\
      	\hline
	100116A 		&  $\gtrsim 6\sigma$ & $29^{\circ}$ & $\sim35$ [long] & $\sim 10$ & 3 &  $\sim 2$ & - & \cite{McEnery4:10} \\	
      	\hline
      \end{tabular}
   \caption{List of GRBs detected by the LAT up to January 2010 along with some general information. The Test-Statistic (TS) is computed via a likelihood fit for a point source ($\dagger$: based on specific analysis requiring non-standard data selection; $\dagger\dagger$: emission is weak for these bursts so the significance was based on an extended emission analysis carefully taking into account background level). $\Theta$ is the angle between the GRB best localization and the LAT boresight at the time of the GBM trigger (* specifies when an ARR has been triggered). $\Delta T$ is an approximate estimate of the duration of the high energy prompt emission (this value is only indicative as the transition between prompt emission and extended emission is extremely subtle to define; **: the signal is too weak to give a rough estimate of the prompt emission). Boxes with no entry means that the information is unknown or extremely difficult to estimate.}
   \label{LATGRBtable}
\end{table}

\begin{figure}
\begin{center}
\includegraphics[ width=.9\linewidth, keepaspectratio]{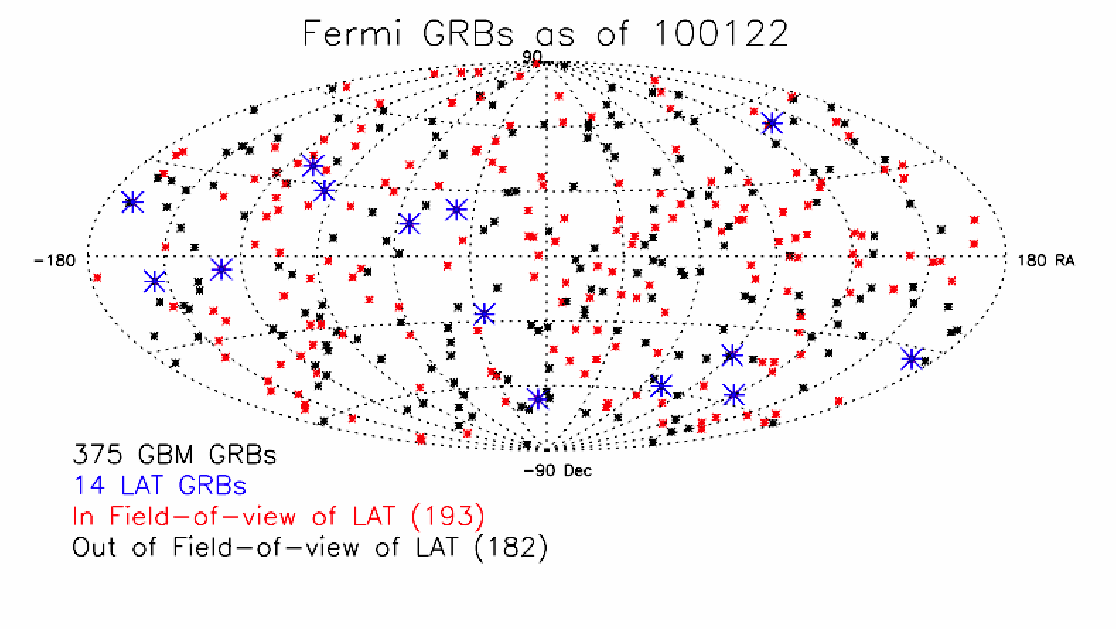}
\end{center}
\caption{Position of GBM and LAT detected GRBs on the celestial sphere. Up to the January $22^{nd}$ 2010, GBM detected 375 GRBs among which 14 were detected by the LAT.  No signs of anisotropy or correlation with the galactic plane can be observed given the level of statistics.}
\label{skymap}
\end{figure}

\begin{figure}
\begin{center}
\includegraphics[ width=.9\linewidth, keepaspectratio]{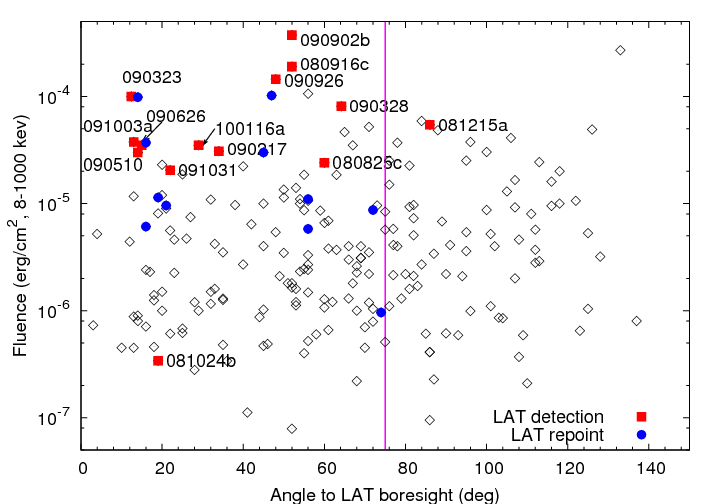}
\end{center}
\caption{Position of GBM and LAT detected GRBs in the instrument coordinates; more precisely with respect to the angle between the GRB best localization and the LAT boresight since the azimuthal angle has fairly low effect on the instrument response.}
\label{fluence_theta}
\end{figure}

Blue stars in figures \ref{skymap} show the position of the LAT detected bursts in the celestial sky. Similarly to the GRBs without any high energy emission detected, the LAT GRBs do not show any signs of anisotropy or correlation with the galactic plane given the level of statistics. Besides their cosmological origin has been verified through redshift measurement or a few LAT bursts.
Figure \ref{fluence_theta} reprents the GBM and LAT detected bursts in the spacecraft coordinate or more precisely to the angle between the GRB best localization and the LAT boresight. All bursts (except GRB 081215A which was quite unusual) are detected  with incidence angle less than $60^{\circ}$ below which the LAT effective area drops rapidely from one-third the on-axis value to zero (see figure \ref{irf diffuse}).
Fig. \ref{fluence_theta} show that the LAT is mainly sensitive to bright GBM bursts although it does not take into account the hardness of a burst or his temporal profile. Hard bursts will of course be easier to detect in the LAT energy range. It was found that the LAT detections do not correlate perfectely with the GBM fluence as GRB 081024B show in figure \ref{fluence_theta}. The hardness of this bursts allowed for its detection compare to other brighter but softer bursts.

\begin{figure}
\begin{center}
\includegraphics[angle=0, width=.9\linewidth, keepaspectratio]{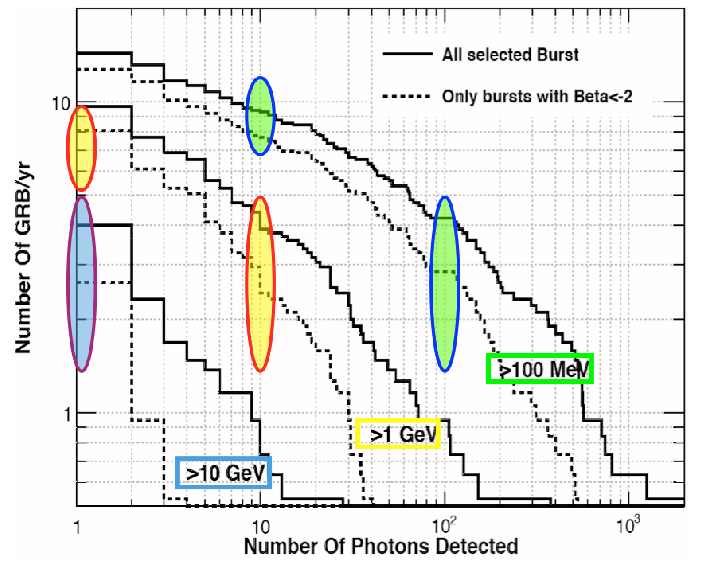}
\end{center}
\caption{LAT GRB detection rates (color ellipses) superimposed on top of pre-launch
expected rates based on the extrapolation of a Band spectrum fit from the BATSE
energy range \cite{Band:09}. The ellipsesÕ inner color indicates the minimal photon energy
(green, yellow and cyan correspond to 0.1, 1 and 10 GeV, respectively), while their
height indicates the uncertainty ($\pm \frac{\sqrt{N}}{\mbox{1.5 year}}$) on the corresponding LAT detection rate ($\frac{N}{\mbox{1.5 year}}$) due to the small number N of detected GRBs.}
\label{GRBrate}
\end{figure}

An interesting question to ask ourselves is whether or not the actual detection rates observed on orbits match with pre-launch expectations \cite{Band:09}. The study performed in \cite{Band:09} was based on a sample of bright BATSE GRBs \citep{Kaneko:06} for which the fit to a Band spectrum over the BATSE energy range [20 keV - 2 MeV] was extrapolated to the LAT energy range without assuming any unusual behavior such as spectral cutoff or rising additional component. Fig. \ref{GRBrate} shows the number of expected bursts per year as a function of the number of photons per burst detected by the LAT. The different lines refer to different energy threshold (100 MeV, 1 GeV, 10 GeV). Dashed lines are a similar extrapolation but only using bursts with high energy index $\beta$ smaller than -2 to exclude cases with a rising $\nu F\nu$ spectrum at high energy which is either indicative of an inaccurate $\beta$  determination or requires a spectral curvature at high energy.

All 14 LAT GRBs had $>10$ photons above 100 MeV which corresponds to a rate of $\sim 9.3$ GRB/yr. 4 bursts where particularly bright
in the LAT with $>1$ photon above 10 GeV, $>10$ photons above 1 GeV,
and $>100$ photons above 100 MeV. This corresponds to a bright LAT GRB
 detection rate of $\sim 2.7$ GRB/yr (with a rather large uncertainty
due to the small number statistics). There were also 11 GRBs
with $>1$ photon above 1 GeV, corresponding to $\sim 7.3$ GRB/yr.
Ellipses in fig. \ref{GRBrate} report those different rates with the vertical size being representative of the uncertainty in this number due to the low statistics of detected GRBs. The agreement found with the expected numbers suggests that, on average, there is no significant and systematic excess or deficit of high-energy emission in the LAT energy range relative to such an extrapolation from lower energies. We note that it is still possible for a limited fraction of bursts to display such significant high energy excess and section \ref{spec} will address cases where this behavior was observed. It is also possible for the majority of the bursts to display a slight excess above low energy extrapolation but this is currently not detectable at the population level given the low statistics of LAT detected bursts.

\begin{figure}
\begin{center}
\includegraphics[angle=0, width=.75\linewidth, keepaspectratio]{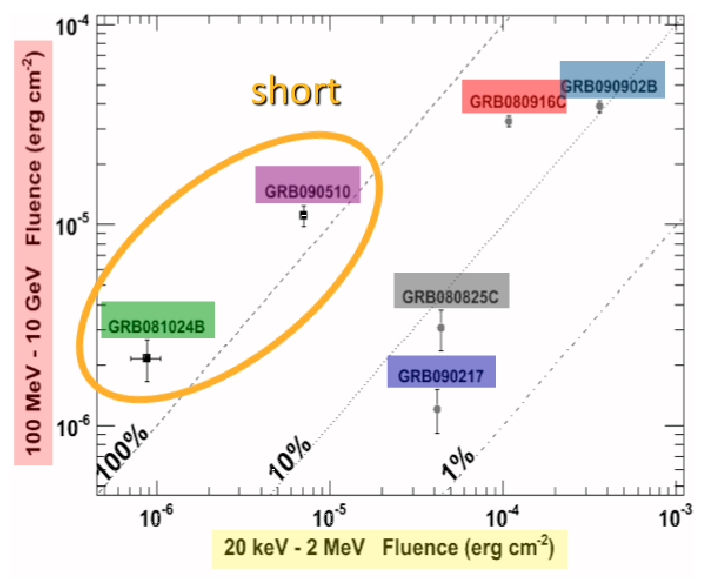}
\end{center}
\caption{The fluence at high (0.1 Ð 10 GeV) versus low (20 keV Ð
2 MeV) energies (from \cite{Abdo_081024B:10}), for 4 long (080825C, 080916C, 090217, 090902B) and
2 short (081024B, 090510) duration LAT GRBs. The diagonal lines indicate high to
low energy fluence ratios of 1\%, 10\%, and 100\%.}
\label{energetics}
\end{figure}

The observed GRB detection rate implies that, on average, only about $\sim 10-20\%$ of the
energy that is radiated during the prompt GRB emission phase is channeled
into the LAT energy range (corresponding to a simple Band function extrapolation), suggesting that in most GRBs the high-energy
radiative output does not significantly affect the total energy budget. This is quite an important fact for the whole issue of looking for correlations of a distance independent parameter with the total energetics of bursts. 
Short GRBs, however, appear to be different in this respect as the high energy output observed on the few short bursts detected by the LAT was on the same order of magnitude as the energy output in the sub-MeV domain (see figure \ref{energetics}). We note however a possible observational bias in this plot as long bursts might be difficult to detect in the lower left corner since the event rate in the NaI detectors might be too faint to trigger GBM and a LAT only detection is difficult due to the low high energy fluence.

Because different fundamental processes dominate the photonic emission at different time and energy regime, it is crucial to investigate the evolutionary behavior of GRB prompt emission both as a function of those two parameters: time and energy. 
In principle, one would like to know the continuous behavior of a source as a function of time and energy. However, this is a very complex problem that is more easily tackled by binning one dimension and investigate the continuous evolution in the other dimensions in each bin. The choice of the binning is usually subjective but it is important to choose it carefully, based on the physics if possible.
This time-energy phase space can be approached with two slightly different approach. A first approach investigates the spectral behavior of the burst as it evolves with time, this is usually refered as time-resolved spectroscopy. Another approach investigates the lightcurve in different energy intervals. This could be refered as 'energy-resolved temporal analysis'.
The following sections will use both approaches in order to look at specific features from different viewing angles.

Although the sample of bursts detected at high energy is still limited - especially bursts with high statistics which are needed to detect temporal and spectral features - some re-occuring features have been observed during the prompt emission of LAT detected GRBs after one year and a half of operation.
We now propose an overall description of the key features of LAT GRBs focusing on the prompt emission.
It is for the first time possible to perform detailed temporal and spectral analysis of the high energy prompt emission thanks to a few bright and hard bursts detected with unprecedented high statistics by the LAT instrument. However these high statistics bursts do not constitute the bulk of the LAT GRB sample with only 4 bursts observed with $> 100$ photons up to January 2010 while the LAT detected a total of 14 bursts. High statistics GRBs in the LAT are extremely useful laboratory to learn about GRB physics as we will present in the following section but one has to be very careful to draw conclusions on the overall GRB population. The diversity of phenomenology observed on the 4 bright LAT bursts is another incentive for great caution in extrapolating the characteristics observed on bright bursts to generic population features.

\section{Delay of the high energy emission}
\label{features}

This section presents a key feature observed on all bright LAT GRBs: a delay of the high energy emission with respect to the onset of the sub-MeV emission (section \ref{brightGRBs}). Section \ref{faintGRBs} addresses the case of faint LAT GRBs where this feature is not strong enough to be detected on individual bursts.  The absence (or lack of significance) of a specific feature in a faint GRB could simply be due to a lack of statistics so section \ref{faintGRBs} estimates the significance of this feature for some faint bursts.

\subsection{Bright bursts: GRB 080916C, 090510, 090902B, 090926A}
\label{brightGRBs}

\begin{figure}
\begin{center}
\includegraphics[width=.9\linewidth, keepaspectratio]{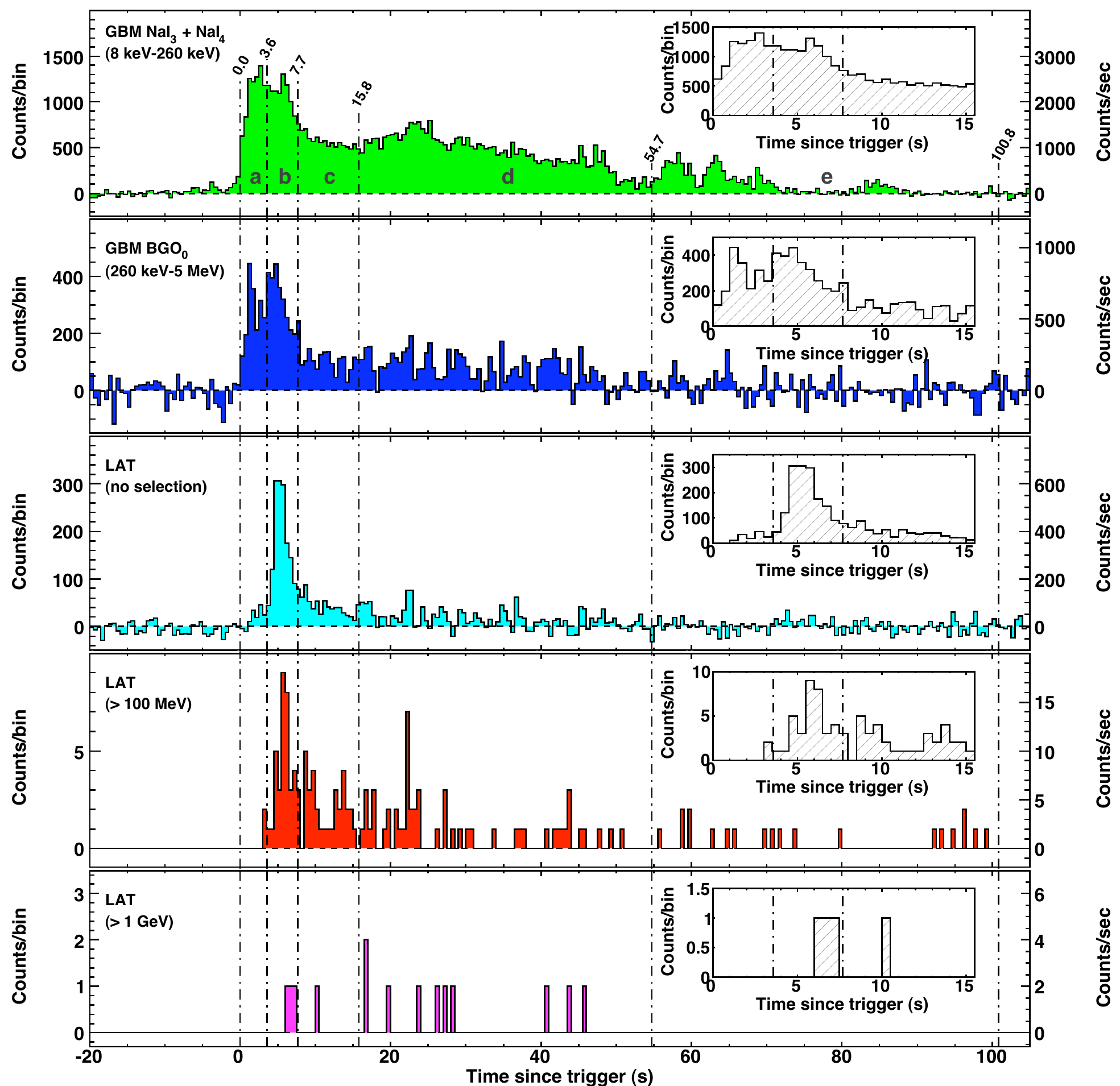}
\end{center}
\caption{Lightcurves for GRB 080916C observed by the GBM and the LAT, from lowest to
highest energies. The inset panels give a view of the first 15 s from the trigger time. In all cases,
the bin width is 0.5s; the per-second counting rate is reported on the right for convenience. The time windows (a) to (e) represent the time selection for time resolved spectroscopy.}
\label{lc_0916C}
\end{figure}

\begin{figure}
\begin{center}
\includegraphics[width=.9\linewidth, keepaspectratio]{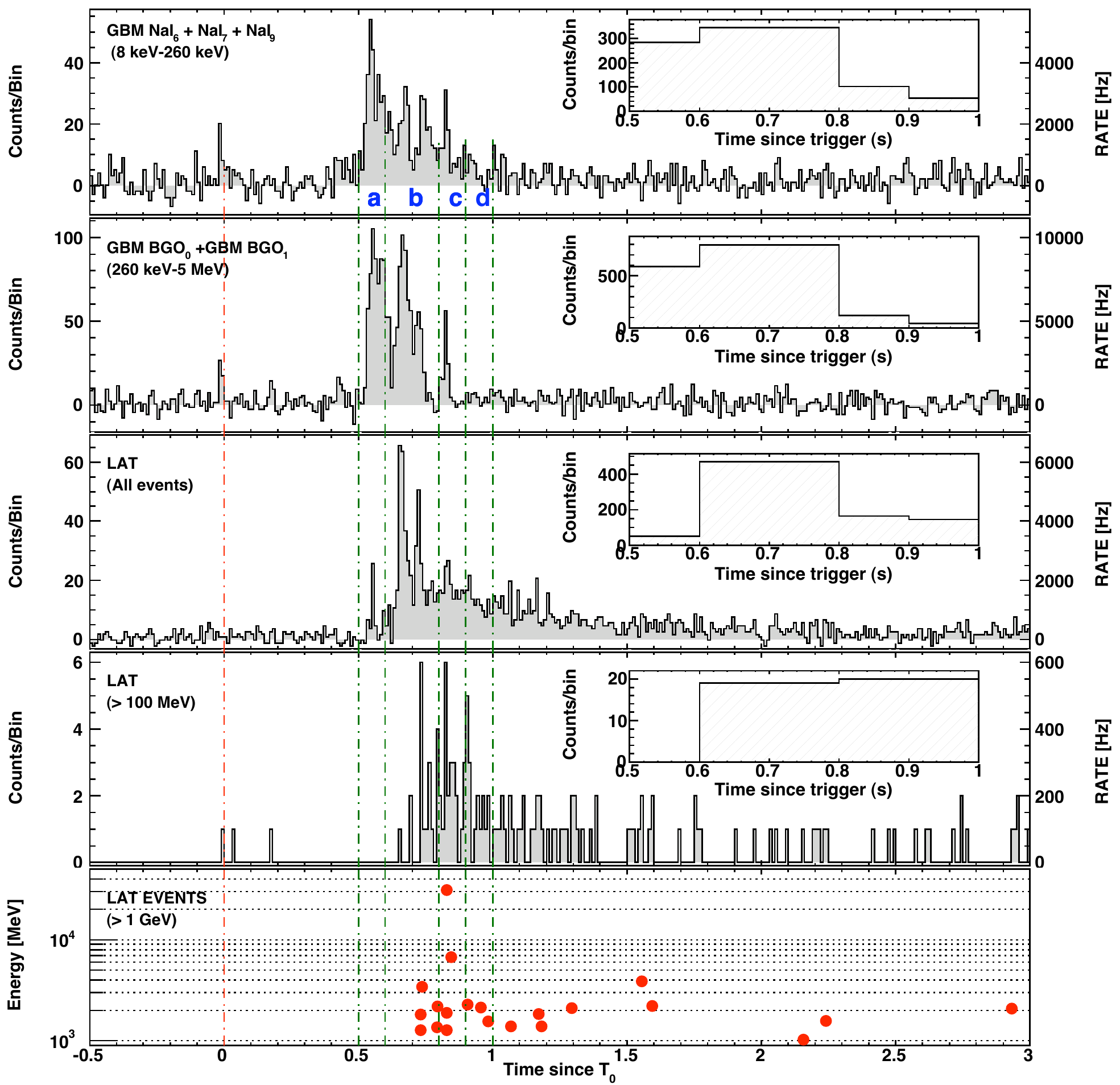}
\end{center}
\caption{Lightcurves for GRB 090510 observed by the GBM and the LAT, from lowest to
highest energies. The inset panels give a view from 0.5 to 1 second after the trigger time. In all cases,
the bin width is 0.2s; the per-second counting rate is reported on the right for convenience. The time windows (a) to (d) represent the time selection for time resolved spectroscopy.}
\label{lc_0510}
\end{figure}

\begin{figure}
\begin{center}
\includegraphics[width=.9\linewidth, keepaspectratio]{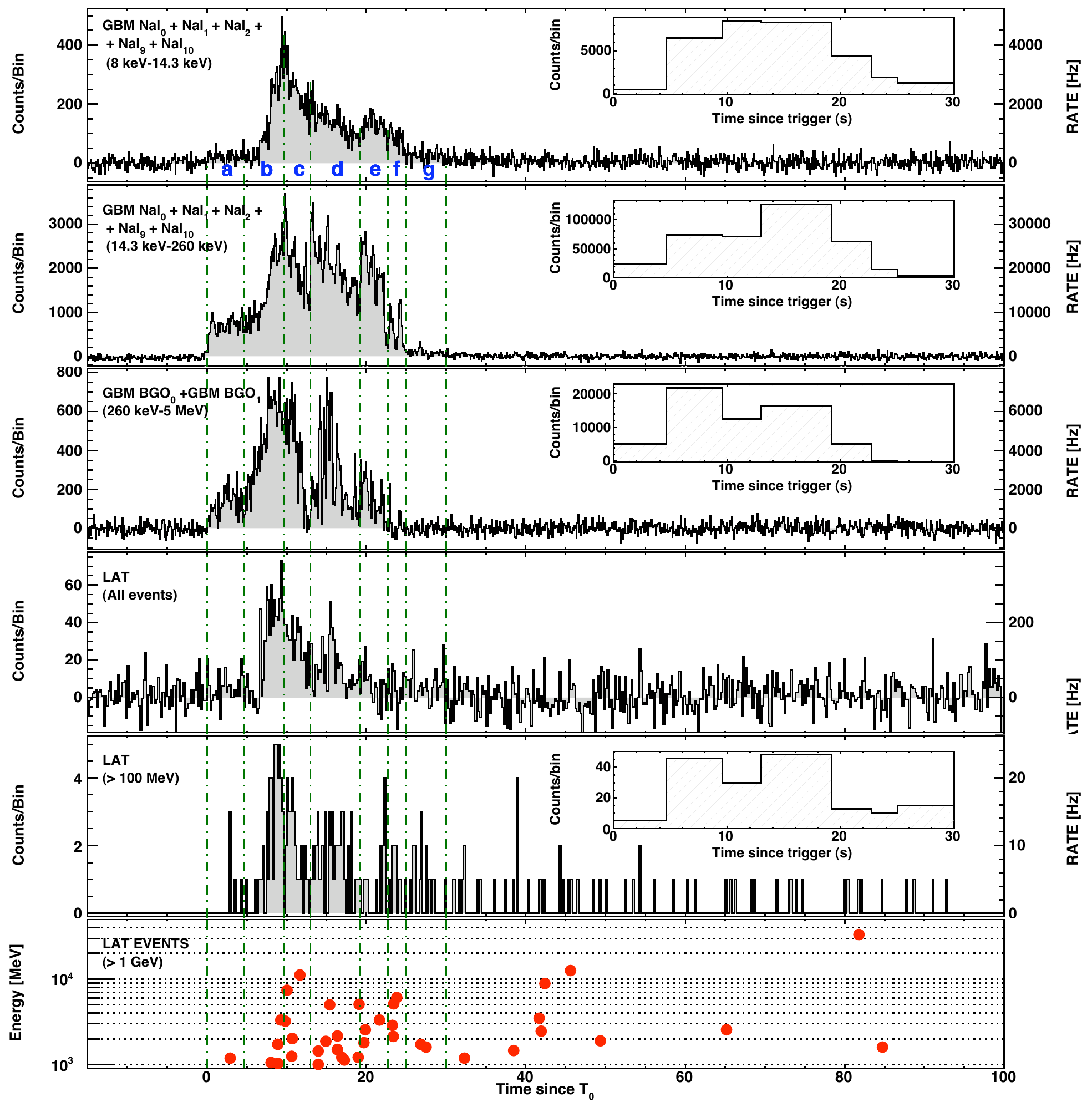}
\end{center}
\caption{Lightcurves for GRB 090902B observed by the GBM and the LAT, from lowest to
highest energies. The inset panels give a view of the first 30 s from the trigger time. In all cases,
the bin width is 0.5s; the per-second counting rate is reported on the right for convenience. The time windows (a) to (g) represent the time selection for time resolved spectroscopy.}
\label{lc_0902B}
\end{figure}

\begin{figure}
\begin{center}
\includegraphics[width=.9\linewidth, keepaspectratio]{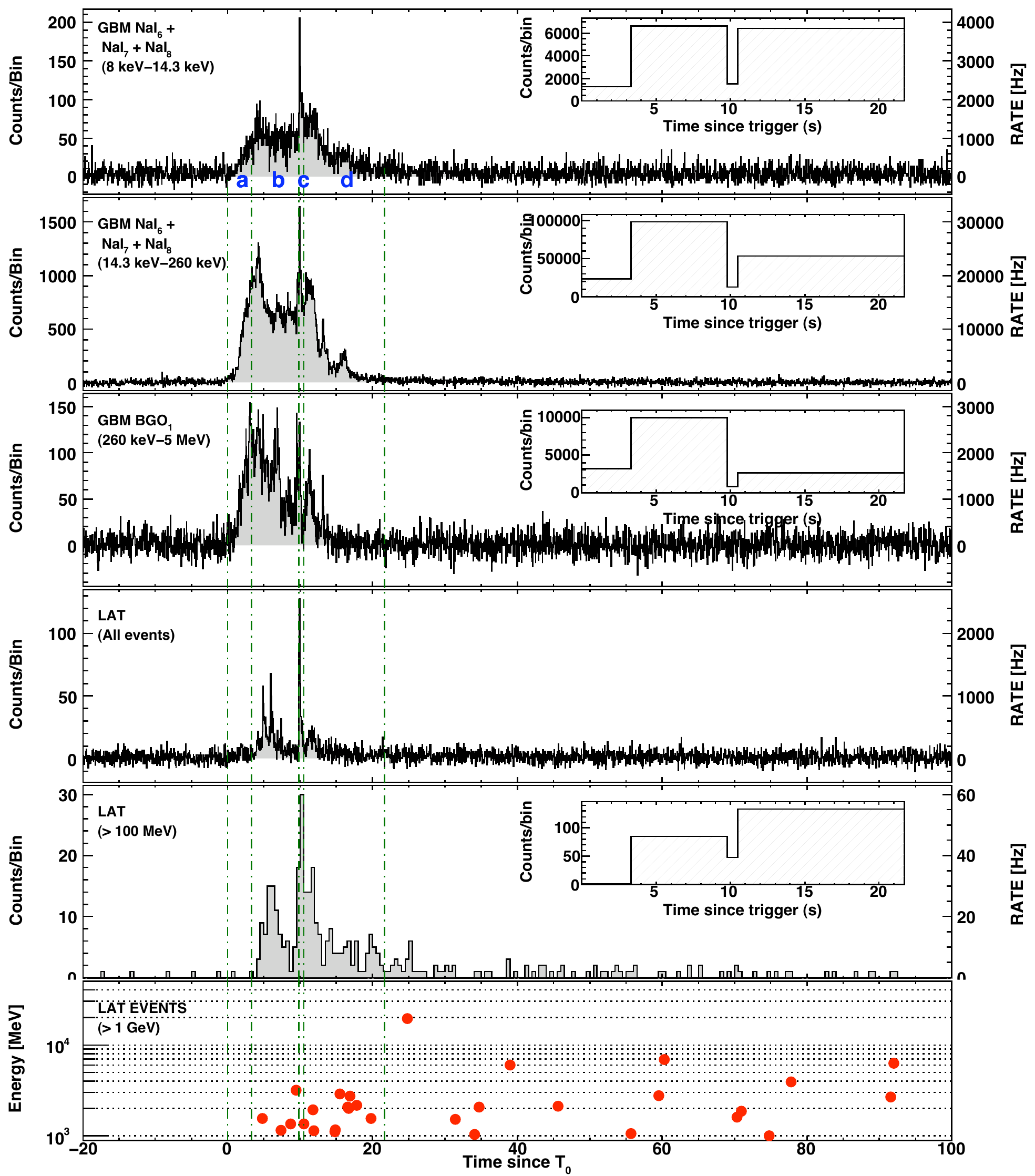}
\end{center}
\caption{Lightcurves for GRB 090926A observed by the GBM and the LAT, from lowest to
highest energies. The inset panels give a view of the first 22 s from the trigger time. In all cases,
the bin width is 0.5s; the per-second counting rate is reported on the right for convenience. The time windows (a) to (d) represent the time selection for time resolved spectroscopy.}
\label{lc_0926A}
\end{figure}

The lightcurves of the 4 brightest bursts observed by the LAT during its first year of operation are shown in figure \ref{lc_0916C} (GRB 080916C), \ref{lc_0510} (GRB 090510), \ref{lc_0902B} (GRB 090902B), and \ref{lc_0926A} (GRB 090926A). In each of these figures, the panel 'LAT (All events)' represents all events passing the LAT onboard filter which have a reconstructed track - and for which we therefore have directional information - while the 'LAT ($>100$ MeV)' shows events for which the energy reconstruction was good enough to be used for spectral analysis. 
The emission seen in the 'LAT (All events)' lightcurves is mostly the result of the detection of low energy (20 MeV$<E<$100 MeV) photons which unfortunately can not be used for spectral analysis as their energy reconstruction is too poor (energy deposit in the calorimeter is low or inexistent).
We futher note that the NaI, BGO and LAT ('all event') lightcurves are background subtracted whereas the LAT ($>100$ MeV) and LAT ($>1$ GeV) are not since they are mostly background free event selection. Vertical dotted-dashed lines are boundaries for the time windows that were used for time-resolved spetroscopy and that will be presented in section \ref{spec}.

A clear feature observed in these bright LAT bursts is a lack of significant LAT emission while the NaI and/or BGO detectors already have significant emission in their detectors. 
This is especially striking when comparing the onset of the emission detected by the NaI and BGO and the emission detected by the LAT above 100 MeV.
The significant increase in the high energy event rate is observed but with a time delay which is much longer than the typical variability time seen during the subsequent prompt emission.
This feature will be refered as a 'delay' of the high energy emission ($>100$ MeV) with respect to the keV-MeV emission, however one should keep in mind that this feature is temporal (as the word 'delay' suggests) as well as spectral in nature.
Indeed, labeling this feature as a 'delay' of a certain energy window ($>100$ MeV) with respect to another energy window ($\sim$ keV-MeV) is somewhat restrictive. A powerful way to look at this feature is to investigate the spectral evolution of the burst during the begining of the prompt emission where the lack of high energy emission followed by its gradual appearance is observed. Such detailed time-resolved spectroscopy has been performed on all these four bright bursts and the details of the methodology for the joint spectroscopy between NaI, BGO and the LAT detectors is presented in appendix \ref{spec_method}. Section \label{spec} presents the main results of the time-resolved but we here lay out the different possible phenomenological effects that would lead to a high-energy 'delay':

\begin{enumerate}
\item An overall increase of the normalization of the Band function which extends to high-energy and accounts for the observed LAT emission
\item A hardening of the Band function ({\it i.e.}, increase of the peak energy Epeak and/or high-energy spectral index) which extends to high-energy and accounts for the observed LAT emission
\item A flux increase of an extra-component at high energy (due du normalization increase and/or index hardening)
\item An increase of the typical energy at which a break or curvature in the spectrum is observed
\item Quantum gravity effects producing an energy dependent delay on the emitted photons
\end{enumerate}

Effect (5) will be described in great details in chapter \ref{lorentz}. Of interest to us here is the fact that this effect is not able to explain the magnitude of this delay as seen on long bursts (on order of a few seconds). The strictest explanation stems from the strongest constraints we were able to put on the quantum gravity mass using GRB 090510 (the interested reader should read section \ref{lorentz}). This constraints do not allow such high energy delay to be of order a a few seconds as is the case for several long bursts: GRB 080916C, GRB 090902B, GRB 090926A. A more straightforward way to convince ourselves that quantum gravity is not the answer (at least for long bursts) is the fact that if such delay would apply, it should be observable on all the temporal structure of the prompt emission. GRB 090926A (see lightcurves \ref{lc_0926A}) has an extremely bright and sharp spike which was detected by all Fermi instruments at different energies. It would be very hard to believe that this spike does not have the same time of emission at the GRB site. If quantum gravity were to operate, it should also operate on this spike which does not present any significant 'delay' in the LAT. The several seconds delay at the onset of long bursts is therefore not compatible with quantum gravity effect.

Effect (4) can be investigated by detailed joint spectral analysis between GBM and the LAT. We indeed searched for possible curvature in the prompt spectrum of LAT GRBs. Our findings are reported in details in section \ref{curvature}. In summary, GRB 090926A was the only burst for which a significant curvature in the spectrum was observed with a cut-off energy around a few GeVs which is much higher than the energy that would be needed to explain a lack of emission above 100 MeV. However this does not rule out effect (4) as the reasons for the high energy delay but we are just unable to detect it with the level of statistics in the data. Such curvature could for example be due to internal $\gamma-\gamma$ absorption or point to a curvature in the energy distribution of the particles responsible for the prompt emission. We note that external $\gamma-\gamma$ absorption through interaction with the extragalactic background light (optical-UV) is however not a possibility as its effect starts being significant only above $\sim 10$ GeV (see section \ref{EBL} for a detailed analysis )

From the detailed time-resolved spectroscopy of the 4 brightest LAT bursts, we derived the following spectral reasons for the high-energy 'delay':
\begin{packed_enum}
\item GRB 080916C: (2). \cite{Abdo_080916C:09}
\item GRB 090510: (2) and (3).  \cite{Ackermann_090510:10}
\item GRB 090902B: (3).  \cite{Abdo_090902B:09}
\item GRB 090926A: (1) and (2).  \cite{Abdo_090926A:10}
\end{packed_enum}

This high energy 'delay' feature is a definite challenge to explain for any theoretical model that aims at describing the GRB prompt emission at least of those bright bursts. Up to now, there has been no complete and self-coherent interpretation of this feature. 
We report in section \ref{interpretation} some possible theoretical interpretation for this feature.

An interesting question is to determine how widespread this feature is among the GRB population and in particular whether it only belongs to a sub-class of GRBs (to which the 4 bright LAT bursts belong) or whether it is a generic feature.
We therefore now address the issue of faint bursts which, although do not have clear detection of this feature, still seem consistent with its presence at the level of statistics provided by the data. In the following section, we quantitatively estimate the significance of this delay on GRB 080825C, the first LAT detected burst.

\subsection{Faint burst GRB 080825C}
\label{faintGRBs}

Bright LAT bursts offer an unprecedented sample of observations where specific temporal and spectral features can be clearly identified even on single bursts. Features are naturally much harder to detect on weaker bursts which already barely passed the $5 \sigma$ threshold for detection of their overall emission. However, it is interesting to try to quantify the possible presence of various features and estimate the level of consistency with what is observed on bright bursts.

In the following sections, we estimate quantitatively a possible delay of the high energy emission onset in GRB 080825C, a weak LAT GRB detected with $\sim 6 \sigma$ confidence. The analysis used is also useful to estimate the possibility for the high energy prompt emission to be significantly more extended than the sub-MeV emission.

\begin{figure}
\begin{center}
\includegraphics[width=.9\linewidth, keepaspectratio]{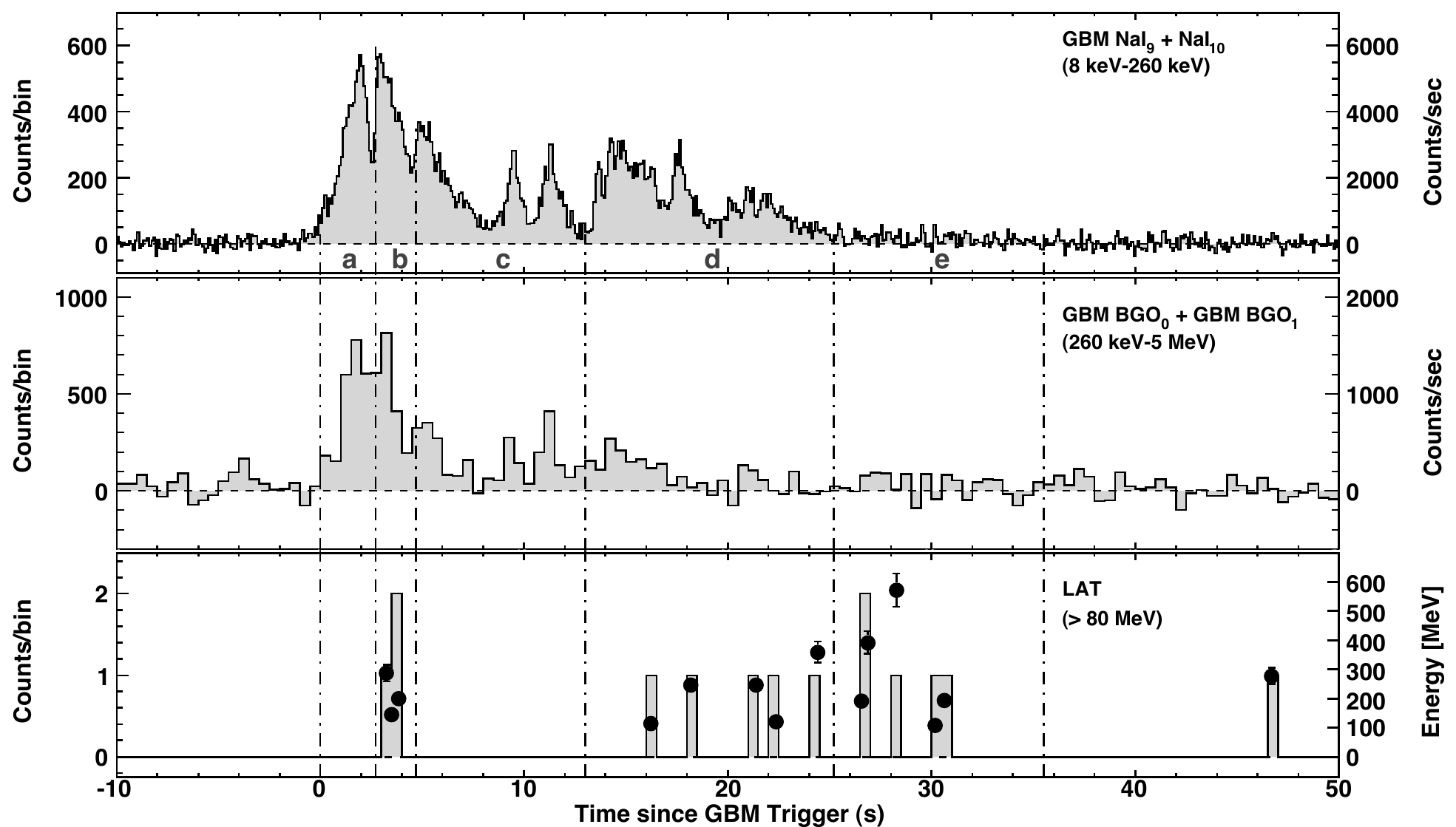}
\end{center}
\caption{Lightcurves for GRB 080825C observed by the GBM and the LAT, from lowest to
highest energies. In all cases, the bin width is 0.5s; the per-second counting rate is reported on the right for convenience. The time windows (a) to (e) represent the time selection for time resolved spectroscopy.}
\label{lc_0825C}
\end{figure}

\subsubsection{Monte-Carlo study}

The NaI, BGO and LAT light curves are shown in
Figure~\ref{lc_0825C}.  No LAT event is seen in coincidence with
the first bright GBM peak which is reminiscent of the high energy 'delay' observed on bright bursts despite the low level of statistics. The first 3 LAT events
are detected in a very short time window, a few seconds after the GBM
trigger, in coincidence with the second GBM peak.  
After a quiet period where no LAT events are detected up to $\sim$$16\;$s, 
4 more events are detected within the GBM $T_{90}$, 
and another 4 events after $T_{90}$ when the
NaI/BGO emissions have faded close to background level. 
Interestingly the highest energy event,
with an energy of $572 \pm 58$~MeV, is detected at fairly late time ($\sim$$T_0+28\;$s).

In order to quantify the different features of this GRB (high energy 'delay' and
extended prompt emission), we have performed Monte-Carlo simulations of the
LAT light curve and estimated the fraction of those simulations that
reproduce these features.  The LAT event distribution was produced
using Poisson statistics for a constant background rate of 0.037$\;$Hz
(the background level was estimated using the method presented in appendix \ref{bkg_est} and consistent with a stable off source region $\sim 600$ sec before the GRB trigger), an estimated detected signal above
$80\;$MeV of 11.7 events (13 events minus 1.3 estimated background
events) and a temporal probability distribution based on the NaI
light curve from $T_0$ to $T_0+35\;$s. 
A total of 30,000 simulations were performed with the exact same number of total LAT events (13).
The probability of the different features were computed as follow:
\begin{itemize}
\item high energy 'delay': the fraction of simulated light curves where the first event arrives later than the first actual observed photon (at $T_0+3.252$ s).
\item extended prompt emission: the fraction of simulated light curves where the last 5 events are detected after the timing of the ninth observed event (detected at $T_0 + 26.570$ s).
\end{itemize}

\begin{figure}
\begin{center} $
\begin{array}{cc}
\includegraphics[ width=.50\linewidth, keepaspectratio]{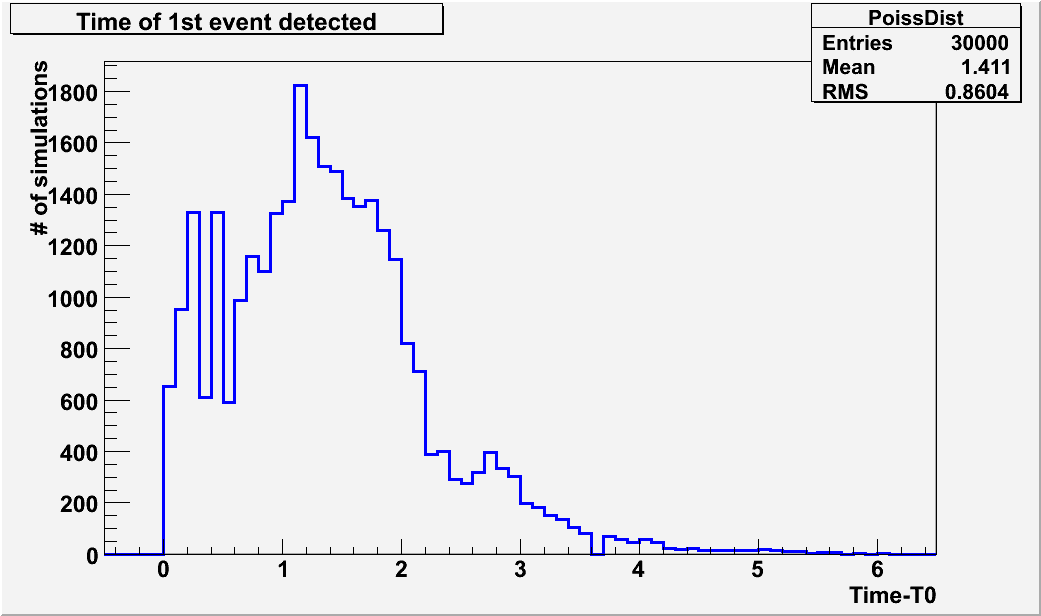}  &
\includegraphics[ width=.44\linewidth, keepaspectratio]{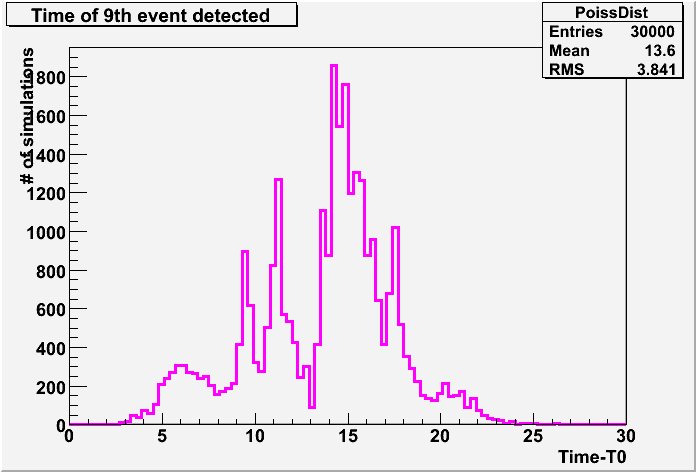} \\
\end{array} $
\end{center}
\caption{The above distributions are based on Monte-Carlo simulations of the LAT lightcurve for GRB 080825C. The LAT event distribution was produced
using a constant background rate of 0.037$\;$Hz
(based on the method presented in appendix \ref{bkg_est}), an estimated detected signal above
$80\;$MeV of 11.7 events (13 events minus 1.3 estimated background
events) and a temporal probability distribution based on the NaI
light curve from $T_0$ to $T_0+35\;$s. {\it Left panel:} Time distribution for the detection of the first simulated events. {\it Right panel:} Time distribution for the detection of the $9^{th}$ simulated events. Based on these distributions, we could derive the chance probability for a high energy 'delay' (3.4\%) and extended prompt emission (3.7$\;\sigma$).}
\label{MC_080825C}
\end{figure}

The distribution of events for those two time definition is given in figure \ref{MC_080825C}.
This analysis finds
weak evidence for the possible high energy delay feature with
a chance probabilities of 3.4\%, and the
evidence for temporally extended emission in the LAT is more significant, with
a 3.7$\;\sigma$ significance.

We now turn to detailed time-resolved spectroscopy that reveals further key features significantly observed in bright LAT GRBs.

\section{Time-resolved spectroscopy of the prompt emission}
\label{spec}

Compared to the $EGRET$ era, the $Fermi$ LAT increased sensitivity, more precise energy resolution, smaller deadtime and new energy window from 10 to 300 GeV is a significicant step forward toward a better understanding of the spectral behavior of high energy sources. A crucial aspect for GRBs is to investigate the relationship of the high energy emission with the sub-MeV emission which is detected onboard $Fermi$ by the Gamm-Ray Burst Monitor (8 keV - 40 MeV). The GBM and LAT instruments combined cover almost 7 order of magnitude in energy ($8 \mbox{ keV} \rightarrow 300$ GeV) with unprecedented sensitivity above 20 MeV.
Precise time-resolved spectrosopy of GRBs in the broad GBM/LAT energy range is key to solve many of the still unanswered questions such as the nature of the jet content, the prevailing emission mechanisms, or the jet bulk lorentz factor.

Because of reasons detailed in section \ref{event_optimization}, spectral studies with the LAT are only performed above 100 MeV. Further analysis are ongoing to open the lower part of the LAT energy range to spectral analysis. Until then, we will restrict ourselves to this lower boundary of 100 MeV in the following analysis.
We also note that because of rapidely decreasing fluxes above $\sim 10$ GeV, only a few photons have been detected above that energy - the highest energy photon is 33 GeV - which therefore limit the LAT capability to derive spectral information of high energy sources above these energies (this limitation can be overcome with huge effective areas which is for example achieved with Air Cherenkov Telescopes).

Besides few cases detected by EGRET with low statistics, the spectral behavior of GRBs in the MeV-GeV range was largely unknown before $Fermi$. We here expose a comprehensive overview of the prompt emission spectrum of the brightest GRB bursts detected by the LAT instrument onboard $Fermi$.
Weak bursts detected by the LAT provide us much less information on the spectral distribution in the MeV-GeV range due to the lack of statistics. Population study might lead to interesting results but require a substantial sample and this work is therefore left for future publications. We therefore mostly concentrate on the analysis of bright LAT GRBs in this section.

In this chapter, we will present the various analysis techniques and results of the joint GBM/LAT spectrosopy on GRBs up to January 2010. The analysis techniques and tools used by the LAT team for fitting and spectral model comparison are presented in appendix \ref{spec_method}. Sections \ref{addcomp} and \ref{curvature} will present the results of our searches for possible departure from the typical Band function which was found to be a good phenomenological model to represent the sub-MeV observations in the BATSE era \citep{Band:93}. Section \ref{addcomp} look for power-law components emerging at high energy and section \ref{curvature} look for curvature of the spectrum at high energy. Finally section \ref{sys} summarize our investigation of systematic uncertainties affecting our spectral results.

\subsection{High energy additional component}
\label{addcomp}

The BATSE era has revealed that a phenomenological model known as the 'Band function' provides an excellent fit to the data in the keV-MeV range \citep{Band:93} - see \citep{Preece:02} and \citep{Ryde:04} however for possible departure from this phenomenological model at low energy. This model is usually interpreted as the synchrotron emission from an energetic electron population energized through internal shocks, although this explanations faces some critical isssues (line of death problem \citep{Preece:00}).
Another possibility is a thermal origin at the photosphere with an underlying power-law component \citep{Ryde:04} although the physical origin of this second component is unclear at the present time.
The detection of an additional component on top of the underlying sub-MeV component would reveal the presence of a new emission at high energy which would be a very valuable clue to derive information on the particle content of the GRB jet as well as the dominant emission mechanisms taking place in various energy regime.

\begin{figure}
\begin{center}
\includegraphics[width=.6\linewidth, keepaspectratio]{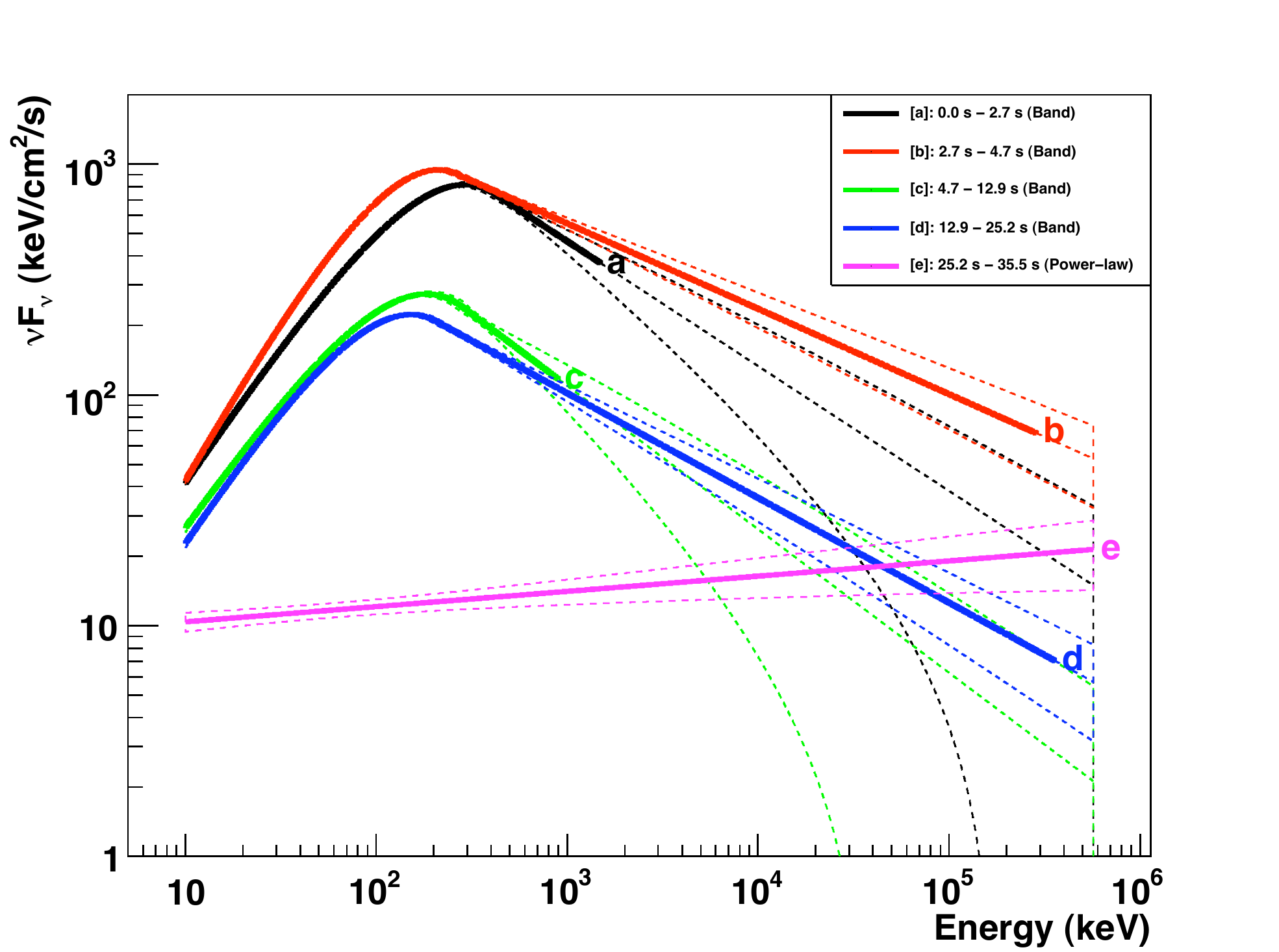}
\end{center}
\caption{Time resolved $\nu F\nu$ spectra for GRB 080825C in time bin (a), (b), (c), (d) and (e) as given in figure \ref{lc_0825C}}
\label{spectra 0825c}
\end{figure}

An additional component at high energy up to 200 MeV was found in one EGRET burst:  GRB 941017 \citep{Gonzalez:03}. We naturally searched for such component in the LAT bursts.
The lack of statistics of most LAT detected bursts prevented us to distinguish between various spectral models so the fact that no signs of departure from a Band function is found in these weak bursts should not be seen as a proof of absence of additional feature but just as a consequence of low statistics in the LAT energy range. It is therefore not a surprise that an additional component was not found significant in any of the faint LAT bursts. An example of such low statistics burst is GRB 080825C. Fig. \ref{spectra 0825c} represents the time-resolved $\nu F\nu$ spectra in the different time bin shown in figure \ref{lc_0825C} (note that a simple power-law is preferred in the last time bin due the very low statistics in the GBM and LAT bands). We now turn to bright LAT GRBs.

\vspace{0.7cm}
\noindent
{\bf GRB 080916C}
\vspace{0.3cm}

\begin{figure}
\begin{center} $
\begin{array}{cc}
\includegraphics[ width=.42\linewidth, keepaspectratio]{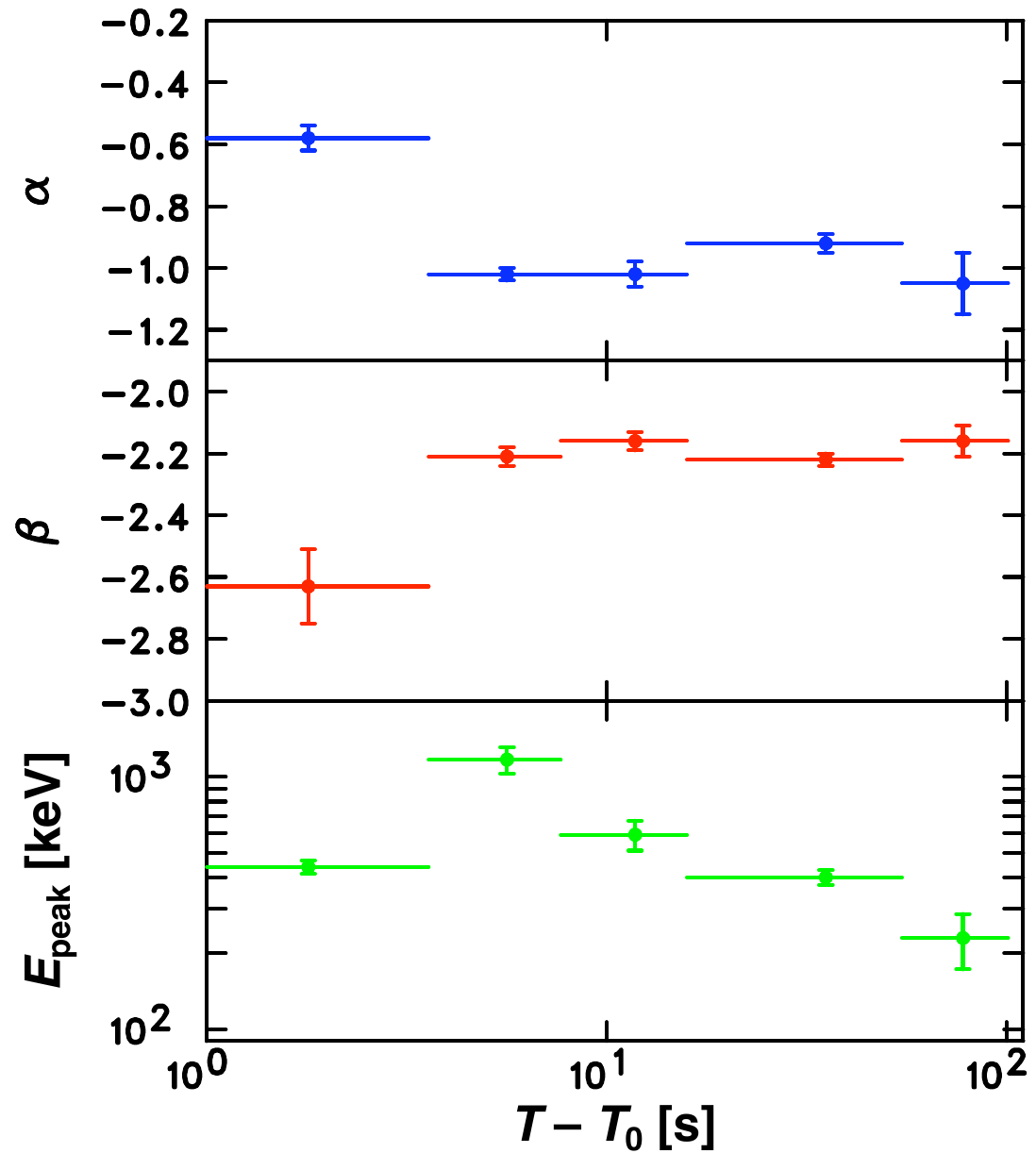} &
\includegraphics[ width=.57\linewidth, keepaspectratio]{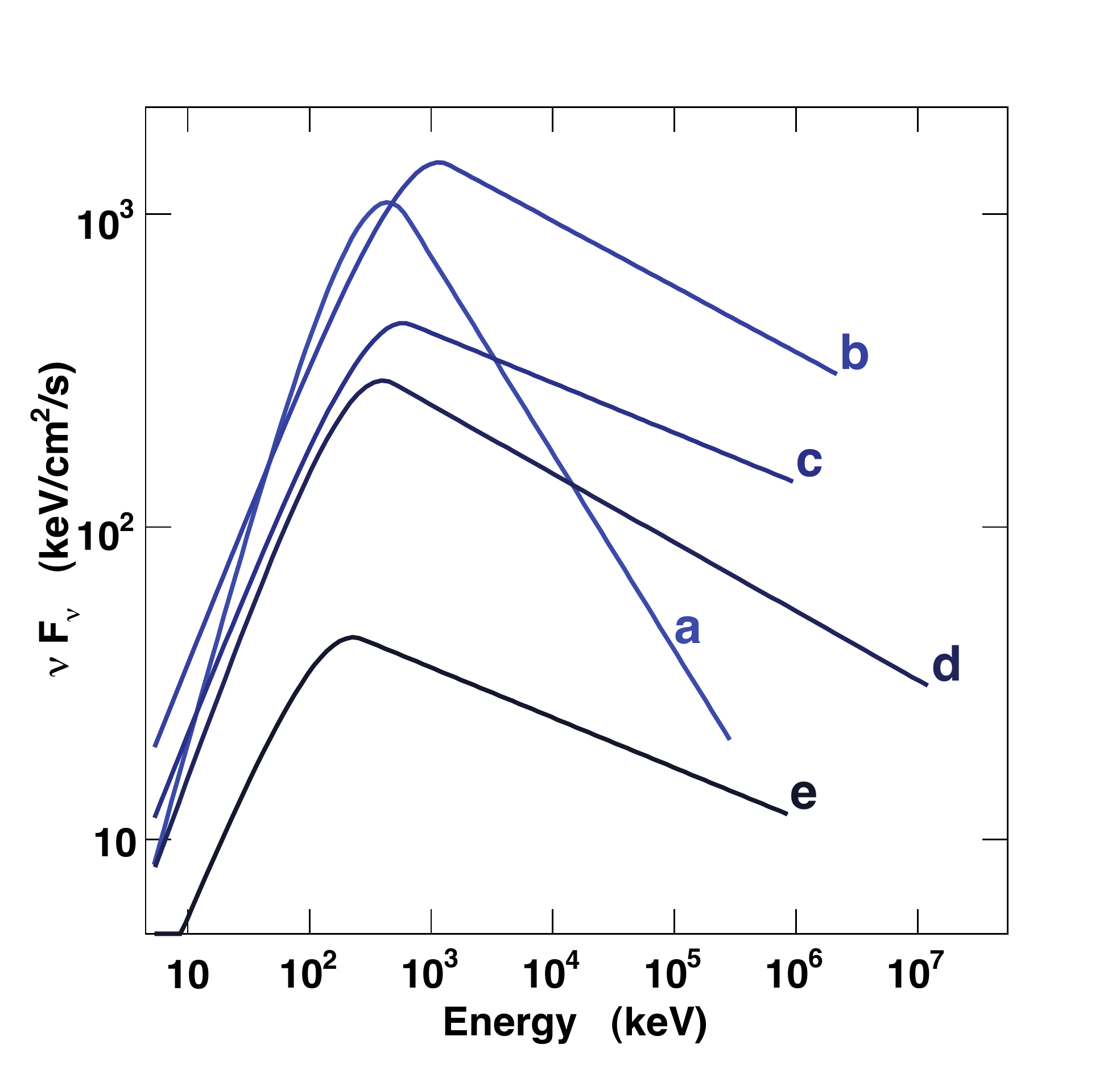}
\end{array} $
\end{center}
\caption{{\it Left}: Fit parameters for the Band function $\alpha$, $\beta$ and $E_{peak}$ as a function of time. {\it Right}: The model spectra for all five time intervals. The curves end at the energy of the highest-energy photon observed in each time interval.}
\label{0916C spectra}
\end{figure}

The first bright burst detected in the LAT was GRB 080916C (figure \ref{lc_0916C}) which had 145 events above 100 MeV associated with the burst during the first 100 seconds after trigger. This allowed a detailed time-resolved spectroscopy to be performed in 5 different time windows (shown in figure \ref{lc_0916C}).
In this burst, we find that the spectra of all five time intervals as well as the time integrated spectrum are well fit by the empirical Band function all the way up to the highest energy observed ($\sim 13$ GeV). The right panel of figure \ref{0916C spectra} shows the spectral models obtained from the fits of the data in each of the five time intervals and the left panel shows the evolution over time of the Band function parameters with their $1-\sigma$ statistical uncertainties. The model curves are shown in $\nu F_{\nu}$ units in which a flat spectrum would indicate
equal energy per decade of photon energy, and the changing shapes show the evolution of the
spectrum over time.
One can for example notice the significant hardening of the spectrum from time bin (a) to time bin (b) - higher $E_{peak}$ and harder $\beta$ index - which is the spectral view of the high energy 'delay' clearly visible on the GBM and LAT lightcurve and which was mentioned in the previous section.

We searched extensively for deviations from the Band function in this bursts. In time bin (d) there are three photons above 6 GeV. We tried modeling
these high-energy photons with a power law as an additional high-energy spectral component.
Starting from the null hypothesis that the data originate from a simple Band GRB function, the
improvement in likelihood obtained by adding the additional power-law component indicates
1\% probability for this time bin that we have no additional spectral component; with 5 time
bins, this is not strong evidence for any additional component. The LAT sensitivity to higher-energy
photons may be reduced at z$\sim 4.2$ through absorption by Extragalactic Background Light
(EBL). Because the effect of various models ranges widely, from leaving the single time bin
probability of an extra component unchanged to decreasing the plausibility of its absence
to 0.03 \%, we do not use EBL absorption effects in our estimation of its significance.

\vspace{0.7cm}
\noindent
{\bf GRB 090510}
\vspace{0.3cm}

\begin{figure}
\begin{center}
\includegraphics[width=.6\linewidth, keepaspectratio]{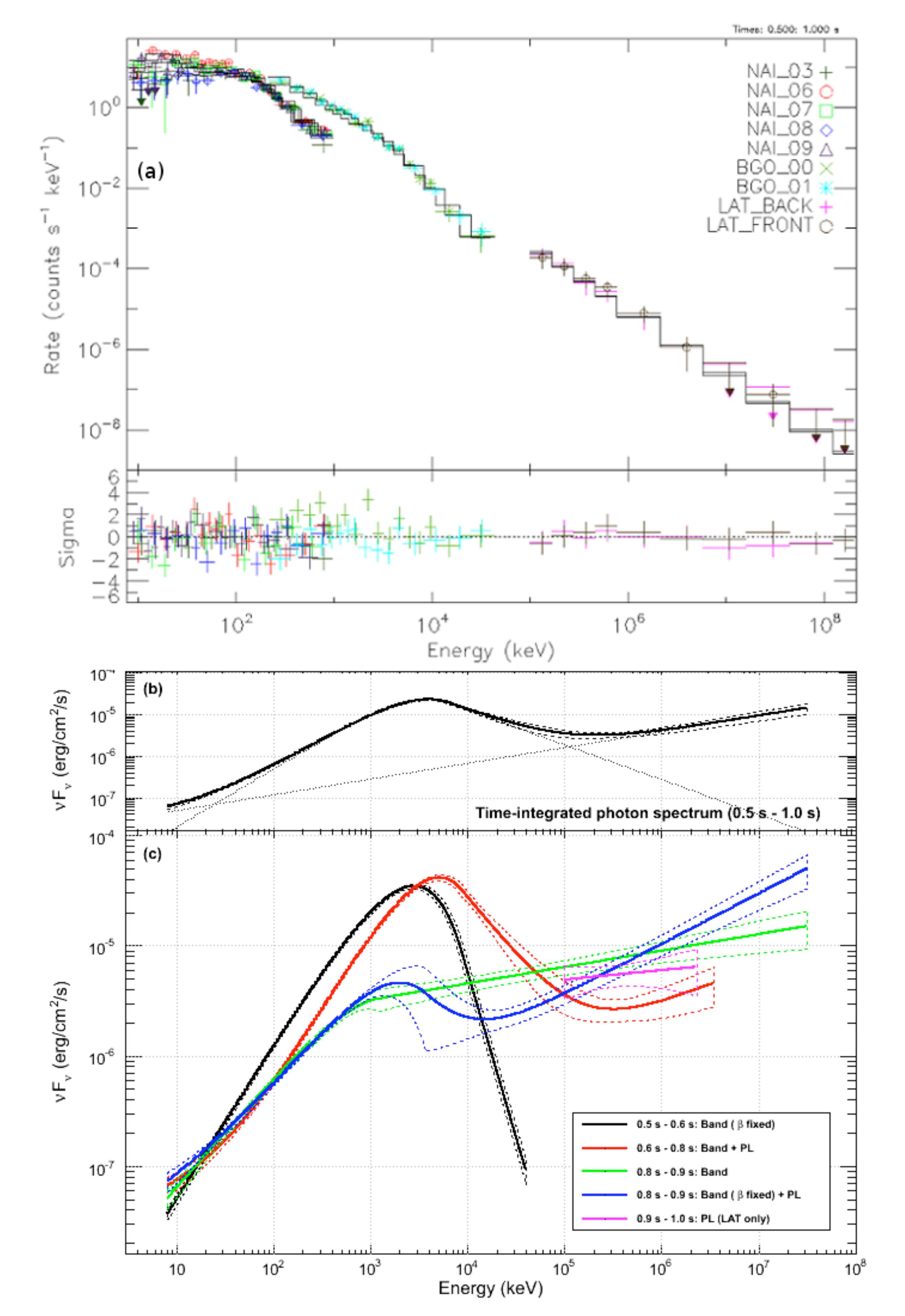}
\end{center}
\caption{{\it Top}: Counts spectrum for the time-integrated ($T_0 + 0.5$, $T_0 + 1.0$ s) data. The Band
+ power-law model has been fit to these data. {\it Middle}: The best-fit Band + power-law model
for the time-integrated data plotted as a $\nu F\nu$ spectrum. The two components are plotted
separately and the sum is plotted as the heavy line. The $\pm 1-\sigma$ error contours derived from
the errors on the fit parameters are also shown. {\it Bottom}: The $\nu F\nu$ model spectra (and $\pm 1-\sigma$ error time resolved spectra in time bin (a), (b), (c) and (d).}
\label{spectra 0510}
\end{figure}

GRB 090510 (figure \ref{lc_0510}) is an extremely intense short burst detected by the LAT instrument. Its count spectra and $\nu F\nu$ spectrum (best fit models for the time-integrated and time-resolved spectrum) can be seen in figure \ref{spectra 0510}.
The peak energy of the Band component for the time-integrated data is $E_{peak} = 3.9 \pm 0.3$ MeV, which is the highest peak energy ever measured in a GRB time-integrated
spectrum. The addition of a power-law component with a photon index of $-1.62 \pm 0.03$
was found to improves the spectral fit with a Test-Statistics of $\sim 36.1$.

\begin{figure}
\begin{center}
\includegraphics[width=.6\linewidth, keepaspectratio]{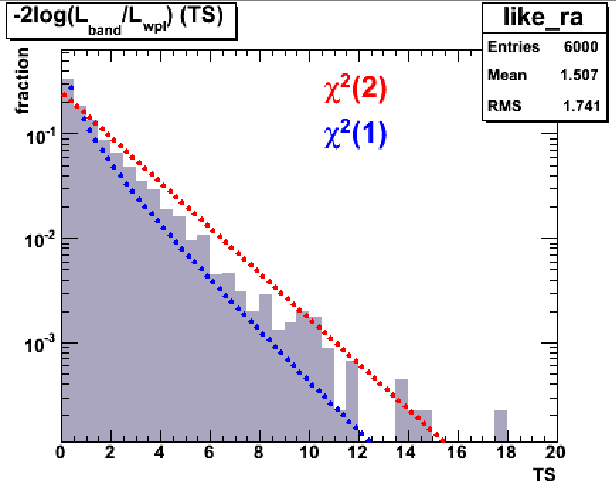}
\end{center}
\caption{TS distribution from simulation of null-hypothesis (Band function) for the time-integrated spectrum of GRB 090510}
\label{ts 0510}
\end{figure}

However, we note that strictly speaking Wilk's theorem does not apply as the null-hypothesis (Band function) and the alternative model (Band function + Power Law) are not fully nested since the null-hypothesis model can be retrieved by simply fixing the power-law normalization to the null value without any constraints on the power-law index.
In order to convince ourselves of the reality of this additional spectral component, we performed simulations of the null-hypothesis (Band function with best fit parameters) to evaluate the distribution of the Test-Statistics (TS) under this hypothesis. The TS distribution is shown on figure \ref{ts 0510} and is close to a chi-square distribution with one degree of freedom (blue dashed curve) as Wilks' theorem would predict. However, we find that a chi-square with two degrees of freedom (red dashed curve) does not underestimate the tail of the distribution which is what matters for the TS conversion into a significance. We therefore use this $\chi^2_{(2dof)}$ parametrization in order to convert the TS value into a significance for the additional component. TS$\sim 36.1$ therefore corresponds to a $ 5.5 \sigma$ discovery of an additional component.
This is the first short burst for which such a hard power-law component has been observed.
Top panel of figure \ref{spectra 0510} shows the counts spectrum of the time-integrated data and
the 'Band function+power-law' best fit. Bottom of figure \ref{spectra 0510} (top panel), we plot this composite model for the time-integrated $\nu F\nu$ spectrum along with the separate contributions for each component.

For the time-resolved spectroscopy, we initially considered the data partitioned into
0.1 s time bins, starting at $T_0 + 0.5$ s. However, for the analyses we present here, we have
combined the data in the $T_0 + 0.6$ s to T0 + 0.8 s interval into a single bin in order to
have sufficient counts at energies $>100$ MeV to constrain the fit of the LAT data. The
Band component undergoes substantial evolution over the course of the prompt phase,
starting out relatively soft with $E_{peak} \approx 3$ MeV, evolving to a very hard spectrum with
$E_{peak} \approx 5$ MeV, accompanied by the appearance of the power-law component at $>100$ MeV, and then becoming softer again with $E_{peak} \approx 2$ MeV. The extra power-law component hints at a similar sort of soft-hard-soft evolution, but these spectral changes do not appear to be commensurate with the Band component evolution - see bottom of figure \ref{spectra 0510} (bottom panel).

\vspace{0.7cm}
\noindent
{\bf GRB 090902B}
\vspace{0.3cm}

\begin{figure}
\begin{center}
\includegraphics[width=.6\linewidth, keepaspectratio]{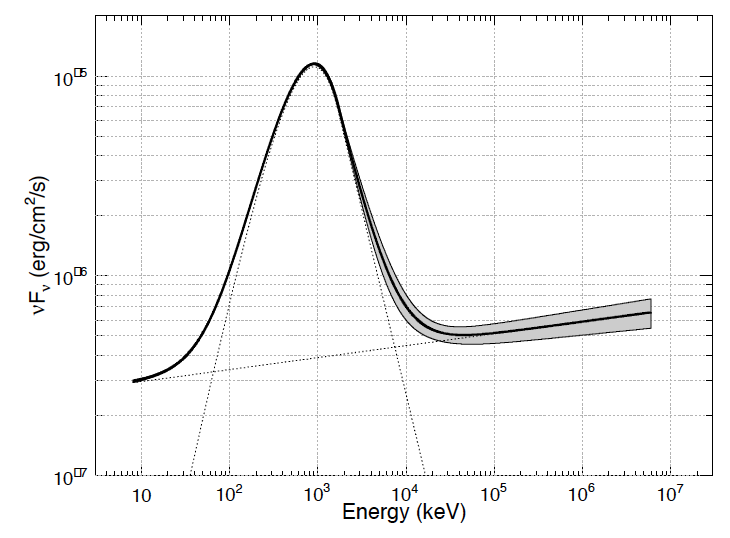}
\end{center}
\caption{Best fit model for time interval (b) (seee fig. \ref{lc_0902B}). $1-\sigma$ error bars with $\nu F\nu$ spectrum are plotted in shaded grey. The extension of the $> 100$ MeV power-law component to the lowest energies ($< 50$ keV) can be clearly seen.}
\label{spec_0902B}
\end{figure}

GRB 090902B (figure \ref{lc_0902B}) is a long burst where the presence of an additional power-law component on top of the Band function was clearly detected in the time-integrated spectrum.
This power-law component
signicantly improves the fit between 8 keV and 200 GeV both in the time-integrated
spectrum and in the individual time intervals where there are sufficient statistics. It is
also required when considering only the GBM data (8 keV - 40 MeV) for the time-integrated
spectrum, as its inclusion causes an improvement of $\sim 1000$ in the Test-Statistic over the
Band function alone. When NaI data below $\sim 50$ keV are excluded, a power-law component can
be neglected in the GBM-only fits. We conclude that this power-law component contributes
a signicant part of the emission both at low ($<50$ keV) and high ($>100$ MeV) energies.
Figure \ref{spec_0902B} shows the unfolded $\nu F\nu$ spectrum for a Band function with a power-law
component fit to the data for interval (b) when the low energy excess is most significant.

Spectral evolution is apparent in the Band function component from the changing
$E_{peak}$ values throughout the burst, while $\beta$ remains soft until interval (e) when it hardens
signicantly. $\beta$ is similarly hard in interval (f), after which the Band function component is
no longer detected. The hardening of $\beta$ is accompanied by an apparent hardening of the
power-law index, $\Gamma$, which until interval (e) does not exhibit much variation. However, this
is not definitive since the flux is too low to constrain $\Gamma$ in intervals e and f separately. A
spectral fit of the sum of these two intervals confirms the presence of both a harder $\beta$ and a
harder $\Gamma$, with a clear statistical preference for the inclusion of the power-law component.
An equally good fit is obtained in the combined (e) + (f) interval if this power-law has an
exponential cut-off at high energies, with the preferred cut-off energy lying above 2 GeV.
Finally, we note that in interval (b), a marginally better fit is achieved using a model with
the additional power-law component having an exponential cut-off at high energies. The
improvement is at the $3 \sigma$ level and indicates weak evidence for a cutoff in the second
component, placing a lower limit on the cutoff energy in this interval of about 1 GeV (we further discuss spectral curvature in section \ref{curvature}).

We note that the
power-law component in GRB 090510 appeared to extrapolate to energies well below $E_{peak}$ and dominates
the Band function emission below $\approx 20$ keV, similarly to the behavior seen in GRB 090902B. However this feature was observed with a much lower significance which is not enough to claim a clear detection on this burst alone.

\vspace{0.7cm}
\noindent
{\bf GRB 090926A}
\vspace{0.3cm}

GRB 090926A is another long burst with a significant additional component observed. This GRB also displayed spectral curvature at high energy as well so we are treating this burst in details in the next section which address exactly this topic.

in the end, all bright LAT GRBs (3 long and 1 short) displayed a significant additional component at high energy at the exception of GRB 080916C which only has a marginal evidence for the presence of such component. It would be tempting to extrapolate this behavior to the whole GRB population however one has to keep in mind that we are only seeing a biased sup-sample of the GRB sample (hard and bright bursts). In section \ref{overviewLATGRBs} (see figure \ref{GRBrate} in particular), it was shown that the current LAT detection rate is in agreement with a simple extrapolation of the sub-MeV behavior of GRBs. This suggests that a fair fraction of the bursts do not display similar additional component as seen in the bright bursts as it would significantly enhance the LAT detection rate compare to predictions from a simple extrapolation of the sub-MeV spectra. More detailed analyses would be needed to further investigate this behavior at the population level.

\subsection{Spectral curvature}
\label{curvature}

Spectral curvature in the prompt emission of GRBs could occur for a multiple of reasons: intrinsic $\g-\g$ opacities, Extragalactic Background Light absorption (see section \ref{EBL}), roll-over of the energy distribution of the emitting particles, inverse compton scattering entering the Klein-Nishina regime.
The fact that the high energy index of many GRBs is harder than -2 is an indication that such curvature in the spectrum needs to occur at a certain energy otherwise the emission would have an infinite amount of energy. Such curvature is also expected for a dense plasma as it becomes optically thick at high energy due to $\gamma-\gamma$ pair annihilation. 

The fact that such energy cutoff has not been observed in the pre-Fermi era has led to the conclusion that the GRB jet must be ultra-relativistic.
Indeed, the non-detection of any spectral signature sign of absorption effect in gamma-ray bursts can be used to place lower limits on the bulk Lorentz factor. We will address this topic in section \ref{lorentz}. At the same time, the detection of such curvature would be an important discovery to understand the physics at the source. GRB 090926A was the first GRB with a clear detection of a spectral curvature at high energy while GRB 080825C is a much more controversial case. We describe our findings on these sources below.

\vspace{0.7cm}
\noindent
{\bf GRB 090926A}
\vspace{0.3cm}

GRB 090926A is a long burst and its lightcurve in different energy bands is presented in figure \ref{lc_0926A}.
Based on the behavior of the light curve, we chose the following time region to perform time-integrated spectroscopy: $T_0 + 3.3$s - $T_0 + 21.6$s.
We jointly fitted a canonical Band function for all time regions and all data sets.
The observed spectra are roughly described by the Band function; however,
large residuals from the fit can be seen in figure \ref{band_pl_cutoff}.
Therefore, we added a power-law component, as in the case of GRB 090902
(Band + PL).
The increase of $107.3$ in Test-Statistics (with 2 degrees-of-freedom) is a significant improvement of the fit equivalent to a $\sim$10~$\sigma$ level; 
the power-law photon index is $\lambda$=$1.80\pm0.01$.
This is now the third case of a LAT detection of an extra component in GRB spectra.
The power-law component dominates below 50~keV and above 10~MeV, similarly to GRB 090902B. Nevertheless, the residuals in the LAT data still remain at high energies (see figure \ref{band_pl_cutoff}).

We then fitted the GBM/LAT spectra with the combinations of the Band function and a power-law model 
with and without a high energy break. For the latter, we tried both a power-law with exponential 
cutoff and a broken power-law. The power-law with exponential cutoff (CUTPL) is modeled as in 
equation \ref{eq:cutoff}, with $A$ the normalization in units of photons s$^{-1}$ cm$^{-2}$ 
keV$^{-1}$, $E_{\rm piv}$ a pivot energy fixed to 1~GeV, $E_F$ a folding energy and $\lambda$ 
is the power-law photon index.

\begin{equation}
\label{eq:cutoff}
 f(E) = A\left(\frac{E}{E_{\rm piv}}\right)^{\lambda}\exp\left(-\frac{E}{E_F}\right)
\end{equation}

The results are summarized in table~\ref{tab:spec} and the best fit count spectra
and residuals are shown on figure~\ref{band_pl_cutoff} for the best model.
The folding energy is $E_F = 1.41_{-0.42}^{+0.22}\;[stat.]\,\pm0.30\;[syst.]$~GeV, while the power-law photon index below the cutoff energy is $\lambda\simeq-1.72_{-0.02}^{+0.10}\;[stat.]\,\pm0.01\;[syst.]$, which is a bit harder than in the (Band+PL) case.
The systematic uncertainties quoted here have been derived using the bracketing instrument response functions, as described in detail in section \ref{sys}.
The parameters of the Band function change little from one fit to another. The C-STAT value for this model is an improvement of 39.5 compared to the Band+PL model (a detailed discussion is provided below on the exact significance of such improvement)

\begin{figure}
\begin{center}
\includegraphics[width=.8\linewidth, keepaspectratio]{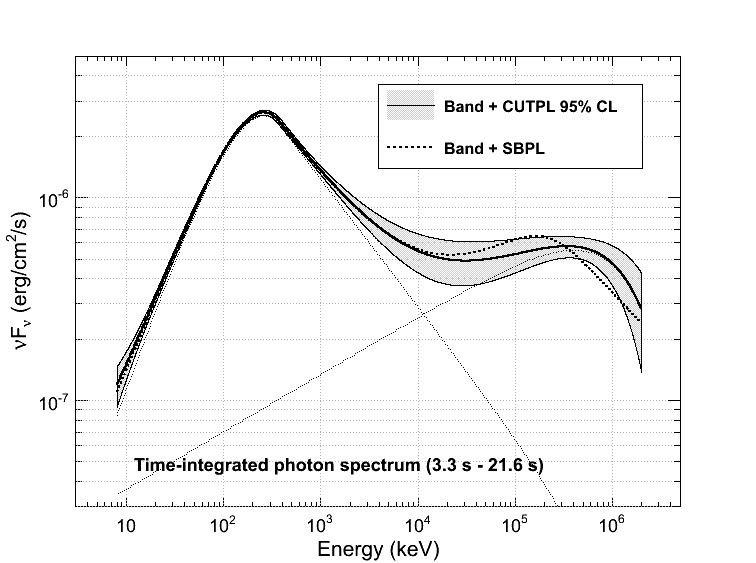}
\end{center}
\caption{Best fit model for the time-integrated data of GRB 090926A. $2-\sigma$ error bars with $\nu F\nu$ spectrum are plotted in shaded grey for the Band+CUTPL model while only the best fit model is plotted for the Band+BPL model.}
\label{0926_nufnu}
\end{figure}

\begin{figure}
\begin{center} $
\begin{array}{cc}
\includegraphics[ width=.48\linewidth, keepaspectratio]{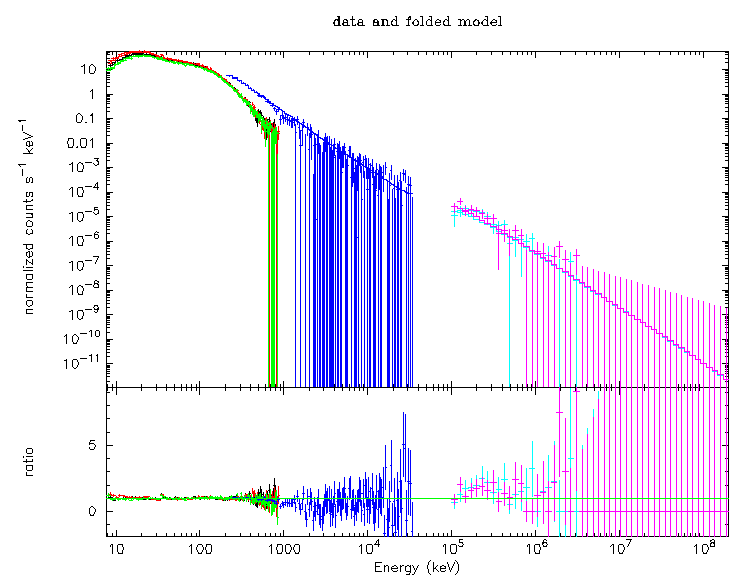} &
\includegraphics[ width=.48\linewidth, keepaspectratio]{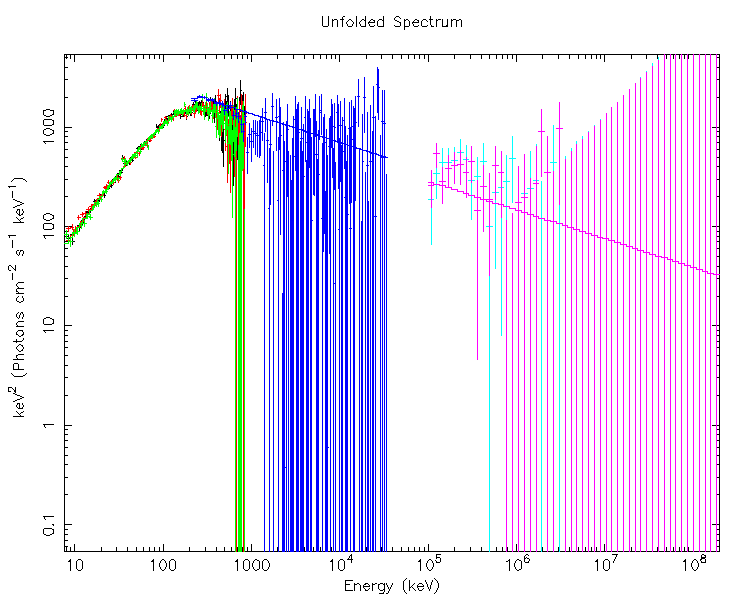} \\
\includegraphics[ width=.48\linewidth, keepaspectratio]{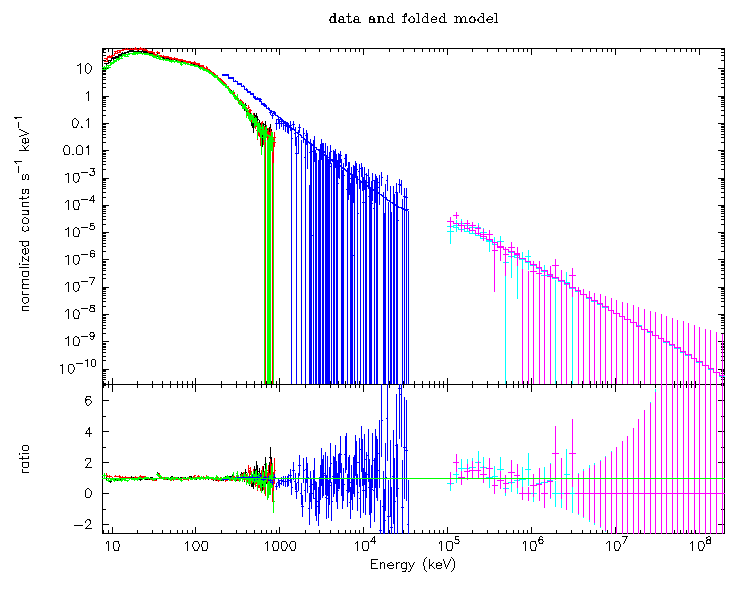} &
\includegraphics[ width=.48\linewidth, keepaspectratio]{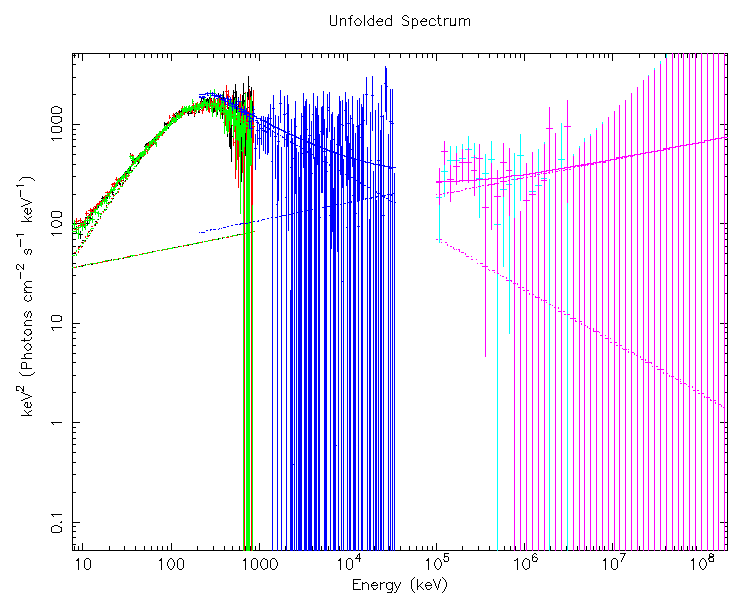} \\
\includegraphics[ width=.48\linewidth, keepaspectratio]{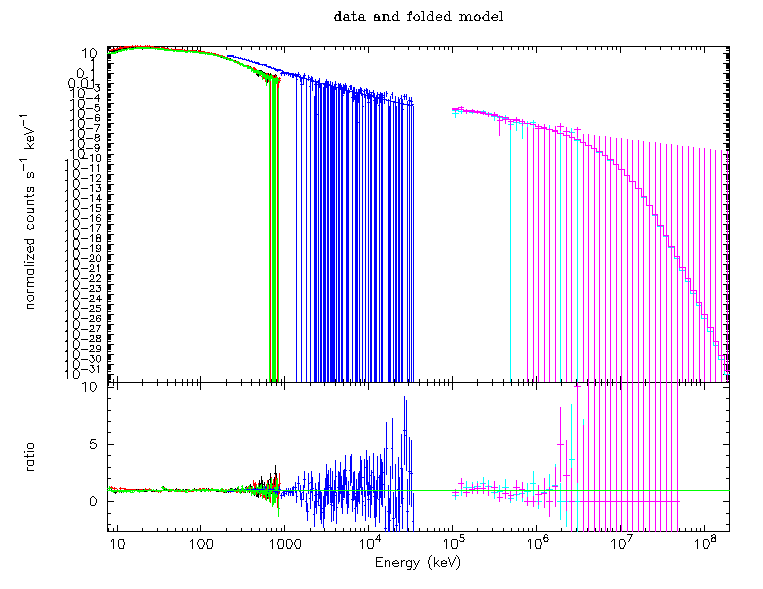} &
\includegraphics[ width=.48\linewidth, keepaspectratio]{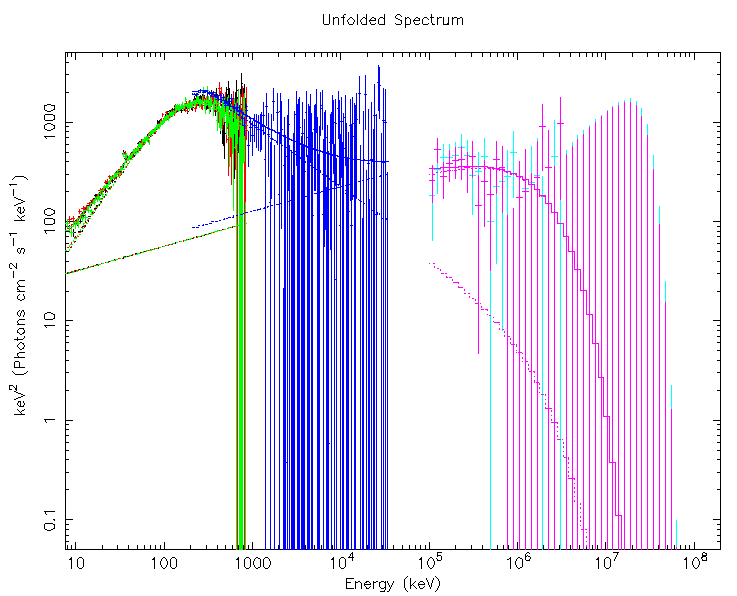} \\
\end{array} $
\end{center}
\caption{Folded and unfolded spectrum for the Band, Band+PL and Band+CUTPL in time-integrated spectrum of GRB 090926}
\label{band_pl_cutoff}
\end{figure}

The fit of the data with the Band plus broken power-law (Band+BPL) model has a significance very close to that of the (Band+CUTPL) model --- the
$\Delta$(C-STAT) between the two models is only 6 --- so 
that we cannot distinguish between them although each of these functions lead to a different interpretation of the data (e.g., see section \ref{interpretation}, and ~\cite{Baring:06, Granot:08}). The broken power-law (BPL) is modeled as

\begin{equation}
\label{eq:bknpo}
f(E)~=~%
\left\{
\begin{array}{clc}
A\left(\frac{E}{E_{\rm piv}}\right)^{\lambda_l}& & {\rm for}\;E\leq E_{\rm break} \\
A\left(\frac{E_{\rm break}}{E_{\rm piv}}\right)^{\lambda_l}&\left(\frac{E}{E_{\rm break}}\right)^{\lambda_h} & {\rm for}\;E>E_{\rm break}
\end{array}
\right\}
\end{equation}
where $\lambda_l$ and $\lambda_h$ are lower and higher power-law photon
indices, $E_{\rm piv}$ is a pivot energy fixed to 1~GeV, $E_{\rm break}$ is the
break energy. For this (Band+BPL) fit, we found a break energy
$E_{\rm break} = 219_{-56} ^{+65}$~MeV and a photon index above $E_{\rm break}$ of 
$\lambda_h=-2.47_{-0.17} ^{+0.14}$. The parameters of the Band function
change little from one fit to another.
We show in figure \ref{0926_nufnu} shows the time-integrated $\nu F\nu$ spectrum for the two best fit: Band + CUTPL and Band +BPL.

We tested the significance of the cutoff via a model comparison test, by differencing C-STAT for the two models, (Band+PL) and (Band+CUTPL).
This is a likelihood ratio test (LRT) since C-STAT is equal to twice the log of the likelihood. It is conventional in astrophysics to calculate the significance of the test by using the $\chi^2$ distribution to convert the LRT value, or C-STAT difference in this case, to a probability. However, certain assumptions are required for the validity of this calculation. For the highest reliability we conducted simulations, creating $2 \times 10^4$ random realizations of the null-hypothesis 
(parameters of (Band+PL) fixed at the best fit values found in the data) and fitting with both models, (Band+PL) and (Band+CUTPL).
We tested for the presence of a statistical-fluctuation exponential cutoff
in the simulations by fitting the simulated counts data with both models and differencing the C-STAT values. In $2 \times 10^4$ simulations the largest C-STAT difference obtained was 16.7, which is much smaller than the value of 39.5 obtained with the actual data (last line of table ~\ref{tab:spec}). We therefore place a firm upper-limit on the probability that the exponential cutoff is a chance probability of $5 \times 10^{-5}$, for a significance that the cutoff is real of at least 99.995\% (Gaussian equivalent 4.05$\sigma$).

\begin{figure}
\begin{center}
\includegraphics[angle=90, width=.6\linewidth, keepaspectratio]{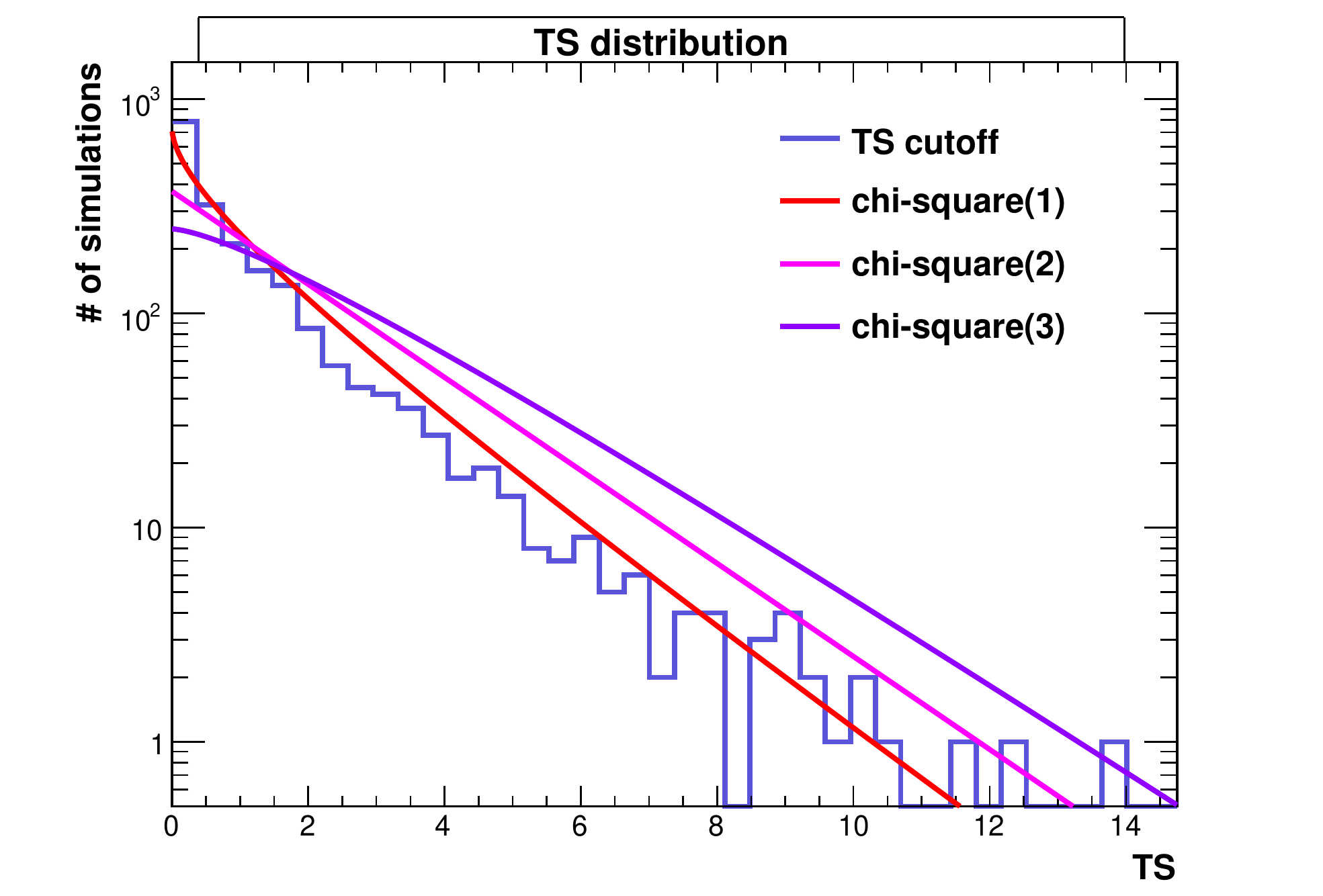}
\end{center}
\caption{TS distribution from simulation of null-hypothesis (BAND+PL) on the time-integrated spectrum of GRB 090926A}
\label{ts 0926}
\end{figure}

According to Wilks' Theorem, the $\Delta$(C-STAT) values should be asymptotically
distributed according to $\chi^2$ of one degree of freedom. The $\Delta$(C-STAT) distribution plotted in figure \ref{ts 0926} from the simulation exhibited a slight excess
at large values -- perhaps the asymptotic distribution has not been reached for those values. Consequently we do not evaluate the significance according to the conventional procedure of using the observed $\Delta$(C-STAT) value of 39.5 and the $\chi^2$ distribution. The number of simulations to determine the significance of the cutoff rather than a limit is prohibitive. The large gap between the largest $\Delta$(C-STAT) value obtained in the simulations, 16.7, and the value of 39.5 for the actual data, suggests that the significance is much better than 99.995\%. 
For the 4 different sets of instrument response functions, we always found the $\Delta$(C-STAT) to be greater than 32. The significance of the spectral cutoff will be hereafter quoted as $>4\,\sigma$.

Using the fit results for the best model (Band+CUTPL), 
we estimate a fluence of 2.07$\pm 0.04 \times 10^{-4}\,{\rm erg}\,{\rm cm}^{-2}$  (10~keV–-10~GeV) from \Tz{3.3} to \Tz{21.6}.
%\Tz{3.328} to \Tz{21.632}.
These data give an isotropic energy
$\mathcal{E}_{\gamma, iso}$ = 2.24 $\pm 0.04 \times 10^{54}$~erg,
comparable to that of GRB~090902B.

\begin{table}
\begin{small}
\caption{Summary of GBM/LAT joint spectral fitting between \Tz{3.3} and 
\Tz{21.6}. The flux range covered by both instruments is 10~keV -- 10~GeV.}
\label{tab:spec}
\begin{center}
\begin{tabular}{p{5cm}cccc}
\hline
Fitting model & Band  & Band+PL &  Band+CUTPL \\ \hline
\multicolumn{4}{l}{\it Band function}\\
A ($\gamma$ cm$^{-2}$ s$^{-1}$ keV$^{-1}$) 
                      & 0.176 +/- 0.002 & 0.173 +/- 0.003   &0.170   $_{-0.004}^{+0.001}$ \\
$E_{\rm{peak}}$ (keV) &   249 +/- 3      & 256 +/-   4      & 259  $_{-2}^{+8}$     \\
$\alpha$ (index 1)    & -0.71 +/- 0.01   & -0.62 +/- 0.03   &-0.64 $_{-0.09}^{+0.02}$  \\
$\beta$ (index 2)     & -2.30 +/- 0.01   &-2.59 $_{-0.05}^{+0.04}$&-2.63 $_{-0.12}^{+0.02}$  \\\hline

\multicolumn{4}{l}{\it Power-law}\\
A ($10^{-10}\;\gamma$ cm$^{-2}$ s$^{-1}$ keV$^{-1}$)&-&3.17 $_{-0.33}^{+0.35}$&5.80  $_{-0.60}^{+0.81}$ \\
$\lambda$ (index)                                           &-& -1.79 +/- 0.02        &-1.72 $_{-0.02}^{+0.10}$ \\ 
$E_{\rm piv}$                                               &-& 1~GeV (fixed)           & 1~GeV (fixed) \\ \hline
\multicolumn{4}{l}{\it High-energy cutoff}\\
$E_F$ (GeV)                    & -                 & -                    & 1.41 $_{-0.42}^{+0.22}$   \\ \hline

\multicolumn{4}{l}{\it Effective area correction}\\
BGO/LAT (back)   &  0.79 (fixed)  &  0.79 (fixed)  &   0.79 (fixed)    \\
\hline
Flux ($\gamma$ cm$^{-2}$ s$^{-1}$)      &  42.2+/-0.1  &  43.5+/-0.3 &  43.3+/-0.2  \\
Flux ($10^{-5}$ erg~cm$^{-2}$ s$^{-1}$) &  1.18+/-0.01 &  1.15+/-0.02 & 1.13+/-0.02  \\
\hline
C-STAT / DOF     &  1395.1 / 579 & 1287.8 / 577 & 1247.3 / 576 \\
\hline
\end{tabular}
\end{center}
\end{small}
\end{table}

\vspace{0.7cm}
\noindent
{\bf GRB 080825C}
\vspace{0.3cm}

\begin{figure}
\begin{center}
\includegraphics[width=.6\linewidth, keepaspectratio]{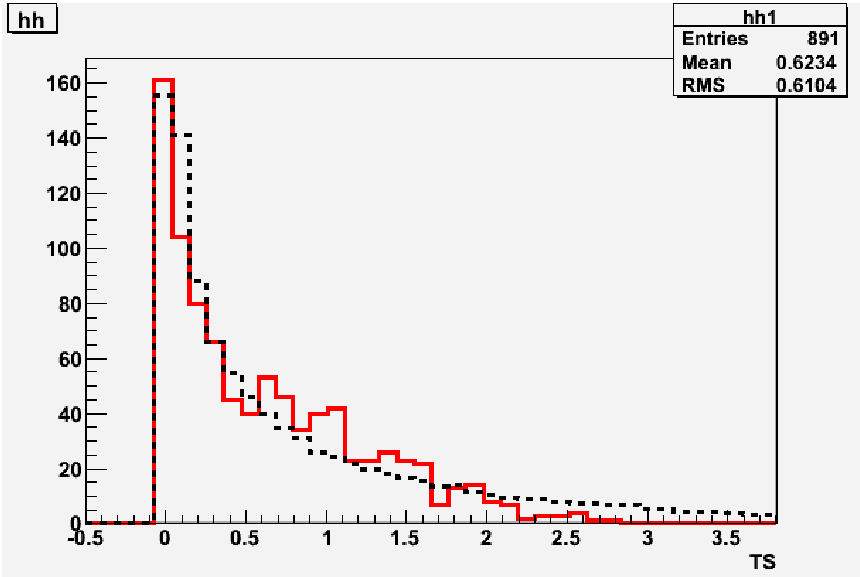}
\end{center}
\caption{TS distribution from simulation of null-hypothesis (Band function) on time bin (a) of GRB 080825C}
\label{ts 0825c}
\end{figure}

GRB 080825C (see lightcurve in figure \ref{lc_0825C}) is the first GRB significantly detected by the LAT instrument.
Following a similar procedure as described for GRB 090926A, we found an exponential cutoff in time bin (a) with a significance of $4.3\,\sigma$ and with a cutoff energy of $E_{\rm{cutoff}} = 1.77^{+1.59}_{-0.56}$ MeV (with following Band function parameters: $\alpha \sim -0.57$, $\beta \sim -1.64$, $E_{peak} \sim 211$ keV). 
We performed Monte-Carlo simulation similar to what was done with GRB 090926A. The Test-Statistics distribution is plotted in figure \ref{ts 0825c} and a $\chi^2_{1 dof}$ function was found to be a good fit to this distribution and was therefore used to derive the above significance.
We investigated the dependence of this significance with the estimated systematics
of the various instruments (see next section) and found the strongest effect to be a $\pm\,15\%$
variation in the BGO effective area which can bring the significance
down to $\sim$$3.7\,\sigma$. With 5 time bins, this is not strong
enough to claim the existence of an exponential cutoff.

\subsection{Investigation of systematics}
\label{sys}

Accurate estimate of the uncertainties in a measurement is as important as the measurement itself. Statistical uncertainties are estimated using appropriate statistical techniques based on the poissonian nature of the background and signal measured. Any other sources of uncertainty are refered to as 'systematic' uncertainties and there estimate is usually less straightforward. 
Systematic errors mostly come from uncertainties in the performance of the instrument, as well as inaccurate or even wrong assumptions made during your analysis. Therefore a reliable estimation of systematic uncertainties can be achieved via an honest evaluation of the degree of knowledge of the instrument used and of the assumptions made throughout the study. Once a list of all the possible uncertainties (besides statistical uncertainties) have been established, one needs to estimate how these uncertainties propagate to the specific measurement that was made. This is usually done by performing the exact same analysis with different instrument description or assumptions, both slightly modified by an amount deemed 'reasonable'. 'Reasonable' here means that is not very unlikely and $1 \sigma$ uncertainties are usually chosen. Each systematic effect is usually investigated separately as even for reasonable amount of systematic uncertainties, the number of combination to analyze rapidly becomes very large and difficult to carry out since the automatization of these investigations is usually not possible.

We list below the instrumental uncertainties that were investigated during the $Fermi$ GRB analysis as well as a 'reasonable' estimate of the level of each of these uncdertainties:

\begin{itemize}

\item {\bf LAT effective area:} the LAT team performed detailed studies of the systematic uncertainties associated with the instrument effective area. This uncertainty was found to be strongly energy dependent. Because a positive fluctuation of the effective area at low and high energy would have a different effect on the spectral index of a source than a positive fluctuation at low energy and a negative fluctuation at high energy, it was decided to create a set of 4 modified Instrument Response Functions which encompass the worst case scenarios in terms of overall normalization and slope of the effective area corrections. 
Systematics on the LAT effective area have
been derived from a study of the Vela pulsar and Earth albedo photons
because it is possible in both cases to extract a $\g$-ray sample extremely pure. Using this sample, the uncertainties on the LAT
effective area have been found to be: 10\%
below 100 MeV, 5\% around 1 GeV and 20\% above 10 GeV. 

\item {\bf Pile-up events:} While examining downlinked events it became clear that some unexpected interactions between background and $\g$-ray events happened, due to the time evolution of the
energy deposition in the detector, the timing of the electronics and of the trigger system, and the details of the reconstruction analysis. This was not observed in Monte
Carlo simulation as each event is generated independently and interactions between subsequent events is not possible. This caused a decrease in the LAT efficiency, as
some valid events are rejected due to pile-up
signals. Furthermore the decrease in efficiency is correlated with the instrument raw trigger rate. 
Corrections were applied to the reconstruction and event analysis routines to recover this loss in order to to use the Monte Carlo simulation to correctly estimate the LAT performance and reduce the systematics
affecting our science analysis. However some systematics remain especially the dependence with trigger rate mentioned above and we computed its effect into all our GRB studies.

\item {\bf LAT energy scale:} the uncertainty in energy measurement for the LAT is estimated to
be of the order of 5\% from our Monte-Carlo simulations. Since this is not included into the description of the LAT instrument while performing GRB analysis, we properly investigated its effect on our measurements.

\item {\bf NaI \& BGO effective area:} We adopted a 10\% uncertainty in the NaI and
BGO effective area (both overall normalization and slope) [provided by the GBM team].

\item {\bf NaI \& BGO background estimation:} we also considered the uncertainty in the choice of off-timing sample for NaI and BGO background subtraction and found the corresponding systematics to be negligible [provided by the GBM team].

\end{itemize} 

We used GRB 080825C as a test case to evaluate how the various systematic uncertainties propagate to the Band spectral parameters.
In each time bin, the error values on
the spectral parameters are found to be similar to or smaller than the
statistical uncertainties, except
for the case of time bin (a) and (b) where the systematics on $E_{\rm
peak}$ and the normalization are found to be about two times and three
times larger, respectively. Table~\ref{tab:sys} reports the
systematics found for each parameter of the time-integrated best fit
Band function of GRB 080825C as well as the
predominant systematic effects.  We note that similar trends for the
systematic uncertainties were found for GRB~080916C,
which had many more LAT events than GRB~080825C.

\begin{table}[htbp]
\begin{center}
\renewcommand{\arraystretch}{1.2}
\begin{tabular}{|p{2.5cm}|p{2.1cm}|p{2.1cm}|p{2.1cm}|p{2.5cm}|} 
\hline
 & Norm & $\alpha$ & $\beta$ & $E_{\rm{peak}}$ (keV) \\
\hline
systematic error & $\pm 15 \%$ & $\pm 0.03$ & $\pm 0.03$ & $\pm 8$ \\
\hline
predominant systematics & NaI eff. area & NaI eff. area & BGO eff. area & NaI \& BGO eff. area \\
\hline
\end{tabular}
\caption{Systematic uncertainties derived for GRB 080825C as well as the predominant effect for each spectral parameter of the time-integrated best fit Band function.}
\label{tab:sys}
\end{center}
\end{table}

\section{Discussion}
\label{discussion}

\subsection{Possible theoretical interpretations of LAT GRB features}
\label{interpretation}

$Fermi$ instruments are discovering multiple new features so far unobserved in the GRB prompt emission and all these features are important clues to unravel the underlying physics behind the GRB prompt emission. We now discuss some of the possible theoretical interpretations of the GRB features observed by $Fermi$.

A delayed onset of the high-energy emission ($>100$ MeV) relative to the low energy
emission ( $< 1$ MeV) appears to be a very common feature in LAT
GRBs. It clearly appears in all 4 of the particularly bright LAT GRBs,
while in dimmer LAT GRBs it is often inconclusive due to poor photon
statistics near the onset time (GRB 090217 being the only case where the data is inconsistent with such feature \cite{Abdo_090217:10}). Besides, the time delay appears to scale with the
duration of the GRB (e.g., it is several seconds in the long GRBs 080916C
and 090902B, while being only $\sim 0.1 - 0.2$ s in the short GRBs 090510).
A leptonic scenario producing the high-energy emission via inverse Compton scattering would have a hard time to produce the delayed onset on a timescale larger than that of individual spikes in the lightcurve (a delay up to this timescale might be due to the
build-up of the seed synchrotron photon field in the emitting region over
the dynamical time).
A hadronic model could explain the late onset with the time it takes to accelerate
protons (or heavier ions) to very high energies (at which they loose
a considerable fraction of their energy on a dynamical time, e.g. via proton
synchrotron \cite{Razzaque:09}, in order to obtain a reasonable radiative efficiency).
If the observed high-energy emission (and in particular the distinct high-energy
spectral component) also involves pair cascades (e.g. inverse-Compton emission
by secondary $e^{\pm}$ pairs \cite{Asano:09} produced through cascades initiated by
photo-hadronic interactions) then it might take some additional time for
such cascades to develop. However, this requires the high-energy emission
to originate from the same physical region for a time of order of the high energy delay which is much larger than the typical variability time scale.
The sharp spikes present in the light curves are a definitive challenge as it is hard
to make the same emission region turn on and off so abruptly. Moreover, the fact that the first spikes are missing while later spikes coincides is problematic in this picture as we would expect a similar 'high energy delay' for those later spikes which is not observed.

Only 3 LAT GRBs so far have shown clear ($>5 \sigma$) evidence for a distinct
spectral component. However, these GRBs are the 3 brightest in the LAT range,
while the next brightest GRB in the LAT (GRB 080916C) showed a hint for
an excess at high energies. This suggests that such a distinct high-energy
spectral component is probably very common, although its clear detection is only possible in particularly bright LAT GRBs since a large
number of LAT photons is needed in order to detect it with $>5 \sigma$ significance.
The distinct spectral component is usually well fit by a hard powerlaw
that dominates at high energies. 
In a leptonic scenario, the high-energy spectral component might be inverse-Compton
emission, and in particular synchrotron-self Compton (SSC) if the usual
Band component is synchrotron. However, the gradual increase in the photon index
of the high-energy power-law component is not naturally expected in such
a model, and the fact that it is different than the Band low-energy photon
index as well as the excess flux (above the Band component) at low energies
are hard to account for in this type of model.
GRB 090902B shows for the first time clear evidence of excess
emission both at low energies ( $\lesssim 50$ keV) and at high energies ($>100$ MeV), while the Band
function alone fits data at intermediate energies adequately (GRB 090510 also shows a hint of such excess at low energy). These excesses are well-fit by
a single power-law component suggesting a common origin. This power-law component
accounts for 24\% of the total 
uence in the 10 keV - 10 GeV range, and its photon index
is hard, with a value $\sim -1.9$ throughout most of the prompt phase. Such a hard component
producing the observed excess at low energies is difficult to explain in the context of leptonic
models by the usual synchrotron self-Compton (SSC) mechanisms.
In the simplest versions of these models, the peak of the SSC emission is expected
to have a much higher energy than the synchrotron peak at MeV energies, and the SSC
component has a soft tail that is well below the synchrotron flux at lower energies and so
would not produce excess emission below 50 keV. However, theoretical extensions which include additional processes such as the photospheric component are promising and might lead to a resolution of this issue in the leptonic scenario \cite{Ryde:10,Toma:10}. Hadronic models, either in the form of
proton synchrotron radiation \citep{Razzaque:09} or photohadronic interactions \citep{Abdo_090902B:09,Asano:09}, can produce a hard component with a similar low energy excess via direct and
cascade radiation (e.g., synchrotron emission by secondary pairs at low energies). However,
the total energy release in hadronic models would exceed the observed gamma-ray energy
significantly and may pose a challenge for the total energy budget.
For GRB 090510 a hadronic model requires a total isotropic
equivalent energy $> 10^2$ times larger than that observed in gamma-rays \cite{Asano:09},
which is a serious challenge for the progenitor of this short GRB.
Collimation into a narrow jet may alleviate the energy requirements, since the actual
energy release can be smaller by a jet beaming factor $> 1/\Gamma_0^2$ from the
apparent isotropic value, where $\Gamma_0$ is the bulk Lorentz factor of the fireball.
A further difficulty for hadronic models is the gradual increase in $\beta$ which is not naturally expected, though it might be mimicked by a time-evolution of a high-energy Band-like spectral component
\cite{Razzaque:09}. 
Altogether, both leptonic and hadronic models still face many challenges, and do not yet
naturally account for all of the Fermi observations.

Finally, the delayed extra component could be emitted from a forward shock that propagates into
the external medium \cite{Meszaros:97,Sari:98}, while the Band component
would have a separate origin. The delay timescale of the extra spectral component
would correspond to the time needed for the forward shock to sweep up material and
brighten \cite{Kumar:09,Ghisellini:10,Razzaque:09}. The forward shock radius in an external shock model is $R_f \sim \Gamma_f^2 c T_{dur} / (1+z)$, where $\Gamma_f$ is the Lorentz factor of the forward shock and $T_{dur}$ is the duration of the burst. 
One major issue with such model though is to explain the rapid variability observed in GRB light curves at various energies. Indeed the external medium needs to be clumpy on length scale
$\sim \Gamma_f c \Delta T / (1+z) \sim 10^{12} (\Gamma_f/10^3) (\Delta T/10^2 \mbox{ s)}$ cm \cite{Dermer:99,Dermer:08}.
This is based on interactions between a very thin shell, prior to
the onset of the self-similar expansion phase, and the external medium with very small
scale clumps. If the extra component is synchrotron emission from the forward shock, then
the synchronization of the pulse peak times of the Band and extra component requires an
explanation. Another possibility is that the extra component arises from inverse Compton
(IC) scattering of the radiation of the Band component by the high-energy electrons in the
forward shock.

Another remarkable feature is the spectral break (or cutoff) of the extra-component that has been measured in GRB 090926A for the whole time range of the prompt emission and for time interval (c) with a high significance. This cutoff may be due to pair production ($\g \g \rightarrow e^+ e^-$) within the emitting region, although we cannot rule out the possibility that there is an intrinsic spectral break related to the energy distribution of the emitting particles or the emission mechanism (e.g., IC scattering with
Klein-Nishina effect). The absorption on the extragalactic background light (EBL) cannot
cause this spectral feature since the opacity at the observed break energy for the redshift of
GRB 090926A is very small for practically all EBL models.
If one presumes that this spectral break is due to pair production attenuation, one can get a direct measurement of the bulk Lorentz factor.
Although the instantaneous emission from a thin shell exhibits a photon spectrum like $f \propto E^{\lambda} exp(-\tau_{\g \g}(E))$, the shape of the time-averaged spectrum of a single pulse may depend on the details of the emission mechanism \cite{Baring:06,Granot:08} which performed a fully time-dependent and self-consistent semi-analytic calculation featuring emission from a very thin spherical shell over a finite range of radii).
Using the former description, we get a bulk Lorentz factor of
$\Gamma \sim 700$, when the more elaborate time-dependent model gives $\Gamma \sim 200$. The motivation to provide both estimation is that in some
models the high-energy photons are expected to be emitted from the bulk
of the shocked region, rather than from a thin cooling layer behind the shock front, in which
case such an intermediate value of the opacity might be expected (see \cite{Abdo_090926A:10} for further explanations).
We now turn to the constraints that could be set using the highest energy emission observed during the prompt emission of other LAT GRBs.

\subsection{Constraints on the bulk Lorentz factor of the jet}
\label{bulk_lorentz_factor}

Short time scale variability and brightness (isotropic equivalent luminosities: $L \sim 10^{50} - 10^{53}$ erg.s$^{-1}$) of GRBs are strong evidence of the ultra-relativistic nature of the outflow where the $\g$-ray emission is produced (bulk lorentz factor $\Gamma >> 1$). Indeed without such relativistic corrections, the plasma would have huge optical depth to pair production ($\g \g \rightarrow e^+ e^-$), which would thermalize the spectrum and thus be at odds with the observed non-thermal spectrum
This ÒcompactnessÓ argument has been used to derive lower-limits, $\Gamma_{min}$, on the
value of $\Gamma$, which were typically $\sim 10^2$ and in some cases as high as a few
hundred (see \cite{Lithwick:01} and references therein). However, the photons that provided
the opacity for these limits were well above the observed energy range,
so there was no direct evidence that they actually existed in the first place.
With Fermi, however, we adopt a more conservative approach of relying
only on photons within the observed energy range. 

Assuming the highest energy observed from a GRB is $E_{max}$ and that this highest energy photon interacts with a powerlaw component  $f(E) = A \times E^{\beta}$ [photons/cm$^2$/keV] (might it be the high energy additional component or the high energy tail of the Band function), the total photon number above $E_0$ is given by:

$$N_{tot} = \frac{4 \pi d_{L}(z)^2}{(1+z)^2} \int^{\infty}_{E_0} dE f(E) = \frac{4 \pi d_{L}(z)^2 f(E_0)}{(1+z)^2 (-\beta-1)} E_0$$

where $d_L (z)$ is the luminosity distance. Defining the emission radius R and the shell width in
the comoving frame $W^{\prime}$, the shell volume becomes $4 \pi R^2 W^{\prime}$. Then, we compute the photon distribution in the shell rest frame:

$$n^{\prime}_{\g}(E^{\prime}) = \Big(\frac{d_L(z)}{R}\Big)^2 \frac{\Gamma f(E_0)}{(1+z)^3 W^{\prime}} \Big(\frac{E^{\prime}}{E^{\prime}_0}\Big)^{\beta}, \mbox{[photons/cm$^3$/keV]}$$

for $E^{\prime} > E^{\prime}_0$, where $E^{\prime}_0 = (1+z)E_0/\Gamma$. We consider photons of energy $E^{\prime}_{max} = (1+z)E_{max}/\Gamma$.
The photoabsorption optical depth of high energy $\g$-rays from lower energy photons emitted
cospatially in the shell is given by

$$\tau_{\g \g}(E^{\prime}_{max}) = \int d\Omega \int_{E^{\prime}_0}^{\infty} dE^{\prime} \frac{n_{\g}^{\prime} (E^{\prime})}{4 \pi} \sigma_{\g \g}(E^{\prime}_{max},E^{\prime},\theta^{\prime}) (1-cos(\theta^{\prime})) W^{\prime}$$

where $\theta^{\prime}$ is the incident angle. The absorption cross section is

$$ \sigma_{\g \g}(y) = \sigma_T g(y), g(y) = \frac{3}{16} (1-y^2) [(3-y^4) ln\frac{1+y}{1-y} - 2y (2-y^2)]$$

where $y^2 = 1-2 (m_e c^2)^2 / [E^{\prime}_{max} E^{\prime} (1-cos(\theta^{\prime}))]$. Then we obtain

$$\tau_{\g \g}(E^{\prime}_{max}) = \sigma_T \Big(\frac{d_L(z)}{R}\Big)^2 \frac{\Gamma E^{\prime}_0 f(E_0)}{(1+z)^3} \Big(\frac{E^{\prime}_{max} E^{\prime}_0}{m_e^2 c^4}\Big)^{-\beta-1} F(\beta)$$

where the dimensionless function $F(\beta) = \frac{4}{1-\beta} \int_0^1 dy (1-y^2)^{-\beta-2} g(y) y \sim 0.1$ for $-2.0> \beta > -2.3$\cite{Asano:03}. Adopting $R = c \Delta t \Gamma^2 /(1+z)$,

$$\tau_{\g \g}(E_{max}) = \sigma_T \Big(\frac{d_L(z)}{c \Delta T}\Big)^2  E_0 f(E_0) (1+z)^{-2(\beta+1)} \Gamma^{2(\beta-1)} \Big(\frac{E_{max} E_0}{m_e^2 c^4}\Big)^{-\beta-1} F(\beta)$$

The fact that the high energy spectrum shows no cutoff implies that the $\g \g$ optical depth of the
maximum observed photon is $\tau_{\g \g} (Emax) < 1$. This allows the minimum Lorentz factor to be
estimated by

$$ \Gamma > \Gamma_{min} = \Big[ \sigma_T \Big(\frac{d_L(z)}{c \Delta T}\Big)^2 E_0 f(E_0) F(\beta) \Big]^{\frac{1}{2(1-\beta)}} (1+z)^{\frac{\beta+1}{1-\beta}} \Big(\frac{E_{max} E_0}{m_e^2 c^4}\Big)^{\frac{\beta+1}{2(\beta-1)}}$$

The main uncertainty in deriving $\Gamma_{min}$ is usually the exact choice for
the variability time. Other uncertainties arise from those on the spectral
fit parameters, or on the degree of space-time overlap between the high-energy
photon and lower energy target photons, in cases where there is
more than one spectral component without conclusive temporal correlation
between their respective light curves. Finally, the fact that our limits rely
on a single high-energy photon also induces an uncertainty, as it might still
escape from an optical depth of up to a few. However, in most cases the
second or third highest-energy photons help to relax the effect this has on
$\Gamma_{min}$ (as the probability that multiple photons escape from $\tau_{\g \g} > 1$ rapidly
decreases with the number of photons). 

\begin{figure}
\begin{center}
\includegraphics[ width=.7\linewidth, keepaspectratio]{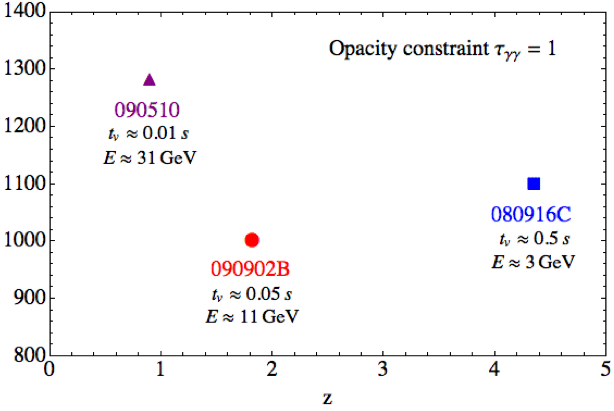}
\end{center}
\caption{Lower limit on the bulk Lorentz factor using GRB 080916C, GRB 090510, GRB 090902B. The energy of the highest energy photon clearly associated with the burst and the time variability $t_{\nu}$ used is indicated for each GRB. We note that more realistic computations \cite{Granot:08}, which account for geometrical and dynamical effects, can lead to smaller opacities and thus to smaller lower limits (by a factor of $\sim 3$) on $\Gamma$.}
\label{lorentz_factor_limit}
\end{figure}

Following the above recipe, we have derived $\Gamma_{min}$ values for three of the brightest LAT GRBs: $\Gamma_{min} \approx 900$ for
GRB 080916C \cite{Abdo_080916C:09}, $\Gamma_{min} \approx 1200$ for GRB 090510 \cite{Ackermann_090510:10}, and $\Gamma_{min} \approx 1000$ for GRB 090902B \cite{Abdo_090902B:09}. This shows that short GRBs (GRB 090510) seem to be
as highly relativistic as long GRBs (GRB 080916C and GRB 090902B). 
Figure \ref{lorentz_factor_limit} shows these limits as a function of the different redshifts measured for these bursts.
Since our highest values of
$\Gamma_{min}$ are derived for the brightest LAT GRBs, they are susceptible to strong
selection effects. It might be that GRBs with higher $\Gamma$ tend to be brighter
in the LAT energy range (e.g. by avoiding intrinsic pair production \cite{Ghisellini:10}).
We also note that more realistic computations \cite{Granot:08}, which account for geometrical and dynamical effects, can lead to smaller opacities and thus to smaller lower limits (by a factor of $\sim 3$) on $\Gamma$.

        	\chapter{Limits on Lorentz Invariance Violation}
\label{lorentz}

A cornerstone of EinsteinÕs theory of special relativity is the concept of Lorentz invariance - the postulate that all observers measure exactly the same speed of light in vacuum, independent of photon-energy. While special relativity does not introduce any fundamental length or energy scales, modern quantum field theory requires several fundamental constants, one of which is the Planck scale, expressed in spatial distance $l_{Planck} \simeq1.62 \times 10^{-33} cm$ or in energy scale $E_{Planck} = M_{Planck} c^2 \simeq 1.22 \times 10^{19}$ GeV, where quantum-effects are expected to strongly affect the nature of space-time. Quantum gravity theory is one of such particle theory in which all four forces (gravity, electromagnetism, weak and strong force) are treated in a unified framework.

There is great interest in the (not yet validated) idea that, in a realization of the quantum gravity theory, Lorentz invariance might break near the Planck scale.  A key test of such violation of Lorentz invariance is possible variation of speed of light with energy \citep{Wheeler:98,Amelino:98,Mattingly:05,Kostelecky:08,Amelino:09}. Even a tiny variation in photon-speed, when accumulated over cosmological light-travel times, may be revealed by studying sharp features in gamma-ray burst light-curves. 

We here focus on measurement of possible time lag between the GRB onset and the arrival of very high energy photons such as a $\sim 13$ GeV photon associated with GRB 080916C and a $\sim 31$ GeV from the distant ($z \sim 0.9$) and short GRB 090510 which help us set the significant constraints on Lorentz Invariance Violation (LIV). 
In summary, we find no evidence for violation of Lorentz invariance, and place a conservative lower limit of $1.2 \times E_{Planck}$ on the energy-scale (or an upper-limit of $l_{Planck}/1.2$ on the length-scale) of a linear energy-dependence (or inverse-linear wavelength-dependence), subject to reasonable assumptions about the emission. Our results support Lorentz invariance, and disfavour a class of quantum-gravity theories \citep{Mattingly:05,Jacobson:06,Amelino:08} in which a postulated granularity of space-time on a very small scale alters the speed of light, and predict that the velocity of light depends linearly on the photon-energy.

By placing an upper limit on the observed time delay between arrival time of simultaneously emitted photons, one can put a lower limit on the quantum gravity mass which is related to a scale factor in the power series expansion of the photon velocity $v_{ph}$.
Gamma-Ray Bursts are excellent sources to perform such study as the onset of their emission is usually sharply defined in time and can therefore be used as the earliest time at which any photon was emitted (detection might be delayed however due to quantum gravity effects).
Naturally, high energy photons detected soon after the GRB onset are the best candidates to put significant constraints on LIV.
In the following analysis, we make use of a $\sim 13$ GeV event associated with GRB 080916C and a $\sim 31$ GeV event associated with GRB 090510.
Of course the association of the high energy event with the GRB is of utmost importance as the basic assumption in our analysis is that this high energy event detected by the LAT was indeed emitted by the GRB no earlier than the emission onset.

Section \ref{method_LIV} describes the methodology used to derive our constraints on Lorentz Invariance Violation. Section \ref{characterization_LATevts} gives a detailed description of analyses used to characterize the highest energy LAT events, establish their chance association with the GRB and their energy uncertainty which is crucial to compute reliable limits on the quantum gravity parameter. Finally, section \ref{LIV_results} summarizes our results concerning Lorentz Invariance Violation that were obtained with GRB 080916C and GRB 090510.

\section{Method}
\label{method_LIV}

Some quantum gravity models \citep{Mattingly:05,Jacobson:06,Amelino:08} allow violation of Lorentz invariance, and in
particular allow the photon propagation speed, $v_{ph}$, to depend on its energy,
$E_{ph}$:

\begin{equation}
v_{ph}(E_{ph}) \neq c \mbox{, where: } c = \lim_{Eph \rightarrow 0} v_{ph}(E_{ph})
\end{equation}

The Lorentz invariance violating (LIV) part in the dependence of the photon momentum, $p_{ph}$, on
its energy, $E_{ph}$, can be expressed as a power series,

\begin{equation}
\frac{p_{ph}^2 c^2}{E_{ph}^2} - 1 = \sum_{k=1}^{\infty} s_k \left(\frac{E_{ph}}{M_{QG,k} c^2} \right)^k
\end{equation}

in the ratio of $E_{ph}$ and a typical energy scale $M_{QG,k} c^2$ for the $k^{th}$ order, which
is expected to be up to the order of the Planck scale, $M_{Planck} = (\hbar c/G)^{1/2} \sim 1.22 \times 10^{19} GeV/c^2$, where $s_k \in \{-1, 0, 1\}$. Since we observe photons of
energy well below the Planck scale, the dominant LIV term is associated
with the lowest order non-zero term in the sum, of order $n = min\{k|s_k  \neq 0\}$,
which is usually assumed to be either linear (n = 1) or quadratic (n = 2).
The photon propagation speed is given by the corresponding group velocity,

\begin{equation}
v_{ph} = \frac{\partial E_{ph}}{\partial v_{ph}} \approx c \left[ 1 - s_n \frac{n+1}{2} \left( \frac{E_{ph}}{M_{QG,n} c^2}\right)^n \right]
\end{equation}

Note that $s_n = 1$ corresponds to the sub-luminal case ($v_{ph} < c$ and a positive
time delay), while $s_n = -1$ corresponds to the super-luminal case ($v_{ph} > c$
and a negative time delay). Taking into account cosmological effects, this
induces a time delay (or lag) in the arrival of a high-energy photon of energy
$E_h$, compared to a low-energy photon of energy $E_l$ (emitted simultaneously
at the same location), of

\begin{equation}
%\begin{displaymath}
\Delta t = s_n \frac{(1+n) (E_h^n -E_l^n)}{2 H_0 (M_{QG,n} c^2)^n} \int_0^z \frac{(1+z^{\prime})^n}{\sqrt{\Omega_m (1+z^{\prime})^3 + \Omega_{\Lambda}}} dz^{\prime}
\label{deltat}
%\end{displaymath}
\end{equation}

\section{Characterization and energy uncertainty of LAT events}
\label{characterization_LATevts}

This section will provide detailed description of the analysis performed in order to establish the significant association of individual high energy events with the GRB with which they were simultaneously detected - we note that similar analysis was performed on the high energy events used to put a lower limit on the bulk Lorentz factor of the jet (see section \ref{bulk_lorentz_factor}).
Section \ref{evt_char} look at the characterization of the event in the detector. Section \ref{assoc} computes the chance probability for this event to originate from a background fluctuation (and therefore not be associated with the GRB). Finally section \ref{energy} evaluates the best energy and energy uncertainty of the high energy event.
We will use the $\sim 31$ GeV event detected during GRB 090510 prompt emission to illustrate the analysis we performed. Similar analysis were performed each time detailed information on a high energy event were needed for particular analyses such as LIV analysis and constraints on bulk Lorentz factor.

The highest energy event detected by the LAT during the prompt emission period and in the vicinity of the GRB 090510 location is a $\sim31$ GeV event which was observed $\sim0.829s$ after the GBM trigger. This event is particularly interesting because it could set very stringent limits on the Quantum Gravity mass scale and the bulk Lorentz factor. It is therefore of great interest to have a closer look to this particular event, evaluate the probability of this event being a photon coming from GRB 090510 and have a precise estimate of its best energy and its energy uncertainty.

\subsection{Characteristics of high energy events}
\label{evt_char}

\begin{figure}
\begin{center} $
\begin{array}{cc}
\includegraphics[ width=.47\linewidth, keepaspectratio]{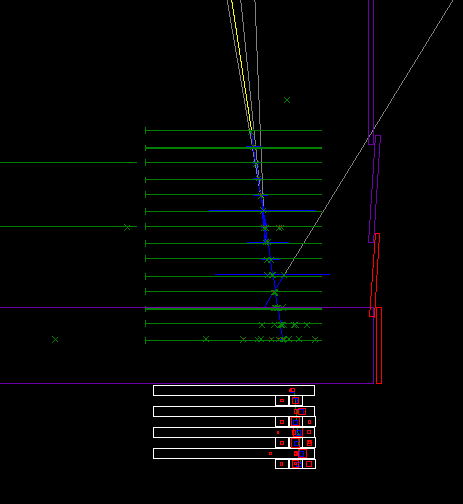} &
\includegraphics[ width=.49\linewidth, keepaspectratio]{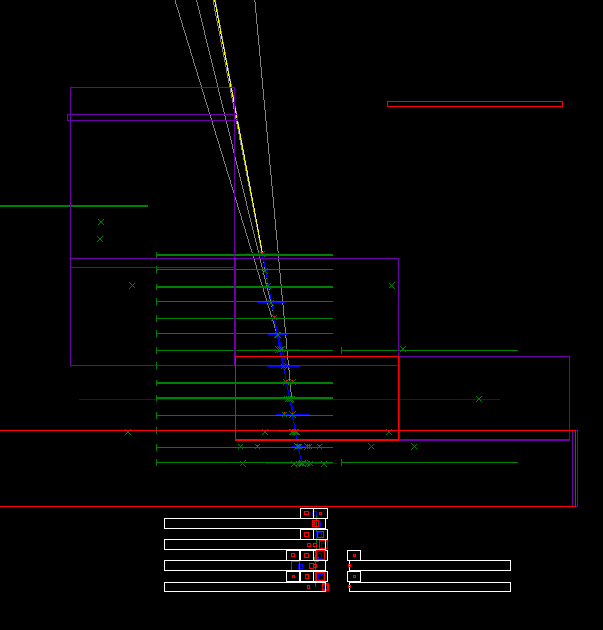}
\end{array} $
\end{center}
\caption{The 31 GeV event as ween with the {\it Wired} tool which allow a visual representation of the event with the LAT detector. Left and right panels represent the XY and YZ views respectively (Z-axis being parallel to the LAT boresight). Green lines and crosses are the silicon strips that triggered on this event. While rectangles are the calorimeter logs where energy deposit have been detected (size of small red squares are proportional to the energy deposit in the calorimeter logs). Red rectangles represents ACD tiles that fired - the two fired tiles at the bottom are due to x-rays which scatter upward in high energy events. A monolitic ACD such as in the EGRET instrument would therefore not have been able to detect such high energy event. Gray lines represent best track directions while the yellow track is the best direction found by the direction reconstruction algorithm.}
\label{event31}
\end{figure}

A powerful way to investigate the event interaction with the detector and therefore access low-level information is through the use of an event display. The LAT team developed such a tool named {\it Wired} which exactly performs this task.
We have looked at the 31 GeV event with the LAT event display (figure \ref{event31}), where it shows the following characteristics:

\begin{itemize}
  \item clean track and few spurious hits in the tracker
  \item reconstructed track well aligned with the calorimeter cluster
  \item no ACD tiles fired along the reconstructed trajectory of the event
  \item several silicon-strip detector layers above the photon vertex shows no signal which is indicative of a neutral particle
  \item calorimeter cluster is compact with no spurious hit far away
  \item electromagnetic shower is not fully contained which is expected given the high energy of this event (correction for non-contained electromagnetic shower is applied by the energy reconstruction algorithm as explained below)
\item electromagnetic shower passes through a crack between two towers at small incidence angle $\sim13^\circ$ so a significant fraction of the energy is "lost" in the crack (we dedicated a special analysis to estimate the energy lost in the crack for such event as explained below)
\end{itemize}

The event identification algorithm for the LAT uses the full low-level characteristics of each event to determine its probability to be a photon or a background event. For that purpose, it uses the standard classification tree that is trained on Monte-Carlo simulations. By this method, it can determine the chance probability that an event has to be a gamma-ray, as well as compute some parameters that describe the quality of the direction and energy reconstruction. It also assigns each event to different event classification which are used for different scientific purpose: 'diffuse', 'source', 'transient' from the purest to the most background contaminated event class.
Our $\sim31$ GeV event was classified as a 'transient' event which raises some concern on its probability for being a photon. However, looking closer at the parameters used for the classification, one notices that the only reason why it is not classified as a 'diffuse' event is that the parameter characterizing the quality of the direction reconstruction is barely below the 'diffuse' threshold (which is probably related to the fact that this event goes through a crack). In particular, parameters describing the probability of this event being a photon from the point of view of the different elements of the LAT (ACD, Tracker, Calorimeter) and the LAT as a whole are all very close to 1.
As a consequence, purely looking at the LAT event classification scheme, we can already attribute a very high confidence level for this event being a gamma-ray photon.

\subsection{Asociation with GRB090510}
\label{assoc}

Now that the event has been characterized as a photon with very high probability, it is crucial to look at the probability it has of being associated with the GRB emission.
For the 31 GeV event of GRB 090510, the angular separation of the reconstructed direction to the best localization found by Swift-UVOT (R.A.=333.55208, Dec.=-26.58311; 90\% confidence error radius=1.5 arcsecond), is found to be 5.8 arcmin which is perfectly consistent with the point spread function of the instrument at this energy (95\% PSF at $\sim30$ GeV is $\sim16$ arcmin for this type of LAT events). 

Another piece of evidence of its possible association with the GRB is the fact that it is detected in coincidence with the low energy emission of the burst. 
A simple estimate of the significance of the 31~GeV photon can be derived by calculating the 
probability that the background can generate at least one event with an energy $E>31\,{\rm GeV}$
during the bulk GeV emission (from 0.65 to $\sim 1.3\,{\rm s}$ post trigger). 
To accomplish that, a background-estimation method (see appendix \ref{bkg_est})
that has an accuracy of $\sim15-20\%$ was used. The expected number of background events passing the transient classification was estimated to be $9\times10^{-6}$, and the associated probability of detecting at least one event when expecting that number of background events is $9\times10^{-6}$ or $4.4\,\sigma$ (for a two-sided
Gaussian distribution). 
Since the 31~GeV event barely missed the 'diffuse' event classification, the associated significance would be higher than $4.4\,\sigma$ since a diffuse-class event can be identified as a photon with a considerably greater certainty.
Therefore, it would be worthwhile to repeat the above calculation using the expected number
of diffuse-class events instead. The result in this case can be used as an upper limit on the 
significance. The actual significance should be closer to the upper limit (since the 31~GeV photon barely did not pass the diffuse-class cuts). We find that the expected number of diffuse-class background events is $1.7\times10^{-8}$ and an associated probability
$1.7\times10^{-8}$ or $5.6\,\sigma$. Based on the above, the significance of the 31~GeV photon shows is between $4.4\,\sigma$ amd $5.6\,\sigma$, with a most likely value towards the upper bound. 

As a conclusion, an association of this high-energy event with GRB 090510 seems very likely given its temporal and spatial coincidence with the burst.

\subsection{Refined energy estimation}
\label{energy}

Precise energy estimation of these high energy event is crucial for LIV or bulk Lorentz factor analyses.
In order to state a confidence interval for the energy, we performed a dedicated analysis of the 31~GeV photon taking into account the most important characteristics of this event (see section \ref{evt_char}). For that, we generated a large Monte-Carlo simulation of events with similar characteristics (angle of incidence and point of impact at the entry of the calorimeter, and of course the energy range) and then applied cuts to select only the events that look like the one we are interested in. This allowed us to estimate the proper energy uncertainty for this specific event.

First of all a remark has to be made on the energy lost in the crack between the two towers and at the back of the calorimeter. Our energy reconstruction algorithm takes into account both kind of leakage (cracks and back), this is why even if the raw energy deposit in the calorimeter is just $\sim16$ GeV, the best reconstructed energy is as high as $\sim 31$ GeV. This kind of large correction is not unusual for the LAT and has been extensively validated with both Monte-Carlo simulations and beam test data.
This said, the bigger the correction, the larger the error on the reconstructed energy, so the ratio of raw energy over real photon energy is a crucial parameter to take into account when trying to state a confidence energy interval.
We have also investigated the importance of other parameters like the quality in the direction and energy reconstruction in order to determine the final best energy estimate and energy uncertainty for this particular event.

Given the specificities of this event and the constraints needed to get a representative sample of this type of event, we have setup a dedicated Monte-Carlo simulation with the following parameters: similar impact point at the entrance of the calorimeter, similar angle of incidence (which will place the event track close to the calorimeter crack where our event lost a significant fraction of its energy) and finally a true energy distribution flat from 10 to 50 GeV.
More than 10 billion events were written to disk which was necessary to nail down the correct true energy range of this event.

Starting from the sample of simulated gamma-ray photons, we applied a number of quality cuts to select the interesting events that are the most representative of our $\sim31$ GeV event: photon classification, gamma-ray conversion in the same tracker layer and reconstructed energy around 31 GeV ($|E_{recon}-31.3107|<2.0$).

\begin{figure}
\begin{center}
\includegraphics[ width=.8\linewidth, keepaspectratio]{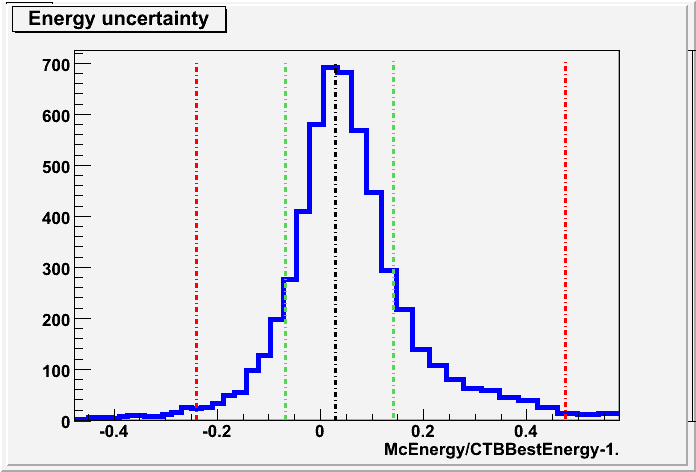}
\end{center}
\caption{Distribution of the reconstructed energy (CTBBestEnergy) with respect to the Monte-Carlo energy (McEnergy): $\frac{\mbox{McEnergy}}{\mbox{CTBBestEnergy}} - 1 $. Starting from the sample of simulated $\g$-ray photons, a number of quality cuts were applied to select events that are most representative of real event for which we are estimating the energy uncertainty: photon classification, gamma-ray conversion in the same tracker layer and reconstructed energy around the measured energy. Vertical dot-dashed lines represent the mean measured energy (black) as well as the $1-\sigma$ (green) and $2-\sigma$ asymetric uncertainties.}
\label{energy_uncertainty}
\end{figure}

Finally from the distribution of the ratio of the reconstructed energy to the true energy ($E_{recon}/E_{true}$) (see figure \ref{energy_uncertainty}), we estimated the correction factor to obtain the best true energy estimate knowing the reconstructed energy and estimated the $68\%$ and $95\%$ confidence interval (symetric with respect to the best value just mentioned). We find the best energy estimate for this particular event to be 30.53 GeV with a $1-~\sigma$ ($2-~\sigma$) confidence interval of [27.97, 36.32] GeV ([25.86, 42.94] GeV).

\section{Limits on possible time delay and on $M_{\rm QG}$}
\label{LIV_results}

\begin{figure}
\begin{center}
\includegraphics[ width=.52\linewidth, keepaspectratio]{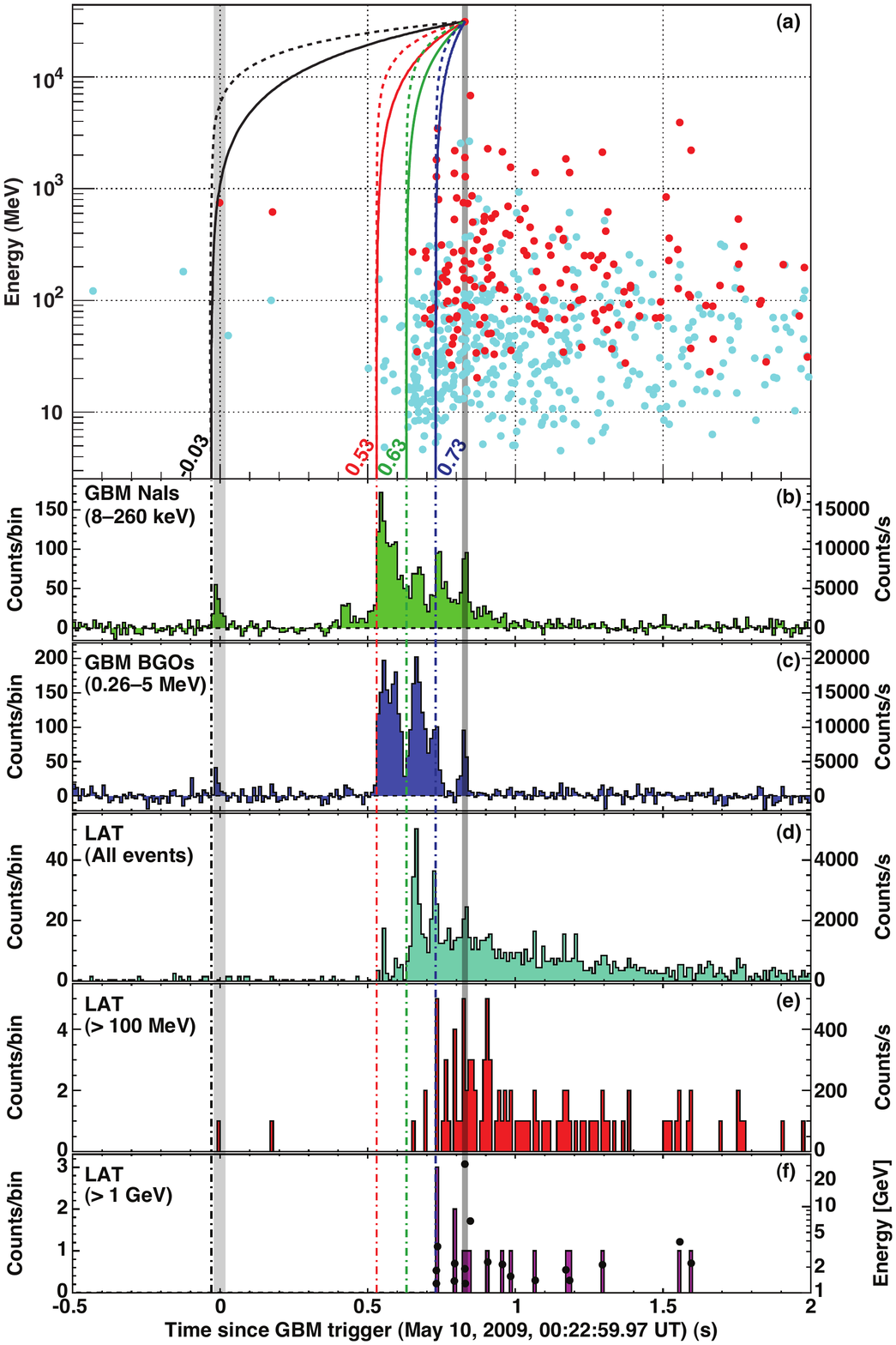}
\end{center}
\caption{Light curves of GRB 090510 at different energies. Panel (a): energy vs. arrival time w.r.t the GBM trigger time for the LAT photons that passed the transient off-line event selection (red) and the photons that passed the onboard $\g$-ray filter (cyan), and are consistent with the direction of GRB 090510. The solid and dashed curves are normalized to pass through the highest energy (31 GeV) photon and represent the relation between a photon's energy and arrival time for linear (n=1, solid) and quadratic (n=2, dashed) LIV, respectively, assuming it is emitted at $t_{start}  = 30$ ms (black; first small GBM pulse onset), 530 ms (red; main $< 1$ MeV emission onset), 630 ms (green; $>100$ MeV emission onset), 730 ms (blue; $> 1$ GeV emission onset). Photons emitted at $t_{start}$ would be located along such a line due to (a positive) LIV induced time delay. Panels (b)Ð(f): GBM and LAT light curves, from lowest to highest energies. Panel (f) also overlays energy vs. arrival time for each photon, with the energy scale displayed on the right side. The dashed-dotted vertical lines show our 4 different possible choices for $t_{start}$. The gray shaded regions indicate the arrival time of the 31 GeV photon $\pm 10$ ms (on the right) and of a 750 MeV photon (during the first GBM pulse) $\pm 20$ ms (on the left), which can both constrain a negative time delay. Panels (b) and (c) show background subtracted light curves for GBM NaI in the 8Ð260 keV band and a GBM BGO in the 0.260Ð5 MeV band, respectively. Panels (d)Ð(f) show, respectively, LAT events passing the onboard $\g$-ray filter, LAT transient class events with E $> 100$ MeV, and LAT transient class events with E $> 1$ GeV. In all light curves, the time-bin width is 10 ms. In panels (b)Ð(e) the per-second count rate is displayed on the right for convenience.}
\label{090510_QG}
\end{figure}

We now report on the constraints we obtain for the linear energy dependence (n = 1), while we do not obtain any significant limit for n=2 or higher orders.
GRB 090816C was the first bright burst detected by the LAT with its highest energy photon detected at $13.22^{+0.70}_{-1.54}$ GeV, and arrived $\sim
16.54$ s after trigger (i.e. the onset of the 100 keV emission).
From equation \ref{deltat}, we could derive lower limits on the quantum gravity mass scale $M_{QG}$ based on various assumptions. 
Since it is hard to associate the highest energy photon with a particular spike
in the low-energy lightcurve, we made the conservative assumption that
it was emitted anytime after the GRB trigger, i.e. $\Delta t \leq t-t_{trigger}$, in order to obtain
a limit for the sub-luminal case ($s_n = 1$). We obtain a limit an order of magnitude below the Planck mass: $M_{QG,1} > 0.1 M_{Planck}$. This was
the strictest limit of its kind at that time.

However, the very bright LAT GRB 090510 would push this limit even further. It is a short burst with very narrow sharp spikes in its light curve with a highest energy photon of $\sim 31$ GeV (see Fig. \ref{090510_QG}). Ground-based optical spectroscopy data exhibited prominent emission lines at a common redshift of $z = 0.903 \pm 0.003$ \citep{Rau:09}, corresponding to a luminosity distance of $dL = 1.8 \times 10^{28}$ cm (for a standard cosmology, [$\Omega_{\Lambda}$, $\Omega_m$, h] = [0.73, 0.27, 0.71]). 
Our main results for GRB 090510 are summarized in table \ref{QG_limits}. 
The first 4 limits are based on a similar method as described above
for GRB 080916C, using the highest energy photon, $E_h = 30.53^{+5.79}_{-2.56}$ GeV.
We do not attempt to associate the relevant high-energy photon with a particular spike in the low-energy light-curve; instead, we simply assume that it was emitted sometime during the relevant lower-energy emission episode, {\it i.e.} we assume that its emission time $t_h$ was after the start of a relevant lower
energy emission episode: $t_h > t_{start}$. We have conservatively used the 31 GeV photon, even if another photon gave a stricter limit, since it is less sensitive to the exact choice of tstart or to intrinsic-lags. And we also conservatively
used the low end of the $1-\sigma$ confidence interval for the highest
energy photon ($E_h = 28$ GeV) and for the redshift (z = 0.900). 
The 4 limits correspond to different
choices of $t_{start}$, which are shown by the vertical lines in Fig. \ref{090510_QG}.
In the following, we describe several possible different assumptions along with the astrophysical reasoning behind them and the corresponding lower limits on $M_{QG}$, starting from the most conservative assumption, and ending with the least conservative assumption (which is still very likely, and with good astrophysical motivation).
our most conservative assumption is based on the fact that no high-energy photon has ever been detected before the onset of the low-energy emission in a GRB. Therefore, it is highly unlikely that the 31 GeV photon was emitted before the observed onset of GRB 090510 and we therefore associate $t_{start}$ with the onset of
any detectable emission from GRB 090510, namely the start of the small
precursor that GBM triggered on, leading to $\xi_1 = M_{QG,1}/M_{Planck} > 1.19$. We consider this as our most conservative lower limit on the quantum gravity mass scale.
Our data can be used to set additional limits, which while not as secure as the one mentioned above, are still very useful.
It is indeed highly unlikely that the 31 GeV photon is indeed associated
with the small precursor. It is much more likely associated with the main
soft gamma-ray emission, leading to $\xi_1 > 3.42$. Moreover, for any reasonable
emission spectrum, the emission of the 31 GeV photon would be accompanied
by the emission of a large number of lower energy photons, which
would suffer a much smaller time delay due to LIV effects, and would therefore
mark its emission time. We could easily detect such photons in energies
above 100 MeV, and therefore the fact that significant high-energy emission
is observed only at later times (see Fig. \ref{090510_QG}) strongly argues that the 31 GeV
photon was not emitted before the onset of the observed high-energy emission.
One could choose either the onset time of the emission above 100 MeV
or above 1 GeV, which correspond to $\xi_1 > 5.12$, and $\xi_1 > 10.0$, respectively.
We note that there is no evidence for LIV induced energy dispersion that might be
expected if indeed the 31 GeV photon was emitted near our choices for tstart, together
with lower energy photons, as can be expected for any reasonable emission spectrum. This
is evident from the lack of accumulation of photons along the solid curves in panel (a) of
Fig. \ref{090510_QG}, at least for the first 3 $t_{start}$ values, and provides support for these choices of $t_{start}$ (i.e. that they can indeed serve as upper limits on a LIV induced energy dispersion).

Finally, one can also set limits on LIV-induced time-delays of either sign based on the temporal association of the 31-GeV photon with the $7^th$ GBM spike, and by associating the 0.75 GeV photon with the first GBM spike, since these photons arrive near the peak of a very narrow GBM spike, which is unlikely due to chance coincidences. These associations would imply $\xi_1 > 102$ and $\xi_1 > 1.33$, respectively. It is important to keep in mind, however, that while these associations are most likely, they are not very secure.

\begin{table}[htbp]
   \begin{tabular}{|p{1.1cm}|p{1.2cm}|p{5.9cm}|p{1.1cm}|p{0.6cm}|p{1.5cm}|}      \hline
      	$t_{start}$ (ms)    & $|\Delta t|$ limit (ms) & Reason for choice of $t_{start}$ or limit on $\Delta t$  & $E_l$ (MeV) & $s_n$ & Lower limit on $M_{QG,1}$ / $M_{Planck}$  \\
      	\hline
      	\hline
	-30 & $<859$ & start of any observed emission & 0.1 & 1 & $> 1.19$ \\
	530 & $<299$ & start of main $<1$ MeV emission & 0.1 & 1 & $> 3.42$ \\
	630 & $<199$ & start of $>100$ MeV emission & 100 & 1 & $> 5.12$ \\
	730 & $<99$ & start of $>1$ GeV emission & 1000 & 1 & $> 10.0$ \\
	\hline
	- & $<10$ & association with $<1$ MeV spike & 0.1 & $\pm 1$ & $> 102$ \\
	- & $<19$ & if $0.75$ GeV $\g$ is from $1^{st}$ spike & 0.1 & $\pm1$ & $> 1.33$ \\
      	\hline
      \end{tabular}
   \caption{Lower-limits on the Quantum Gravity mass scale associated with a possible
linear (n = 1) variation of the speed of light with photon energy, that we can place from the
lack of time delay (of sign $s_n$) in the arrival of high-energy photons relative to low-energy
photons, from our observations of GRB 090510.}
   \label{QG_limits}
\end{table}

Using the so-called DisCan method \cite{Scargle:08}, \cite{Abdo_090510:09} were able to get another independent limit on the quantum gravity mass scale of $\xi_1 > 1.22$ which confirms the conservative limit obtained in the method presented above. 
This method relies on very different and largely independent analysis, but yet still give a very similar limit. This lends considerable support to this result, and makes it more robust and secure than for each of the methods separately.
We refer the interested reader to \cite{Abdo_090510:09} for more details on this alternative method.

In the end, our most secure and conservative new limit, $\xi_1 > 1.2$, is much stronger than the previous best limit of this kind ($\xi_1 > 0.1$ from GRB080916C) and fundamentally more meaningful. Since, in most quantum gravity scenarios, $M_{QG,n} \lesssim M_{Planck}$, even our most conservative limits greatly reduce the parameter space for n = 1 models \citep{Ellis:08,Zloshchastiev:10}. Our other limits, and especially our least conservative limit of $\xi_1 > 102$, make such theories highly implausible (Models with $n > 1$ are not significantly constrained by our results). Thus, it is unlikely that other predictions of such n = 1 models would be observed. Our stringent photon dispersion limit greatly solidifies the conclusion that violation of Lorentz invariance implied by a presumed lumpiness of space-time at the Planck scale is not supported by astrophysical observations.

       	\chapter{\label{EBL}LAT Constraints on the Gamma-ray Opacity of the Universe}

The so-called Extragalactic Background Light (EBL) includes photons from ultraviolet to 
infrared, which are effective in attenuating high energy $\g$-rays through pair creation ($\g \g \rightarrow e^+ e^-$) during 
propagation from sources at cosmological distances.
This results in a redshift- and
energy-dependent attenuation of the $\g$-ray flux of extragalactic
sources such as blazars and Gamma-Ray Bursts (GRBs).  
A major science goal of {\em Fermi} is to probe the opacity of the
Universe to high-energy (HE) $\g$-rays as they propagate from their
sources to Earth.
The LAT onboard {\em Fermi} detects a sample of
$\g$-ray blazars with redshift up to $z\sim 3$, and GRBs with redshift
up to $z\sim 4.3$.  The LAT is a unique instrument for such study in the sense that it observes both the unabsorbed part of the spectrum (typically $\lesssim 10$ GeV) as well as the energy range where EBL absorption is significant (typically $>> 10$ GeV)
Using photons above 10 GeV collected by {\em Fermi} over more than one year of
observations for these sources, we investigate the effects of the EBL on the $\g$-ray flux.  
We constrain  the $\g$-ray opacity of the Universe at various
energies and redshifts, and compare this with predictions from
well-known EBL models.  We find that an EBL intensity in the 
optical--ultraviolet wavelengths as great as predicted by the
``fast-evolution'' and ``baseline'' models of \citet{Stecker06}
can be ruled out with high confidence.
After a brief introduction in section \ref{intro_ebl},  section \ref{sec:analysis} describes the sample of blazars and GRBs detected by the LAT along with its various biases. Section \ref{sec:methods} gives the detailed of the various methods used as well as their results. Finally we discuss the implication of our results in section \ref{sec:discussion}.

\section{Introduction}
\label{intro_ebl}

A presentation of the EBL and its absorbing effect of $\g$-rays have been presented in chapter \ref{chap:introEBL}.
We now focus on the use of $Fermi$ data to constrain different models of EBL that have been proposed in the literature. 
The models encompass different degrees of
complexity, observational constraints and data inputs. Unfortunately, 
the available direct EBL measurements do not constrain these
models strongly at optical-UV wavelengths due to the large scatter in the data points. 
Detailed description
of these models is beyond the scope of this work; we refer the reader to 
the original works on the various EBL models 
\citep[e.g.,][]{Salamon98, Stecker06, Kneiske02, Kneiske04, Primack05, Gilmore09, Franceschini08, 
Razzaque09, Finke09_model}.

For the analyses presented in this work we have made use of the optical depth values $\tau(E,z)$ provided by 
the authors of these EBL models. These models are available via webpages 
\footnote{ \url{http://www.physics.adelaide.edu.au/~tkneiske/Tau_data.html} for Kneiske 2004; \url{http://www.phy.ohiou.edu/~finke/EBL/index.html} for Finke et al. 2010}, analytical 
approximations (as in, e.g., \citet{Stecker06}), published tables (as in, e.g., \citet{Franceschini08}) 
or via private communications (which is the case for, e.g., \citet{Salamon98, Primack05, Gilmore09, Finke09_model} for this work).
Since the optical depth values are usually available in tabular form, for exact values of observed energy $E$ 
and redshift $z$ a linear interpolation of $\tau(E,z)$ is used for arbitrary values of $E$ and $z$ in our 
calculations below.

The range of predictions by these EBL models is illustrated in Figure~\ref{fig:tau_vs_energy} as a 
function of observed $\g$-ray energy for sources at different redshifts.  
The Universe is optically thin ($\tau_{\g\g} < 1$) to 
$\gamma$-rays with energy below $\simeq 10$ GeV up to redshift $z\simeq 3$, independently of the model 
(see also \citet{dieter}). This is due to 
the rapid extinction of EBL photons shortwards of the Lyman limit. 
Gamma rays below $\sim 10$~GeV are not attenuated substantially because far-UV and X-ray diffuse backgrounds is weak.

The primary sources of HE extragalactic $\g$-rays are blazars and gamma ray bursts (GRBs). 
Blazars are active galactic nuclei (AGN) with relativistic plasma outflows (jets) 
directed along our line of sight. Long GRBs are believed to be associated with
the core collapse of massive stars, while short GRBs might be caused by binary
mergers of neutron stars or neutron star - black hole systems.
Some GRBs produce beamed high-energy radiation similar to the case of blazars but lasting for a short period of time.
GRBs have not been used to constrain EBL absorption during
the pre-{\em Fermi} era mainly because of a lack of sensitivity to transient
objects above 10 GeV. 
The sensitivity of EGRET decreased significantly above 10
GeV, and the field-of-view (FoV) of TeV instruments is small (typically $2-4^\circ$) to
catch the prompt phase where most of the HE emission occurs. The new
energy window ($10- 300$ GeV) accessible by {\em Fermi}, and the wide FoV of the LAT,
makes GRBs interesting targets to constrain EBL absorption in this
energy band.

Evaluating the ratio of the absorbed to unabsorbed fluxes from a
large number of distant blazars and GRBs observed by {\em Fermi} could
result in interesting EBL constraints, as proposed by \citet{chen04},
although intrinsic spectral curvature \citep[e.g.,][]{Massaro06} or
redshift dependent source internal absorption \citep{Reimer07} could make
this, or similar techniques, less effective. \citet{georgan08} have
proposed that Compton scattering of the EBL by the radio lobes of
nearby radio galaxies such as Fornax A could be detectable by the {\em
Fermi}-LAT.  If identified as unambiguously originating from such process, a LAT
detection of Fornax A could constrain the local EBL intensity.

Because the e-folding cutoff energy, $E(\tau_{\g\g}=1)$, from $\g\g$ pair production in
$\g$-ray source spectra decreases with redshift, modern Cherenkov $\g$-ray
telescopes are limited to probing EBL absorption at low redshift due
to their detection energy thresholds typically at or below 50~GeV to 100~GeV \citep{Hinton09}.
Ground-based $\g$-ray telescopes have detected 35 extragalactic
sources to date\footnote{e.g.,\url{http://www.mpi-hd.mpg.de/hfm/HESS/pages/home/sources/}, \url{http://www.mppmu.mpg.de/~rwagner/sources/}}, mostly of the high-synchrotron peaked (HSP) BL Lacertae objects
type.  The most distant sources seen from the ground with a confirmed
redshift are the flat spectrum radio quasar (FSRQ) 3C~279 at $z=0.536$ \citep{Albert08}
and PKS~1510-089 at $z=0.36$ \citep{PKS1510}.  Observations
of the closest sources at multi-TeV energies have been effective in
placing limits on the local EBL at mid-IR wavelengths, while spectra
of more distant sources generally do not extend above 1 TeV, and
therefore probe the optical and near-IR starlight peak of the
intervening EBL \citep[e.g.,][]{Stecker93, Stanev98, Schroedter05,Aharonian99,Aharonian02,Costamante04,
Aharonian06a, Mazin07, Albert08, Krennrich08, Finke09}.

The starting point for constraining the EBL intensity from observations
of TeV $\g$-rays from distant blazars with atmospheric Cherenkov
telescopes is the
assumption of a reasonable intrinsic blazar spectrum, which, in the case of a power law, $dN/dE \propto
E^{-\Gamma_{int}}$ for example, that is not harder than a pre-specified
minimum value, e.g., $\Gamma_{int}\geq\Gamma_{min}=0.67$ or 1.5. Upper limits on the
EBL intensity are obtained when the reconstructed intrinsic spectral
index from the observed spectrum, $\Gamma_{obs}$, presumably softened by EBL absorption of
very high energy (VHE) $\g$-rays, is required to not fall below $\Gamma_{int}$.  The
minimum value of $\G$ has been a matter of much debate, being reasoned
to be $\G_{int}=1.5$ by \citet{Aharonian06a} from simple shock
acceleration theory and from the observed spectral energy distribution (SED) properties of blazars, while 
\citet{stecker07} argued for harder values (less than 1.5)
under specific conditions based on more detailed shock acceleration simulations.
\citet{Katarzynski06} suggested that a spectral index as hard as
$\Gamma_{int}=0.67$ was possible in a single-zone leptonic model if
the underlying electron spectrum responsible for inverse-Compton
emission had a sharp lower-energy cutoff. \citet{boett08} noted that
Compton scattering of the cosmic microwave background radiation by
extended jets could lead to harder observed VHE $\g$-ray spectra, and
\citet{aharonian08} have argued that internal absorption could, in
some cases, lead to harder spectra in the TeV range as well.

An approach less dependent on specifics of models uses the (unabsorbed) photon index as
measured in the sub-GeV range as the intrinsic spectral slope
at GeV-TeV energies. This method has recently been applied
to PG~1553+113 \citep{lat1553} and 1ES~1424+240 \citep{lat1424,prandini10} to derive upper
limits on their uncertain redshifts, and to search for EBL-induced spectral softening
in {\em Fermi} observations of a sample of TeV-selected AGN \citep{TeVselected}. 

Attenuation in
the spectra of higher redshift objects ($z \gtrsim 1$) may be 
detectable at the lower
energies that are accessible to the {\em Fermi}-LAT, i.e., at $E\approx
10-300$ GeV. Gamma rays at these energies are attenuated mainly
by the UV background, which is produced primarily by young
stellar populations and closely traces the global star-formation rate.
{\em Fermi} observations of high redshift sources can
therefore reveal information about the star-formation history of the
early Universe, as well as the uncertain attenuation of UV starlight by
dust.

In this section, we present constraints on the EBL intensity of the Universe derived from {\em Fermi}-LAT observations of blazars and GRBs.
The highest-energy $\g$-rays
from high redshift sources are the most effective probe of
the EBL intensity, and consequently a powerful tool for investigating
possible signatures of EBL absorption.  In contrast to ground-based
$\g$-ray detectors, {\em Fermi} offers the possibility of probing the
EBL at high redshifts by the detection of AGN at $\gtrsim 10$ GeV
energies out to $z>3$, and additionally by the detection of GRB
080916C at a redshift of $\sim 4.35$ \citep{abdo09,Greiner09}. GRBs
are known to exist at even higher redshifts \citep[GRB 090423 is the current
record holder with $z\sim$ 8.2]{Tanvir09}. Therefore
observations of these sources with {\em Fermi} are promising
candidates for probing the optical-UV EBL at high redshifts that are
not currently accessible to ground-based (Cherenkov) telescopes.

In Section~\ref{sec:analysis} we describe our data selections,
the {\em Fermi} LAT AGN and GRB observations during the first year of
operation and analysis, and we discuss potential biases in the selection.  Our
methodology and results are presented in Section~\ref{sec:methods}.  Finally we
discuss implications of our results in Section~\ref{sec:discussion}.

In the following, energies are in the observer frame except where noted otherwise.

\section{Observations and Data Selection} 
\label{sec:analysis}

The data set used for the analysis of the AGNs includes
LAT events with energy above $100$ MeV that were
collected between 2008 August 4 and 2009 July 4.  LAT-detected GRBs are considered up to 2009 September
30.  A zenith angle cut of $105^\circ$ was applied in
order to greatly reduce any contamination from the Earth albedo.
Blazars and GRBs have different emission characteristics,
which result in different analysis procedures here. The event rate detected by
the LAT in the vicinity (68\% confidence radius) of a blazar is largely background dominated and only continuous
observations over long time scales allow the detection of the underlying blazar emission.
To minimize the background contamination when analyzing blazar data we use the
``diffuse'' class events,
which provide the purest $\g$-ray sample and
the best angular resolution.  
GRBs, on the other hand, emit most of their radiative $\g$-ray power on very short time scales (typically on the order of seconds) 
where the event rate can be considered mostly
background free (at least during the prompt emission of bright
bursts).  It is therefore possible to loosen the event class selection
to increase the effective area at the expense of a higher background
rate which is still small on short time scales for bright bursts.  The
``transient'' class was designed for this specific purpose and we use
these events for GRB analysis.\footnote{see \cite{FERMI} for further details on LAT event selection.}

\subsection{AGN sample and potential biases} 
\label{sec:agn_sample}

We use blazars extracted from the First LAT AGN Catalog
\citep[1LAC;][]{1LAC} as the AGN source sample to probe the UV
through optical EBL.  This catalog contains 671 sources at high Galactic latitude
($|b|>10^\circ$) associated with high-confidence with blazars and
other AGNs that were detected with a Test Statistic\footnote{The test statistic (TS) is defined as
$TS = -2 \times (log(L_{0})-log(L_{1}))$ with $L_0$ the likelihood of the
Null-hypothesis model as compared to the likelihood of a
competitive model, $L_1$, (see Section 3.2.2).}
$TS>25$ during the first 11 months of
science operation of the LAT. Detection of correlated
multiwavelength variability was required in order to establish a
source as being identified.

Source associations were made with a
Bayesian approach \citep[similar to][]{mattox01}. 
The Bayesian approach for source association
implemented in the {\it gtsrcid} tool of the LAT {\it ScienceTools}
package\footnote{http://fermi.gsfc.nasa.gov/ssc/data/analysis/scitools/overview.html} uses only spatial coincidences between LAT
and the counterpart sources. 
Candidate source catalogs used for this procedure include CRATES
\citep{healey07}, CGRaBS \citep{healey08} and the Roma-BZCAT
\citep{Massaro09}, which also provide optical classifications and
the latter two provide also spectroscopic redshifts for the sources.  
See \citet{1LAC} for further
details on the source detection and association algorithims refered to
here.

As discussed below, some methods applied here require one to distinguish
among the different blazar source classes. Flat-spectrum radio quasars (FSRQs) 
and BL Lacs are discerned by their observed optical emission line equivalent widths
and the Ca II break ratio \citep[e.g.,][]{stocke91,mar96} following
the procedure outlined in \citet{LBAS}. The BL Lac class itself is
sub-divided into Low-, Intermediate-, and High-Synchrotron peaked BL
Lacs (denoted as LSP-BLs, ISP-BLs and HSP-BLs, respectively) by
estimating the position of their synchrotron peak, $\nu_{\rm peak}^s$,
from the indices $\alpha_{ox}\simeq 0.384\cdot\log(f_{\rm
5000A}/f_{\rm 1keV})$ and $\alpha_{ro}\simeq 0.197\cdot\log(f_{\rm
5GHz}/f_{\rm 5000A})$ determined by the (rest frame) optical $(f_{\rm
5000A})$, X-ray $(f_{\rm 1keV})$ and radio $(f_{\rm 5GHz})$ flux
densities listed in the online version of the Roma-BZCAT blazar
catalog \citep{Massaro09}, and using an empirical relationship
between those broadband indices and $\nu_{\rm peak}^s$ as derived in
\citet{SEDpaper}. LSP-BLs have their synchrotron peak at $\nu_{\rm
peak}^s < 10^{14}$~Hz, ISP-BLs at $10^{14}$~Hz$\leq \nu_{\rm peak}^s \leq
10^{15}$~Hz and HSP-BLs at $\nu_{\rm peak}^s > 10^{15}$~Hz.  This is
found to be in agreement with the classifications used in
\citet{spec_paper,SEDpaper}.  Nearly all the 296 FSRQs are of
LSP-type, while only 23\% of the 300 BL Lacs are LSP-BLs, 15\% are
ISP-BLs, and 39\% are HSP-BLs, 72 AGNs could not be classified, and 41
AGNs are of other type than listed above.

Redshift information on the sources is extracted from the counterpart
source catalogs (CRATES, CGRaBS, Roma-BZCAT).  While all the redshifts
of the 1LAC FSRQs are known, only 42\% of the high-confidence BL Lacs have measured redshifts. 
Obviously, AGN without redshift information are not used in the present work.

The intrinsic average photon indices of {\em Fermi} blazars in the LAT energy range indicate a
systematic hardening with source type from $\sim 2.5$ for FSRQs via
$\sim 2.2$ for LSP- and $\sim 2.1$ for ISP-, to $\sim 1.9$ for HSP-BLs
\citep{spec_paper}.  On the other hand, their redshift distributions
systematically decrease from the high-redshift (up to $z\sim 3.1$)
FSRQs, via LSP-BLs located up to a redshift $z\sim 1.5$, down to the
mostly nearby HSP-BLs at $z<0.5$.  This mimics a spectral softening
with redshift if blazars are not treated separately by source type. A search
for any systematic spectral evolution must therefore differentiate
between the various AGN sub-classes (see below).

To detect absorption features in the HE spectra ($>10$~GeV; see Fig.~\ref{fig:tau_vs_energy}) of {\em Fermi} blazars, 
a thorough understanding of their intrinsic spectra,
including variability, and source internal spectral features is required.  Most
blazars do not show strong spectral variability in the LAT energy
range on $\gtrsim$ week scales \citep{spec_paper}, despite often strong
flux variability \citep{variability_paper}.  Indeed at least three blazars
which turn out to constrain the UV EBL the most (see
Section~\ref{sec:HEP method}, \ref{sec:likelihood}), show a $>99\%$
probability of being variable in flux (using a $\chi^2$ test) with a
normalized excess variance of $\sim 0.02-0.2$ on time scales of hours to weeks.  
PKS~1502+106 (J1504+1029) is one of
the most constraining sources in the sample. It displayed an exceptional flare in
August 2008 with a factor $\sim 3$ increase in flux
within $\sim 12$~hours  \citep{lat1502}.  During this flare a flatter (when brighter)
spectral shape was evident. The spectral curvature at the high energy end increased 
with decreasing flux level.
  If the high energy ($\gtrsim
10$~GeV) photons are emitted during such flare activity, the
constraints on the $\g$-ray optical depth would be tighter if only the 
flare-state spectral data were used.  Because of limited photon statistics
during the flare, however, we use the more conservative time averaged
spectrum in the present analysis.

Absorption in radiation fields internal to the source (e.g., accretion
disk radiation, photon emission from the broad line region) may cause
a systematic break in the $\g$-ray spectra that coincidentally mimics
EBL attenuation \citep{Reimer07}.  In the case where such internal absorption
occurs, its redshift dependence is guaranteed, even in the absence
of accretion evolution.  This is because of the redshifting of that
energy where the interaction probability is maximum
\citep{Reimer07}.  Any technique that explores systematic variaton of
observables (e.g., changes in spectral slope, flux ratios, e-folding
cutoff energy) with redshift to single out EBL-induced absorption
features in blazars with luminous accretion disk radiation (possibly
indicated by strong emission lines) might therefore suffer from such
a bias.

All bright strong-line {\em Fermi} blazars (i.e., {\em Fermi}-FSRQs and some LSP-
and ISP-BLs), however, have been found to show spectral breaks already
in the 1-10 GeV (source frame) range \citep{spec_paper}. This is too low in
energy to be caused by EBL-attenuation for their redshift range $\lesssim 3$ 
(see Fig.~\ref{fig:tau_vs_energy}).  
Although it is not clear if these breaks are due to internal absorption, 
the spectral softening results
in low photon counts at energies $\gtrsim 10$~GeV where EBL
absorption is expected.  Spectra of all bright HSP-BLs and some
ISP-BLs, on the other hand, can be well represented by simple power
laws without any signs of curvature or breaks.  
This indicates not only do they
not have significant internal absorption in the $\gamma$-ray band, but
also the absence of significant EBL absorption, which is expected to be
beyond the LAT energy range for this nearby ($z\lesssim 0.5$) blazar
population.

Consequently, as we show in Section~\ref{sec:flux ratios}, it remains challenging to quantify EBL absorption effects
in the LAT energy range based on population studies.  On the other hand, the determination of
the EBL-caused absorption features from individual blazars requires
bright, high redshift objects with spectra extending to $\gg10$ GeV (Fig.~\ref{fig:tau_vs_energy}), 
and we focus on these blazars in
Section~\ref{sec:individual sources}.

\subsection{GRB sample and potential biases}
\label{sec:grb_sample}

The {\em Fermi} LAT has detected 11 GRBs from the beginning of its science
operation (August, $4^{th}$ 2008) until 30 September 2009,
6 of which have redshift measurements. Figure \ref{fig:emax_vs_redshift}
shows the redshift and highest energy event associated with each of these
GRBs.  The
probability of non-association is extremely small (see Table~1).

GRB prompt emission is highly variable and shows signs of spectral
evolution, a source of systematics to be considered carefully.  Our
approach in this paper is to restrict ourselves to the analysis of small
time windows during the GRB emission where the temporal behavior does
not seem to change significantly.

The GRB spectral behavior is well-represented by the Band function
(Band et al. 1993) in the keV--MeV range.  An additional hard, $\G
\sim 1.5$--2, power-law component, dominating at $\gtrsim 100$~MeV,
has now been firmly identified in a few GRBs:
GRB090510 \citep{grb090510}, GRB090902B, \citep{grb090902B}, GRB090926A \citep{grb090926A}.
Its absence in other LAT
bursts could well be due to limited photon statistics.  We assume that
the power-law component extends well beyond 100~MeV up to $\sim
10$~GeV, below which EBL absorption is negligible (see Fig.~\ref{fig:tau_vs_energy}).  EBL
absorption is then expected to soften the power-law spectra from the
extrapolation of the intrinsic/unabsorbed spectra beyond $\sim
10$~GeV.  Systematic effects will, of course, occur when an intrinsic
spectrum at high energies differs from this extrapolation.  Source
internal and/or intrinsic absorption via pair creation, e.g., would
produce a curvature of the spectrum at higher energy which could be
misinterpreted as an EBL absorption effect.  
Such a spectral break, which could be due to intrinsic pair creation, was detected 
in the LAT data from GRB 090926A \citep{grb090926A} but we note that a corresponding 
roll-off in the intrinsic spectrum can only make our limits on the 
$\g$-ray optical depth more conservative.
By contrast, a rising spectral component above $>$10 GeV would make our limits less constraining, but in the absence of any evidence for inverted gamma-ray spectra in GRBs, we consider this possibility unlikely.

\section{Analysis of $\g$-ray flux attenuation and results} 
\label{sec:methods}

Assuming that high-energy photon absorption by the EBL is the sole
mechanism that affects the $\g$-ray flux from a source at redshift
$z$, the observed (i.e. absorbed) and unabsorbed fluxes at the observed energy $E$ 
can be related by the optical depth,
$\tau_{\g\g}(E,z)$, as
\be
F_{obs}(E) = \exp[-\tau_{\g\g}(E,z)]F_{unabs}(E).
\label{flux_relation}
\ee
This is the primary expression that we use to (i) explore $\g$-ray
flux attenuation in the EBL from AGNs by means of
a redshift-dependent flux ratio between a low- and a high- energy band; 
(ii) constrain EBL
models which predict $\tau_{\g\g}(E,z)$ values much higher than the
optical depth that would give the observed fluxes from individual blazars
and GRBs; and (iii) put upper limits on the $\g$-ray optical depth calculated
from the observed flux of individual blazars and GRBs, and the extrapolation
of the unabsorbed flux to high energies.  We discuss these methods
and the results from our analysis below.

\subsection{Flux ratios - a population based method}
\label{sec:flux ratios}

Because of inherent uncertainties in the determination of the intrinsic
spectrum ($\G_{int}$) for any given blazar in the pre-{\em Fermi} era,
\citet{chen04} proposed the average ratio $F(>10~{\rm GeV})/F(>1~{\rm
GeV})$ for all blazars with significant detections above 1 GeV, weighted
according to the errors in $F(>1~{\rm GeV})$, as a
redshift-dependent tracer of the EBL attenuation of $\g$-ray flux.
The average flux ratio could then be compared with the predictions of
the EBL models, taking selection effects into account.  This approach
assumes that the blazars are sampled from a homogeneous distribution
with a single redshift-dependent luminosity function and a single
intrinsic spectral index distribution.  Preliminary results from 
{\em Fermi} \citep{LBAS} indicate that this assumption is inadequate.
Consequently, we have calculated the average flux ratios for the
different classes of blazars and discuss the results below.

Among the AGN sample described in Section~\ref{sec:agn_sample} we find
that 237 FSRQs, 110 BL Lacs and 25 other AGNs are clean\footnote{i.e., its 
association probability is at least 80\%, it is the sole AGN associated with the 
corresponding $\gamma$-ray source, 
and it is not flagged to have problems that cast doubt on its detection \citep{1LAC}} 
1LAC associations 
with known redshift and detectable fluxes at energies
$\ge 1$~GeV. There are 30 LSP-, 18 ISP- and 60 HSP- BL Lacs in
this sub-sample. 

Of these AGN, only 22 FSRQs, 49 BL Lacs, and 1 other AGN have flux detections
rather than upper limits above 10~GeV, including 10 LSP-, 6 ISP-, and 33 HSP-BL Lacs.
For each of these BL Lacs and FSRQs, we calculated the ratio
between the fluxes above 10~GeV and 1~GeV and their corresponding
statistical errors following \citet{chen04}.

\begin{figure}[ht]
\begin{centering}
\includegraphics[width=0.75\columnwidth]{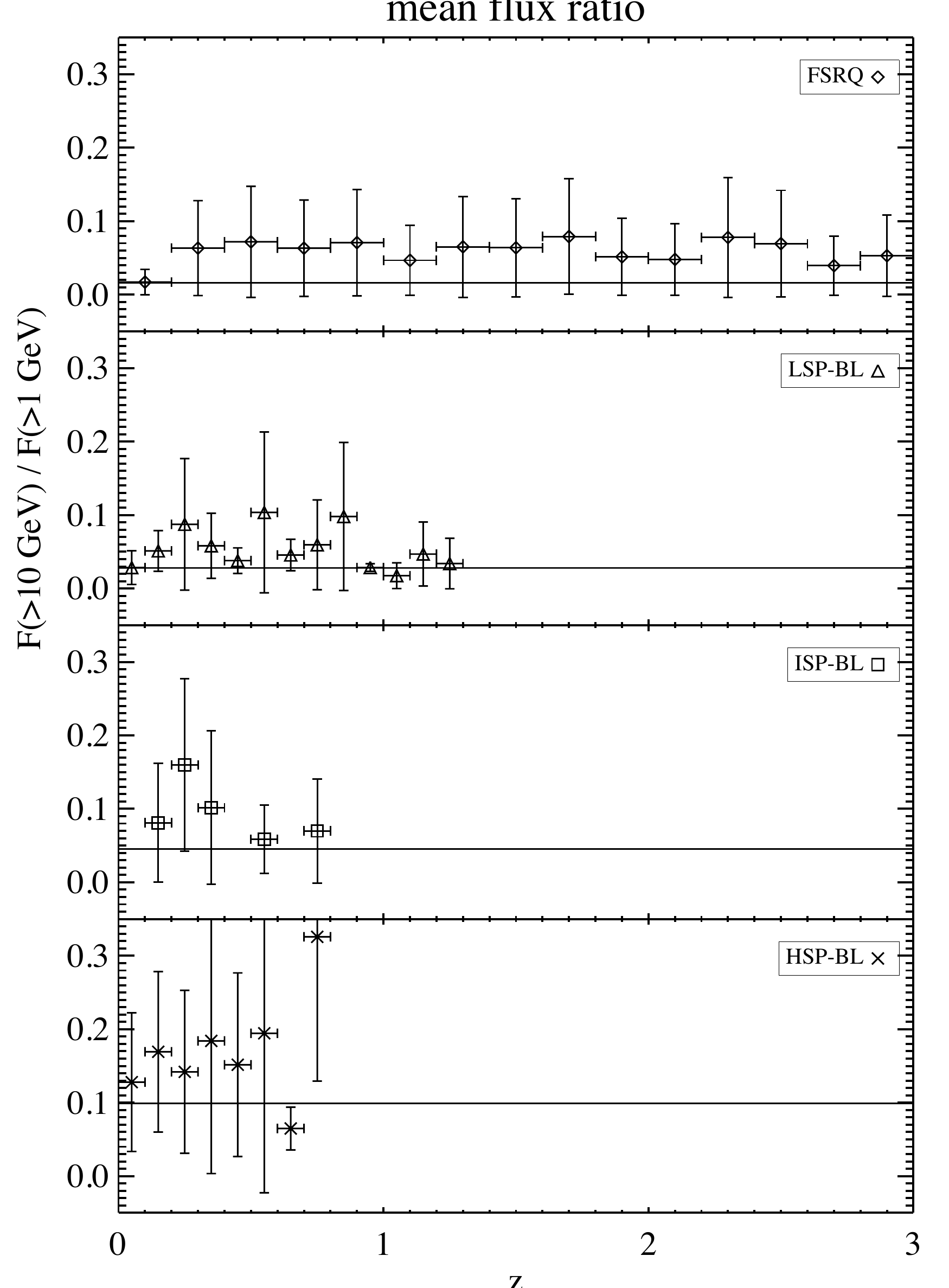}
\par\end{centering}
\caption{\label{fig:ratio} Flux ratio $F(\geq 10~{\rm GeV}) / F(\geq 1~{\rm GeV})$ as
function of redshift, in the {\em Fermi} LAT energy range, for FSRQs and BL
Lac populations.  The horizontal solid lines correspond to the mean
ratios (0.016 (FSRQ), 0.028 (LSP-BL), 0.045 (ISP-BL) and 0.099
(HSP-BL)) which were obtained by considering the individual data points rather that the binned averages.}
\end{figure}

\begin{table}[htdp]
\begin{center}
\begin{tabular}{|p{1.65cm}|p{0.7cm}|p{1.5cm}|p{1.8cm}|p{2.4cm}|p{0.8cm}|p{2cm}|}
\hline
Blazar type & Num & $\Gamma$ & ratio (pred) & mean ratio (obs) & red. $\chi^2$ & prob\\
\hline\hline
FSRQ & 22 & $2.3 \pm 0.1$ & $0.04 \pm 0.01$ & $0.014 \pm 0.001$ & 4.38 & $1.8\times 10^{-10}$ \\
LSP-BL & 10 & $2.2 \pm 0.1$ & $0.07 \pm 0.01$ & $0.028 \pm 0.003$ & 1.65 & 0.11\\
ISP-BL & 6 & $2.1 \pm 0.1$ & $0.08 \pm 0.02$ & $0.048 \pm 0.008$ & 1.86 & 0.11\\
HSP-BL & 33 & $1.9 \pm 0.1$ & $0.12 \pm 0.03$ & $0.100 \pm 0.005$ & 1.29 & 0.13\\
\hline
\end{tabular}
\end{center}
\caption{Spectral indices, mean predicted and observed flux ratios, and reduced $\chi^2$ and probability for blazar sub-populations}
\label{table:ratio}
\end{table}%

Figure~\ref{fig:ratio} shows the observed flux ratios for the FSRQ population and
BL Lac sub-populations as well as the ratios predicted according to the
1FGL spectral index of each blazar, assuming an unbroken power law and
no EBL attenuation. Table~\ref{table:ratio} shows the mean spectral index, mean
flux ratios observed and expected, and the reduced $\chi^2$ and associated probability
given a parent distribution with constant flux ratio. As the blazar classes progress from
FSRQ through LSP-BL, ISP-BL, and HSP-BL, 
\begin{enumerate}
\item the range of redshifts becomes narrower;
\item on average, the spectra become harder;
\item both the predicted and observed mean flux ratios increase;
\item the difference between the predicted and observed flux ratios decreases.
\end{enumerate}

The trend in the predicted flux ratios is a direct consequence of the hardening of the spectra
as a function of source class, while
the difference between the predicted and observed flux ratios is due to the fact that the curvature
of the spectra decrease as the HE peak of the SED moves through the Fermi-LAT energies.
The apparent discrepancies between the flux ratios for different
blazar sub-populations arise from the fact that the LAT samples
different parts of the blazar SED for these sub-classes.  Indeed, a
redshift distribution of the flux ratios for the combined blazar
populations would show a strong, apparently decreasing trend, giving the
appearance of an EBL absorption effect. When we separate the blazars into 
sub-populations, we find no significant redshift dependence of the flux ratios 
within each sub-population. The dearth of sources at high redshift
and the large spread of spectral indices make it difficult to use the
mean trend in the flux ratio as a function of redshift. To set
upper limits on the $\gamma$-ray optical depth, we need to rely on the 
spectra of individual blazars, despite the increased dependence on
the blazar emission model this entails.

The flux ratio versus redshift relationship for BL Lacs is therefore
primarily due to the differing intrinsic spectral characteristics of
BL Lacs, rather than from EBL absorption.  This test is a reminder of
the importance of a careful consideration of the intrinsic spectral
characteristics of the source populations chosen to probe EBL
absorption.

\subsection{Constraints on EBL models from individual source spectra}
\label{sec:individual sources}

The sensitivity of the LAT over a broad energy range provides a unique
opportunity to probe $\g$-ray spectra from AGNs and GRBs at $< 10$~GeV
where EBL absorption is negligible and at $\gtrsim 10$~GeV where EBL
absorption can be substantial (see Fig.~\ref{fig:tau_vs_energy}).  Thus extrapolations of the unabsorbed
flux at low energies from individual sources to high energies, and
assuming that the intrinsic spectra do not become harder at high
energies, allows us to derive a measure of the total absorption
(source in-situ and in EBL).  We note that this is the only assumption
made for the following methods.
Furthermore, since any intrinsic
spectral curvature or internal absorption effects can not be decoupled
from EBL-caused curvature, the constraints derived below shall be
considered as conservative upper limits on the EBL-caused opacity.
These are then confronted with various EBL models. Clearly, high EBL
density models possess a higher probability of being constrained by these
methods than low density ones. In the following, we use two methods:
the highest energy photon (Section~\ref{sec:HEP method}) and the
likelihood (Section~\ref{sec:likelihood}) methods.

\begin{figure}[ht]
\begin{centering}
  \begin{minipage}[b]{15cm} 
     \includegraphics[width=12cm]{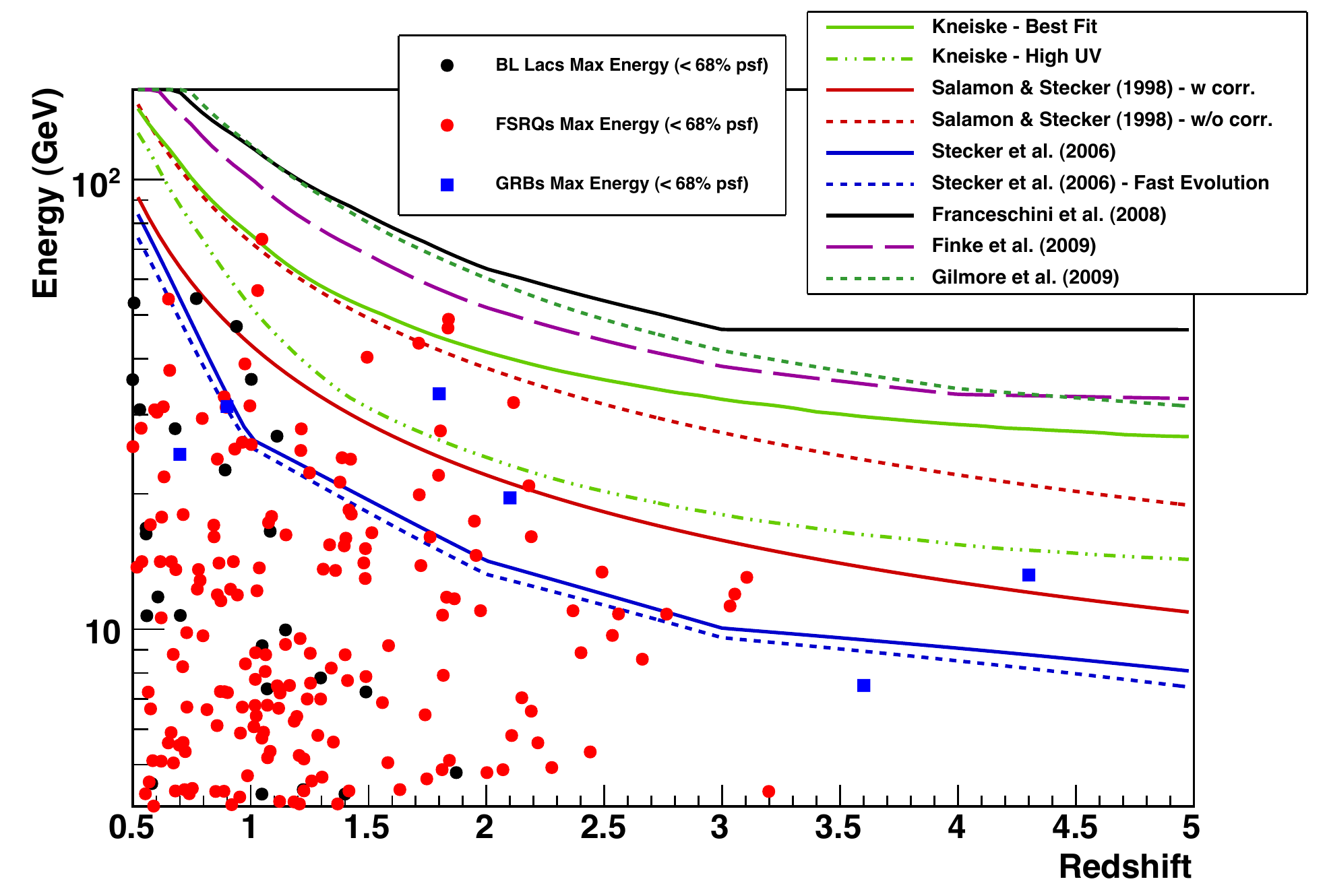}
  \end{minipage}
  \begin{minipage}[b]{15cm}
     \includegraphics[width=12cm]{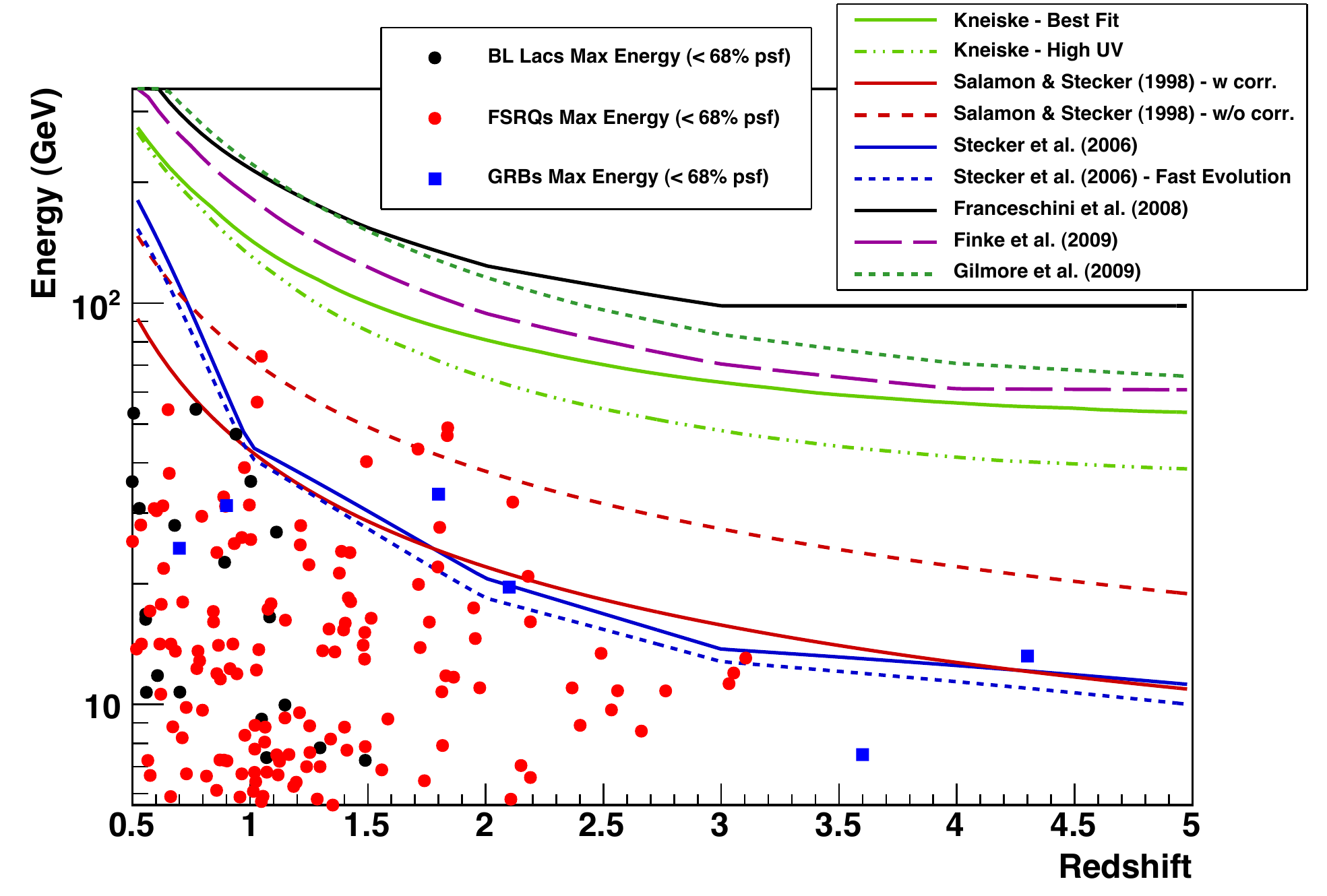} 
  \end{minipage}
\par\end{centering}
\caption{\label{fig:emax_vs_redshift}Highest-energy photons from blazars and GRBs from different
redshifts.  Predictions of $\g\g$ opacity $\tau_{\g\g} = 1$ (top
panel) and $\tau_{\g\g} = 3$ (bottom panel) from various EBL models
are indicated by  lines. Photons above model predictions in this figure traverse an EBL medium with a high $\gamma$-ray opacity. The likelihood of detecting such photon considering the spectral characteristics of the source are considered in the method presented in section \ref{sec:HEP method}.}
\end{figure}

\begin{table}
\begin{scriptsize}
\begin{center}
\renewcommand{\arraystretch}{1.2}
%\centering
%\topcaption{Table captions are better up top} 
%requires the topcapt package
\begin{tabular}{|p{2.3cm}|p{0.8cm}|p{0.7cm}|p{1cm}|p{1cm}|p{1cm}|p{1.2cm}|p{1.5cm}|} 
% Column formatting, @{} suppresses leading/trailing space
\hline
Source & $z$ & $E_{max}$ (GeV) & Conv.\ Type & $\Delta E/E$  &  68\% radius &  Separation & Chance Probability \\
\hline\hline
J1147-3812 & 1.05 & 73.7 & {\em front} & 10.7 \% & $0.054^\circ$ & $0.020^\circ$ & $7.0\times10^{-4}$ \\
(PKS 1144-379) & & & & & & &\\
\hline
J1504+1029 & 1.84 & 48.9 & {\em back} & 5.4\% & $0.114^\circ$  & $0.087^\circ$ & $5.6\times10^{-3}$ \\
(PKS 1502+106) &      & 35.1 & {\em back} & 12.4\% & $0.117^\circ$ & $0.086^\circ$& $9.8\times10^{-3}$\\
         &      & 23.2 & {\em front} & 7.2\% & $0.072^\circ$ &  $0.052^\circ$& $5.6\times10^{-3}$\\
\hline
J0808-0751 & 1.84 & 46.8 & {\em front} & 9.7\% & $0.057^\circ$ & $0.020^\circ$ & $1.5\times10^{-3}$ \\
(PKS 0805-07)&      & 33.1 & {\em front} & 5.9\% & $0.063^\circ$ & $0.038^\circ$ &$2.7\times10^{-3}$\\
         &      & 20.6 & {\em front} & 8.9\% & $0.075^\circ$ & $0.029^\circ$ &$6.9\times10^{-3}$\\
\hline
J1016+0513 & 1.71 & 43.3 & {\em front} & 11.4\% & $0.054^\circ$ & $0.017^\circ$ & $1.2\times10^{-3}$ \\
(CRATES J1016+0513) &      & 16.8 & {\em front} & 6.3\%  & $0.087^\circ$ & $0.035^\circ$ &$8.2\times10^{-3}$\\
         &      & 16.1 & {\em front} & 7.6\% & $0.084^\circ$ & $0.018^\circ$ &$8.2\times10^{-3}$\\
\hline
J0229-3643 & 2.11 & 31.9 & {\em front} & 10.7\% & $0.060^\circ$ & $0.035^\circ$ & $1.7\times10^{-3}$ \\
(PKS 0227-369) & & & & & & &\\
\hline
GRB 090902B & 1.82 & 33.4 & {\em back} & 10.5\% &  $0.117^\circ$ &  $0.077^\circ$ & $6.0\times 10^{-8}$ \\
\hline
GRB 080916C & 4.24 & 13.2 & {\em back} & 11.6\% &  $0.175^\circ$ & $0.087^\circ$ & $2.0\times 10^{-6}$ \\
\hline
\end{tabular}
\end{center}
\end{scriptsize}
\caption{List of blazars and GRBs detected by the LAT which have
redshift measurements, and which constrain the EBL opacity the most. For each source, J2000 coordinate based name (other name), the energy of the 
highest-energy photon (HEP), the conversion type of the event ({\em front} or {\em back}) of the instrument, the energy 
resolution, $\Delta E/E$, for 68\% containment of the reconstructed incoming photon energy, and the 68\%
containment radius based on the energy and incoming direction in instrument coordinates of the event, the separation from the source and the chance probability of the HEP being from
the galactic diffuse or isotropic backgrounds 
are also listed. The energy resolution for the GRB HEP events is taken from \citet{grb090902B} and \citet{grb080916C} using
the respective lower energy bounds.
The three highest-energy photons are listed for those sources that have multiple constraining photons.}
\label{high-energy-photons}
\end{table}

\subsubsection{Highest energy photons}
\label{sec:HEP method}

A simple method to constrain a given EBL model is to calculate the
chance probability of detecting a photon with energy $E \geq E_{max}$, where
$E_{max}$ is the energy of the most energetic photon that we would expect when the source intrinsic spectrum is folded with the
optical depth from the specific EBL model we want to test. We derive a conservative estimate of the
 intrinsic flux of the source by
extrapolating the unabsorbed spectrum at low energies to high
energies.  We consider the LAT spectrum to be representative of the
intrinsic spectrum at energies where the EBL is supposed to absorb
less than $\sim 1 \%$ of the photons for the most opaque models. This
corresponds to an energy of around $10$~GeV  
(down to $\sim 6$ GeV for GRB 080916C at $z \sim 4.3$). 
Best fit spectral
parameters of this ``low-energy'' unabsorbed spectrum were derived for
all sources of the HEP set (see Table~\ref{tab:fit}).  The spectrum is
assumed to be a power law unless a significant deviation from this
shape is measured at $\lesssim 10$~GeV  (as is indeed observed from, e.g.,
FSRQs at GeV energies). This is the case for source J1504+1029, for which a log parabola 
model provides the best fit.

Iterating through the source list described in
Section~\ref{sec:agn_sample} and Section~\ref{sec:grb_sample} we find the
energy $E_{\rm max}$ of the highest-energy photon detected within the 68\%
containment radius (using the specific P6\_V3\_DIFFUSE instrument response
functions for {\em front} and {\em back} events) of each source position.  The resulting $E_{\rm max}$
versus source redshift is shown in Figure~\ref{fig:emax_vs_redshift}
for sources with $z>0.5$, and compared to the energy at which the optical depth $\tau_{\gamma\gamma}$ is 
equal to 1 and 3 according to the various EBL models. As shown in this figure, 5 AGN have $E_{\rm max}$ that is 
significantly greater $(\gtrsim 2)$ than the energy at which $\tau_{\gamma\gamma}=3$ for the 
``baseline EBL model'' of \citet{Stecker06}. These 5 AGN (and 2 comparable GRBs) have emitted a number 
of events (hereafter {\em highest-energy photons} or HEP) that populate a region of the $E_{max}-z$ phase
space where EBL attenuation effects are predicted to be
significant. These (henceforth called ``HEP set'') will be used in the following sections to 
constrain EBL models and to calculate the
maximum amount of EBL attenuation that is consistent with the LAT
observations\footnote{Only the highest energy photon from each source is shown in Figure~\ref{fig:emax_vs_redshift}.  
There are a few sources however with more than one constraining photon as indicated in Table~\ref{high-energy-photons}.}.

It is possible that the high-energy  photons considered here may not be emitted in the high-redshift source and instead are originated in any of the following background sources:  Galactic $\gamma$-ray diffuse,  isotropic (Extragalactic $\gamma$-ray plus charged-particle residuals) or  a nearby point source.  The likelihood of detecting any of these background photons within the 68\% containment radius used to find the HEP set  is quantified by computing the number of expected events within the 68\% containment radius at the location of the source as determined by the best fit background model (Galactic and isotropic diffuse + point sources) and the instrument acceptance.  The last column of Table~\ref{high-energy-photons} shows such probability for  photons in the HEP set. These chance probabilities, although being fairly small, are non-negligible (at least in the case of blazars) if one would like to set significant constraints on specific EBL models by using this HEP. We later describe how this probability for the HEP to be a background fluctuation was incorporated in our final results for this method. For now we will assume that this HEP is indeed from the source and we will first derive the type of constraints it allows us to set on different EBL models.
We also note that a stricter set of cuts ({\em extradiffuse}) has been
developed by the LAT team to study the Extragalactic $\g$-ray
background \citep{EGBpaper}. Despite the decreased $\g$-ray acceptance
we find all photons in the HEP set to be retained when using these
selection cuts.

Monte-Carlo simulations are performed in order to test a particular EBL model with the derived intrinsic
spectrum absorbed by the EBL as the Null-hypothesis.   The simulations were performed using {\em gtobssim}, one of the science tools distributed by the {\em Fermi} science support center and the LAT instrument team. For each simulation we define the  unabsorbed spectrum of the source as a power law (or log parabola in the case of  J1504+1029) with spectral parameters drawn randomly from the best-fit values (and corresponding uncertainty)  shown in Table~\ref{tab:fit}. EBL absorption is applied according to the optical depth values of the considered model, and finally, the position and orientation of the {\em Fermi} satellite  during the time interval described in Section~2  is used to account for the instrument acceptance that corresponds to the observations. The highest-energy photon from the simulated data is obtained following the exact same cuts and analysis procedure that was used for the data. 

\begin{table}
\begin{center}
\renewcommand{\arraystretch}{1.2}
\begin{tabular}{|p{2.5cm}|p{3.2cm}|p{2.6cm}|p{2.3cm}|p{1cm}|c|c|cl} 
\hline
Source & normalization $N_0$ ($10^{-7} \mbox{ph}$& photon index $\Gamma$  & $\Gamma$, $\beta$, $E_b$/GeV & TS \\
& $\mbox{ cm$^{-2}$ s$^{-1}$ MeV$^{-1}$}$)& (PL) & (LP) & \\
\hline\hline
J1147-3812 & $0.570 \pm 0.081$ & $2.38 \pm 0.09$ & ... & 221 \\
\hline
J1504+1029 & $(1.84 \pm 0.23)\cdot10^{-4}$ & ... & $2.36 \pm 0.03,$ & 34638\\
&&& $0.09 \pm 0.01,$ & \\ 
&&& $2.0 \pm 0.1$& \\
\hline
J0808-0751 & $1.212 \pm 0.078$ & $2.09 \pm 0.04$ & ... & 1498 \\
\hline
J1016+0513 & $1.183 \pm 0.078$ & $2.27 \pm 0.05$ & ... & 1220 \\
\hline
J0229-3643 & $0.789 \pm 0.075$ & $2.56 \pm 0.07$ & ...& 394\\
\hline
J1012+2439 & $0.552 \pm 0.058$ & $2.21 \pm 0.07$ & ... & 443\\
\hline
GRB 090902B  & $146 \pm 56$ & $1.40 \pm 0.37 $ &... & 1956 \\
\hline
GRB 080916C  & $1146 \pm 199$ & $2.15 \pm 0.22 $ &... & 1398 \\
\hline
\end{tabular}
\caption{Parameter values of the power law (PL) fits $dN/dE = N_0(-\Gamma+1) E^{-\Gamma}/[E_{max}^{-\Gamma+1}-E_{min}^{-\Gamma+1}]$ 
in the range $E_{\rm min}=100$~MeV to $E_{\rm max}=10$~GeV of the sources (AGN and GRBs) listed in
Table~\ref{tab:spectres} except for source J1504+1029 where a log parabolic parametrization (LP) 
$dN/dE = N_0 (E/E_b)^{-(\Gamma+\beta \log(E/E_b))}$
has been found to be preferable over a power law fit (with $\Delta TS=71.9$).
The spectral fits for the GRBs are performed below 6 GeV and 3 GeV for GRB 090902B and GRB 080916C, respectively. The TS values 
are obtained through a likelihood ratio test comparing a model with background only and a model where a point source was added.}
\label{tab:fit}
\end{center}
\end{table}

\begin{figure}[ht]
\begin{centering}
\includegraphics[width=0.75\columnwidth]{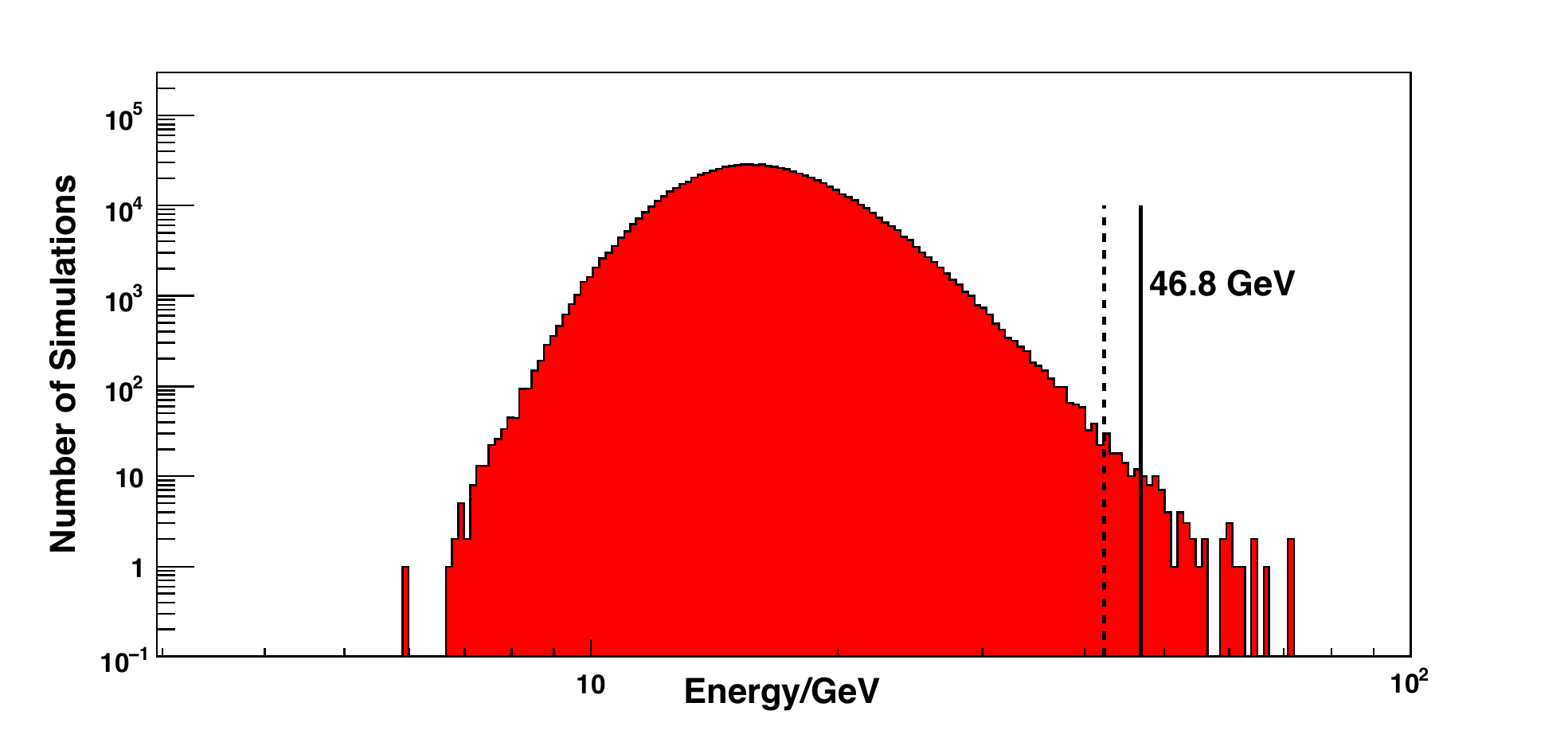}
\par\end{centering}
\caption{\label{fig:hep_mcsim} Distribution of high-energy photons obtained from Monte-Carlo
simulations of the source J0808-0751 with the EBL attenuation by
\citet{Stecker06}. $E_{max}$ and $E_{max} - \sigma_{Emax}$ (where  $\sigma_{Emax}$ is the energy uncertainty) are indicated by a solid and dotted vertical black lines, respectively. The probability of detecting a photon with energy equal or greater to  $E_{max} - \sigma_{Emax}$ is equal to
$6.8\times10^{-5}$.}
\end{figure}

\begin{figure}[ht]
\begin{centering}
  \begin{minipage}[b]{8cm}
      \includegraphics[angle=90, width=8cm]{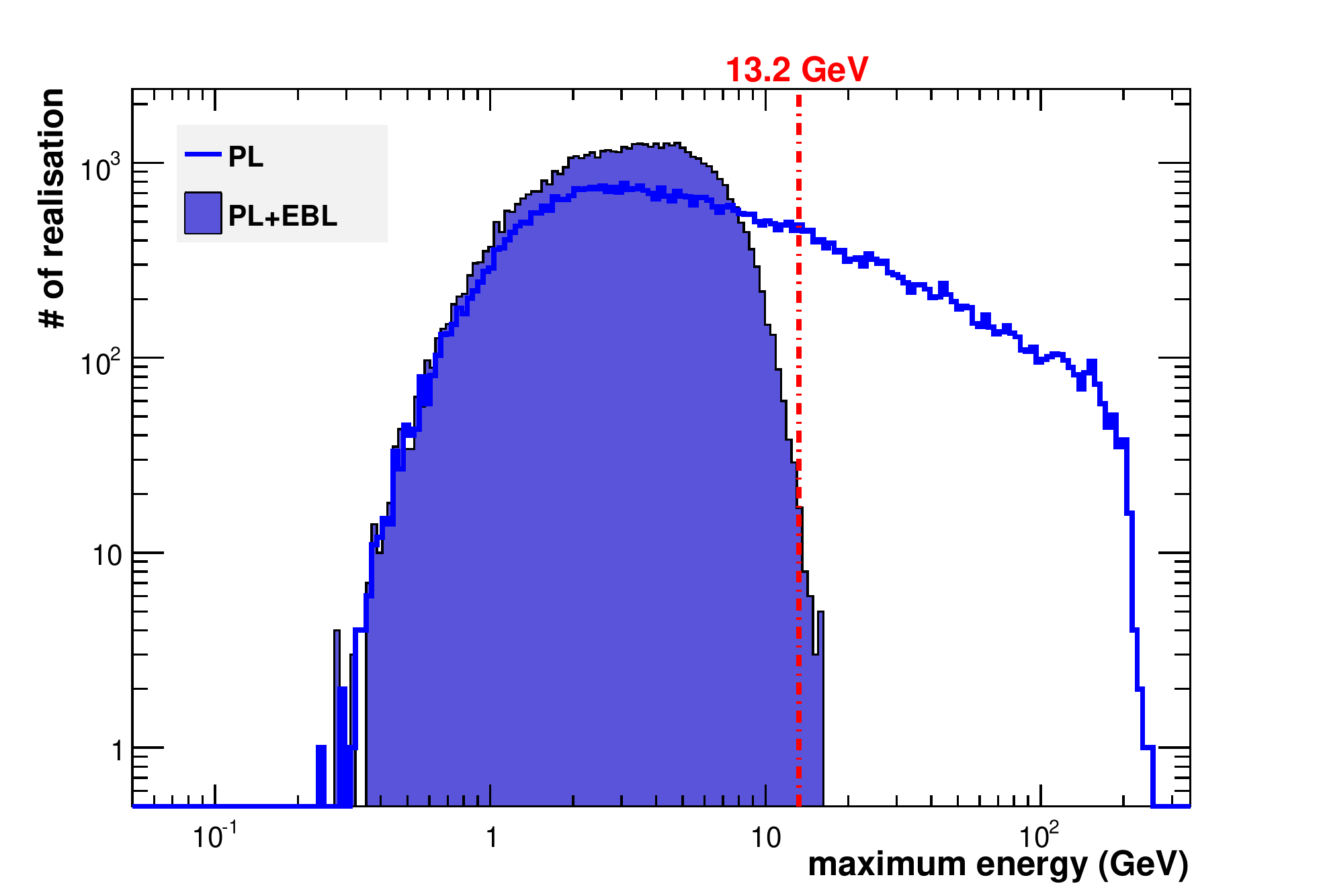}
  \end{minipage}
  \begin{minipage}[b]{7.5cm}
     \includegraphics[angle=90, width=8cm]{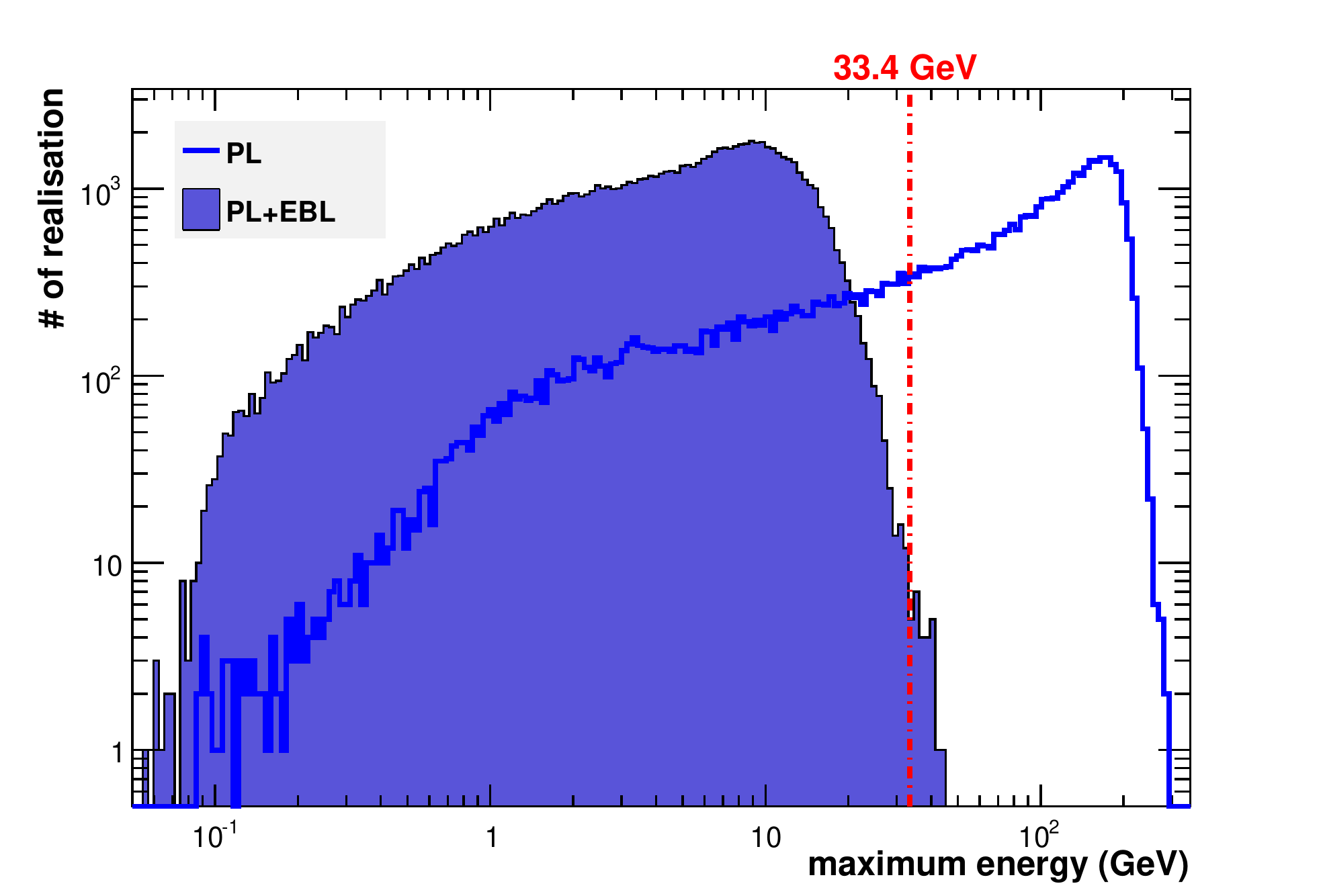} 
  \end{minipage}
\par\end{centering}
\caption{\label{fig:heevt} Distributions of the highest energy photons from simulations
performed with estimates of our intrinsic spectra for GRB 080916C
(top panel) and GRB 090902B (bottom panel), folded with EBL
attenuation calculated using the \citet{Stecker06} baseline model.
The total number of realizations ($10^5$) in both the power-law and power-law convolved with the EBL cases is the same.}
\end{figure}

The resulting distribution of the HEP simulated in each case (see
e.g., Figure~\ref{fig:hep_mcsim}) is used to estimate the chance
probability of detecting a photon from the source with energy equal or greater than  $E_{\rm max}$.
We produced 
$\sim 800,000$ and $\sim 100,000$ simulations for each of the HEP sets for
AGN and for GRBs, respectively. Assuming the HEP is indeed from the source, the probability of observing such high energy photon 
given the specific EBL model tested (called $P_{HEP}$) is calculated as the ratio between the number of cases where the HEP energy
 is above $E_{max}$ (actually $E_{max} - \sigma_{Emax}$  given the energy dispersion) and the total number of simulations performed. 
The number of simulations in each case was chosen to reach sufficient statistics at the tail of the distribution where the energy 
of the HEP is measured.
Distributions of the HEP events from these MC simulations for
GRB 080916C and GRB 090902B are shown in Figure~\ref{fig:heevt}.  
The open and filled histograms correspond to the
distributions using the GRB spectra without and with EBL absorption
using the ``baseline model'' of \citet{Stecker06}.

To compute the final probability of rejection for the specific EBL model tested (called $P_{rejection}$), one needs 
to consider the 
fact that the HEP could be a background photon. We compute the probability for this to happen in Table~2 ($P_{bkg}$). 
In the end, 
one can fail to reject the EBL model because the HEP might be a background event or because there is a chance for a 
source photon 
with energy $E_{max}$ not to be absorbed by the EBL so that 
\begin{equation}
P_{rejection} = P_{bkg} + P_{HEP} \times (1-P_{bkg}).
\label{Prej}
\end{equation} 
%\sim max(P_{bkg},P_{HEP})$. 
\noindent In Table~\ref{tab:prob}, we list these 3 probabilities for each of our
most constraining sources. When more than one photon is available for
a given source, the probabilities are combined resulting in
a stronger rejection.  Although $P_{HEP}$ can be quite constraining,
our final significance of rejection is limited by $P_{bkg}$ which is
non-negligible in the case of blazars and which depends on the size of
the region around each source defined {\em a priori} to
look for associated high-energy events (68\% PSF containment radius in
this analysis). A larger  HEP acceptance region (90\% or 95\%
containment radius instead of 68\%) would increase the background
probability $P_{bkg}$ while also adding constraining photons to the
HEP set.  On a source-by-source basis, the rejection probability goes
up or down with increasing radius  depending on the number and
energy of these additional photons, but our overall result remains the same. 
The unbinned likelihood method,
which we describe in the next subsection (\ref{sec:likelihood}),
does not make use of an acceptance radius, and instead makes full use
of available information in the data to systematically calculate
a model rejection probability.

The analysis described in this section was applied to all sources from the HEP set. We find the ``baseline'' 
model of \citet{Stecker06} to be significantly constrained by our observations. Column 5 of
Table~\ref{tab:spectres} shows the optical depth of the ``baseline'' model of \citet{Stecker06} for the HEP
events. Since the
``fast evolution'' model\footnote{The ``baseline'' model considers the case where all galaxy $60\mu$m luminosities
evolved as $(1+z)^{3.1}$ up to $z\leq 1.4$, non-evolving between
$1.4 < z < 6$ and no emission at $z>6$. In contrast, the ``fast evolution'' model assumes a more rapid galaxy luminosity 
evolution: $\propto (1+z)^4$ for $z<0.8$, $\propto (1+z)^2$ for $0.8<z<1.5$, no evolution
for $1.5<z<6$, and no emission at $z>6$. Consequently, for a given redshift the ``fast evolution'' model predicts a higher 
$\gamma$-ray attentuation than the ``baseline'' model.}
of \citet{Stecker06} predicts higher opacities in the LAT energy range at all redshifts, our constraints on this model will naturally be 
higher than the ones found in Table~\ref{tab:spectres} for the ``baseline'' model.

\begin{table}
\begin{center}
\renewcommand{\arraystretch}{1.2}
\begin{tabular}{|p{2.3cm}|p{0.8cm}|p{1.1cm}|p{1.9cm}|p{2.9cm}|p{2.5cm}|p{3cm}l} 
\hline
Source & z & $E_{max}$ & $\tau (z,E_{max})$ & $\tau (z,E_{max})$ & Number of photons \\ 
       &   & (GeV)     & (F08)       & (St06, baseline)    &  above 15 GeV    \\
\hline\hline
J1147-3812 & 1.05 & 73.7 & 0.40 & 7.1 & 1 \\
J1504+1029 & 1.84 & 48.9 & 0.56 & 12.2  & 7\\
J0808-0751 & 1.84 & 46.8 & 0.52 & 11.7  & 6\\
J1016+0513 & 1.71 & 43.3 & 0.39 &  9.0  & 3\\
J0229-3643 & 2.11 & 31.9 & 0.38 & 10.2 & 1\\
GRB 090902B & 1.82 & 33.4 & 0.28  & 7.7 & 1\\
GRB 080916C & 4.24 & 13.2 & 0.08  & 5.0 & 1\\
\hline
\end{tabular}
\caption{Gamma-ray optical depth to HEP calculated using the EBL model of Franceschini et al (2008; F08)
in comparison to the ``baseline" model of \citet{Stecker06} (St06). Also listed are the number of photons associated to the source which have $\geq 15$ GeV energy and which can potentially constrain EBL models.}
\label{tab:spectres}
\end{center}
\end{table}

\begin{table}
\begin{center}
\begin{scriptsize}
\begin{tabular}{|p{1.55cm}|p{0.4cm}|p{1.cm}|p{1.5cm}|p{1.6cm}|p{1.6cm}|p{2cm}l} 
\hline
 &  &  &  & \multicolumn{2}{|c|}{HEP method applied to Stecker 06}  & HEP Rejection \\ 
Source & $z$ & Energy (GeV) & $P_{bkg}$&$P_{HEP}$ & $P_{rejection} $ & Significance\\ 
\hline\hline
J1147-3812 & 1.05 & 73.7&  $7.0\times10^{-4}$ &$1.2\times10^{-4}$ & $8.1\times10^{-4}$  & 3.2 $\sigma$ \\
\hline
J1504+1029 & 1.84 & 48.9 &  $5.6\times10^{-3}$ &$6.7\times10^{-5}$ & $5.7\times10^{-3}$ &  \\
                        &          &  35.1 &  $9.8\times10^{-3}$ & $6.8\times10^{-3}$ & $1.7\times10^{-2}$ & \\
                        &          &  23.2 &  $5.6\times10^{-3}$ & $1.8\times10^{-1}$ & $1.9\times10^{-1}$ & \\
                        &          &           &                                     &  \multicolumn{2}{c|}{Combined $P_{rej}$ = $1.7\times10^{-5}$} & 4.1 $\sigma$                 \\
\hline
J0808-0751 & 1.84 & 46.8 & $1.5\times10^{-3}$ &$1.9\times10^{-4}$ & $1.7\times10^{-3}$&  \\
                        &          &  33.1 &  $2.7\times10^{-3}$ & $3.7\times10^{-3}$ & $6.4\times10^{-3}$ & \\
                        &          &  20.6 &  $6.9\times10^{-3}$ & $2.5\times10^{-1}$ & $2.6\times10^{-1}$ & \\
                        &          &           &                                     &  \multicolumn{2}{c|}{Combined $P_{rej}$ = $2.8\times10^{-6}$} & 4.5 $\sigma$                  \\
\hline
J1016+0513 & 1.71& 43.3& $1.1\times10^{-3}$ &$5.4\times10^{-4}$ & $1.6\times10^{-3}$& \\
                        &          &  16.8 &  $8.2\times10^{-3}$ & $4.9\times10^{-1}$ & $4.9\times10^{-1}$ & \\
                        &          &  16.1 &  $8.2\times10^{-3}$ & $6.5\times10^{-1}$ & $6.5\times10^{-1}$ & \\
                        &          &           &                                     &  \multicolumn{2}{c|}{Combined $P_{rej}$ = $5.3\times10^{-4}$} & 3.3 $\sigma$                 \\
\hline
J0229-3643 & 2.11 & 31.9 & $1.7\times10^{-3}$ &$8.9\times10^{-5}$ & $1.8\times10^{-3}$ & 2.9 $\sigma$\\
\hline
GRB 090902B & 1.82 &33.4 & $2 \times 10^{-6}$ &$2.0\times10^{-4}$ & $2.0\times10^{-4}$  & 3.7 $\sigma$ \\
\hline
GRB 080916C & 4.24 &13.2 & $8 \times 10^{-8}$ & $6.5\times10^{-4}$ & $6.5\times10^{-4}$  & 3.4 $\sigma$  \\
\hline
\end{tabular}
\end{scriptsize}
\caption{Listed are the significance of rejecting the ``baseline'' model (\citet{Stecker06}), calculated using the HEP method as described in Section
\ref{sec:HEP method}.  For completeness, we also report individually the probability of the HEP to be a background event ($P_{bkg}$) and the probability for this HEP not to be absorbed by the EBL if it were emitted by the source ($P_{HEP}$) following Eq.~\ref{Prej}. For those sources with more than one constraining photon, the individual and combined $P_{rejection}$ are calculated.  The ``fast evolution" model by \citet{Stecker06} is more opaque and leads to an
even higher significance of rejection. Applying this method to less opaque models leads to no hints of rejection since the probability $P_{HEP}$ is large in those cases (e.g.~ $\gtrsim 0.1$ for the \citet{Franceschini08} EBL model). Note that a log parabola model was used as the 
intrinsic model for source J1504+1029 since evidence of curvature is observed here 
even below 10 GeV (see Table~\ref{tab:fit}).
}
\label{tab:prob}
\end{center}
\end{table}

\subsubsection{Likelihood method}
\label{sec:likelihood}

This second method to constrain specific EBL models makes use of a
Likelihood-Ratio Test (LRT) technique. This approach compares the likelihood of the
Null-hypothesis model ($L_0$) to best represent the data with the likelihood of a
competitive model ($L_1$).  The test statistic (TS) is defined as
$TS = -2 \times (log(L_{0})-log(L_{1}))$. Following Wilks' theorem
\citep{Wil38}, the TS is asymptotically distributed as $\chi^2_n$
(with $n$ the difference in degrees of freedom between the two models) if the two models
under consideration satisfy the following two conditions \citep{Pro02}:
1) the models must be nested and 2) the null-values of the additional
parameters are not on the boundary of the set of possible parameter
values.

For the LRT we use the power law intrinsic spectrum convolved with the EBL
absorption predicted by the model ($\tau_{mod}$) we are testing,
$F_{unabs}(E) \times \exp[-\alpha\tau_{mod} (E,z)]$, as the observed flux.
For the Null-hypothesis we set $\alpha = 1$ and we compare it to an
alternative model where $\alpha$ is left as a free parameter, which
therefore has one more degree of freedom than the Null-hypothesis.  In
the absense of any flux attenuation by the EBL, $\alpha = 0$.  Note that
we allow the normalization parameter, $\alpha$ to go to negative
values. This choice, although not physically motivated, allows us to
satisfy the second condition mentioned above. As a consequence the Test-Statistics can simply be converted into a significance 
of rejecting the Null-hypothesis by making use of Wilks' theorem. Because of the lack of information on the intrinsic spectrum 
of a distant source above 10 GeV, we use the (unabsorbed) $\leq 10$ GeV observed spectrum as a reasonable assumption for the
functional shape of the intrinsic source spectrum. A simple power law was found to be a good fit to the $\leq 10$ GeV data for 
the sources listed in Table~\ref{tab:spectres} except in the case of J1504+1029 where a log-parabolic spectrum was preferred. 
We note that if the actual intrinsic curvature is more pronounced than the one found with the best fit below 10 GeV, this would 
only make the results more constraining.

\begin{figure}[ht]
\begin{centering}
\includegraphics[width=0.75\columnwidth]{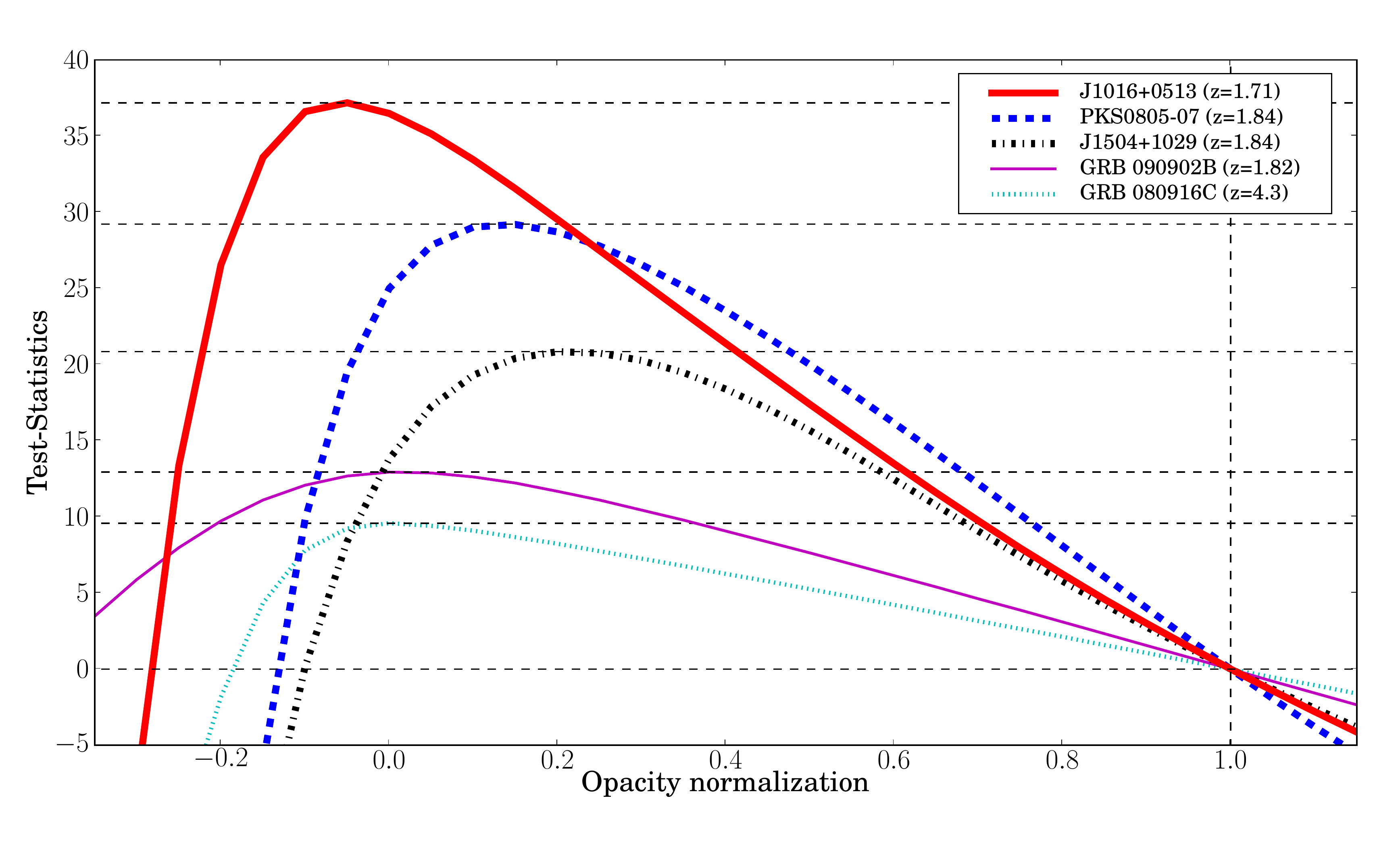}
\par\end{centering}
\caption{\label{fig:TS} Test statistic (TS) as a function of the opacity normalization
parameter calculated from the likelihood ratio test (LRT) for
J1016+0513, J0808-0751, J1504+1029 and GRBs 090902B and 080916C. The ''baseline'' model of \citet{Stecker06} has been used and the rejection for this model can be directly read out as $\Delta TS$ between $\alpha = 1$ and the best fit $\alpha$ for the source (horizontal dashed line). The confidence interval for the normalization parameter can be obtained using $\Delta TS = CL^2$ where $CL$ is the confidence level.}
\end{figure}

As we mentioned earlier, although we are considering all EBL models in the literature, we find that our observations are only constraining the most opaque ones.
Figure~\ref{fig:TS} shows the TS value as a function of the optical depth
normalization parameter $\alpha$, for the three most constraining blazars (J1016+0513, J0808-0751, J1504+1029) and the two GRBs (GRBs 090902B and 080916C) when considering the ``baseline'' model of \cite{Stecker06} with the LRT method.  All sources are found to have an optical depth normalization parameter that is consistent with $\alpha \geq 0$ at the $1\sigma$ level which is reassuring as we do not expect a rise in the spectrum on a physical basis. $\sqrt{TS_{max}}$ for $\alpha=1$ corresponds to the rejection significance for the specific model considered.
The most constraining source, J1016+0513,  rejects the Null-hypothesis ($\alpha = 1$, corresponding to the ``baseline'' model of \cite{Stecker06} in this case) with a significance of $\sim 6.0\sigma$.
This source could also constrain the ``high UV model" of Kneiske et al. (2004) with a significance of $3.2\sigma$ although multi-trials effect substantially reduce this significance (see Section~\ref{sec:combined_P}).

\begin{table}{}
\begin{center}
\begin{small}
\begin{tabular}{|c|c|c|c|}
\hline
 & & \multicolumn{2}{|c|}{LRT Rejection Significance} \\
Source & $z$  & pre-trial & post-trial \\
\hline\hline
J1147-3812 & 1.05 & 3.7$\sigma$ & 2.0 $\sigma$ \\
\hline
J1504+1029 & 1.84 & 4.6$\sigma$ & 3.3 $\sigma$ \\
\hline
J0808-0751 & 1.84   & 5.4$\sigma$ & 4.4 $\sigma$ \\
\hline
J1016+0513 & 1.71 & 6.0$\sigma$ & 5.1 $\sigma$ \\
\hline
J0229-3643 & 2.11  &3.2$\sigma$ & 1.2 $\sigma$ \\
\hline
GRB 090902B & 1.82 & 3.6$\sigma$ & 1.9 $\sigma$ \\
\hline
GRB 080916C & 4.24  & 3.1$\sigma$ & 1.0 $\sigma$ \\
\hline
\end{tabular}
\end{small}
\caption{Significance of rejecting the ``baseline'' model (\citet{Stecker06}), calculated using the LRT method described in 
Section \ref{sec:likelihood}.  Again,  the ``fast evolution" model by \citet{Stecker06}  leads to a  high rejection significance 
with two sources (J0808-0751 and J1016+0513) with $> 4 \sigma$ post-trial significance. The post-trial significance is computed 
by taking into account the fact that our analysis is considering $\sim 200$ independent sources.
}
\label{tab:lrt}
\end{center}
\end{table}

As compared to the HEP method, the LRT method incorporates the
possibility of each photon being from the background into the unbinned
maximum likelihood computation.  Thus separate calculations of the
background probablity and corresponding rejection probablity are not
needed.  Also since the LRT method takes into account all high-energy
photons rather than the highest-energy ones in the HEP method, it
gives more constraining results for the EBL model rejection with the
exception of 2 GRBs where the HEP method gives slightly more constraining
results.
Finally, we note that the {\em a priori} choice of the size of the region around each source defined to look for associated high-energy events is a source of systematics for the HEP method ( which uses 68\% PSF containment radius) while it does not affect the LRT method.

\subsubsection{Multi-trial effects and combined probabilities}
\label{sec:combined_P}

Because the search for EBL signatures or rejection of specific EBL
models is performed on all blazars and GRBs detected by the LAT, one
has to consider multi-trials, which is potentially affecting our
analysis.  For independent searches, as is the case here, the
post-trial probability threshold for obtaining a $4\sigma$ result is
$P_{\rm post-trial} = 1-(1- P_{4\sigma})^{1/N_{\rm trials}}$, where
$N_{\rm trials}$ is the number of trials and $P_{4 \sigma}$ is the $4
\sigma$ probability threshold for a single search ($\approx 6.3 \times
10^{-5}$).  In the present case, the LAT AGN catalog that we have
used \citep{1LAC} includes 709 AGNs of which $\sim 200$ have a
sufficiently high redshift ($\sim 100$ with $\gtrsim 10$~GeV photon)
to allow for the testing of EBL attenuation models with their $\g$-ray
spectra.  Only a handful of LAT GRBs were observed with sufficient
statistics to hope to constrain the EBL. In the end, we have
$N_{trials} \sim 200$ which corresponds to a post-trial probability
for a $4 \sigma$ result of $P_{4\sigma,post-trial} \approx 3.17
\times 10^{-7}$. This corresponds to a significance of $\approx 5.11
\sigma$ on an individual source which we will therefore consider as
our threshold for a $4\sigma$ post-trials rejection significance for
any specific EBL model.  This $P_{4\sigma,post-trial}$ threshold was
reached in case of the ~\citet{Stecker06} ``baseline" model for sources
J0808-0751 and J1016+0513 using the LRT method.  Note that J1504+1029 is only
slightly below this threshold.

Combining specific EBL model rejection probabilities from multiple
sources\footnote{since the spectral fits of all the sources we
considered in this analysis are independent to each other.} we get a
much higher rejection significance. For HEP probabilities the
combined rejection significance for the ~\citet{Stecker06} ``baseline"
model is $\approx 8.9\sigma$ ($\approx 7.7\sigma$ without the 2 GRBs)
using Fisher's method in order to combine results from independent  
tests of the same Null-hypothesis \citep{fisher}. For
the LRT method, we add the individual likelihood profiles to derive an
overall profile from which $\sqrt{TS_{max}}$ gives an overall significance of $11.4 \sigma$ for the same EBL model.
Therefore both methods give very large rejection significances even
after taking multi-trial effects into account. Since the~\citet{Stecker06}
``fast-evolution" model gives opacities larger than the ``baseline
model" in the LAT range, both models can be rejected by our analysis
with very high confidence level.  All other models can not be
significantly rejected even after such stacking procedure is applied.

\subsection{Opacity upper limits}

Upper limits on the $\g$-ray optical depth have been evaluated with a method based on the comparison between the 
measured energy spectrum of the source and the unabsorbed spectrum above 10~GeV.
The unabsorbed spectrum, F$_{unabs}$, is assumed to be the extrapolation of the low-energy part, 
 E$<$10~GeV, of the spectrum (F$_{E<10}$), where EBL attenuation is negligible (see Fig.~\ref{fig:tau_vs_energy}), to higher energies.
F$_{E<10}$ is fitted with a power-law or log-parabola function, according to the best TS value.
At high energies, if 
no intrinsic hardening of the spectrum is present, the measured spectrum, F$_{obs}$, at (observed) energy E and the unabsorbed spectrum, F$_{unabs}$, are related by
Eq.~\ref{flux_relation}. The $\gamma$-ray optical depth can therefore be estimated at any given energy as
\be
\tau_{\g\g}(E,z) = \ln[ F_{unabs}(E)/F_{obs}(E) ].
\label{eq:tau}
\ee 
Since F$_{unabs}$ is evaluated assuming no EBL attenuation, it gives a maximum value. 
Therefore the
optical depth, $\tau_{\g\g}(E,z)$ given by Eq.~\ref{eq:tau} could already be considered as an upper limit, 
assuming that the difference between $F_{unabs}(E)$ and $F_{obs}(E)$ is only due to EBL effects.
The fit of both $F_{obs}$ and F$_{E<10}$ are carried out with a maximum likelihood analysis \citep{MattoxLikelihood}
\footnote{{\it gtlike} tool
in the standard {\em Fermi} LAT {\it Science Tools} package provided by the
{\em Fermi} Science Support Center (FSSC)}.

To evaluate $F_{E<10}$ we have assumed a background model including all the point-like sources within $15^\circ$ from 
the source under study and two diffuse components (Galactic and extra-galactic).  
The Galactic diffuse emission is
modeled using a {\it mapcube} function, while a tabulated model is used for
the isotropic component, representing the extragalactic emission as
well as the residual instrumental
background\footnote{http://fermi.gsfc.nasa.gov/ssc/data/access/lat/BackgroundModels.html}.
Both diffuse components are assigned a free normalization for the
likelihood fit.  
In the fit we have considered all the nearby point sources within a 10$^\circ$ radius, 
modeled with a power-law with the
photon index fixed to the value taken from the 1FGL catalog \citep{1FGL} and the
integral flux parameter left free.
The remaining point sources are 
modeled with a power-law with all spectral parameters fixed.  

The source
under study has been fitted with a power-law and a log-parabola with all spectral parameters free.
Among the two, we have chosen the fitted function showing the best TS value. The result is that
for all the sources except J1504+1029 a power-law fit is preferred. 
From the fit results of $F_{E<10}$ we have extrapolated the spectral shape to obtain $F_{unabs}(E)$ above 10~GeV.

A different method has been used to derive the measured flux $F_{obs}$ in selected energy bins.
The whole energy range from 100~MeV to 100~GeV is divided in
equal logarithmically spaced bins requiring in each energy bin a TS value greater than 10: 
2 bins per decade above 10 GeV for J0229-3643,  J1016+0513 and
J1147-3812, 4 bins per decade for J0808-0750 and 5 bins per decade for J1504+1029.
In each energy bin the standard \emph{gtlike} tool has been applied assuming for all the 
point-like sources a simple power law spectrum with photon index fixed to 2.0\footnote{since the energy bin is small enough to assume a flat spectrum.}
The integral fluxes of all point-like sources within $10^{\circ}$
are left as free parameters in the fitting procedure, while the diffuse background components are modeled as described in the previous paragraph.
In this way,
assuming that in each energy bin the spectral shape can be
approximated with a power law, the flux of the source
in all selected energy bins is evaluated.

Once both $F_{unabs}$ and $F_{obs}$ are determined, the maximum $\g$-ray optical depth
in each energy bin can be estimated from Eq.~(\ref{eq:tau}).  
An upper limit on $\tau_{\g\g}(\langle E \rangle,z)$ with 95\% CL in a constraining
energy bin with mean energy $\langle E \rangle$ is then calculated by
propagating the parameter uncertainties in the fitted flux\footnote{It has been veryfied that the statistical errors follow a Gaussian
distribution. The standard error propagation formula has then been
applied.}:

\be
\tau_{\g\g,UL95\%CL}(\langle E \rangle,z) = \ln[ F_{unabs}(\langle E \rangle)/F_{obs}(\langle E \rangle) ] + 2\sigma.
\label{eq:tauUL}
\ee

\noindent We compare these limits with the $\g$-ray optical depths predicted by various EBL models.

\begin{figure}[ht]
\begin{centering}
  \begin{minipage}[b]{6.5cm} 
     \includegraphics[angle=90, width=6.5cm]{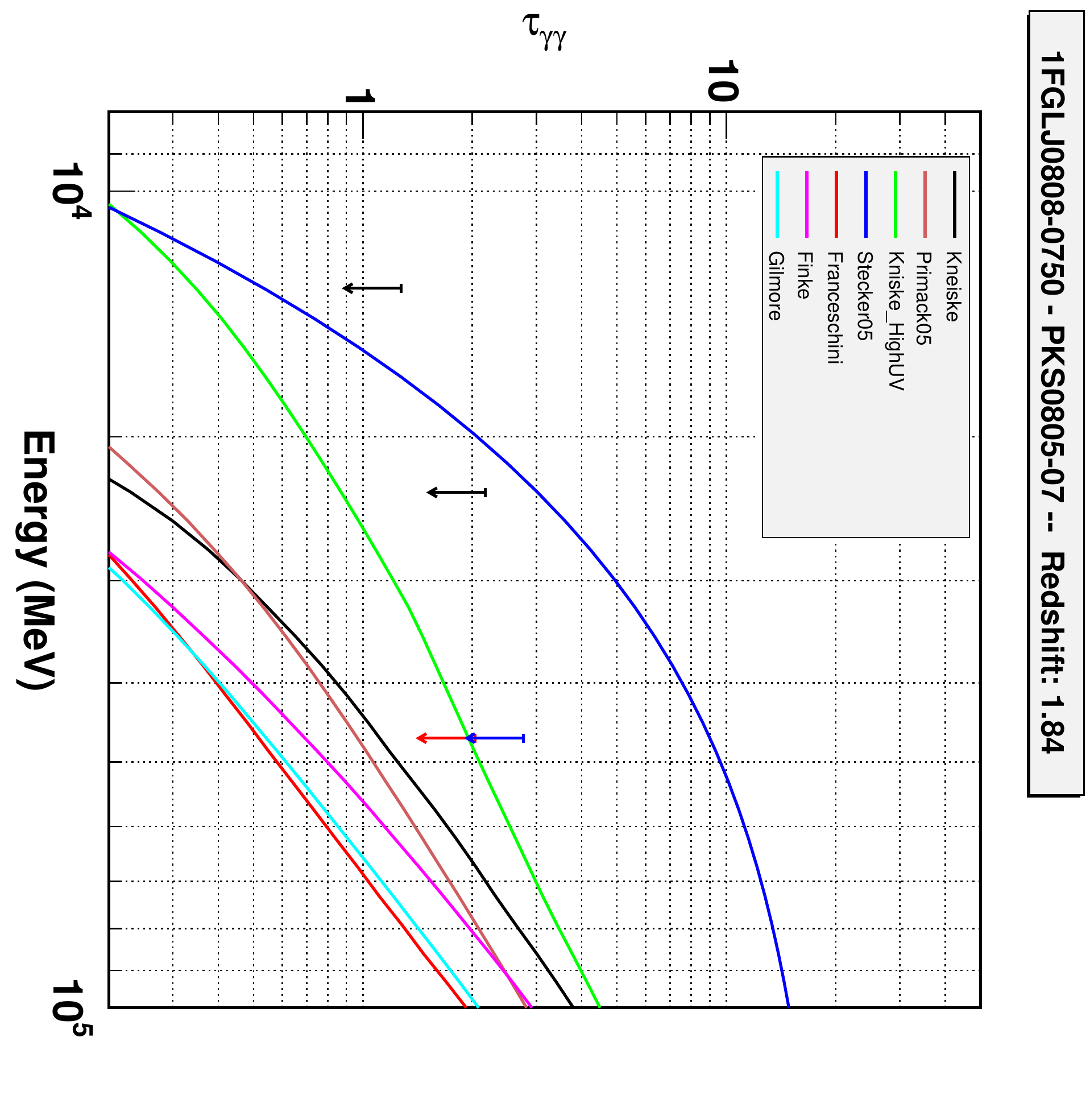}
  \end{minipage}
  \begin{minipage}[b]{6.5cm}
     \includegraphics[angle=90, width=6.5cm]{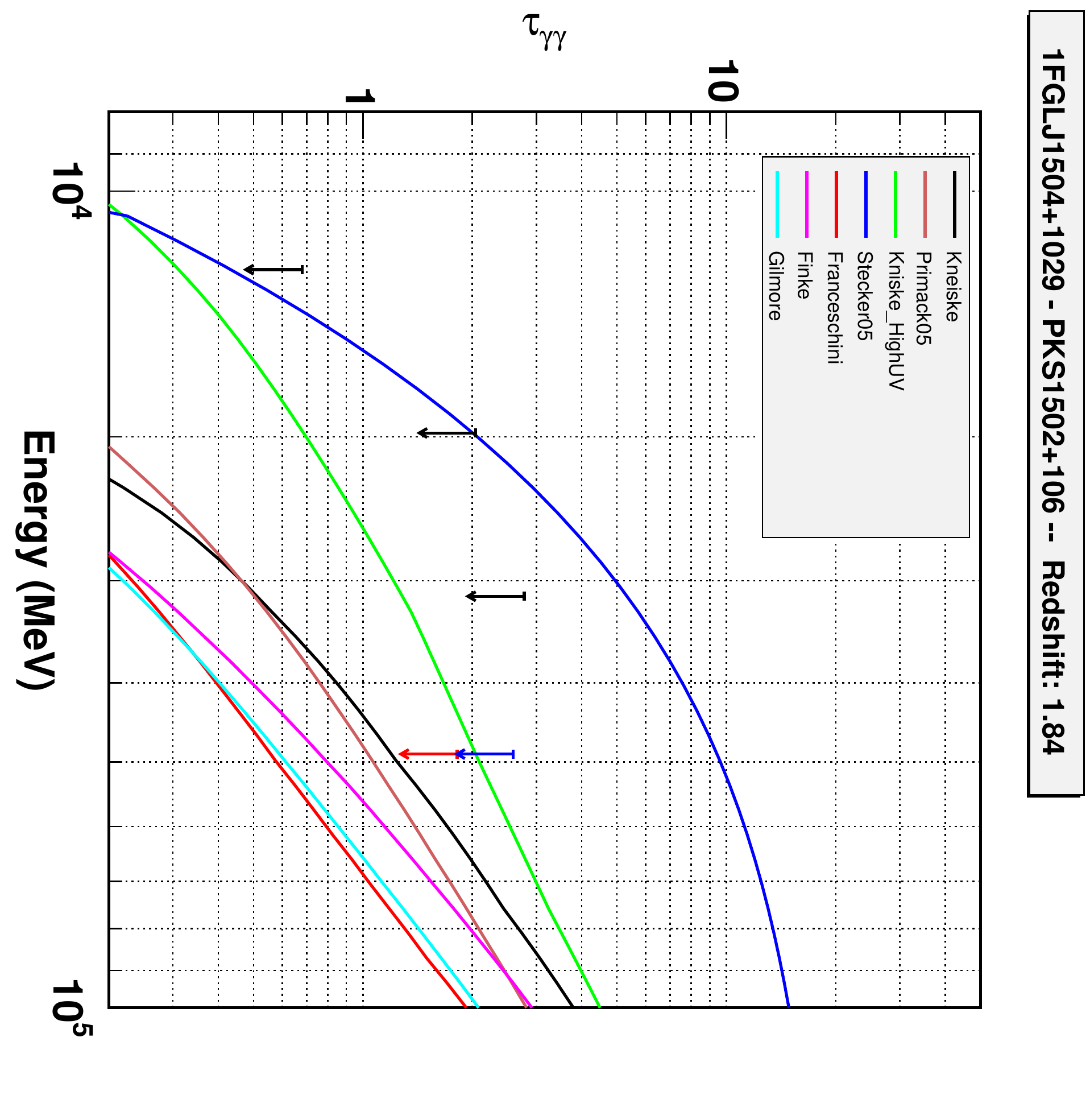} 
  \end{minipage}
  \begin{minipage}[b]{6.5cm}
      \includegraphics[angle=90, width=6.5cm]{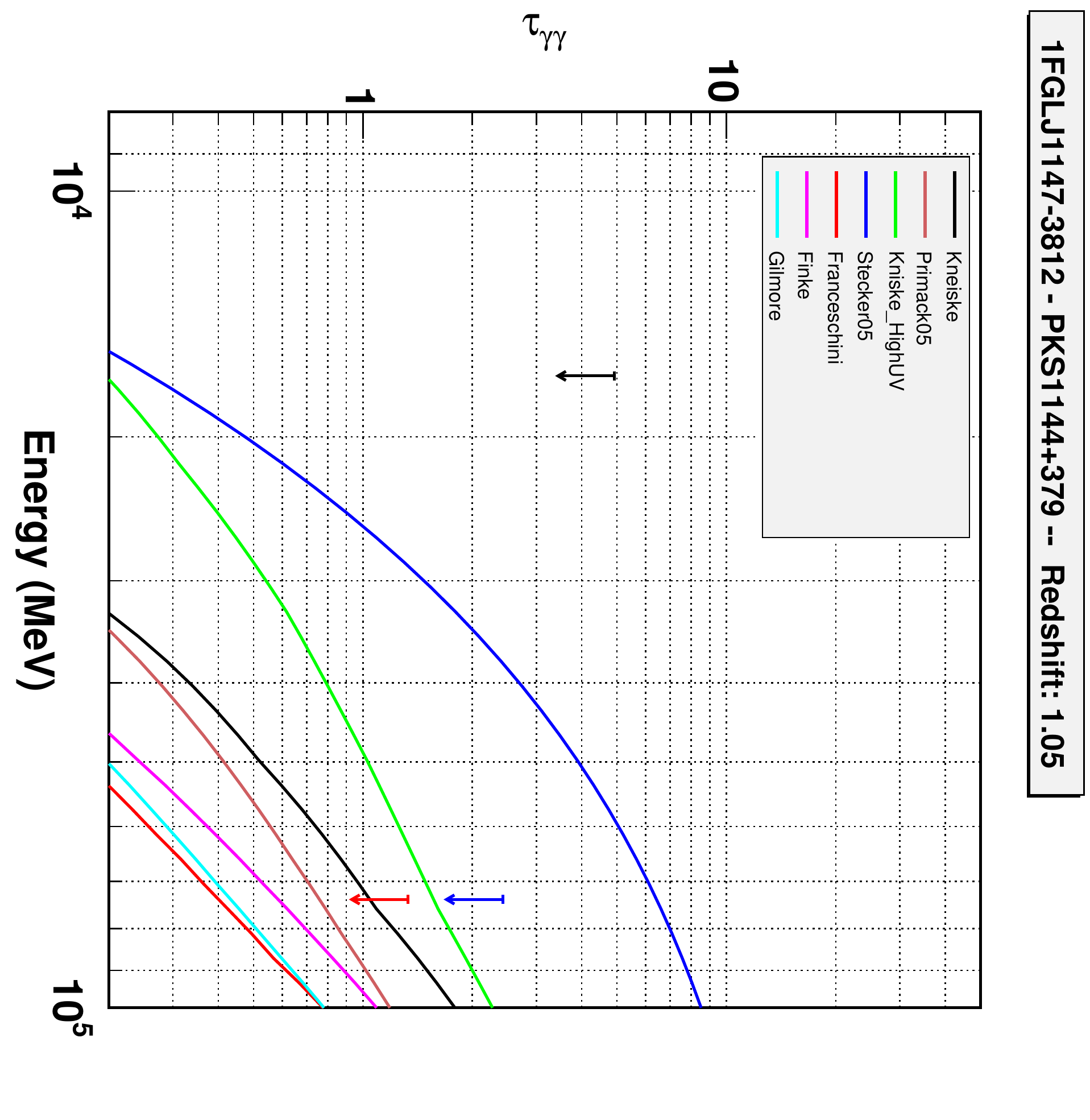}
  \end{minipage}
  \begin{minipage}[b]{6.5cm}
     \includegraphics[angle=90, width=6.5cm]{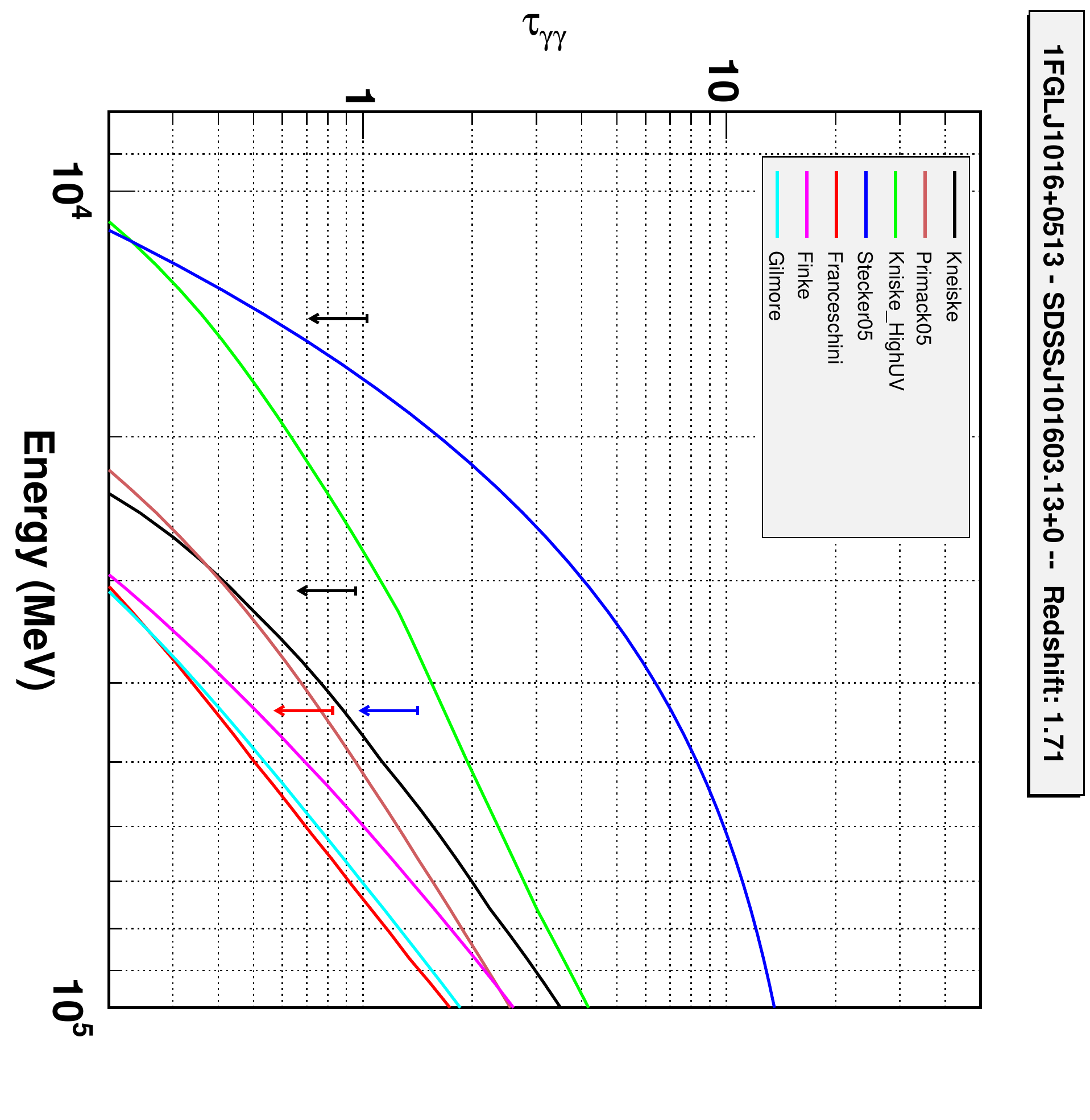} 
  \end{minipage}
\par\end{centering}
\caption{\label{fig:tau_limits} Derived upper limits for the optical depth of $\g$-rays emitted at z=1.84 (J0808-0751, J1504+1029),
z=1.05 (J1147-3812) and z=1.71 (J1016+0513).
 Black arrows: upper limits at 95\% c.l. in all energy bins used to determine
 the observed flux above 10 GeV. Red arrow: upper
 limits at 95\% c.l. for the highest energy photon. Blue arrow: upper limit
 at 99\% c.l. for the highest energy photon. The upper limits are
 inconsistent with the EBL models that predict the strongest opacity.}
\end{figure}

In Figure~\ref{fig:tau_limits} we show the upper limits ($95\%$
CL) derived at the mean energy of the bins above 10 GeV for various objects. 
In the highest energy bin the optical depth UL has been evaluated at the highest photon energy
as reported in Table~\ref{tab:spectres}. 
At this energy, the results of the optical depth UL at 99\% CL are also reported (blue arrow).
As an example, consider blazar J0808-0751 at $z=1.84$ shown in the upper left plot: a larger optical depth would require an
intrinsic spectrum that at high energies lies significantly above the
extrapolation obtained from the low energy spectrum.
The figure shows that 
the upper limit rules out those EBL models
that predict strong attenuation. This result is consistent with all
other upper limits derived with this method (see the other plots in Figure~\ref{fig:tau_limits} and summary of the results in Table~\ref{tab:opacityUL}).

\begin{table}
\begin{center}
\renewcommand{\arraystretch}{1.2}
\begin{tabular}{|c|c|c|c|c|cl} 
\hline
Source & z & $E_{max}$ & $\tau_{UL} (z,E_{max})$ & Energy bins \\
 & & & & 10~GeV$-$100~GeV \\
\hline\hline
J1147-3812 & 1.05 & 73.7 &  1.33 & 2 bins/dec \\
J1504+1029 & 1.84 & 48.9 &  1.82 & 5 bins/dec \\
J0808-0751 & 1.84 & 46.8 &  2.03 & 4 bins/dec \\
J1016+0513 & 1.71 & 43.3 &  0.83 & 2 bins/dec \\
J0229-3643 & 2.11 & 31.9 &  0.97 & 2 bins/dec \\
J1012+2439 & 1.81 & 27.6 &  2.41 & 2 bins/dec \\
\hline
\end{tabular}
\caption{Upper limits (95\% c.l.) on the $\g$-ray optical depth for AGN in Table~\ref{tab:spectres}.
The first and second column report the name of the sources and their redshift, 
the third
column the maximum photon energy and the fourth column the optical depth UL
evaluated at 95\% c.l. as $\tau_{UL} = ln[F_{unabs}(E)/F_{obs}(E)] + 2\sigma$., the fourth column the number of energy bins/dec (for E$>$10~GeV) used to evaluate $F_{obs}(E)$.}
\label{tab:opacityUL}
\end{center}
\end{table}

\section{Discussion}
\label{sec:discussion}

Studies with the highest energy extragalactic photons seen by the
{\em Fermi} LAT primarily probe the UV and optical components of the EBL.
The background light responsible for $\g$-ray attenuation can
evolve strongly with redshift.  In many of the models analyzed in this
paper the EBL intensity can exceed the local value by a factor of 10
or more at redshifts near the peak of star-formation rate density.
The optical depth to $\gamma$ rays from extragalactic sources
 is determined by integrating the EBL intensity
along the line of sight to the source from the observer.

The value of the longest
wavelength photon (source frame) with which a $\g$-ray emitted at $z_{\rm src}$ with
observed energy $E_{\rm obs}$ can interact is given by setting $\theta = \pi$ in equation \ref{eqEBL}:

\begin{equation}
\lambda_{\rm max}=47.5 \: (1+z_{\rm src}) 
\; \left[{\frac{E_{\rm obs}}{\mbox{GeV}}}\right] \; \mbox{\AA~.}  
\label{max_wavelength}
\end{equation} 

\noindent The equation is used to set an upper limit on background photon wavelengths
that can contribute to the $\g$-ray optical depth.  Limits for
$\lambda_{\rm max}$ are 
7175 \AA~for blazar J1147-3812, and 4474 \AA~for GRB 090902B and 3286
\AA~for GRB 080916C, based on the highest-energy $\g$-rays seen from
these sources.  In reality, interactions with shorter wavelength
photons are more likely and will contribute more to the optical depth
because the $\g$-ray redshifts during propagation to Earth.
The cross section peaks at approximately twice the threshold energy which corresponds to an interaction angle of $\theta \sim
90^\circ$.

The results of our analysis of the highest energy $\g$-rays from
blazars and GRBs detected by the {\em Fermi} LAT disfavor a UV background intensity
at the level predicted both by the baseline and fast-evolution models of
\citet{Stecker06}. The LAT observations discussed here however do not
constrain the predictions of the two models at longer wavelengths.  The two
models are based upon a backwards evolution model of
galaxy formation.  In this scenario, the IR SED of a galaxy is
predicted from its luminosity at 60 $\mu$m.  The locally-determined IR SED (at 60
$\mu$m) luminosity function is then assumed to undergo pure luminosity
evolution following a power law in $(1+z)$.  Optical and UV
luminosities, relevant to {\em Fermi}'s extragalactic observations, are then
determined by analytic approximation to the SEDs in \citet{Salamon98}
and are normalized to the short wavelength portions of the IR SEDs.
These models do not include absorption of UV light by dust in
star-forming regions and the interstellar medium of galaxies, which
may partially account for the high background in this model.  While
this model does account for redshift evolution in the UV-optical SEDs
of galaxies, it does not allow for any evolution in the IR emission to
which these SEDs are normalized.  As mentioned by \citet{Stecker06},
this is another factor which could result in overpredicted UV
emission.
Emissivity at UV wavelengths is closely tied to the global
star-formation rate density.  Because the models of \citet{Stecker06}
are not derived from an assumed function for the star-formation rate,
limits on the UV emissivity in this case cannot be used to directly
constrain star-formation.  

We do not find that our results
are conclusively in disagreement with the `best-fit' model of \citet{Kneiske04}. 
In this models, the
optical-UV EBL is based upon a Salpeter IMF and a star-formation rate
density that peaks at $z\sim$1.25, with a value of $\sim 0.2$
M$_\odot$ Mpc$^{-3}$ yr$^{-1}$, and falls slowly towards higher
redshift.  In the high-UV model, ultraviolet flux is boosted by a
factor of 4 above the level of the best-fit model, greatly enhancing
the opacity for $\g$-rays at energies below about 200 GeV.  A star
formation history of the magnitude required to produce the background
in the high-UV model would be above essentially all estimates of the
global star formation rate (see for example
\citet{Hopkins&Beacom06}). All other EBL models are of such low density in the
UV range that they can not be constrained by the data presented in this work.
Although the results
of our analysis can not yet place any stringent upper limits on the cosmological
star-formation history that are competitive with current observational
estimates, future prospects for probing low density UV models of the EBL 
by means of improved methods and enlarged GeV photon data sets
may be promising.

High-energy $\g$ rays that are absorbed by the EBL photons can also
initiate a pair cascade by subsequent Compton scattering of the CMB
photons by the pairs.  In case the intergalactic magnetic field (IGMF)
is very weak, so that the pairs do not deflect out of our line of
sight, this cascade radiation component can be
detectable~\citep{Plaga95}.  Calculations of such cascade signatures have been
carried out for AGNs~\citep[see e.g.][]{Dai02,Murase08,Essey10} and
for GRBs~\citep[see e.g.][]{DL02,Razzaque04,Takahashi08} and found to
compensate for a large portion of the flux that is absorbed in the EBL.
If blazars or GRBs are sources of ultrahigh energy cosmic rays (UHECRs; \citet{Waxman}), then photohadronic 
interactions by protons during their propagation in the background light can also 
induce a high-energy cascade signature that would 
form appreciable high-energy emission, provided the IGM magnetic field is sufficiently small
\citep{Essey10}.
However recent flux upper-limits calculated in the {\em
Fermi} LAT range from TeV blazars 1ES 0347-121 and 1ES 0229+200
constrain the IGMF to be $> 3\times 10^{-16}$~G~\citep{Neronov10}.
Such a strong field reduces the cascade flux significantly (because the emission
becomes essentially isotropic due to the large deflection angles) and the
contribution to the observed flux is likely to be small.  
Furthermore, since the constraining blazar sample consists of FSRQs only, which seem weak TeV emitters,
any of their reprocessed emission can only be small also.

An interesting possibility to reduce $\g$-ray attenuation by EBL has been proposed through 
exotic scenarios involving oscillation between $\g$-rays and axionlike
particles, while propagating in the Galactic magnetic field, from
distant sources and may produce observable signatures in the TeV
range~\citep[see e.g.][]{axions1,axions2}.  However the effect may not
set in for typically assumed IGMF values or likely to be too small to
make up for EBL flux attenuation in the $\lesssim 100$~GeV
range~\citep[see e.g.][]{axions3}

In the end, using the high-energy 11-month photon data set collected by {\em Fermi}
from distant blazars and two GRBs, we have (i) placed upper limits on the opacity of
the Universe to $\g$ rays in the $\sim$10--100~GeV range coming
from various redshifts up to $z\approx 4.3$; and (ii) ruled out an EBL
intensity in the redshift range $\sim$ 1 to 4.3 as great as that predicted by
\citet{Stecker06} in the ultraviolet range  
at more than $4\sigma$ post trials in two independent sources (blazars). The overall rejection significance is found to be $>10 \sigma$ post trials therefore making this result very robust. Our most 
constraining sources are blazars
J1504+1029, J0808-0751 and J1016+0513 with $(z, \langle E_{max} \rangle)$
combinations of (1.84, 48.9~GeV), (1.84, 46.8 GeV) and (1.71, 43.3 GeV), respectively.
Although a likelihood ratio analysis of the latter source indicates
that the sensitivity of our analysis method is approaching the EBL flux level of
the ``high UV model'' of Kneiske et al (2004), multi-trial effects markedly reduced
the rejection significance.
The two most constraining GRBs are GRB 090902B and GRB 080916C, both
of which rule out the ``baseline'' EBL model of \citet{Stecker06} in the UV energy range at more than
3$\sigma$ level. The ``fast evolution'' model of \citet{Stecker06} predicts higher opacities in the LAT energy range at
all redshifts, and therefore is also ruled out. Together with the results from VHE observations (e.g., \citet{aharonian07,Mazin07})
the models by \citet{Stecker06} seem now disfavored in the UV and mid-IR energy range.
We have also calculated model-independent optical depth
upper-limits $\tau_{\g\g, \rm UL} (z, \langle E_{max} \rangle)$
at 95\% CL in the redshift $z\simeq 1-2.1$ and $E_{\rm max}\approx 28-74$ GeV ranges.

As the high-energy photon data set collected by {\em Fermi} grows in the future
and more blazars and GRBs are detected at constraining energies, the
$(E, z)$ phase space that constrains $\tau_{\g\g}$ will become more
populated.  
This will provide us with unique opportunities to
constrain the opacity of the Universe to $\g$-rays over a large energy
and redshift range, and eventually help us further understand the evolution of the
intensity of the extragalactic background light over cosmic time.

       	\chapter{\label{Conclusion}Conclusions and future prospects}

The $Fermi$ Large Area Telescope, jointly with the Gamma-ray Burst Monitor, has opened a new era for high-energy astrophysics and in particular for the study of Gamma-Ray Bursts. This thesis provides a comprehensive overview of the $Fermi$ observations of the prompt emission of fourteen GRBs detected by the LAT instrument during its first year and a half of operation. In this sample, key temporal and spectral features were revealed, in particular through the detailed time-resolved spectroscopy performed on the four brightest bursts in the LAT. These new observations stand as serious challenges for theories attempting to provide a complete and coherent picture of the GRB emissions. As a consequence, $Fermi$ LAT observations triggered a substantial amount of theoretical work in order to tackle specific issues faced by the standard scenario where GRB prompt emission is produced by the acceleration and emission of high energy particles at internal shocks in a relativistic outflow while the afterglow emission results from its deceleration by the circumburst medium. While no single model can provide a satisfactory explanation for all observed features, various revisions of the internal/external shock scenario are being investigated under the light of the GRB temporal and spectral properties revealed by $Fermi$ with the goal of eventually achieving a unified pictures of the GRB phenomenon.

The most striking feature uncovered by $Fermi$ is the time delay of the onset of the high energy $\g$-ray emission (typically $\gtrsim 100$ MeV) with respect to the onset of the lower energy emission (typically $\lesssim 1$ MeV). This feature was clearly observed in all four LAT bright bursts while faint LAT bursts observations are consistent with such feature at the level of statistics detected (GRB 090217 being the only exception \cite{Abdo_090217:10}). Hadronic models ({\it e.g.} proton synchrotron, photo-hadron electromagnetic cascades), which investigate possible evidence of Ultra-High Energy Cosmic Rays in GRBs, could explain the late onset of the $\gtrsim 100$ MeV emission by the time needed to accelerate protons (or heavier ions) and to develop the subsequent electromagnetic cascades. However this class of models has difficulty to reproduce the correlated variability between the low and high energies which is observed in the $Fermi$ lightcurves. On the contrary, leptonic scenarios ({\it e.g.} electron synchrotron or jitter radiation \cite{Medvedev:00} in the sub-MeV regime and inverse Compton or self-Compton processes above $\gtrsim 100$ MeV energies) naturally predict the temporal correlation but have issues producing a delayed onset of the high energy emission which is longer than a spike width (which could be the accumulation time for a target photon field). This feature may also originate from internal absorption in the jet which would imply a spectral curvature between the BGO and the LAT energy ranges. Unfortunately, such curvature has not been detected or ruled out so far mainly because the BGO detectors are background dominated at the highest energy channels and because the LAT lacks photons with reliable energy estimation below $\sim 100$ MeV. Such spectral diagnosis would however be crucial to pin down the underlying process responsible for this feature. Future improvement of the LAT systematics at low energy (currently being developed in the collaboration \cite{Pelassa:10}) will hopefully allow such detailed spectroscopy to be performed.

A second important feature is an additional high energy component (modeled by a power- law) which was first detected by EGRET in GRB 941017. This additional component was clearly found in all bright LAT GRBs at the exception of GRB 080916C which only had a marginal detection. Particularly interesting is the fact that this power-law component was found to extrapolate in the NaI detectors below an energy of $\sim 50$ keV. This is a real challenge to explain for leptonic models under the usual Inverse Compton scattering scenario although models including additional processes such as a photospheric component may provide better agreement with these observed properties. Hadronic models however could produce such low energy extension via synchrotron emission from secondary $e^+ e^-$ pairs produced via photo- hadron interactions.

The detection of such additional component requires a significant signal both in the LAT and in the high energy channels of the BGO in order to constrain the shape of the spectrum properly. The absence of such component in faint LAT bursts is non-conclusive as it could simply be that the level of statistics is not sufficient to detect this additional component. However, one has to be very careful in generalizing features observed in bright LAT GRBs to the whole GRB population as these burst represent a biased sample - brightest and hardest GRBs - which could display peculiar features not present in the rest of the GRB population. At the current level of statistics, the LAT detection rate is consistent with a simple extrapolation from the sub-MeV behavior observed by BATSE. This seems to suggest that a fair fraction of GRBs do not display additional component that would enhance the high energy emission significantly (as is the case for three of the four bright LAT GRBs). However, detailed population studies will be necessary to quantify this statement in the close future.

A third feature observed for the first time by $Fermi$ is the clear signature of a spectral curvature in GRB 090926A (at an energy of $\sim 1$ GeV). This type of spectral feature is even more difficult to detect than the hard powerlaw component as it requires high statistics in the LAT up to high energies where the curvature occurs. Distinguishing the exact shape of the curvature (exponential cutoff, broken power-law, log-parabola...) would require even more statistics which are not available in this burst. However resolving the shape of the spectral curvature is key to unravel the underlying physical process. If due to internal absorption, this would be the first indirect measurement of the bulk Lorentz factor of the jet during the prompt emission phase. A value of $\Gamma \sim 200-700$ was derived while for other bright LAT GRBs, the absence of such curvature up to the highest energy detected allowed us to set stringent constraints on the bulk Lorentz factor: $\Gamma_{min} \gtrsim 300-1000$. The uncertainties in the $\Gamma$ measurement and lower limits come from theoretical uncertainties on the evolution and homogeneity of the colliding shells.

In the end, $Fermi$ LAT revealed several clear properties of the high energy prompt emission of GRBs (chapter \ref{HEprompt}). Although a unified picture of GRBs has not yet emerged, the $Fermi$ observations provide very important clues on what some features of the model might be and are a crucial help to rule out some theoretical interpretations and thus reduce the phase space of possible explanations of the GRB phenomenon. Answers to the pending open questions, however, still require further observations with the $Fermi$ LAT instrument of brighter GRBs as well as a sizable sample of GRBs for reliable population studies of the high energy emission of GRBs. It is the hope that these future observations along with theoretical efforts will lead to a better understanding of Gamma-Ray Busts in the near future.

With their very high energy emission (typically $\gtrsim 10$ GeV), GRBs make very powerful tools to constrain interesting physics. The detection of a 31 GeV photon associated with the short GRB 090510 allowed us to set the most constraining limits on Lorentz Invariance Violation so far (chapter \ref{lorentz}). In particular, a linear dependence of the speed of light with energy (positive delay) can be ruled out with high confidence ($M_{QG} \gtrsim 1.2 M_{Planck}$ using the most conservative assumption) by our observations while higher orders are still far from being significantly constrained. Thanks to distant and bright sources in the $\g$-ray regime such as GRBs and blazars, we constrained the opacity of the optical-ultraviolet extragalactic background light (chapter \ref{EBL}). While no signature of EBL absorption was found in the LAT data, we set constraints on the total opacity at different redshift (up to $z \sim 4.3$). In particular, we found our data inconsistent at a high significance ($\gtrsim 10 \sigma$) with the two most opaque models tested which are SteckerÕs ÔbaselineÕ and Ôfast evolutionÕ predictions. We also set upper-limits on the total opacity as a function of energy for our most constraining sources (see fig. \ref{fig:tau_limits}).

Detecting GRBs at very high energy with ground-based Cherenkov telescopes has proved very difficult. However, a single such observation would be extremely valuable in many respects: investigate the effect of EBL, probe the highest energy emitted by the prompt and afterglow emission where the LAT is statistically limited (crucial to get a good handle on the bulk Lorentz factor of the jet for example), perform time-resolved spectroscopy with enhanced statistics at very high energy (lack of statistics currently preventing us to resolve specific spectral shape: power-law component, exponential cutoff, broken power-law...). Better constraints of the underlying physics of GRB emission (internal absorption, particle energy distribution, emission processes...) would result from such VHE GRB detection. Although the chance of placing a more sensitive GeV telescope in orbit in the near future are small, the prospects for very high-energy astronomy on the ground are definitely brighter. Imaging Air Cherenkov Telescopes (IACTs) in particular with the next generation instrument CTA (Cherenkov Telescope Array) probably hold the best chances of GRB detection thanks to its enhanced effective area and a low energy threshold of a few tens of GeVs. Moreover, this instrument is scheduled to operate during the lifetime of $Fermi$. Finally, the High Altitude Water Cherenkov Experiment (HAWC), another ground TeV telescope, probably has better chances to catch the prompt emission of GRBs thanks to its wide field-of-view and high duty cycle compared to IACTs, although its much smaller effective area would only make it sensitive to an extremely bright event.

	\appendix
       	\chapter{Background estimation for transient events}
\label{bkg_est}

Having a good handle on the background is a crucial part of any instrument. A first step is to achieve minimum background contribution to the overall measurement while still keeping as much of the desired signal as possible. This was the goal of section \ref{event_optimization} where a set of different event classes where recommended in the case of specific applications ({\it e.g.} detection, spectral analysis of high or low flux sources) as good compromise between event rejection and signal efficiency.
As a second step (once a we are happy with the event selections), one needs a precise background estimation of the experiment in order to properly extract the signal information. 

In the context of studying GRB (and more generally transient objects) with the LAT, a good estimation of the background is useful for many different reasons:

\begin{packed_enum}
\item Estimation of the significance of the detection of a transient event such as the prompt or afterglow emission of a GRB, flares
from sources such as Soft Gamma-ray Repeaters or blazars (see section \ref{significance}).
\item Computation of upper-limits in the case of non-detection by the LAT
\item Perform spectral analysis with a proper description of the background level as a function of energy.\item Perform blind searches for transient emissions.
\end{packed_enum}

These are critical tasks that cannot be performed in any accurate way without any proper description of the background level for a particular transient observation (at the exception of the spectral analysis of bright bursts for which background contamination is fairly negligible).
At the begining of the mission, much effort was put into developing a tool that could accurately estimate the background for short time periods, for any position of the spacecraft and at any position in the sky. I myself pioneered a work in this area and provided help in the development of the subsequent project. The goal of this section is to describe the tool that is used to perform accurate estimation of transient background\footnote{Vlasios Vasileiou is the person that developped the background estimator tool which is now a standard tool for LAT transient analysis in the $Fermi$ team. The description of the inner working of the background estimator in this appendix is actually mostly based on Vlasios' own writing.}.

\section{Estimating background level during a transient emission}

\begin{figure}
\begin{center}
\includegraphics[ width=1\linewidth, keepaspectratio]{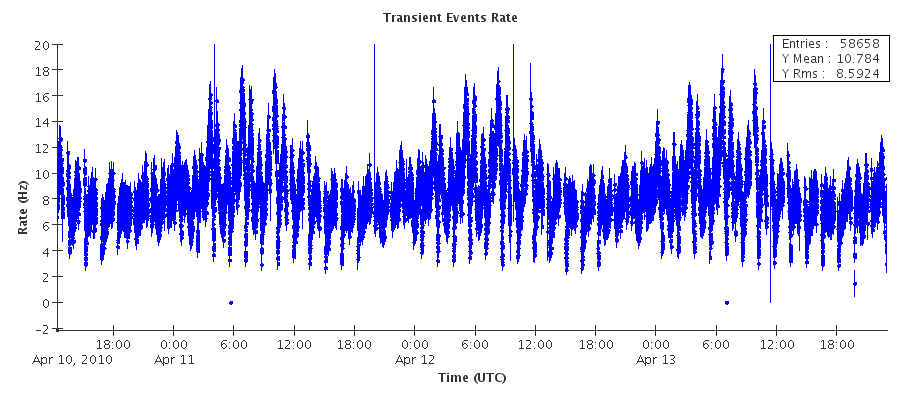}
\end{center}
\caption{All sky transient background as a function of time. Notice the large fluctuations (up to a factor of two) as the spacecraft orbits around the earth and changes geomagnetic latitude.}
\label{transient_bkg}
\end{figure}

An issue with the estimation of background on short time-scale occurs when the background is significantly fluctuating with time. And this is very much the case for the LAT instrument in particular for the transient event selection which is used for observation of short transient objects.
The instantaneous rate of LAT background events from a source depends strongly
on the geomagnetic coordinates at the location (see figure \ref{transient_bkg}) of the spacecraft and the off-axis
angle of the source (see figure \ref{irf diffuse}). These dependencies can cause short-term fluctuations of the
background rate that can be as large as a factor of two. For long-term observations
these fluctuations average out, however for short-term observations (seconds to
$\sim$ several tens of minutes) they are important.

One possible approach to derive background during transient sources is the so called 'on-off technique'. The idea behind it is to measure the background level at the source location just before and after the transient emission where the absence of signal is certain or alternatively during the transient emission but in a slightly offset location of the sky (notice that the instrument needs to have localization capabilities to do that). And one could of course perform both of these approach at the same time.
Let us know mention some potential issues with this 'on-off technique':

\begin{packed_enum}
\item The background level in the off-source region needs to be a good representation of the background level in the on-source region (after proper re-scaling of the off-source background measurement to the on-source window with the factor $\frac{\Delta T_{on} \times \Omega_{on}}{\Delta T_{off} \times \Omega_{off}}$ where ($\Delta T_{on}$, $\Omega_{on}$) and ($\Delta T_{off}$, $\Omega_{off}$) are the time window and solid angle of the on-source and off-source region respectively)
\item Enough statistics needs to be accumulated to have an accurate measurement of the background level. This is an issue in particular for low-background measurement
\item The background region should not be contaminated by the signal you are trying to measure as this would result in an overestimation of your background level and thus an underestimation of the measured signal.
\end{packed_enum}

The first bullet is particularly difficult when the background level is varying rapidly. A way around it is to establish a proper extrapolation of the background level in the off-source region into the on-source region. This is in-fact the strategy adopted for the GBM detectors for which a measurement of the varying background before and after the burst is performed and a polynomial fit to the data allows to establish a good estimation of the time-varying background in the on-source region.

In the case of the LAT instrument, the background rate is a function of many parameters and can vary by
more than a factor of two, depending on the observational
conditions. For example, there is a strong dependence of the
CR background rate on the geomagnetic coordinates at the
location of the spacecraft. Furthermore, the
background rate also depends on the burst position in instrument
coordinates, because the LAT's acceptance
varies strongly with the inclination angle.
For these reasons, it is not straightforward to estimate
the expected amount of background during the GRB emission using
off-source regions around the trigger time, since the spacecraft will
have moved to regions of different geomagnetic coordinates,
and the inclination angle of the region of interest will have
changed significantly.  

In the case of the LAT instrument, the background level is very low with an all sky rate of $\sim$ 2 Hz for transient events which corresponds to 1 background event in a $10^{\circ}$ region-of-interest every $\sim$10 sec (this is very crude estimate as it strongly depends on the location of the spacecraft and the source location in the instrument coordinate). In order to accumulate enough statistics, we would need to integrate over long period of times or/and in large region of the sky surrounding the source position. However, the $Fermi$ observatory is moving at a rate of $\sim 4^{\circ}$/ minute around the earth which is also the rate of rotation of the pointing direction of the LAT boresight. The effective area being strongly dependent with the angle to the LAT boresight (even for cosmic-rays which constitute a large fraction of the transient background), the time of integration can not be arbitrarily large. And in most situation (see GRB 080825C in section \ref{significance} for a lucky situation), it will not be possible to apply this technique in order to derive the background level during the time of source emission.

And alternative approach is to make use of past data where the instrument was in simular conditions. If enough data have been stored, one can accumulate enough statistics in order to reach the desired level of statistics for accurate background estimation. This is the approach that we used to develop our background estimator tool.

\section{Modelisation of the LAT background for transient events}

The background-estimation tool can estimate the background for observations of
any duration and from any direction in the celestial sphere (including the galactic
plane). It can provide estimates for both past and future observations and works with
the S3, Transient and Diffuse LAT data classes. The estimates are provided versus
the energy and have an accuracy of ~15\%. The tool is currently available to only the
members of the LAT collaboration, however it is planned to be included in the
Science Tools and released to the public.

The gamma-ray component of the background from some direction in the sky
depends only on the exposure at that direction, while the Cosmic-Ray (CR)
component of the background also depends on the geomagnetic coordinates of the
spacecraft during that observation. For that reason, these two components are
estimated separately using two different techniques.

\subsection{'Cosmic-ray' background}

An integral part of the background estimation is the calculation of the all-sky cosmic-ray background for each second of the observation under consideration. The all-sky background is composed of multiple components; cosmic rays, gamma-rays from point and extended sources, gamma-rays from the diffuse galactic and extra-galactic emissions, etc. The relative contributions of these component may vary depending on which part of the galaxy the LAT is observing, and on the properties of the earth's magnetic field at the location of the spacecraft.

The cosmic-ray all-sky background rate ($R_{CR,all-sky}$) depends on the properties of the earth's magnetic field at the location of the spacecraft. A parameter that describes the local magnetic field and is appropriate for estimating the cosmic-ray background rate is McIlwain's L parameter (McIlwainL). The backgroung estimator uses the pre-calculated function $R_{CR,all-sky}$ (McIlwainL) to estimate the all-sky cosmic-ray rate $R_{CR,all-sky,obs}(t)$ for each second of an observation obs described by a $McIlWainL_{obs}(t)$.

Special care has been taken to avoid the inclusion of gamma-rays from the galactic plane during this calculation. Only data taken when the LAT +z axis was pointing to a galactic latitude $|PtBz| > 70^{\circ}$ were used. This way the galactic plane was observed from an off-axis angle that was large enough (or equivalently the effective-area was small enough) that the contamination from galactic-plane gamma-rays to be considerably reduced. 

As mentioned above, the all-sky background is composed of many components with the cosmic-ray component dominating. The second strongest component is the gamma-ray component from the galactic plane. The galactic-latitude cut applied above ($|PtBz|>70^{\circ}$) did reduce that second largest (galactic-plane gamma-ray) component, but did not reduce the gamma-ray component that is present at high galactic latitudes (extra-galactic point sources, high-latitude galactic sources, galactic and extra-galactic diffuse). For that reason, when I refer to the "cosmic-ray all-sky background" I should more precisely refer to the "all-sky background minus a big fraction of the galactic-plane background". The difference between these two backgrounds is small.

The $R_{CR,all-sky}$ (McIlwainL) fits for each of the 15 energy bins are shown below. The first (underflow) histogram corresponds to all the $E<20$ MeV events, and the last (overflow) histogram corresponds to all the $E>300$ GeV events. The overflow and underflow histograms are only shown here for reference; estimates for these two energy ranges are produced by the estimator but are not shown in the final result.

\begin{figure}
\begin{center} $
\begin{array}{cc}
\includegraphics[ width=.5\linewidth, keepaspectratio]{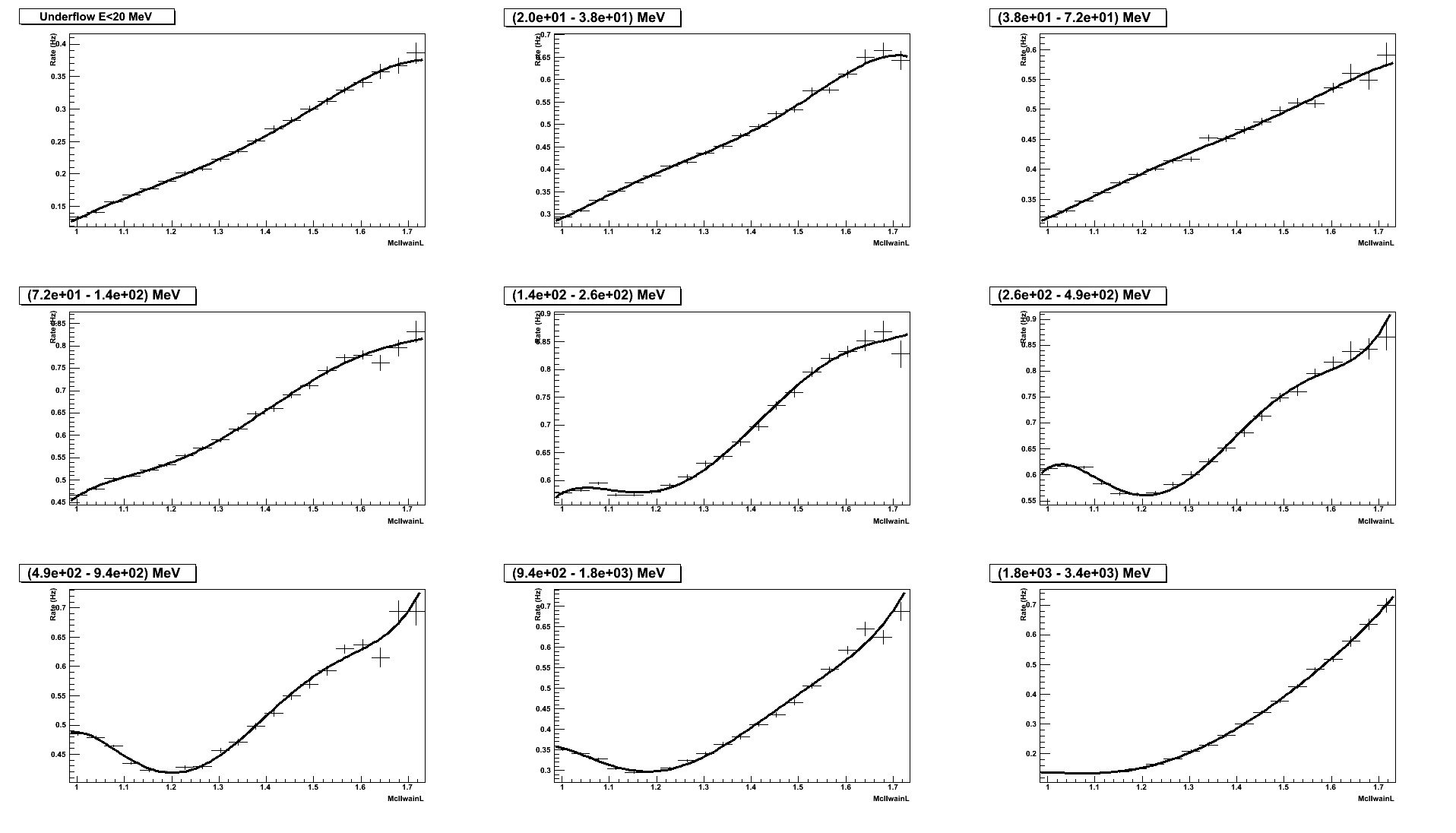} &
\includegraphics[ width=.5\linewidth, keepaspectratio]{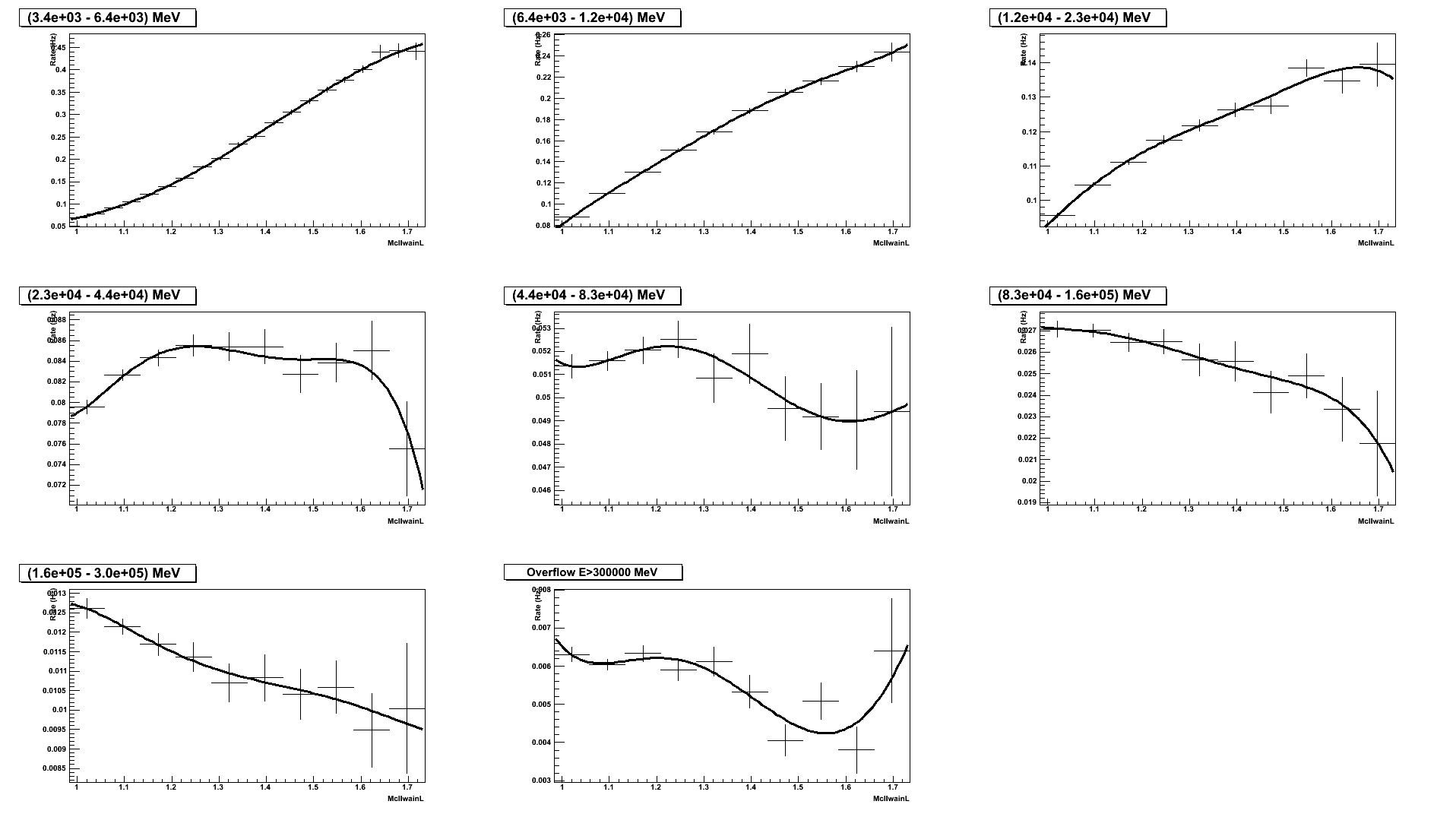}
\end{array} $
\end{center}
\caption{All-sky CR-background rate for transient class events with energies $\sim 1$GeV versus the
McIlwain L coordinate of the location of the spacecraft.}
\label{Rates}
\end{figure}

Using the function $R_{CR,all-sky}$ (McIlwainL), the code can estimate the all-sky cosmic-ray background of an arbitrary observation. The next step is to calculate from which part of the celestial sphere these cosmic-ray events came from. Here, an assumption is made that the theta and phi distributions of cosmic-ray events are constant independently on the McIlwainL parameter, or the direction the LAT is pointing to. Based on that assumption, these distributions produced from a sample of the data can be used as probability distribution functions for a generic cosmic-ray event to have a specific theta and phi. Because the theta and phi distributions also depend on the energy, these distributions are produced for different energy bins. Also, because the cross section of the LAT is square (instead of circular), the theta distribution also depends on the phi angle (the theta distribution extends to higher values for phis that point towards the corners of the LAT). For that reason the theta distributions are also produced versus theta.

For the theta distribution, in order to reduce the contamination from galactic-plane gamma-rays, only data taken when the LAT was pointing to a galactic latitude $|PtBz|>70^{\circ}$ were used. With such a cut, a small amount of galactic-plane gamma-rays is still present at the tail of the zenith-angle distribution.

The way a (theta/phi) pair is generated by the code is the following:

A phi is sampled from the phi distribution of the energy bin under consideration
A theta is sampled from a theta distribution appropriate for the sampled phi and the energy under consideration.
The zenith angle of the (theta/phi) sampled pair is calculated. If the zenith angle is over the zenith-angle cut then another pair is sampled. The zenith-angle calculation uses the directions of the Earth's Zenith and the LAT's +Z axis for the specific second of the observation simulated.

\subsection{$\gamma$-ray background}

Using the information calculated above the code can calculate the amount of cosmic-ray background arriving from each direction of the celestial sphere during an arbitrary observation. The next step is to calculate how much gamma-ray background is also expected to be detected during that observation. To accomplish that, a map that contains the amount of gamma-ray background per unit of exposure is needed. First, a skymap is created by the reconstructed directions of the events included in the months-long analysed dataset. Then, a map containing the simulated directions of cosmic-ray events for the same dataset is created. The difference of these two maps (the "residual map") shows the amount of accumulated gamma-ray signal during that period. By dividing the residual map with a map that shows the exposure of the analysed data set, a new map called the "residual-over-exposure" map is created. The gamma-ray background of an arbitrary observation can be calculated by multiplying the residual-over-exposure map with the exposure that corresponds to that observation.

\begin{figure}
\begin{center} $
\begin{array}{cc}
\includegraphics[ width=.5\linewidth, keepaspectratio]{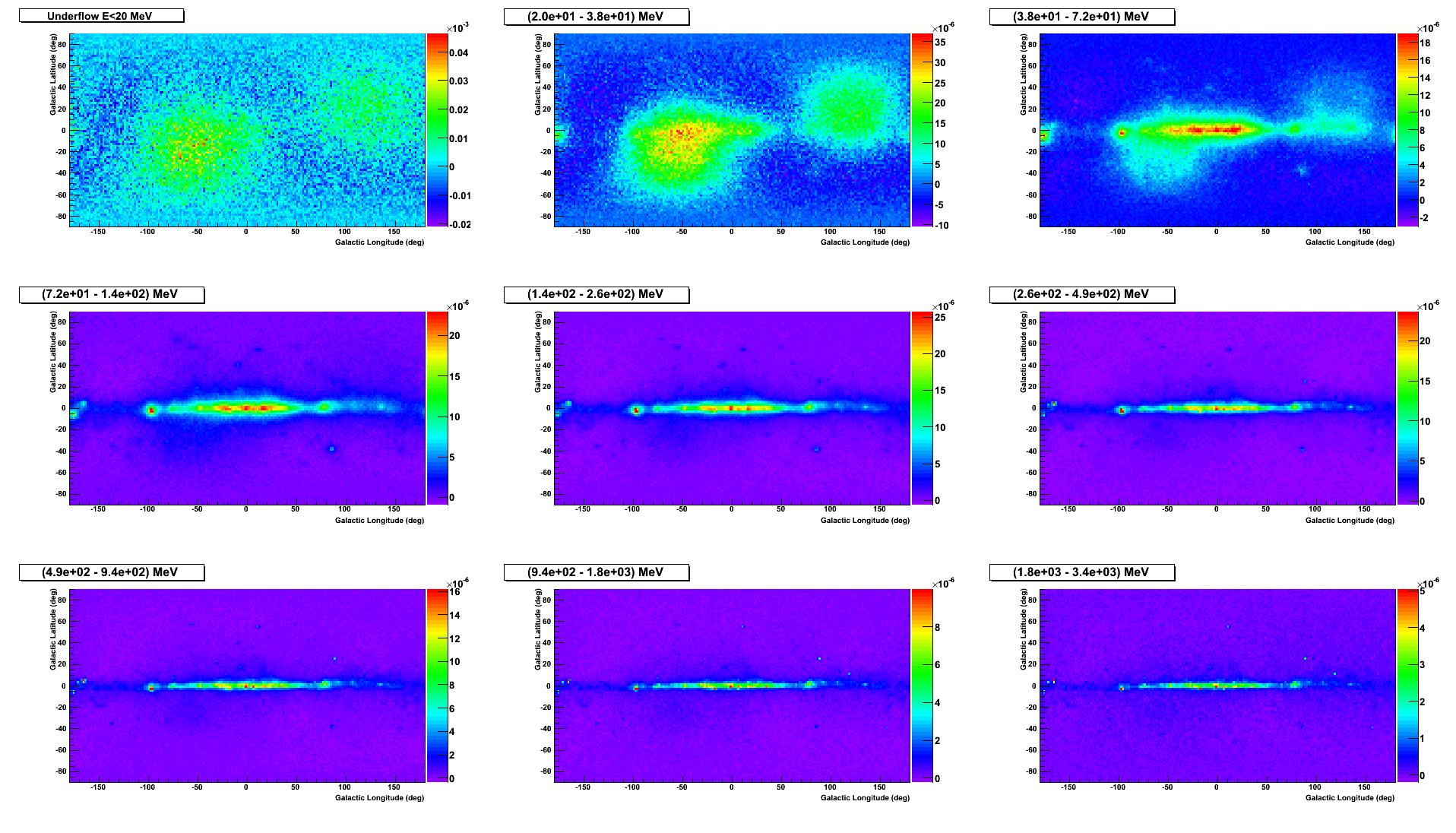} &
\includegraphics[ width=.5\linewidth, keepaspectratio]{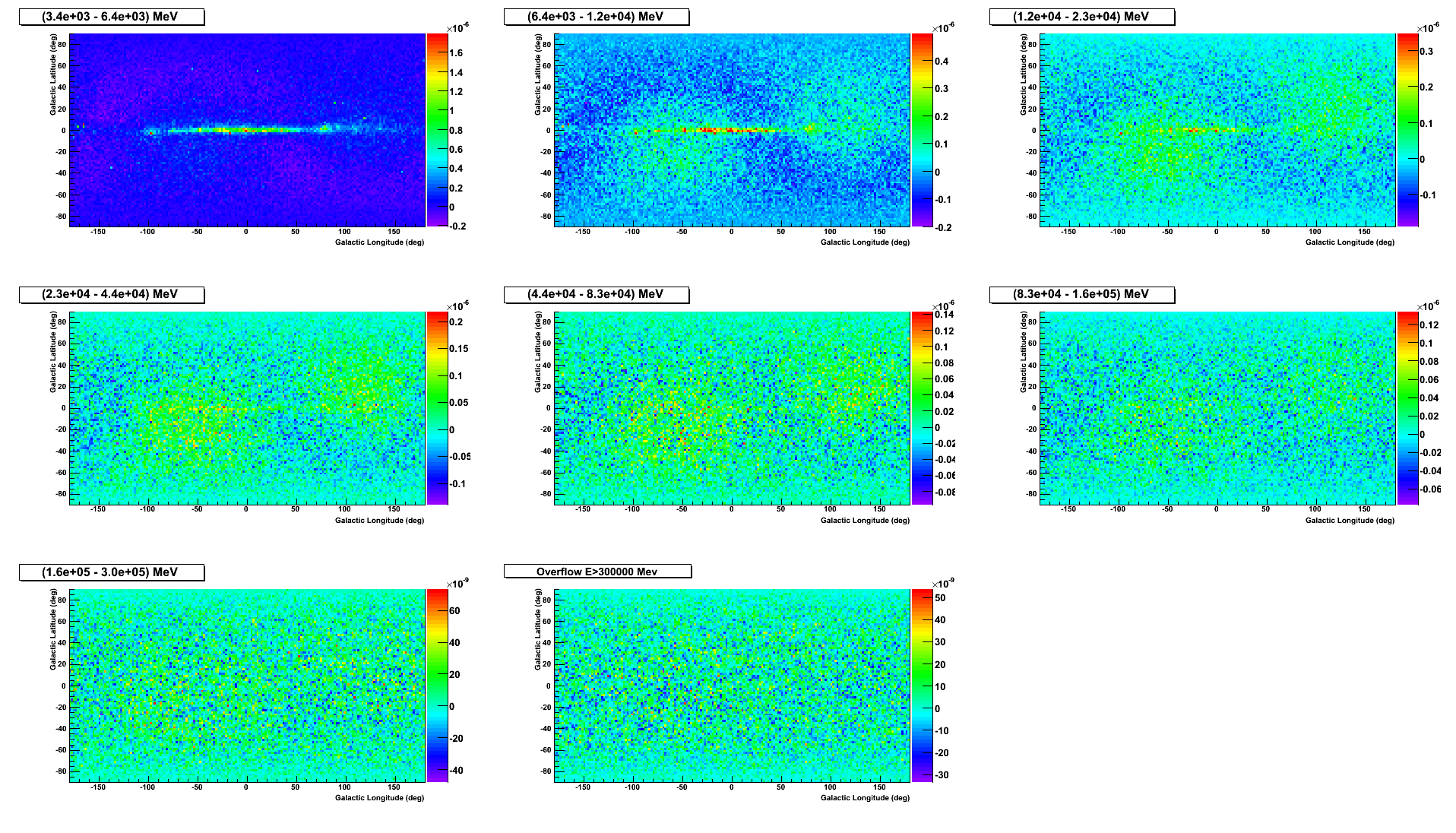}
\end{array} $
\end{center}
\caption{Residual-over-background maps for each of the 15+2 energy bins. The first (underflow) map corresponds to all the $<20$ MeV events, and the last (overflow) map corresponds to all the $>300$ GeV events.}
\label{residual}
\end{figure}

The residual-over-background maps for each of the 15+2 energy bins are shown in figure \ref{residual}. The first (underflow) map corresponds to all the $E<20$MeV events, and the last (overflow) map corresponds to all the$E>300$GeV events. The overflow and underflow maps are only shown here for reference; estimates for these two energy ranges are produced by the estimator but are not shown in the final result.

\subsection{Putting it all together}

The estimation in the end is the sum of a cosmic-ray and a gamma-ray component:

\begin{packed_enum}
\item The estimation of the first component initially involves the calculation of the all-sky cosmic-ray rate, for each second of an observation, using its dependence on the McIlwain L parameter at the location of the spacecraft at that second. Then, the amount of cosmic-rays from each direction of the celestial sphere is estimated using the theta and phi distributions of cosmic rays (calculated from real data) as probability distribution functions.
\item The second gamma-ray component is estimated by multiplying a pre-calculated map that contains the amount of gamma-ray background per unit of exposure by the exposure of the observation under consideration.
\end{packed_enum}

The above process is repeated for a number of (30-30-20) energy bins logarithmically spaced in the (20-50-80)MeV-300GeV range for the (S3, transient, diffuse) class. The calculation requires some pre-calculated data that come with the code. Details on how these data were produced can be found in the next chapter. The detailed steps involved in the background estimation are the following:

\begin{packed_enum}

\item {\bf Step 1. Calculation of the all-sky cosmic-ray rate  $R_{CR,all_sky}$ (McIlwainL):}
The code breaks the observation under consideration in one-second time segments. For each segment, it calculates the all-sky cosmic-ray rate based on the value of McIlwain L (for that segment) and the pre-calculated dependence of the all-sky cosmic-ray rate on that parameter. Because the code uses a ZenithTheta$<90^{\circ}$ cut to reject the earth's albedo, a fraction of the sky is almost always (depending on the rocking angle) excluded by that cut. For that reason the "all-sky" rate at each moment is somewhat smaller than the true all-sky rate of the case of no ZenithTheta cut. The amount of rate reduction depends on the rocking angle (higher rocking angle $\rightarrow$ more rate decrease) and the distribution of inclination angles for CR background events (see map below). The all-sky rate used by the code is the correct reduced rate that takes into account the fraction of the FOV removed by the ZenithTheta cut.

\item {\bf Step 2. Calculation of the cosmic-ray rate from a specific direction of the celestial sphere:}
For each one-second time segment, the cosmic-ray background from each direction of the celestial sphere is calculated. The code uses the pre-calculated distributions of theta and phi for cosmic-ray background events as probability distribution functions, and the pointing information of the LAT (latitude and longitude that the LAT +z axis points to - calculated from PtRaz/PtDecz) to distribute the cosmic-ray background events to directions on the celestial sphere. The ZenithTheta angle of all generated events is also calculated and a ZenithTheta$<90^{\circ}$ cut is applied. A skymap in galactic coordinates is filled with the resulting simulated events. That map contains the estimated amount of cosmic-ray background for the simulated observation.

\item {\bf Step 3. Calculation of the gamma-ray background:}
The exposure of the observation under consideration is calculated using the Science Tools. Then, a map that contains the amount of gamma-ray background per unit of exposure is multiplied by the exposure of the observation. The produced map contains the estimated amount of gamma-ray background for that observation.

\item {\bf Step 4. Calculation of the background map:}
A background map is created by adding the cosmic-ray and gamma-ray background maps calculated above.

\item {\bf Step 5. Calculation of the background from some area of the celestial sphere:}
The contents of the background map are integrated over a circular ROI. The radius of the ROI can be either provided by the user or calculated automatically. For the second case, the ROI radius depends on the PSF and the localization error of the ROI's center. The variables Containment and LocalizationError found in the GRBConfig files set these two parameters. The automatically-calculated ROI radius is equal to the radius that contains a fraction Containment of the LAT's transient-class PSF and the LocalizationError added in quadrature.
The integration does not happen by simply approximating the circular ROI with the rectangular bins of the skymap. Instead, the amount of background per steradian from each direction on the celestial sphere is initially calculated, and a non-binned integration over the ROI is then performed.

\end{packed_enum}

%\section{Evaluation of systematics}

%The background estimate is sensitive to different kind of systematics at long and at short time scales. For short time scales, it is sensitive to errors in the phi/theta distributions, and in the dependence of the all-sky rate on the McIlwain L parameter. For longer time durations the directionality information is washed out, since we are integrating over many overlapping orbits, and the background estimate is mostly sensitive to errors in the all-sky rate calculation.

%Test on ARR observation:

%\begin{figure}
%\begin{center}
%\includegraphics[ width=1\linewidth, keepaspectratio]{Plots/Background_estimation/bkg_ARR.png}
%\end{center}
%\caption{Background estimation for an ARR of 20 ks which did not yield any positive detection of signal from the GRB tracked}
%\label{bkg_ARR}
%\end{figure}

       	\chapter{Methodology for joint LAT/GBM spectroscopy}
\label{spec_method}

\section{Forward-folding technique}

The interaction of a gamma-ray with the NaI, BGO or LAT instruments (through photoelectric effect, compton scattering and pair production) is a complicated process where it is not possible to have a one-to-one correspondence between the energy of an incident photon and the energy measured by the detector. Indeed given the true energy and direction of a gamma-ray photon, the observed energy (derived from the instrumental observables in the detector) will have a statistical dispersion with a width that represents the finite energy resolution of the instrument. It is important to note that this dispersion originates from the statistical nature of the interaction of each photon with the detector and is therefore a fundamental limitation of any instrument. In the case of $Fermi$ instruments, the energy resolution is found at a level of $\sim 10\%$ for the NaI, BGO and LAT instruments\footnote{The energy dependent resolution of the LAT instrument can be found at the following adress: $http://www-glast.slac.stanford.edu/software/IS/glast\_lat\_performance.htm$}.
As a consequence, the reconstruction of the true incoming energy spectrum of a source from the interaction parameters measured by the instrument is a definite challenge. The detected count spectrum cannot be simply inverted into a measured photon spectrum since the energy of a single photon cannot be deduced from the energy of its observed properties of interaction.

In case a lot of counts are observed for a particular source, the incident spectrum
can be unfolded with certain limitations.
Several direct inversion techniques have been proposed
\citep{Loredo:89}, but practitioners have generally not found them to be very useful.
Sometimes the fitted photon function is used to scale the count rate data
points into what appear to be photon flux points. But these must be treated with great caution since what appear to be data points are actually model dependent.
For this reason, the reality of a spectral feature should never be judged on a plot of deconvolved
photon flux.

Instead, the `forward-folding' deconvolution procedure is generally recommended procedure \citep{Briggs:99}. This technique assumes a photon spectral form and compares it to the data by propagating it through a reliable model of your instruments.
A model photon spectrum is represented with a parameterized function and is converted into a model count spectrum in the instrument using the model of the detector which is usually given the name of 'Instrument Response Function' (IRF). This 'forward-folding' of the source model into the IRF provides an expectation for the number of counts in the different instrument channels which allows direct comparison of the model and the observed count spectra.

A reliable description of the instrument response to incoming gamma-rays can be obtained via Monte-Carlo simulations and on-the-ground (before launch) and in-flight calibrations. Continuous in-flight verifications are performed to make sure that the instrument description used for $Fermi$ analysis are reliable. 

The model description of the detector response to an incident gamma-ray spectrum are contained into the IRF which can be factored into 3 different components:

\begin{eqnarray*} 
R(\epsilon',\hat{p}',\epsilon,\hat{p}) = A_{eff}(\epsilon',\hat{p}') \times PSF(\hat{p}',\epsilon,\hat{p}) \times D(\epsilon',\epsilon,\hat{p})
\end{eqnarray*}

where $\epsilon$ ($\epsilon'$), $\hat{p}$ ($\hat{p}'$) are the true (measured) energy and direction of the incoming photon respectively. $A_{eff}$ is called the effective area and is the cross-section of the LAT for detecting an incident photon with energy $\epsilon$ and direction $\hat{p}$. $PSF$ and $D$ represents the point spread function and energy dispersion which are both probability distribution functions. Note that the $PSF$ term is only relevant for an instrument with event-by-event localization capability like the LAT. Therefore only the LAT IRF contains this term as the GBM instruments only record counts in energy channels.

Assuming an input photon spectrum $S(\epsilon,\hat{p})$ one can derive the expected measured count spectrum using the knowledge on the instrument response:

\begin{eqnarray*} 
S_j = \int_{j} d\epsilon' \int_{ROI} d\hat{p}' \int_ d\epsilon d\hat{p} R(\epsilon',\hat{p}',\epsilon,\hat{p}) S(\epsilon,\hat{p})
\end{eqnarray*}

where $S_j$ is the observed number of counts in the energy channel j. 
This predicted value of observed counts in channel j can be directly compared to the actual observed counts in the data $N_j$.

Note that the source spectrum and instrument response have been integrated over the spatial Region-Of-Interest (ROI) used for the analysis. This integration is not needed for the GBM which does not have event-by-event spatial reconstruction capability. Because the LAT $PSF$ is strongly energy dependent (see fig. \ref{irf diffuse}), the region-of-interest have been chosen to be also strongly energy dependent and based on the 95\% containment radius. Details on this spatial selection are given in section \ref{detLATGRB}.

\section{Statistical hypothesis testing}

From experimental data, one is interesting in deriving a theoretical understanding of the particular phenomenon under investigation. It is therefore crucial to develop a reliable framework in order to quantitatively evaluate the capacity of specific theoretical models to provide a satisfying representation of the our observation.
Statistical hypothesis testing is a key technique of frequentist statistical inference to approach this issue. The idea revolves around the assumption of a null-hypothesis which will be confronted to the data. A test statistics will serve as a tool to determine the degree of similarity between the null-hypothesis model and the data. 

$\chi^2$ and likelihood are well-known example of such test-statistics. Because of the low detection rate and the extent of the LAT point-spread function, a specific statistical technique is required to analyse LAT data. Likelihood (first intruduced by \citep{Fisher:25}) is the predominant LAT point-source analysis method and is used to quantify the relative extent to which the LAT data support a statistical hypothesis. In case of joint LAT/GBM analysis we still prefered to use a maximum likelihood procedure as the $\chi^2$ test statistics usually breaks down for low number of counts\footnote{$https://astrophysics.gsfc.nasa.gov/XSPECwiki/low\_count\_spectra$}.

Specifically the likelihood is the joint probability of the observed data given the hypothesis. In case of binned spectral fitting, this probability is easy to compute:

$$Likelihood(Data|Model) = \Pi_i  \frac{M_i^{N_i}e^{-M_i}}{N_i}$$

where $M_i$ and $N_i$ are the predicted number of counts and the actual number of counts in energy channel i.

Parameter estimation is performed by estimating the value of the parameters that maximizes the likelihood. Hence the expression 'method of maximum likelihood'. The free parameters of the model considered are varied until the likelihood is maximized. This provides the best fit function for the specific spectral form assumed. In order to derive uncertainties on the free spectral parameters, one can use the profile likelihood method which will be described after the likelihood ratio (LRT) test has been introduced.

The idea behind the LRT is to compare quantitatively how well different models can account for the observed data. One naively would think that the model with the highest likelihood should be prefered but if this was the case, models with numerous degrees of freedom would have a definitive advantage.
The LRT (introduced by \cite{Neyman:28}) is specifically designed for quantitative model comparison. The likelihood ratio is the likelihood  of the null hypothesis for the data divided by the likelihood of the alternative hypothesis for the same data:

$$LRT = L_{H_{1}} / L_{H_{0}}$$

\cite{Wilks:38} established a very useful expression for the likelihood ratio in the null-hypothesis case (meaning if the null-hypothesis is true) that is asymptotically exact. He demonstrated that $TS = -2 \times log(L_{H_{1}} / L_{H_{0}})$ (where the likelihood are computed for the best fit models) is asymptotically distributed as $\chi_n^2$ in the null-hypothesis  with n the number of additional degrees of freedom that are optimized for the alternative hypothesis. The application of likelihood to photon counting experiments was described by \citep{Cash:79}. The likelihood method was applied to many experiments and in particular to EGRET data \citep{Mattox:96}.

In order for the TS to distribute as $\chi_n^2$ the two models under consideration must satisfy the following 2 conditions \citep{Protassov:02}: 1) the null-hypothesis model must be nested within the more complicated alternative model meaning that the simpler model must be obtainable by setting the extra n parameters of the alternative model to specific values and 2) the null-values of the additional parameters should not be on the boundary of the set of possible parameter values. It is not unusual for the alternative and null hypothesis not to follow one or both of these conditions in which case, strictly speaking, Wilks' theorem cannot apply, despite its common use in the literrature.

If the required conditions cannot be matched in order to simply apply Wilks' theorem to convert the test statistics into a quantitative probability for the alternative hypothesis to better describe the data than the null hypothesis, another procedure needs to be adopted wich is quantitatively correct although usually more intensive to carry out. This procedure makes use of Monte Carlo simulations of the particular experiments you are interested in. Using reliable description of your experiments one simulates the null hypothesis considered in order to derive the distribution of the test statistics used (in this case the likelihood ratio). The number of times the test statistics is higher than the measured value will provide an estimate of the significance of the alternative model over the null-hypothesis model. Not to be statistically limited, enough simulations needs to be produced in order to have a reliable estimate of the chance probability to obtain the test statistics measured on the actual data. For high confidence detection of spectral feature, this requires extremely intensive simulations which is usually not possible to perform. A $5 \sigma$ detection would for example correspond to a $5.73 \times 10^{-7}$ probability therefore requiring several tens of millions of simulation to reach the desired level of accuracy. In case one would like to use the Wilks' theorem anyway, one safety cross-check that should be performed is that the test statistics distribution is indeed fairly close to a $\chi^2_n$ distribution. This is of course not sufficient since one is usually interested in the tail of this distribution which requires intensive simulations as mentioned above but is at least better than nothing.

It is important to realize here a limitation of the LRT method in that it will never tell you if a spectral model is the most appropriate model to describe your data. The only thing it can do is compare models and since the range of possible models is infinite, the task of demonstrating the superiority of one model over all others is hopeless.
If we can never prove a fit to be superior in an absolute sense, how is one suppose to approach the model fitting problem?
The answer is two-fold: by viewing the problem as one of statistical
model comparison and by using scientific judgment.
The statistical approach used is model comparison as described above and a potential new spectral feature is tested against the conventional spectral model adopted and we ask whether the fitting statistic (maximum likelihood in our case) is sufficiently improved to convince us of the reality of the spectral feature.
In principle, to demonstrate the existence of a spectral feature one must show that it is significant over any continuum model. But scientific judgment can be used to limit the space of trial continuum
models to all 'reasonable' models.
Currently, the most popular continuum model adopted to describe GRB prompt emission is the four-parameter 'Band' function \citep{Band:93}. This function  fits
the keV-MeV gamma-ray data quite well. Most of the popular simpler
models can be represented as special cases of the Band function;
depending on the signal-to-noise ratio and energy range
of a spectrum, it may not be possible to demonstrate that the four-parameter
Band function provides a statistically better fit compared to a simpler model.
And therefore it may be sufficient to use just the Band function as the set of
all reasonable continuum models. We will therefore use it as our null-hypothesis in the subsequent analysis.

Interestingly, an analysis of data extending below 20 keV indicates that in some cases the GRB  function may fail in the X-ray range (Preece et al. 1996):  additional low-energy spectral component below 20\,keV for $\sim 15$\% of BATSE bursts.. We will see that the Fermi GBM/LAT data help shed some light on this finding. Also worthwhile to mention is the work by \citep{Ryde:04} which found a significant improvement in the spectral fitting of BATSE bursts when using a blackbody spectrum overlayed by a powerlaw component dominant in the low and high energy regime.

A final consideration when considering the possibility of a new spectral feature on a large number of spectra from a large number of GRBs is to correct the significance for the possibility to mimick the spectral feature purely by chance.   This must be allowed for in calculating the overall
significance of a possible spectral feature by adjusting the threshold of the minimum test statistics required to be convinced of the reality of the spectral feature (multi-trial effect already discussed in section \ref{significance}).

The preceding discussion is based on the assumption that the instrument
works correctly, that the computer model of the detector is sufficiently
accurate, and that the errors in the observed spectrum arise solely
from Poisson fluctuations in the arrival of photons and their interaction
in the detector. Systematic errors are always a concern in such study and a precise investigation of their effect must be carried out. We describe our investigation of systematic uncertainties for spectral analysis in section \ref{sys}.

In summary, the steps to demonstrate the existence of an unknown spectral feature are:
%\begin{enumerate}
\begin{list}{$\bullet$}{\setlength{\itemsep}{0pt} \setlength{\parsep}{0pt}}
\item Convolve the model spectra using the forward-folding technique to show the model fit and the data in units of counts per energy channels.
\item Test the necessity of a spectral feature by comparing fits with and without
spectral feature using the LRT
\item Test the significance of the spectral feature against all 'reasonable' continuum models
\item If a large number of spectra have been examined, consider the
increased probability of a chance fluctuation and appropriately
degrade the significance of the detection of a spectral feature on a candidate
%\end{enumerate}
\end{list}

\section{Spectral fitting tools for joint GBM/LAT analysis}

We performed the joint spectral fitting of GBM and LAT, using the tool
RMFIT version 3.2 \citep{Kaneko:06,Abdo_080825C:09}.

\begin{figure}
\begin{center}
\includegraphics[angle=90, width=.6\linewidth, keepaspectratio]{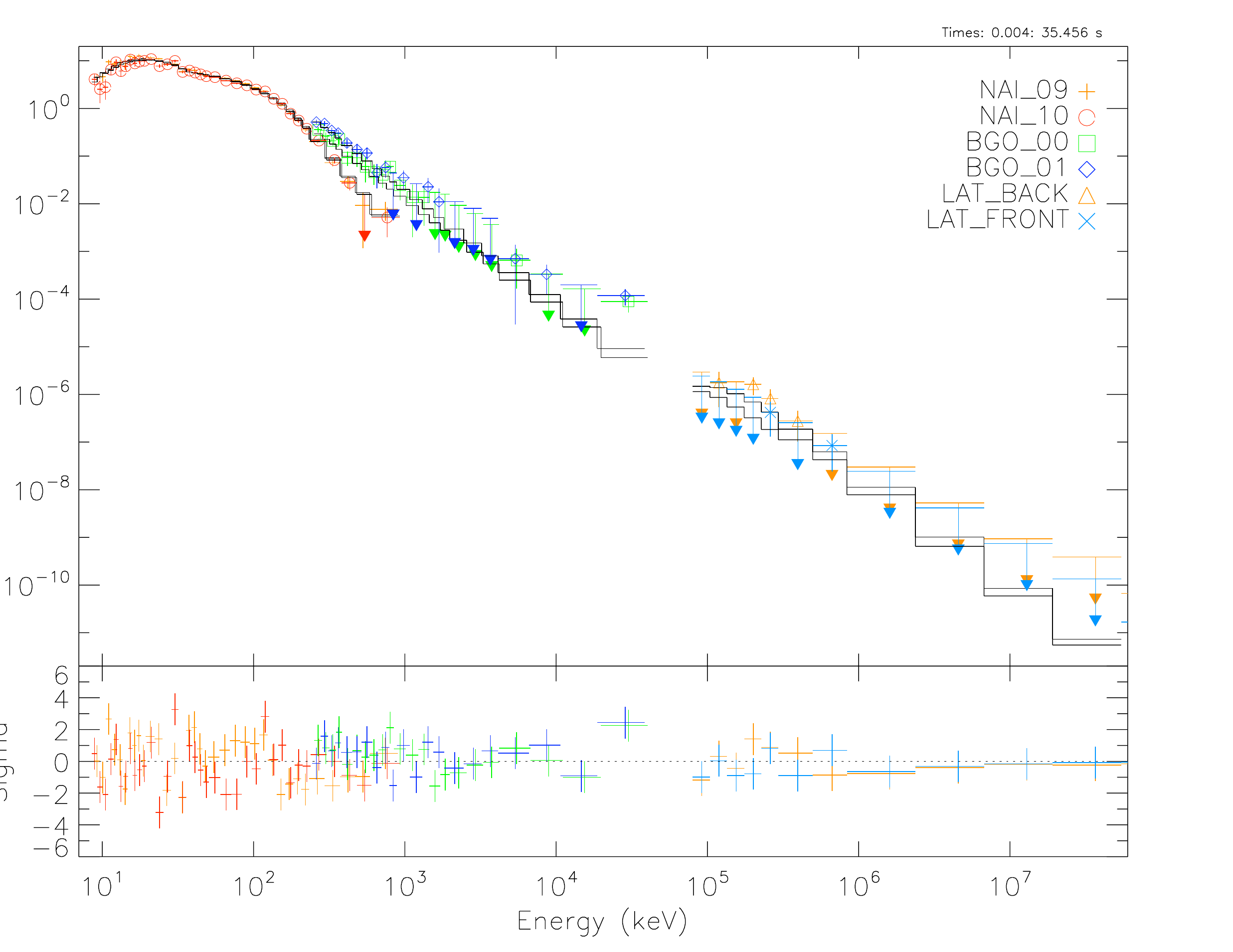}
\end{center}
\caption{GRB 080825C time-integrated spectrum}
\label{rmfit spec}
\end{figure}

The Fermi GRB group makes use of RMFIT as its designated tool to carry out detailed spectral analysis of the prompt gamma-ray emission observed by the NaI, BGO and LAT instruments. 
This tool has been designed by the GBM team building upon the thorough experience aquired with the spectral analysis of BATSE and EGRET bursts. It is written in Fortran/IDL and is used through a Graphical User Interface which allows to interactively perform all the needed steps involved in detailed spectroscopy of the NaI, BGO and LAT data and perform joint fit analysis.
RMFIT uses the forward-folded technique described above and both the chi-square and likelihood statistics have been implemented. The likelihood statistics being used for joint fitting between the GBM and the LAT instruments because of the inherent low statistics of the LAT data.
RMFIT provides a variety of spectral models like simple power-law, Band function, blackbody spectrum, Comptonized model, exponential cut-off which can be additively and multiplicatively combined.
As an example, figure \ref{rmfit spec} represents the time-integrated spectrum of GRB 080825C along with the best fit Band function found (solid line).

Because RMFIT does not support easy scriptability and simulation capability is not fully operational yet, the spectral fitting tool XSPEC\footnote{$http://heasarc.gsfc.nasa.gov/docs/xanadu/xspec/$} has been used for specific analysis involving heavy simulation and fitting of multiple spectra. We have found necessary to use this tool as the use of the GUI of RMFIT for such analysis would have been an endless task. RMFIT developpers are working on making available a working scriptability and simulation capability in the close future which will be quite useful. In each of those cases, the consistency of RMFIT and XSPEC results was of course verified.
        	\chapter{\label{GRB_cosmology}Gamma-ray Bursts as Cosmological Tools}

\section{Introduction}

The change in our understanding of gamma-ray bursts (GRBs) in less than a decade
has been
unprecedented. We have gone from groping for ways to determine their distances
(from solar system to cosmological scales) to attempts to use them  as
cosmological probes. Observations by instruments on board a series of satellites
starting
with {\it BeppoSAX} and continuing with {\it HETE}, {\it INTEGRAL} and
{\it Swift}, have been the primary source of this change. The higher spatial
resolution of these instruments has allowed the measurement of redshifts of many
well-localized GRBs, which has in turn led to several
attempts to discover some emission characteristics  which appears to be
a ``standard candle" ({\bf SC} for short), or shows a well defined
correlation (with a small dispersion) with another distance independent
measurable characteristic. One can use such relations to determine the distances
to
GRBs in a manner analogous to the use of the Cepheid variables.
Example of this are the lag-luminosity and variability-luminosity
relations (Norris et al 2000, Norris 2002, Fenimore \&  Ramirez-Ruiz
2000, Reichart et al 2001) which were exploited for determining some
cosmological aspects of these sources (Lloyd, Fryer \& Ramirez-Ruiz 2002,
Kocevski \& Liang 2006) using the methods developed by
Efron \& Petrosian (1992, 1994, 1999). More recently there has
been a flurry of activity dealing with the observed relation between the
peak energy $\ep$ of the $\nu F_\nu$ spectrum and the total (isotropic)
gamma-ray  energy output
$\eiso$ ($\ep\propto\eiso^{\eta}$, $\eta \sim 0.6$) predicted by Lloyd {\it et al.}
(2000, {\bf LPM00}) and established to be the case by Amati {\it et al.} (2002) (see also
Lamb {\it et al.} 2004, Attiea et al 2004). Similarly Ghirlanda {\it et al.} (2004a) has shown a
correlation with smaller dispersion between $\ep$ and the beaming corrected
energy $\egamma$ ($E_p \propto \egamma^{\eta'}$, $\eta' \sim 0.7$) where:

\be
\egamma=\eiso \times \frac{(1-cos(\tjet))}{2}
\ee

and $\tjet$  is the half width of the jet. 

These are followed by many attempt to use these relations
for determining  cosmological parameters (Dai {\it et al.} 2004,Ghirlanda {\it et al.} 2004b, 2005
[Gea05],  Friedman \& Bloom 2005, [FB05]). There are, however, many
uncertainties associated with the claimed relations and even more with
the suggested cosmological tests. The purpose of this paper is to investigate
the utility of GRBs as cosmological tools either as SCs or via some correlation.
In the next section we review the past and current status of the first
possibility
and in \S3 we discuss the energy-spectrum correlation and whether it can be used
for  cosmological model parameter determination. Finally in \S4, we address the
question of cosmological luminosity and rate density evolution of GRB based on
the existing sample with known redshifts.

\section{Standard Candle?}

The simplest method of determining cosmological parameters is through SCs. Type
Ia supernovae (SNIa) are a good example of this. But currently their
observations
are limited to relatively nearby universe (redshift $z<2$). Galaxies and active
galactic nuclei (or AGNs), on the other hand, can be seen to much higher
redshifts ($z>6$) but are not good SCs. GRBs are observed to similar redshifts
and can be detected to even higher redshifts by current instruments. So that if
there were SCs they can complement the SNIa results. In general GRBs show
considerable dispersion in their intrinsic characteristics. The first indication
that GRBs might be SCs came from Frail {\it et al.} (2001) observation showing that for a
sample of 17 GRBs  the
dispersion of the distribution of $\egamma$ is significantly smaller than that
of $\eiso$.
The determination of $\egamma$ requires a well defined light curve with a
distinct steepening. The jet opening $\theta_{jet}$ depends primarily on the
time of the steepening and the bulk Lorentz factor, but its exact value is model
dependent and depends also weakly on $\eiso$ and the density of the background
medium

\be
\label{theta}
\theta_{jet}=0.101 \mbox{ radian} \times \left( \frac{t_{break}}{1 \mbox{ day}}
\right)^{3/8} \left( \frac{\eta}{0.2} \right)^{1/8} \left( \frac{n}{10 \mbox{
cm}^{-3}} \right)^{1/8} \left( \frac{1+z}{2} \right)^{-3/8} \left( \frac{{\cal
E}_{iso}}{10^{53} \mbox{ ergs}} \right)^{-1/8},
\ee

\begin{figure}[htbp]
\begin{center}
\includegraphics[width=14cm]{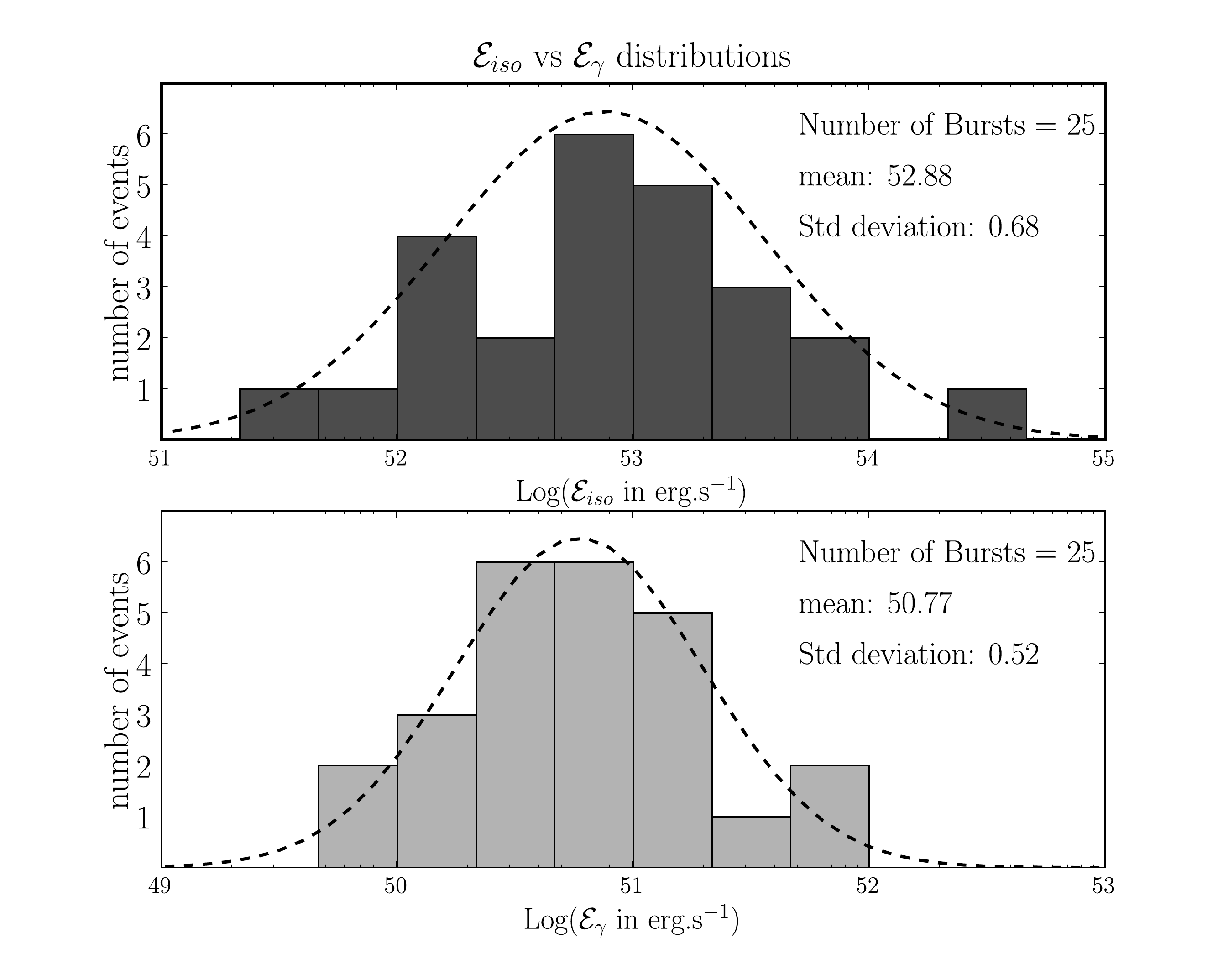}
\end{center}
\caption{Distribution of $\eiso$ and $\egamma$ for 25 pre-Swift GRBs with
evidence for a jet-break and beaming. $\egamma$ distribution is shifted by about
two orders of magnitudes compared to $\eiso$ distribution due to the beaming
factor correction. However the dispersion of the two distribution are very
similar ($\sigma_{\eiso} =0.68$, $\sigma_{\egamma} =0.52$) and broad indicating
that GRBs cannot be assumed to be {\bf SCs}}
\label{1}
\end{figure}

In Figure \ref{1} we show the distribution of $\eiso$ and $\egamma$ for  25
pre-{\it Swift} GRBs (mostly compiled in FB05). As evident, there is little
difference between the two distributions (except for their mean values) and
neither characteristics is anywhere close to being a SC.

We have calculated the jet angle $\tjet$ (from equation \ref{theta}) for 25
pre-{\it Swift} GRBs with relatively well defined $t_{break}$. Assuming a
gamma-ray efficiency $\eta = 0.2$ and a value of circum burst density estimated
from broadband modeling of the lightcurve when available (otherwise we use the
default value of $n=10$ cm$^{-3}$).

Unfortunately fewer than expected {\it Swift} GRBs  have optical light curves
and their X-ray light curves show considerable structure (several breaks and
flaring activity)
with several GRBs showing no sign of jet-break or beaming (Nousek {\it et al.} 2006).
This has brought the whole idea of jet breaks and calculating $\egamma$ into
question.
The upshot of this is that {\it Swift} $\egamma$, like $\eiso$ also has a broad
distribution extending
over two decades. Thus,  any cosmological use of GRBs must
include the effects of the breadth of the distributions%
\footnote{It should be also noted that there is an observational
bias in favor of detecting smaller jet angles (\ie earlier jet-breaks), so that
the population as a whole (including those with very late jet-breaks)
will have even a broader distribution. Note also that a {\bf SC} $\egamma$ means
that the Ghirlanda relation would have an index $\eta' \sim 0$}.

\section{Correlations}

When addressing the correlation between any two variables, one should
distinguish between a
one-to-one relation and a statistical correlation. In general the correlation
between two variables (say, $\ep$ and $\eiso$) can be described by a
bi-variate distribution  $\psi(\ep, \eiso)$. If this is a
separable function, $\psi(\ep, \eiso) = \phi(\eiso)\zeta(\ep)$, then the two
variables are
said to
be uncorrelated. A correlation is present if some characteristic (say the mean
value)
of one variable depends on the other:
\eg $\langle\ep\rangle=g(\eiso)$. Only in the absence of dispersion there will
be
a one-to-one relation;
$\zeta(\ep)=\delta(\ep-g(\eiso))$.

In general, the determination  of the exact
nature of the correlation is complicated by the fact that the extant data
suffers from many observational selection biases and truncations. An obvious
bias is that most sample are limited to GRBs with peak fluxes above some
threshold. There are also biases in the
determination of $\epo$ (see \eg Lloyd \& Petrosian 1999; {\bf LP99}). The
methods devised by Efron \& Petrosian (1992, 1994) are particularly suitable for
determination of correlations in such complexly truncated data.

The first indication of a correlation between
the energetics and spectrum of
GRBs came from Mallozzi et al (1995), who reported a
correlation between observed peak flux $f_p$ and $\epo$.
A more comprehensive analysis  by {\bf LPM00}, using the above mentioned
methods, showed that a
similar correlation also exist  between the observed total energy fluence
$F_{\rm tot}$ and $\epo$. Both these quantities depend on the redshift $z\equiv
Z-1$;
\be
\label{eiso}
F_{\rm tot}=\eiso/(4\pi d_m^2 Z), \,\,\,\,\,\, {\rm and} \,\,\,\,\, \epo=\ep/Z,
\ee
Here $d_m$ is the metric distance, and for a flat universe
\be
\label{dm}
d_m(Z) = (c/H_0)\int^Z_1 dZ'(\Omega (Z'))^{-1/2}, \,\,\,\,\, {\rm with}
\,\,\,\,\, \Omega (Z) = \rho(Z)/\rho_0,
\ee
describing the evolution of the
total energy density $\rho(z)$  of
all substance (visible and dark matter, radiation, dark energy or the
cosmological
constant).
{\bf LPM00} also showed that the correlation expected from these
interrelationships is not sufficient to account for the observed correlation,
and  that there must be an
intrinsic correlation between $\ep$ and $\eiso$. Without knowledge of
redshifts LPM00 predicted the relation ${\cal E}_{iso} \propto E_p^{0.5}$ which
is very close
%\be\label{corr}
%\langle\ep\rangle\propto\eiso^\eta \,\,\,\,\, {\rm with} \,\,\,\,\, \eta\sim
0.5, which is very similar
to the so-called Amati relation obtained for GRBs with known
redshifts. However, it should be emphasized that the LPM00 result implies a
statistical correlation  and
not a one-to-one relation needed for using GRBs as a reliable distance
indicator.
Nakar \& Piran (2004, 2005) and Band \& Preece
(2005) have shown convincingly that  the claimed tight one-to-one
relations cannot be valid for all GRBs. We believe that the
small dispersion seen in GRBs with known redshifts is due to
selection effects arising in the localization
and redshift determination processes: {\it e.g.}, these GRBs
may represent the upper envelope of the
distribution. A recent analysis  by Ghirlanda {\it et al.}
(2005) using
pseudo-redshift  shows a much broader dispersion (as in LPM00).
The claimed  tighter Ghirlanda relation,
could be due to additional correlation between the jet opening angle $\tjet$
and $\eiso$, $\ep$, or both. However, as mentioned above the picture of jet
break and measurements of $\theta_{jet}$ and ${\cal E}_\gamma$ is a confusing
state in view of {\it Swift} observation.

\begin{figure}[htbp]
\begin{center}
\includegraphics[width=15cm]{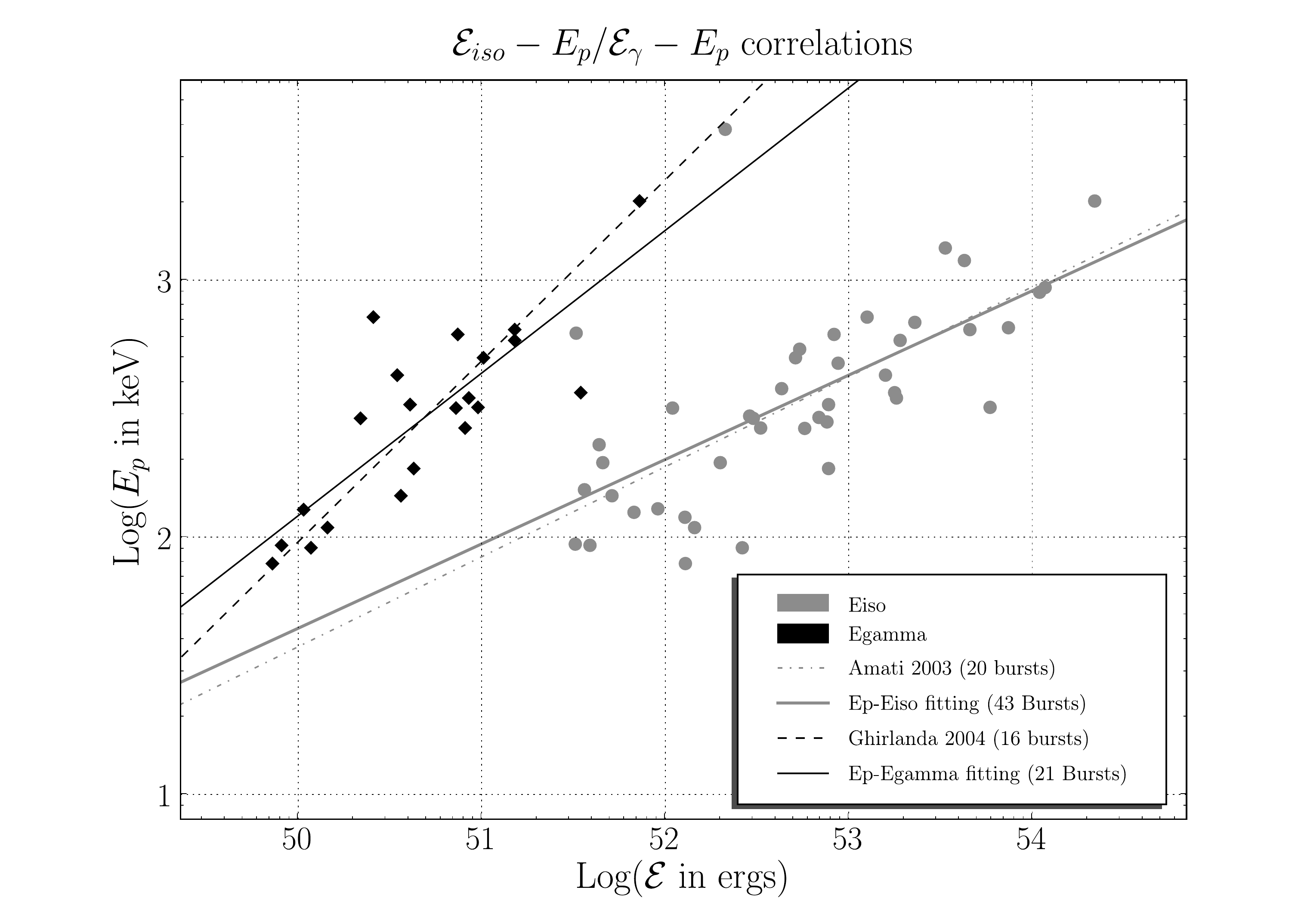}
\end{center}
\caption{$E_p-{\cal E}_{iso}$ and $E_p-{\cal E}_{\gamma}$ correlations. The 43
gray circles are all the bursts from our sample that had good enough  spectral
observations to find the energy peak of the $\nu F_{\nu}$ spectrum, and the 21
black diamonds are a subset of those bursts with a jet break found in their
optical lightcurve. Solid lines are the best fit we find for the two
correlations ($E_p \propto {\cal E}_{iso}^{\eta}$, $\eta \sim 0.328 \pm 0.036$
and $E_p \propto \egamma^{\eta'}$, $\eta' \sim 0.555 \pm 0.089$) and dashed
lines are the best fit found by Amati {\it et al.} 2003 with 20 bursts ($E_p \propto
{\cal E}_{iso}^{0.35}$) and Ghirlanda {\it et al.} 2004 with 16 bursts ($E_p \propto
{\cal E}_{\gamma}^{0.70}$).}
\label{2}
\end{figure}

We have reanalyzed the existing data and determined the parameters of the Amati
and Ghirlanda relations. In Figure \ref{2} we show the the $\eiso$ (and
$\egamma$ excluding some outliers) vs $\ep$ for all GRBs with known redshifts
(and $\tjet$). We compute best power law fit for both these correlations and we
describe the dispersion around it by the standard deviation. For the $E_p -
{\cal E}_{iso}$ we find: $E_p \propto {\cal E}_{iso}^{\eta}$, $\eta \sim 0.328
\pm 0.036$ and $\sigma_{iso} = 0.286$. For $E_p - \egamma$ correlation we find:
$E_p \propto \egamma^{\eta'}$, $\eta' \sim 0.555 \pm 0.089$ and $\sigma_{\gamma}
= 0.209$. We find that additional data has reduced the significance of the
correlations or has increased the dispersions (compared to values obtained by Amati {\it et al.}
and Ghirlanda {\it et al.}). This is contrary to what one would expect for a sample with
a true correlation.
From this we conclude that, as predicted by LPM00, there is a strong correlation
between $\ep$ and $\eiso$ (or $\egamma$), but for the population  GRBs
as a whole both variables have a broad distribution and most
GRBs do not obey the tight relations claimed earlier.

\subsection{Correlations and Cosmology}

Attempts to use observations of extragalactic sources for
cosmological studies have shown us that
extreme care is required. All observational biases must be accounted
for and theoretical ideas tested self-consistently, avoiding circular
arguments. This is especially true for GRBs at this stage of our
ignorance about the basic processes involved in their creation,
energizing, particle acceleration and radiation production.
Here we
outline some of the difficulties and how one may address and possibly
overcome them.

Let us assume that there
exists a one-to-one but unknown relation between
$\eiso$ and $\ep$, $\eiso = {\cal E}_0 f(\ep/E_0)$, and that we have a measure
of  $F_{\rm tot}$ and  $z$. Here ${\cal E}_0$ and $E_0$
are some constants, and for convenience we have defined $f(x)$ which is the
inverse of the
function $g$ introduced above. The

From equations  (\ref{eiso}) and (\ref{dm}) we can write
\be\label{omega}
\int^Z_1 dZ'[\Omega (Z')]^{-1/2} =
\left({{f(\epo Z)/E_0)}\over{ZF_{\rm tot}/F_0}}\right)^{1/2}\,\,\,\,\,{
\rm
with}\,\,\,\,\, F_0={{{\cal E}_0}\over{4\pi (c/H_0)^2}}.
\ee
For general equations of
state $P=w_i \rho$, $\Omega(Z)=\sum_{i} \Omega_i Z^{3(1+w_i)}$.
The aim of any cosmological test is to determine
the values of different $\Omega_i$  and their evolutions (\eg changes in $w_i$)
.
If we make the somewhat questionable assumption of complete
absence of cosmological evolutions  of $\eiso, \ep$ and the
function $f(x)$, then this equation involves two unknown
functions $\Omega(z)$ and $f(x)$. {\underline{In principle,}} if the
forms of these functions are known, then one can rely on
some kind of minimum $\chi^2$ method to determine the parameters of
both functions, assuming that there is sufficient data to overcome the
degeneracies inherent in dealing with large number of
parameters.
By now the parametrization of $\Omega(Z)$
has become standard. However,
the form and parameter values of $f(x)$ is
based on poorly understood data and theory,
and currently requires an assumed cosmological
model. Using the {\underline{form}}  ({\it e.g.} the power law used by Gea05)
derived based on an assumed cosmological model to carry out such
a test  is strictly speaking circular. (It is even
more circular to fix the value of  {\underline{parameters}}, in this case the
index $\eta$, obtained in one cosmological model to test others as
done by Dai {\it et al.} 2004). Even though different models yield
results with small differences, this does not
justify the  use of circular logic. The differences
sought in the final test using equation \ref{omega} will be of the same order.
The situation is even more difficult because
as stressed above the correlation  is not a
simple one-to-one relation but is a statistical one.
Finally, the most important unknown which plagues all cosmological tests using
discrete sources is the possibility of the existence of an a priory unknown
evolution in one or all of the relevant characteristics. For example, the
intrinsic luminosity $L_{iso}$ might suffer large evolution which we refer to as
luminosity evolution. The value of $E_p$ can also be subject to selection
effects, or the correlation function f(x) may evolve with redshift, i.e.
$\eta=\eta(z)$. For such a general case we are dealing with 4 unknown of the
two above. Moreover the rate function of GRBs most likely is not a constant and
can influence the result s with a broad distribution. We address some of these
questions now.

\section{GRB Evolutions}

For a better understanding of GRBs themselves and the possibility of their use
for cosmological tests we need to know whether characteristics such as $\eiso$,
$\tjet$, $\ep$, the correlation function $f(x)$ and the occurrence rate ${\dot
\rho_{\rm GRB}}$ (number of GRBs per unit co-moving volume and time) change with
time or $Z$. For example to use the $E_p - \eiso$ correlation for cosmological
purposes, one need to first establish the existence of the correlation and
determine its form locally (low redshift).One then has to rely on a theory or
non-circular observations to show that either this relation does not evolve or
if it does how it evolves.The existing GRB data is not sufficient for such a
test. In factthere seems to be some evidence that there is evolution. Lie has
shown by subdividing the data into 4 z-bins, he obtained different index $\eta$
which change significantly, rendering previous use of this relation for
cosmological test invalid. This emphasizes the need for a solid understanding of
the evolution of all GRB characteristics. Two of the most important
characteristics are the energy generation $\eiso$ and the rate of GRBs. These
are also two characteristics which can be determined more readily and with
higher uncertainty. In what follows we address these two questions. We will use
all GRBs with known $Z$ irrespective of whether we know the jet angle $\tjet$
because this gives us a larger sample and because in view of new {\it Swift}
observations (Nousek {\it et al.} 2006) the determination of the latter does not seem to
be straightforward. Also since it is often easier to determine the peak flux
$f_p$ rather than the fluence threshold, in what follows we will use the peak
bolometric luminosity $L_p=4\pi d_m^2 Z^2 f_p$ instead of $\eiso=\int L(t)dt$.

\subsection{Evolution with Pseudoredshifts}

Before considering GRBs with known redshifts we briefly mention that there has
been two indications of strong evolutionary trends from use of pseudo redshifts
based on the so-called luminosity-variability and lag-luminosity correlations
(Lloyd {\it et al.} 2002, Kocevski \& Liang 2006) using the methods developed by Efron \&
Petrosian (1992, 1994). These works show existence of a relatively strong
luminosity evolution $L(z)=L_0 Z^{\alpha}$ ($\alpha= 1.4 \pm 0.5, 1.7 \pm 0.3 $)
from which one can determine a GRB formation rate which also varies with
redshifts and can be compared with other cosmological rates such as the star
formation rate.

\subsection{Evolution with Measured Redshifts}

\subsubsection{Description of the Data}

We have compiled the most complete list of GRBs with known redshift. Since the
launch of the {\it Swift} satellite, this list has become significantly larger.
We include only bursts with good redshift determination meaning that GRBs with
only upper or lower limits on their redshift are not in our sample. 
On total, our sample contains 86 bursts, triggered by 4 different instruments: 
BATSE on board CGRO (7 bursts), BeppoSAX (14), HETE-2 (13), and Swift (52).
For each burst we collected fluence and peak flux in the energy bandpass of the
triggering instrument, as well as the duration of the burst. When available we
have also collected spectral information namely the parameters that define the
Band function; the energy peak ($E_p^{obs}$) as well as the low ($\alpha$) and
high ($\beta$) energy indexes of the $\nu F_{\nu}$ spectrum. When a good
spectral analysis was not available, we took as default values the mean of the
BATSE distributions based on large number of bursts: $<\alpha>=-1.0$,
$<\beta>=-2.3$ and $<E_p^{obs}>=250$ keV.
For non-Swift bursts, all this information was mostly extracted from FB05 data
set to which we added results from recent spectral analysis released in GCN. For
Swift GRBs,
redshift, duration, fluence and peak flux were compiled from the Swift
Information webpage (http://swift.gsfc.nasa.gov/docs/swift/archive) and spectral
information have been retrieved from GCN releases and we have also looked at
spectral analysis ourselves for some of them.
We assumed the following cosmological model:

$$\Omega_M=0.3, \Omega_{\lambda}=0.7, H_0=70 \mbox{ km s$^{-1}$ Mpc$^{-1}$}$$

\noindent
in order to determine the intrinsic properties ({\it e.g.} ${\cal E}_{iso}$ see
eq. [\ref{eiso}]). 
${\cal E}_{iso}$ is here calculated for a rest-frame bandpass [20,2000] keV.
Note that K-correction due to the shift of the photons into the instrument
bandpass has been properly taken into account for the ${\cal E}_{iso}$
calculation (see Bloom {\it et al.} 2001). From this, we calculate the average
isotropic-equivalent luminosity as:

$$L_{iso}=\frac{\eiso}{T_{90}}$$

\noindent
when $T_{90}$ is the duration of the burst that includes 90$\%$ of the total
counts.

\begin{figure}[htbp]
\begin{center}
\includegraphics[width=14cm]{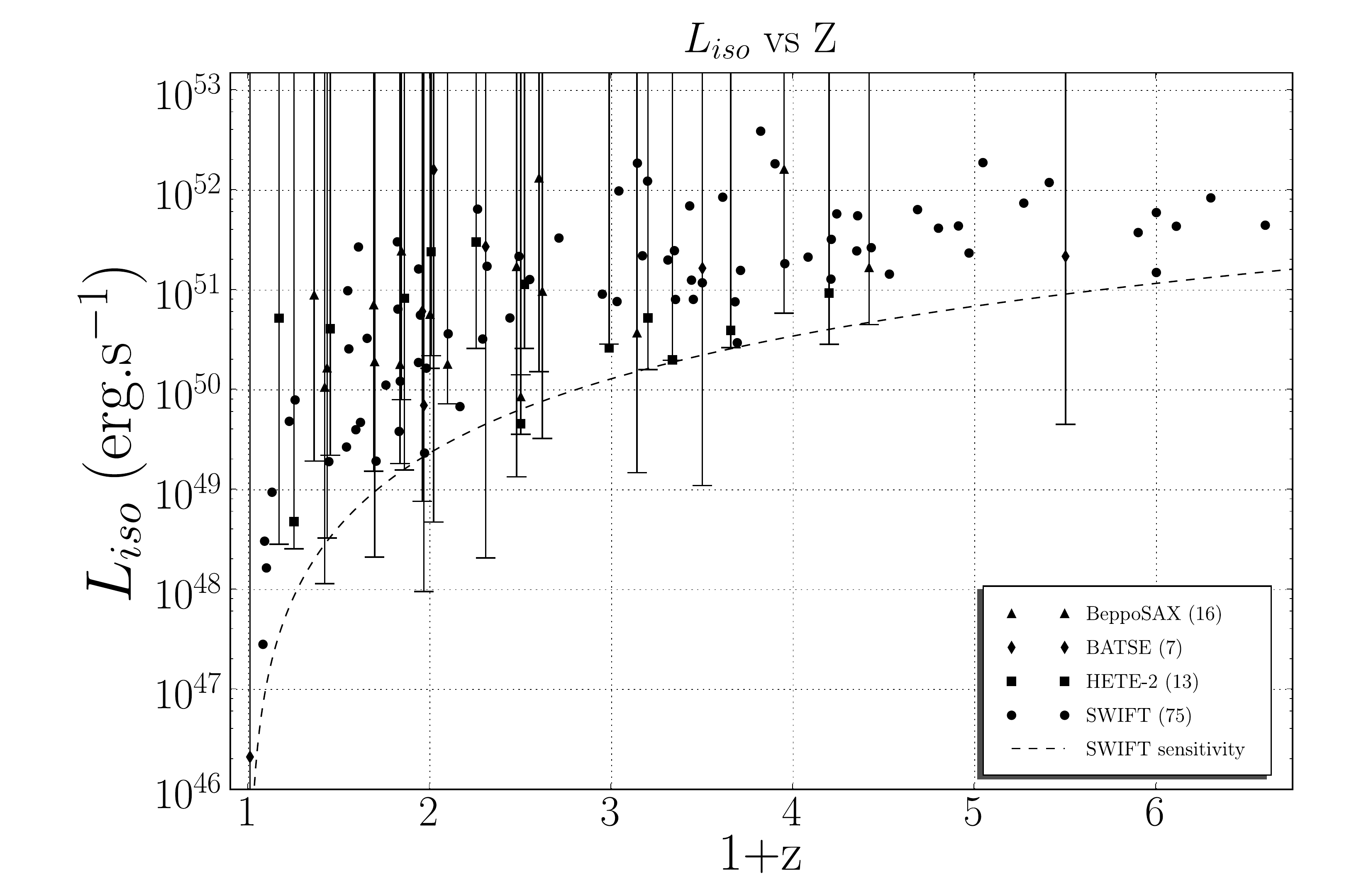}
\end{center}
\caption{Isotropic average luminosity versus redshift for all bursts in our
sample (86). Different symbols represent bursts observed by different
instruments: BATSE(7), Beppo-SAX (14), HETE-2 (13), {\it Swift} (52). For all
non-{\it Swift} burst, a vertical line is plotted representing the range of
isotropic luminosity in which the burst would still have been observable by the
instrument keeping all its others parameters fixed. Using the work of Lamb \&
al. 2005, the limiting luminosity is taken to be only dependent on the energy
peak $E_p$ of the bursts.  For {\it Swift} bursts, a conservative threshold flux
of 0.8 ergs s$^{-1}$ cm$^{-2}$ has been chosen. This limit is shown as a dashed
line.}
\label{3}
\end{figure}

Because different instruments have been used to collect this information, the
sample is very heterogeneous and suffers from various selection and truncation
effects that vary from burst to burst. The most simple of these truncation
effects is due to the limiting sensitivity of the instruments. A GRB trigger
will occur when the peak flux of the burst exceeds the average background
variation by a few sigmas (depending on the setting of the instrument).
In an attempt to carefully take into account this effect into our study, we used
the analysis carried out by Lamb {\it et al.} 2005 for pre-{\it Swift} instruments. In
this analysis, they computed the sensitivity for each instrument depending on
the spectral parameters of the bursts. Therefore, for each specific burst of our
data set, from its spectral parameters it is possible to determine the limiting
photon flux of the instrument (for specific GRB with its specific spectral
parameters). From these, we can easily compute the limiting peak flux
$f_{p,lim}$ for our burst (assuming the Band function for our spectrum).
Finally, we can determine the detection threshold of the observed energy fluence
$F_{obs}$. This lower limit $F_{obs,lim}$ is obtained via the simple
relationship (Lee \& Petrosian 1996):

$$ \frac{F_{obs}}{F_{obs,lim}} =  \frac{f_{p}}{f_{p,lim}} $$

\noindent
Using the same reasoning we can obtain the limiting values for the intrinsic
quantities meaning the intrinsic values that a given burst needs to have in
order to be detected:

$$ \frac{\eiso}{{\cal E}_{iso,lim}} =\frac{L_{iso}}{L_{iso,lim}} = 
\frac{f_{p}}{f_{p,lim}} $$

\noindent
Those limiting average luminosities for each bursts of our sample are
represented in Figure \ref{3}. This analysis was not carried out for BAT
instrument on board Swift therefore we used a conservative threshold of $0.8$ erg
s$^{-1}$ cm$^{-2}$ for all of Swift bursts.

\subsubsection{Analysis and results}

\begin{figure}[htbp]
\begin{center}
\includegraphics[width=14cm]{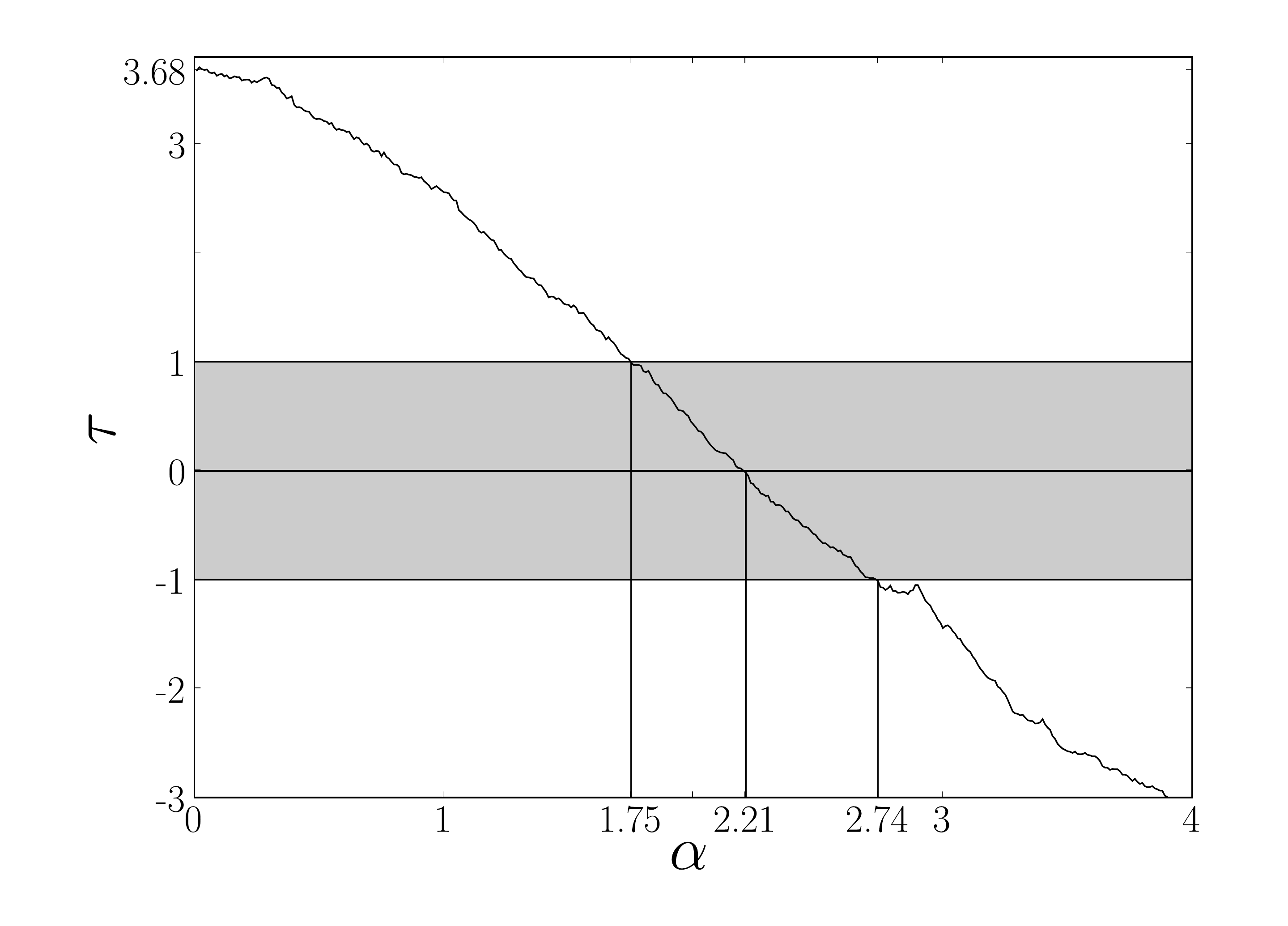}
\end{center}
\caption{Variation of the $\tau$ parameter with the power law index $\alpha$ of
the luminosity evolution. $\tau=0$ means no correlation which gives the best
value of $\alpha = 2.21$ for the assumed power law form with a one sigma range
of $1.75$ to $2.74$.}
\label{5}
\end{figure}

We now describe our determination of luminosity and density rate evolution of
the parent population of our GRB sample. Our analysis is based on the work done
by Efron \& Petrosian (1992, 1994). We refer the reader to these two papers for
details. We will here simply describe the most important steps of the analysis
and what it allows us to infer on our data sample. 
This method has been developed in order to take into account effects of data
truncation and selection bias on a heterogeneous sample from different
instruments with different sensitivities as described above and shown in Figure
\ref{3}. The method corrects for this bias by applying a proper rankings to
different subset of our sample.
The first step is to compute the degree of correlation between the isotropic
luminosity and redshift. For that we use the specialized version of Kendell's
$\tau$ statistics. The parameter $\tau$ represents the degree of correlation
found for the entire sample with proper accounting for the data truncation.
$\tau=0$ means no correlation is found between the two parameters being
inspected (luminosity and redshift in our case). Any other specific value
$\tau_0$ implies presence of a correlation with a significance of $\tau_0 
\sigma$. 
With this statistic method in place, we can calculate the parametrization that
best describe the luminosity evolution. To establish a functional form of the
luminosity-redshift correlation, we assume a power law luminosity evolution:
$L(z)=L_0 Z^{\alpha}$.  We then remove this dependency from the observed
luminosity: $L'  \rightarrow L_{observed}/(1+z)^{\alpha}$ and calculate the
Kendell's $\tau$ statistics as a function of $\alpha$. Figure \ref{5} shows the
variation of $\tau$ with $\alpha$.

Once the parametric form for the luminosity evolution have been determined, this
nonparametric maximum likelihood techniques can be used to determine the
cumulative distribution for luminosity and redshift (see Efron \& Petrosian
1994) say $\Phi(L)$ and $\sigma(z)$, which gives the relative number of bursts
under a certain redshift z. From this last function, we can easily draw the
comoving rate density $\dot{n}(z)$, which is the number of GRB per unit comoving
volume and unit time: 	

\be
\label{prout1}
\dot{n}(Z)=\frac{d \sigma(Z)}{dZ} \frac{Z}{dV/dZ}
\ee

where the factor $Z$ is to take the time dilatation into account. Note that this
method do not provide any constrain on the normalization of any of the quantity
mentioned above. Normalization will therefore be set arbitrarily on all our
figures representing these functions.

We find a 3.68 $\sigma$ evidence for luminosity evolution (see the $\tau$ value
at the onset of Figure \ref{5} when $\alpha=0$). From this figure, we can also
infer that $\alpha = 2.21$ for value obtained when $\tau = 0$ gives the best
description of the luminosity evolution for the assumed form has a one sigma
range of $[1.75,2.74]$. Constrain on the $\alpha$ parameter is not very tight
off course since the size of our sample is still limited. While current
satellites accumulate more data, we will be able to increase our data set and
further constrain this parameter in the future.

\begin{figure}[htbp]
\begin{center} $
\begin{array}{cc}
\includegraphics[ width=.43\linewidth, keepaspectratio]{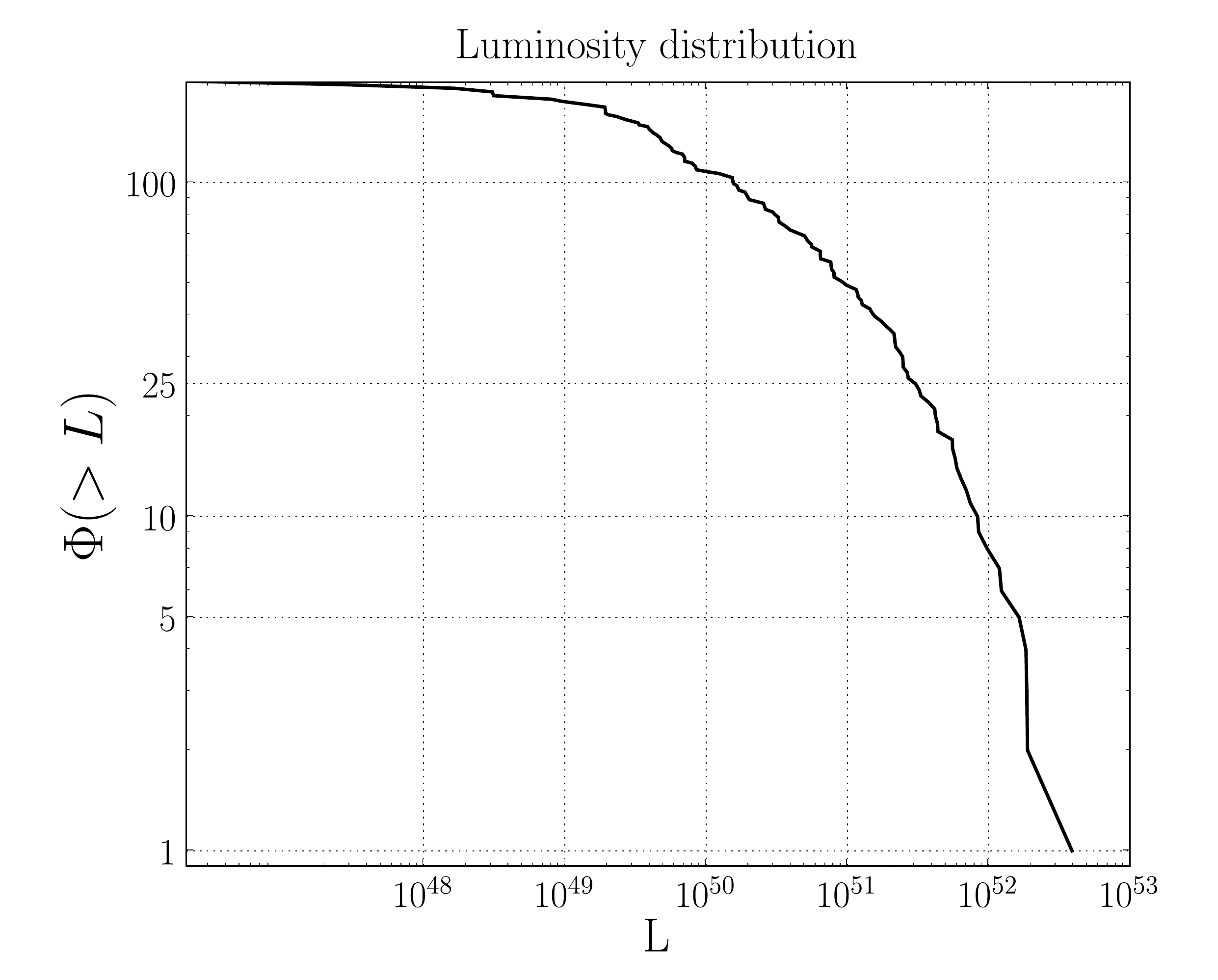} &
\includegraphics[ width=.52\linewidth, keepaspectratio]{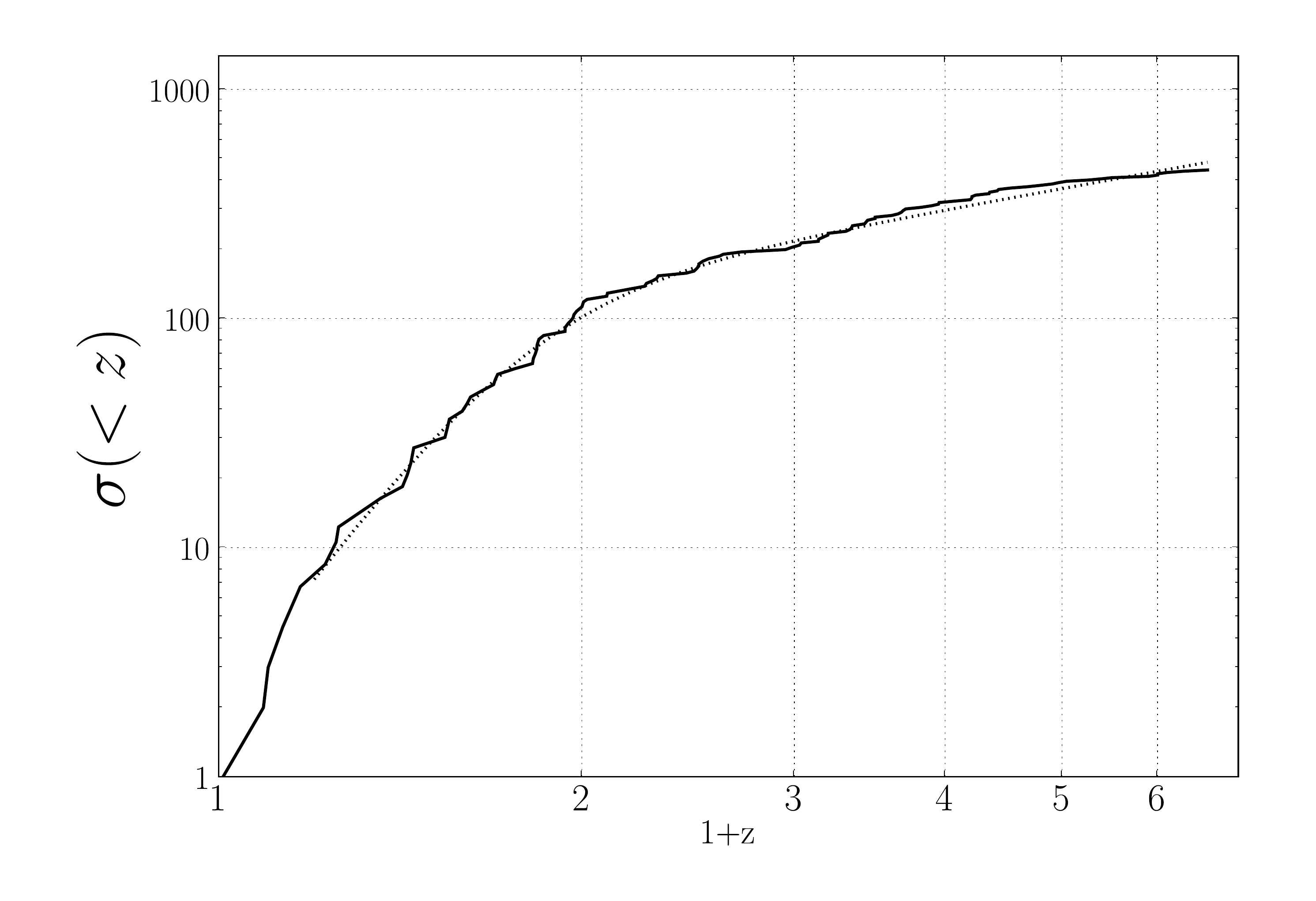}
\end{array} $
%\plottwo{Plots/GRB_cosmology/cumulative_luminosity3}{Plots/GRB_cosmology/cumulative_density_withfit3}
\end{center}
\caption{The cumulative luminosity distribution $\Phi(L)$ (left panel) and the
cumulative redshift distribution $\sigma(z)$ (right panel). Fit to the cumulative
density rate is represented with a point line.}
\label{4}
\end{figure}

\begin{figure}[htbp]
\begin{center}
\includegraphics[width=14cm]{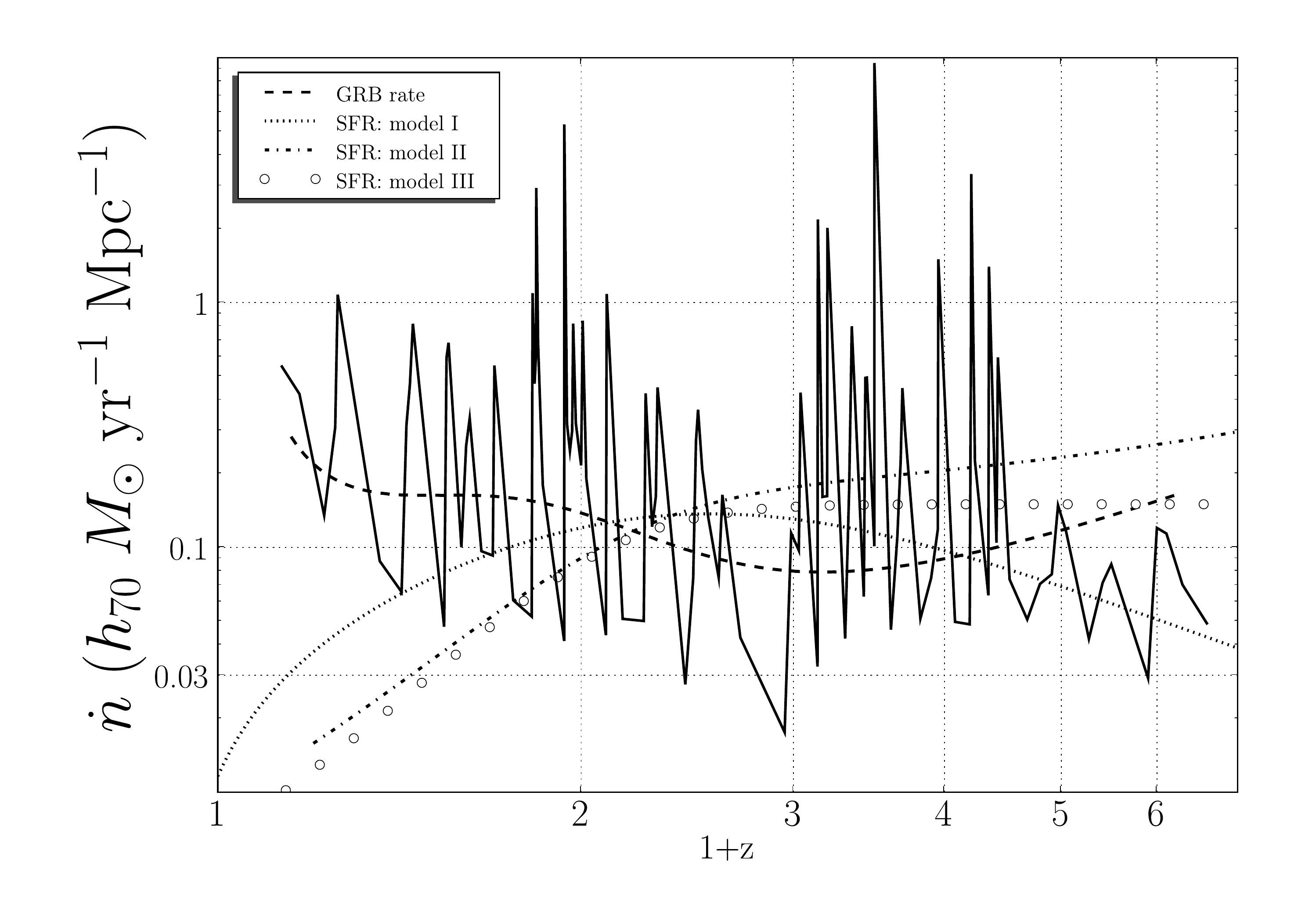}
\end{center}
\caption{The comoving rate density $\dot{n}(z)$. The dashed line was obtained by
fitting the cumulative density distribution by the parametrized smooth function
of equation \ref{fitting}. We also show comparison of the density rate (from
Figure \ref{6}) with three different SFR scenarios taken from literature. No SFR
scenario seems to match the density rate deduced from our analysis.}
\label{6}
\end{figure}

\begin{figure}[htbp]
\begin{center}
\includegraphics[width=14cm]{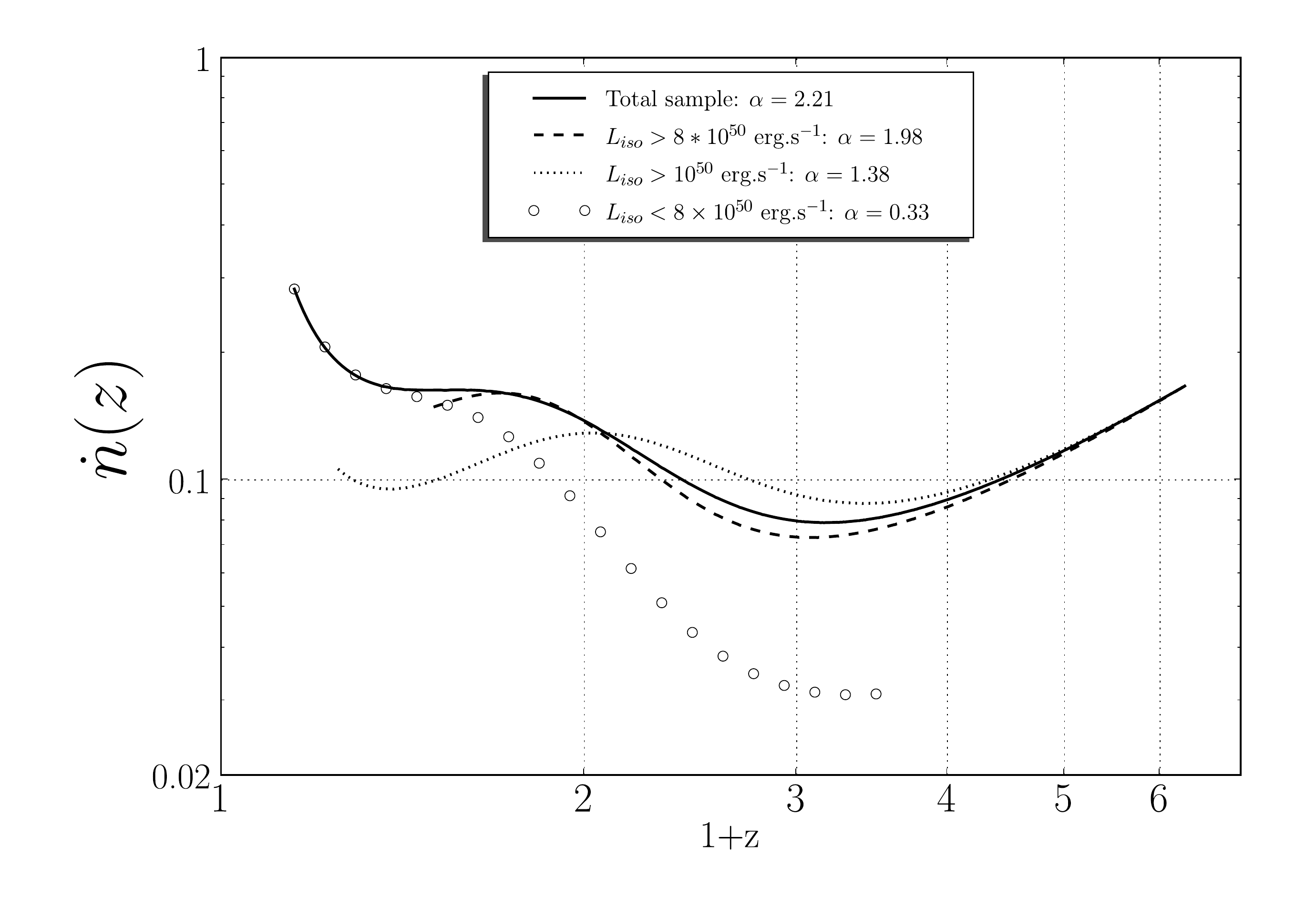}
\end{center}
\caption{Comparison of the comoving density rate evolution of the total sample
with that of several sub-sample where we impose three different luminosity
thresholds: $L_{iso}>10^{49}$, $>10^{50}$, and $>8 \times 10^{50}$
ergs.s$^{-1}$. We also looked at a low luminosity population where we imposed of
maximal luminosity of $8 \times 10^{50}$ ergs.s$^{-1}$. Each sub-sample is
subject to the same analysis and has provided different luminosity evolution as
evident from the different values of $\alpha$. As expected, the rate at low
redshift decreases with increasing values of threshold.}
\label{7}
\end{figure}

The cumulative functions are both shown in Figure \ref{4} and the estimated
comoving density rate is shown by the jagged curve in Figure 6. Most of the high
frequency variation is not real and is due to taking the derivative of a noisy
curve ($\sigma (Z)$). The dashed line was obtained by fitting the cumulative
density distribution by the following parametrized function:

\be
\label{fitting}
\sigma(Z) \propto \frac{(Z/Z_0)^{p_1}}{(1+Z/Z_0)^{p_1-p_2}}
\ee

\noindent
with the following values for the parameters: $Z_0=1.8$, $p_1=7.1$, and
$p_2=0.95$

These results are still very preliminary as more data become available, accuracy
of the density function will increase constraining further the evolution rate of
long bursts. By tackling this problem for the first time we hope to set the
ground for further analysis in the future.

The behavior of the comoving density rate for our sample of long bursts is quite
peculiar with a significant rate increase happening at low redshifts. This
effect might be due to some selection effects that we have not included in our
analysis. For instance, it might be a consequence of the fact that instruments
detect more easily low-redshifts host galaxies and therefore create a bias
toward low redshifts GRBs. Another interesting feature is the steady increase we
obtain in the GRB rate at high redshifts ($z > 3$). Figure 7 compares the
estimated comoving rate evolution with different models of Star Formation Rates
(SFRs I, II, III).

For comparison with Star Formation history, we used three different models taken
from the literature:

- Steidel et al. 1999:

\be
\dot{n}=0.16 h_{70} \frac{e^{3.4 z}}{e^{3.4 z} +22} M_{\odot} \mbox{yr}^{-1}
\mbox{Mpc}^{-3}
\ee

- Porciani \& Madau 2000:

\be
\dot{n}= 0.22 h_{70} \frac{e^{3.05 z-0.4}}{e^{2.93 z}+15} M_{\odot}
\mbox{yr}^{-1} \mbox{Mpc}^{-3}
\ee

- Cole {\it et al.} 2001:

\be
\dot{n}= \frac{(a + b z) h_{70}}{1+(z/c)^d} M_{\odot} \mbox{yr}^{-1}
\mbox{Mpc}^{-3}
\ee

with $(a, b, c, d) = (0.0166, 0.1848, 1.9474, 2.6316)$

As evident, no SFR scenario seems to match the density rate evolution deduced
from our analysis, specially at low redshift. How much of this difference is
real and how much is due to other selection effects that we have not quantified
is unclear. Because of increasing difficulty of identifying the host galaxy with
increasing redshift one would expect some bias against detection of high
redshift bursts. But the largest densities for SFR seems to be in the
intermediate redshift range.

An other possibility is that there may exist subclasses of GRBs such as low or
high luminosity classes. In order to test this eventuality we have defined
several subsets of our total GRB sample carried out the above analysis for each
subsamples, determining a new luminosity evolution (a new $\alpha$) and then
proceeding to obtain $\dot{n} (z)$ from the smooth function fitting $\sigma
(z)$. We impose different luminosity thresholds for the different subsamples.
Three different threshold have been chosen: $L_{iso}>10^{49}$ ergs.s$^{-1}$,
$L_{iso}>10^{50}$ ergs.s$^{-1}$, and $L_{iso}>8 \times 10^{50}$ ergs.s$^{-1}$.
We also looked at a low luminosity population where we imposed of maximal
luminosity of $10^{50}$ ergs.s$^{-1}$. Figures \ref{6} and \ref{7} compare the
new rates with that of the total sample. As expected high luminosity samples
contribute less to the rate at low redshifts. But the general trend and the
differences with SFR are essentially still present. However the method and
framework we presented would be a very valuable tool when enough data has been
accumulated.

\section{Summary and Conclusion}

We have considered GRBs as cosmological tools. We find that GRBs are not SCs and
the correlations found so far are statistical in nature and too broad to be very
useful for cosmological model parameter determination. In addition we have shown
that there is strong evidence for evolution of the peak luminosity of GRBs. This
indicate a very likely possibility that $\eiso$ and $\egamma$ may also have
undergone comparable changes. There may also be evolution of $\ep$, $\tjet$ or
other relevant characteristics. This makes use of GRBs as cosmological tools
more difficult.

We may therefore ask is this process hopeless. Strictly speaking the answer is
no. Some
broad brush conclusion can already be reached.
For example, one can test the relative merits of different forms for $f(x)$.
As shown by FB05 the SC assumption
($\phi(\eiso)\rightarrow\delta(\eiso - {\cal E}_0)$)
gives unacceptable fit to essentially all cosmological models, but the
use of the power-law form agrees with  Lemaitre type model (FB05, Gea05)
with a relatively long quasi-static
phase (refereed to as a loitering model). Such models, which were in vogue some
time
ago (see Petrosian 1974), are currently unacceptable because of their
low (baryonic plus dark) matter  density and large curvature.
This indicates that the form of the correlation and/or other assumptions ({\it
e.g.} no
evolution) are not correct. In this paper, we have shown that there is strong
evolution of luminosity and $\eiso$. Lei {\it et al.} have also shown that the form of
the $E_p-\eiso$ correlation may evolve. These are tentative results and more
data are required to determine  these evolution trend and their meaning.

The relevant variables in addition to redshift are $\ep^{obs}=\ep/(1+z)$ and
$F_{tot}$, determination of which requires a good description of the
total spectrum. We need to
know the observational selection biases for all the variables and
their parameters, and use accurate statistical methods to account for the
biases and data truncations. From these one can learn about the distributions of
$\eiso$ and $\ep$ and their correlations.

Obviously {\it Swift} observations will be extremely helpful
and eventually may provide the data required for this complex
task. In the near future, however, from analysis of the incoming and
archived data we will (and need to) first learn more about the nature of the
GRBs than
cosmological models.  Eventually we may have
enough information to construct a well defined ``SC", which can be used for
global cosmological tests as is
done using type Ia supernovae.   The immediate situation may be more analogous
to
galaxies where the cosmological tests
are rendered complicated because of the multivariate situation and broad
distributions of the relevant variables.
%(see, {\it e.g.} Petrosian1998 and references cited therein).
Consequently, over the years the
focus of activity has shifted
from the determination of the few global cosmological parameters to the
investigation of structure formation, the building
process of the black holes  and the star formation  rate (SFR). Similarly, we
expect that
from investigation of GRBs we will learn about the evolution, distributions and
correlations of their intrinsic characteristics, and the
relationship of these with the evolutionary rates of other
cosmological sources and the formation rates of stars, supernovae and black
holes.

In summary, on the long run cosmological test with
GRBs may be possible, either carried out with clever statistical
methods, or by identification of  a subclass of ``SCs''. On
a shorter time scale, we need to learn more about the intrinsic
characteristics of GRBs, and provide a reasonable
theoretical interpretation for them and their cosmological evolution. As an
example, as shown here, one outcome of our analysis is the determination of the
evolution rate of GRBs. We have shown that for all GRBs with known redshifts
this rate appears to be different from the SFR. These differences seems to be
present with different subdivision of the sample and may be consequences of
other selection biases not included in our analysis. We postpone a proper treatment of these biases to futur work.

	\bibliographystyle{plainnat}
%	\bibliography{}
	\bibliography{bibliography/mybib.bib}
		
\end{document}